\documentclass[11pt]{book}
\usepackage{graphicx} % Required for including pictures
\graphicspath{{Pictures/}} % Specifies the directory where pictures are stored

\usepackage{lipsum} % Inserts dummy text

\usepackage{tikz} % Required for drawing custom shapes

\usepackage[english]{babel} % English language/hyphenation

\usepackage{enumitem} % Customize lists
\setlist{nolistsep} % Reduce spacing between bullet points and numbered lists

\usepackage{booktabs} % Required for nicer horizontal rules in tables

\usepackage{xcolor} % Required for specifying colors by name
\definecolor{ocre}{RGB}{0, 57, 189} % Define the orange color used for highlighting throughout the book

\usepackage{geometry} % Required for adjusting page dimensions and margins

\usepackage{avant} % Use the Avantgarde font for headings
\usepackage{mathptmx} % Use the Adobe Times Roman as the default text font together with math symbols from the Sym­bol, Chancery and Com­puter Modern fonts

\usepackage{microtype} % Slightly tweak font spacing for aesthetics
\usepackage[utf8]{inputenc} % Required for including letters with accents
\usepackage[T1]{fontenc} % Use 8-bit encoding that has 256 glyphs

\usepackage[style=numeric,citestyle=numeric,sorting=nyt,sortcites=true,autopunct=true,babel=hyphen,hyperref=true,abbreviate=false,backref=true,backend=biber]{biblatex}
\defbibheading{bibempty}{}

\usepackage{calc} % For simpler calculation - used for spacing the index letter headings correctly
\usepackage{makeidx} % Required to make an index
\makeindex % Tells LaTeX to create the files required for indexing

\usepackage{titletoc} % Required for manipulating the table of contents

\contentsmargin{0cm} % Removes the default margin

\titlecontents{part}
	[0cm] % Left indentation
	{\addvspace{20pt}\bfseries} % Spacing and font options for parts
	{}
	{}
	{}

\titlecontents{chapter}
	[1.25cm] % Left indentation
	{\addvspace{12pt}\large\sffamily\bfseries} % Spacing and font options for chapters
	{\color{ocre!60}\contentslabel[\Large\thecontentslabel]{1.25cm}\color{ocre}} % Formatting of numbered sections of this type
	{\color{ocre}} % Formatting of numberless sections of this type
	{\color{ocre!60}\normalsize\;\titlerule*[.5pc]{.}\;\thecontentspage} % Formatting of the filler to the right of the heading and the page number

\titlecontents{section}
	[1.25cm] % Left indentation
	{\addvspace{3pt}\sffamily\bfseries} % Spacing and font options for sections
	{\contentslabel[\thecontentslabel]{1.25cm}} % Formatting of numbered sections of this type
	{} % Formatting of numberless sections of this type
	{\hfill\color{black}\thecontentspage} % Formatting of the filler to the right of the heading and the page number

\titlecontents{subsection}
	[1.25cm] % Left indentation
	{\addvspace{1pt}\sffamily\small} % Spacing and font options for subsections
	{\contentslabel[\thecontentslabel]{1.25cm}} % Formatting of numbered sections of this type
	{} % Formatting of numberless sections of this type
	{\ \titlerule*[.5pc]{.}\;\thecontentspage} % Formatting of the filler to the right of the heading and the page number

\titlecontents{figure}
	[1.25cm] % Left indentation
	{\addvspace{1pt}\sffamily\small} % Spacing and font options for figures
	{\thecontentslabel\hspace*{1em}} % Formatting of numbered sections of this type
	{} % Formatting of numberless sections of this type
	{\ \titlerule*[.5pc]{.}\;\thecontentspage} % Formatting of the filler to the right of the heading and the page number

\titlecontents{table}
	[1.25cm] % Left indentation
	{\addvspace{1pt}\sffamily\small} % Spacing and font options for tables
	{\thecontentslabel\hspace*{1em}} % Formatting of numbered sections of this type
	{} % Formatting of numberless sections of this type
	{\ \titlerule*[.5pc]{.}\;\thecontentspage} % Formatting of the filler to the right of the heading and the page number

\titlecontents{lchapter}
	[0em] % Left indentation
	{\addvspace{15pt}\large\sffamily\bfseries} % Spacing and font options for chapters
	{\color{ocre}\contentslabel[\Large\thecontentslabel]{1.25cm}\color{ocre}} % Chapter number
	{}  
	{\color{ocre}\normalsize\sffamily\bfseries\;\titlerule*[.5pc]{.}\;\thecontentspage} % Page number

\titlecontents{lsection}
	[0em] % Left indentation
	{\sffamily\small} % Spacing and font options for sections
	{\contentslabel[\thecontentslabel]{1.25cm}} % Section number
	{}
	{}

\titlecontents{lsubsection}
	[.5em] % Left indentation
	{\sffamily\footnotesize} % Spacing and font options for subsections
	{\contentslabel[\thecontentslabel]{1.25cm}}
	{}
	{}

\usepackage{fancyhdr} % Required for header and footer configuration

\fancypagestyle{plain}{% Style for when a plain pagestyle is specified
	\fancyhead{}%
}

\makeatletter
\renewcommand{\cleardoublepage}{
\clearpage\ifodd\c@page\else
\hbox{}
\vspace*{\fill}
\thispagestyle{empty}
\newpage
\fi}

\usepackage{amsmath,amsfonts,amssymb,amsthm} % For math equations, theorems, symbols, etc

\newtheoremstyle{ocrenumbox}% Theorem style name
{0pt}% Space above
{0pt}% Space below
{\normalfont}% Body font
{}% Indent amount
{\small\bf\sffamily\color{ocre}}% Theorem head font
{\;}% Punctuation after theorem head
{0.25em}% Space after theorem head
{\small\sffamily\color{ocre}\thmname{#1}\nobreakspace\thmnumber{\@ifnotempty{#1}{}\@upn{#2}}% Theorem text (e.g. Theorem 2.1)
\thmnote{\nobreakspace\the\thm@notefont\sffamily\bfseries\color{black}---\nobreakspace#3.}} % Optional theorem note

\newtheoremstyle{blacknumex}% Theorem style name
{5pt}% Space above
{5pt}% Space below
{\normalfont}% Body font
{} % Indent amount
{\small\bf\sffamily}% Theorem head font
{\;}% Punctuation after theorem head
{0.25em}% Space after theorem head
{\small\sffamily{\tiny\ensuremath{\blacksquare}}\nobreakspace\thmname{#1}\nobreakspace\thmnumber{\@ifnotempty{#1}{}\@upn{#2}}% Theorem text (e.g. Theorem 2.1)
\thmnote{\nobreakspace\the\thm@notefont\sffamily\bfseries---\nobreakspace#3.}}% Optional theorem note

\newtheoremstyle{blacknumbox} % Theorem style name
{0pt}% Space above
{0pt}% Space below
{\normalfont}% Body font
{}% Indent amount
{\small\bf\sffamily}% Theorem head font
{\;}% Punctuation after theorem head
{0.25em}% Space after theorem head
{\small\sffamily\thmname{#1}\nobreakspace\thmnumber{\@ifnotempty{#1}{}\@upn{#2}}% Theorem text (e.g. Theorem 2.1)
\thmnote{\nobreakspace\the\thm@notefont\sffamily\bfseries---\nobreakspace#3.}}% Optional theorem note

\newtheoremstyle{ocrenum}% Theorem style name
{5pt}% Space above
{5pt}% Space below
{\normalfont}% Body font
{}% Indent amount
{\small\bf\sffamily\color{ocre}}% Theorem head font
{\;}% Punctuation after theorem head
{0.25em}% Space after theorem head
{\small\sffamily\color{ocre}\thmname{#1}\nobreakspace\thmnumber{\@ifnotempty{#1}{}\@upn{#2}}% Theorem text (e.g. Theorem 2.1)
\thmnote{\nobreakspace\the\thm@notefont\sffamily\bfseries\color{black}---\nobreakspace#3.}} % Optional theorem note
\makeatother

\newcounter{dummy} 
\numberwithin{dummy}{section}
\theoremstyle{ocrenumbox}
\newtheorem{theoremeT}[dummy]{Theorem}

\newtheorem{exerciseT}{Exercise}[chapter]
\theoremstyle{blacknumex}
\newtheorem{exampleT}{Example}[chapter]
\theoremstyle{blacknumbox}

\newtheorem{definitionT}{Definition}[section]
\newtheorem{corollaryT}[dummy]{Corollary}
\theoremstyle{ocrenum}

\RequirePackage[framemethod=default]{mdframed} % Required for creating the theorem, definition, exercise and corollary boxes

\newmdenv[skipabove=7pt,
skipbelow=7pt,
backgroundcolor=black!5,
linecolor=ocre,
innerleftmargin=5pt,
innerrightmargin=5pt,
innertopmargin=5pt,
leftmargin=0cm,
rightmargin=0cm,
innerbottommargin=5pt]{tBox}

\newmdenv[skipabove=7pt,
skipbelow=7pt,
rightline=false,
leftline=true,
topline=false,
bottomline=false,
backgroundcolor=ocre!10,
linecolor=ocre,
innerleftmargin=5pt,
innerrightmargin=5pt,
innertopmargin=5pt,
innerbottommargin=5pt,
leftmargin=0cm,
rightmargin=0cm,
linewidth=4pt]{eBox}	

\newmdenv[skipabove=7pt,
skipbelow=7pt,
rightline=false,
leftline=true,
topline=false,
bottomline=false,
linecolor=ocre,
innerleftmargin=5pt,
innerrightmargin=5pt,
innertopmargin=0pt,
leftmargin=0cm,
rightmargin=0cm,
linewidth=4pt,
innerbottommargin=0pt]{dBox}	

\newmdenv[skipabove=7pt,
skipbelow=7pt,
rightline=false,
leftline=true,
topline=false,
bottomline=false,
linecolor=gray,
backgroundcolor=black!5,
innerleftmargin=5pt,
innerrightmargin=5pt,
innertopmargin=5pt,
leftmargin=0cm,
rightmargin=0cm,
linewidth=4pt,
innerbottommargin=5pt]{cBox}

\newenvironment{theorem}{\begin{tBox}\begin{theoremeT}}{\end{theoremeT}\end{tBox}}
\newenvironment{exercise}{\begin{eBox}\begin{exerciseT}}{\hfill{\color{ocre}\tiny\ensuremath{\blacksquare}}\end{exerciseT}\end{eBox}}				  
\newenvironment{definition}{\begin{dBox}\begin{definitionT}}{\end{definitionT}\end{dBox}}	
		
\newenvironment{corollary}{\begin{cBox}\begin{corollaryT}}{\end{corollaryT}\end{cBox}}	

\newenvironment{remark}{\par\vspace{10pt}\small % Vertical white space above the remark and smaller font size
\begin{list}{}{
\leftmargin=35pt % Indentation on the left
\rightmargin=25pt}\item\ignorespaces % Indentation on the right
\makebox[-2.5pt]{\begin{tikzpicture}[overlay]
\node[draw=ocre!60,line width=1pt,circle,fill=ocre!25,font=\sffamily\bfseries,inner sep=2pt,outer sep=0pt] at (-15pt,0pt){\textcolor{ocre}{R}};\end{tikzpicture}} % Orange R in a circle
\advance\baselineskip -1pt}{\end{list}\vskip5pt} % Tighter line spacing and white space after remark

\makeatletter
\renewcommand{\@seccntformat}[1]{\llap{\textcolor{ocre}{\csname the#1\endcsname}\hspace{1em}}}                    
\renewcommand{\section}{\@startsection{section}{1}{\z@}
{-4ex \@plus -1ex \@minus -.4ex}
{1ex \@plus.2ex }
{\normalfont\large\sffamily\bfseries}}
\renewcommand{\subsection}{\@startsection {subsection}{2}{\z@}
{-3ex \@plus -0.1ex \@minus -.4ex}
{0.5ex \@plus.2ex }
{\normalfont\sffamily\bfseries}}
\renewcommand{\subsubsection}{\@startsection {subsubsection}{3}{\z@}
{-2ex \@plus -0.1ex \@minus -.2ex}
{.2ex \@plus.2ex }
{\normalfont\small\sffamily\bfseries}}                        
\renewcommand\paragraph{\@startsection{paragraph}{4}{\z@}
{-2ex \@plus-.2ex \@minus .2ex}
{.1ex}
{\normalfont\small\sffamily\bfseries}}

\newcommand{\@mypartnumtocformat}[2]{%
	\setlength\fboxsep{0pt}%
	\noindent\colorbox{ocre!20}{\strut\parbox[c][.7cm]{\ecart}{\color{ocre!70}\Large\sffamily\bfseries\centering#1}}\hskip\esp\colorbox{ocre!40}{\strut\parbox[c][.7cm]{\linewidth-\ecart-\esp}{\Large\sffamily\centering#2}}%
}

\newcommand{\@myparttocformat}[1]{%
	\setlength\fboxsep{0pt}%
	\noindent\colorbox{ocre!40}{\strut\parbox[c][.7cm]{\linewidth}{\Large\sffamily\centering#1}}%
}

\newlength\esp
\newlength\ecart
\def\@part[#1]#2{%
\ifnum \c@secnumdepth >-2\relax%
\refstepcounter{part}%
\addcontentsline{toc}{part}{\texorpdfstring{\protect\@mypartnumtocformat{\thepart}{#1}}{\partname~\thepart\ ---\ #1}}
\else%
\addcontentsline{toc}{part}{\texorpdfstring{\protect\@myparttocformat{#1}}{#1}}%
\fi%
\startcontents%
\markboth{}{}%
{\thispagestyle{empty}%
\begin{tikzpicture}[remember picture,overlay]%
\node at (current page.north west){\begin{tikzpicture}[remember picture,overlay]%	
\fill[ocre!20](0cm,0cm) rectangle (\paperwidth,-\paperheight);
\node[anchor=north] at (4cm,-3.25cm){\color{ocre!40}\fontsize{220}{100}\sffamily\bfseries\thepart}; 
\node[anchor=south east] at (\paperwidth-1cm,-\paperheight+1cm){\parbox[t][][t]{8.5cm}{
\printcontents{l}{0}{\setcounter{tocdepth}{1}}% The depth to which the Part mini table of contents displays headings; 0 for chapters only, 1 for chapters and sections and 2 for chapters, sections and subsections
}};
\node[anchor=north east] at (\paperwidth-1.5cm,-3.25cm){\parbox[t][][t]{15cm}{\strut\raggedleft\color{white}\fontsize{30}{30}\sffamily\bfseries#2}};
\end{tikzpicture}};
\end{tikzpicture}}%
\@endpart}
\def\@spart#1{%
\startcontents%
\phantomsection
{\thispagestyle{empty}%
\begin{tikzpicture}[remember picture,overlay]%
\node at (current page.north west){\begin{tikzpicture}[remember picture,overlay]%	
\fill[ocre!20](0cm,0cm) rectangle (\paperwidth,-\paperheight);
\node[anchor=north east] at (\paperwidth-1.5cm,-3.25cm){\parbox[t][][t]{15cm}{\strut\raggedleft\color{white}\fontsize{30}{30}\sffamily\bfseries#1}};
\end{tikzpicture}};
\end{tikzpicture}}
\addcontentsline{toc}{part}{\texorpdfstring{%
\setlength\fboxsep{0pt}%
\noindent\protect\colorbox{ocre!40}{\strut\protect\parbox[c][.7cm]{\linewidth}{\Large\sffamily\protect\centering #1\quad\mbox{}}}}{#1}}%
\@endpart}
\def\@endpart{\vfil\newpage
\if@twoside
\if@openright
\null
\thispagestyle{empty}%
\newpage
\fi
\fi
\if@tempswa
\twocolumn
\fi}

\newif\ifusechapterimage
\usechapterimagetrue
\newcommand{\thechapterimage}{}%
\newcommand{\chapterimage}[1]{\ifusechapterimage\renewcommand{\thechapterimage}{#1}\fi}%
\newcommand{\autodot}{.}
\def\@makechapterhead#1{%
{\parindent \z@ \raggedright \normalfont
\ifnum \c@secnumdepth >\m@ne
\if@mainmatter
\begin{tikzpicture}[remember picture,overlay]
\node at (current page.north west)
{\begin{tikzpicture}[remember picture,overlay]
\node[anchor=north west,inner sep=0pt] at (0,0) {\ifusechapterimage\includegraphics[width=\paperwidth]{\thechapterimage}\fi};
\draw[anchor=west] (\Gm@lmargin,-9cm) node [line width=2pt,rounded corners=15pt,draw=ocre,fill=white,fill opacity=0.8,inner sep=15pt]{\strut\makebox[22cm]{}};
\draw[anchor=west] (\Gm@lmargin+.3cm,-9cm) node {\huge\sffamily\bfseries\color{ocre}\thechapter\autodot~#1\strut};
\end{tikzpicture}};
\end{tikzpicture}
\else
\begin{tikzpicture}[remember picture,overlay]
\node at (current page.north west)
{\begin{tikzpicture}[remember picture,overlay]
\node[anchor=north west,inner sep=0pt] at (0,0) {\ifusechapterimage\includegraphics[width=\paperwidth]{\thechapterimage}\fi};
\draw[anchor=west] (\Gm@lmargin,-9cm) node [line width=2pt,rounded corners=15pt,draw=ocre,fill=white,fill opacity=0.8,inner sep=15pt]{\strut\makebox[22cm]{}};
\draw[anchor=west] (\Gm@lmargin+.3cm,-9cm) node {\huge\sffamily\bfseries\color{ocre}#1\strut};
\end{tikzpicture}};
\end{tikzpicture}
\fi\fi\par\vspace*{270\p@}}}

\def\@makeschapterhead#1{%
\begin{tikzpicture}[remember picture,overlay]
\node at (current page.north west)
{\begin{tikzpicture}[remember picture,overlay]
\node[anchor=north west,inner sep=0pt] at (0,0) {\ifusechapterimage\includegraphics[width=\paperwidth]{\thechapterimage}\fi};
\draw[anchor=west] (\Gm@lmargin,-9cm) node [line width=2pt,rounded corners=15pt,draw=ocre,fill=white,fill opacity=0.8,inner sep=15pt]{\strut\makebox[22cm]{}};
\draw[anchor=west] (\Gm@lmargin+.3cm,-9cm) node {\huge\sffamily\bfseries\color{ocre}#1\strut};
\end{tikzpicture}};
\end{tikzpicture}
\par\vspace*{270\p@}}
\makeatother

\usepackage{hyperref}
\hypersetup{hidelinks,backref=true,pagebackref=true,hyperindex=true,colorlinks=false,breaklinks=true,urlcolor=ocre,bookmarks=true,bookmarksopen=false}

\usepackage{bookmark}
\bookmarksetup{
open,
numbered,
addtohook={%
\ifnum\bookmarkget{level}=0 % chapter
\bookmarksetup{bold}%
\fi
\ifnum\bookmarkget{level}=-1 % part
\bookmarksetup{color=ocre,bold}%
\fi
}
}
\usepackage[utf8]{inputenc}
\usepackage{mathrsfs}

\DeclareUnicodeCharacter{0301}{í}
\DeclareUnicodeCharacter{043E}{************043E*************}
\DeclareUnicodeCharacter{041D}{************041D*************}
\DeclareUnicodeCharacter{0430}{************0430*************}
\DeclareUnicodeCharacter{0443}{************0443*************}
\DeclareUnicodeCharacter{043A}{************043A*************}
\DeclareUnicodeCharacter{0432}{************0432*************}
\DeclareUnicodeCharacter{0438}{************0438*************}
\DeclareUnicodeCharacter{0439}{************0439*************}
\DeclareUnicodeCharacter{0456}{************0456*************}
\DeclareUnicodeCharacter{0441}{************0441*************}
\DeclareUnicodeCharacter{043D}{************043D*************}
\DeclareUnicodeCharacter{0446}{************0446*************}
\DeclareUnicodeCharacter{043B}{************043B*************}
\DeclareUnicodeCharacter{044C}{************044C*************}
\DeclareUnicodeCharacter{0433}{************0433*************}
\DeclareUnicodeCharacter{0440}{************0440*************}
\DeclareUnicodeCharacter{043D}{************043D*************}
\DeclareUnicodeCharacter{0447}{************0447*************}
\DeclareUnicodeCharacter{0435}{************0435*************}
\DeclareUnicodeCharacter{0442}{************0442*************}
\DeclareUnicodeCharacter{044B}{************044B*************}
\DeclareUnicodeCharacter{0444}{************0444*************}
\DeclareUnicodeCharacter{0437}{************0437*************}
\DeclareUnicodeCharacter{0445}{************0445*************}
\DeclareUnicodeCharacter{043C}{************043C*************}
\DeclareUnicodeCharacter{044F}{************044F*************}
\DeclareUnicodeCharacter{0412}{************0412*************}
\DeclareUnicodeCharacter{0425}{************0425*************}
\DeclareUnicodeCharacter{041A}{************041A*************}
\DeclareUnicodeCharacter{0413}{************0413*************}
\DeclareUnicodeCharacter{0415}{************0415*************}
\DeclareUnicodeCharacter{041F}{************041F*************}
\DeclareUnicodeCharacter{0406}{************0406*************}
\DeclareUnicodeCharacter{0434}{************0434*************}
\DeclareUnicodeCharacter{0436}{************0436*************}
\DeclareUnicodeCharacter{0431}{************0431*************}
\DeclareUnicodeCharacter{1EF3}{************1EF3*************}
\DeclareUnicodeCharacter{0421}{************0421*************}
\DeclareUnicodeCharacter{7530}{************7530*************}
\DeclareUnicodeCharacter{587E}{************587E*************}
\DeclareUnicodeCharacter{5927}{************5927*************}
\DeclareUnicodeCharacter{5B66}{************5B66*************}
\DeclareUnicodeCharacter{7D00}{************7D00*************}
\DeclareUnicodeCharacter{8981}{************8981*************}
\DeclareUnicodeCharacter{6D25}{************6D25*************}
\DeclareUnicodeCharacter{2264}{************2264*************}

\DeclareUnicodeCharacter{C218}{************C218*************}
\DeclareUnicodeCharacter{D559}{************D559*************}
\DeclareUnicodeCharacter{AD50}{************AD50*************}
\DeclareUnicodeCharacter{C721}{************C721*************}
\DeclareUnicodeCharacter{D559}{************D559*************}
\DeclareUnicodeCharacter{C5F0}{************C5F0*************}
\DeclareUnicodeCharacter{AD6C}{************AD6C*************}

\makeatletter
\g@addto@macro\@floatboxreset\centering
\makeatother

\usepackage{etoolbox}
\usepackage{amssymb}
\usepackage{nomencl}
\makenomenclature
\usepackage[utf8]{inputenc}
\usepackage{amsmath}
\usepackage{subfig}
\usepackage{mathtools}
\usepackage{float}
\usepackage{amsfonts}
\usepackage{booktabs}
\usepackage{siunitx}
\usepackage{stackengine}
\usepackage{stmaryrd}
\usepackage{txfonts}
\usepackage{graphicx}
\usepackage{titlesec}
\usepackage{xcolor}
\usepackage{undertilde}

\usepackage{etoolbox}
\makeatletter
\patchcmd\@makeschapterhead
  {\includegraphics[width=\paperwidth]{\thechapterimage}}
  {\includegraphics[width=\paperwidth,height=4.13333in]{\thechapterimage}}
  {}{}
\makeatother

\titleformat{\paragraph}
{\normalfont\normalsize\bfseries}{\theparagraph}{1em}{}
\titlespacing*{\paragraph}
{0pt}{3.25ex plus 1ex minus .2ex}{1.5ex plus .2ex}

\stackMath

\newtheorem{thm}{Theorem}[section]
\newtheorem{clm}[thm]{Claim}

\newtheorem{exm}[thm]{Example}

\newtheorem{lem}[thm]{Lemma}
\newtheorem{crl}[thm]{Corollary}

\newcommand{\highlight}[1]{%
  \colorbox{yellow!100}{$\displaystyle#1$}}
\begin{document}

%----------------------------------------------------------------------------------------
%	TITLE PAGE
%----------------------------------------------------------------------------------------

\begingroup
\thispagestyle{empty} % Suppress headers and footers on the title page
\begin{tikzpicture}[remember picture,overlay]
\node[inner sep=0pt] (background) at (current page.center) {\includegraphics[width=\paperwidth]{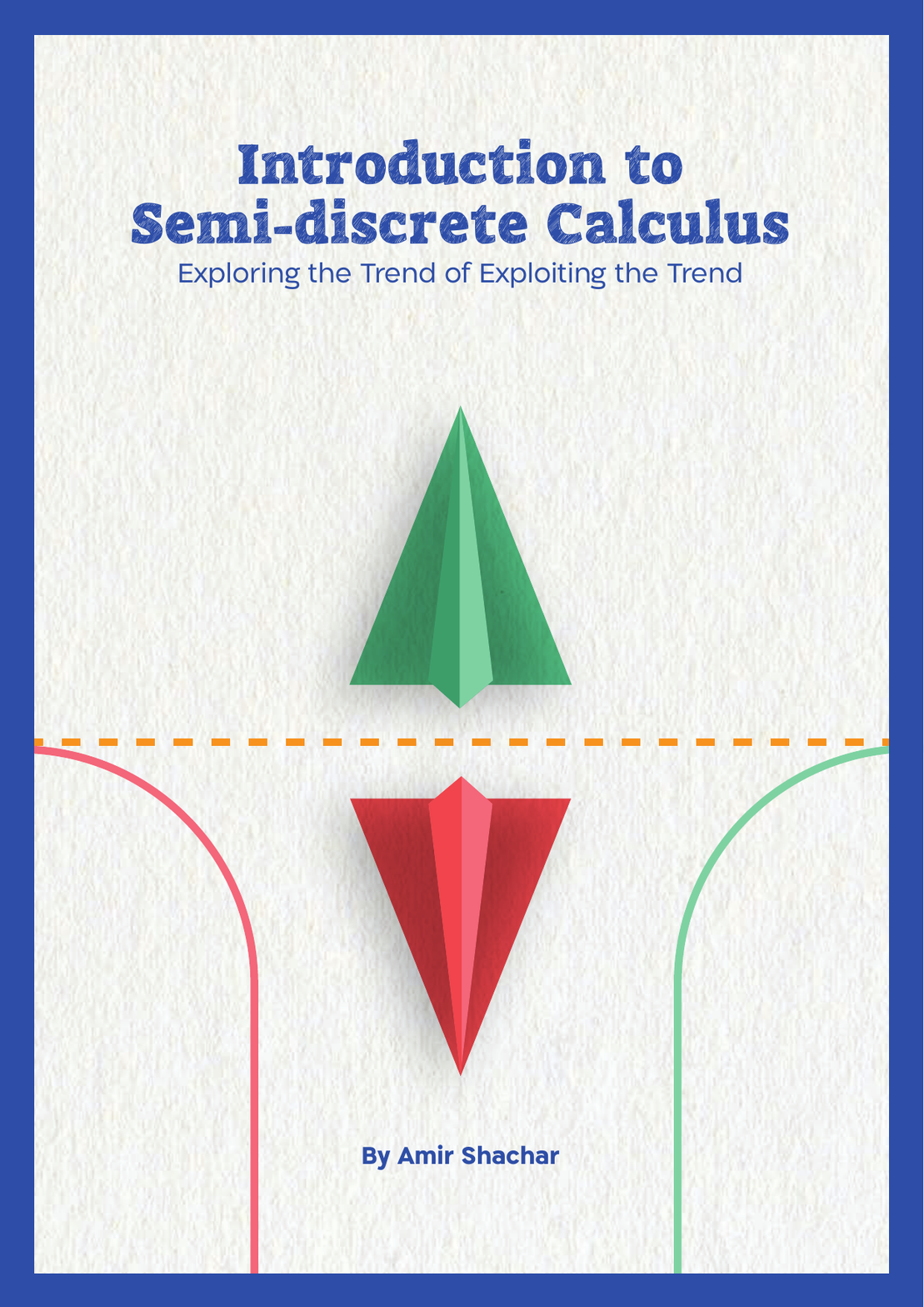}};
\end{tikzpicture}
\vfill
\endgroup

%----------------------------------------------------------------------------------------
%	COPYRIGHT PAGE
%----------------------------------------------------------------------------------------

\newpage
~\vfill
\thispagestyle{empty}

\noindent Copyright \copyright\ 2022 Amir Shachar\\ % Copyright notice

\noindent \textsc{Published by Amir Shachar}\\

\noindent \textsc{https://www.amirshachar.com}\\

\noindent All rights reserved. No portion of this book may be reproduced in any form without permission from the publisher, except as permitted by U.S. copyright law.\\

\noindent For permissions contact: contact@amirshachar.com\\

\noindent Cover by Trio Studio \\

\noindent \textit{First printing, April 2022} % Printing/edition date

%-------------------------------------------------------
%   DEDICATION
%--------------------------------------------------------
\clearpage
\begin{center}
    \thispagestyle{empty}
    \vspace*{\fill}
    To my loved parents:\linebreak
Sarit, who always believed;\linebreak
Yaron, who used to doubt;\linebreak
Both ingredients were equally essential.
    \vspace*{\fill}
\end{center}
\clearpage

%----------------------------------------------------------------------------------------
%	TABLE OF CONTENTS
%----------------------------------------------------------------------------------------

%\usechapterimagefalse % If you do not want to include a chapter image, use this to toggle images off - it can be enabled later with \usechapterimagetrue

\chapterimage{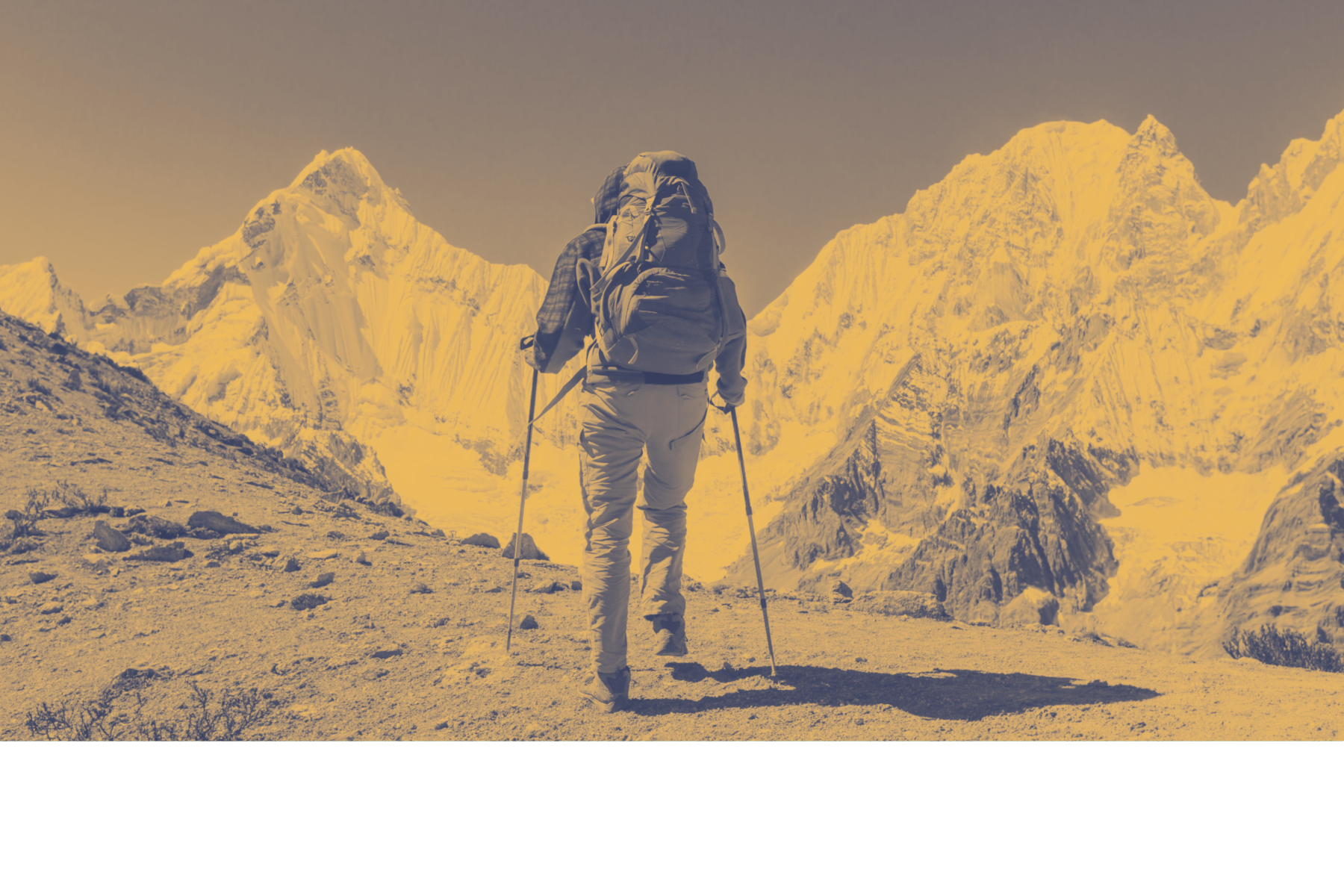} % Table of contents heading image

\pagestyle{empty} % Disable headers and footers for the following pages

\tableofcontents % Print the table of contents itself

\cleardoublepage % Forces the first chapter to start on an odd page so it's on the right side of the book

\pagestyle{fancy} % Enable headers and footers again

\chapterimage{chapter_head_2.pdf} % Chapter heading image

\index{Prologue}
\part{Prologue}

\renewcommand\nomgroup[1]{%
  \item[\bfseries
  \ifstrequal{#1}{P}{Physics constants}{%
  \ifstrequal{#1}{N}{Number sets}{%
  \ifstrequal{#1}{M}{Mathematical Operators}{%
  \ifstrequal{#1}{O}{Other symbols}{}}}}%
]}
% -----------------------------------------

\chapterimage{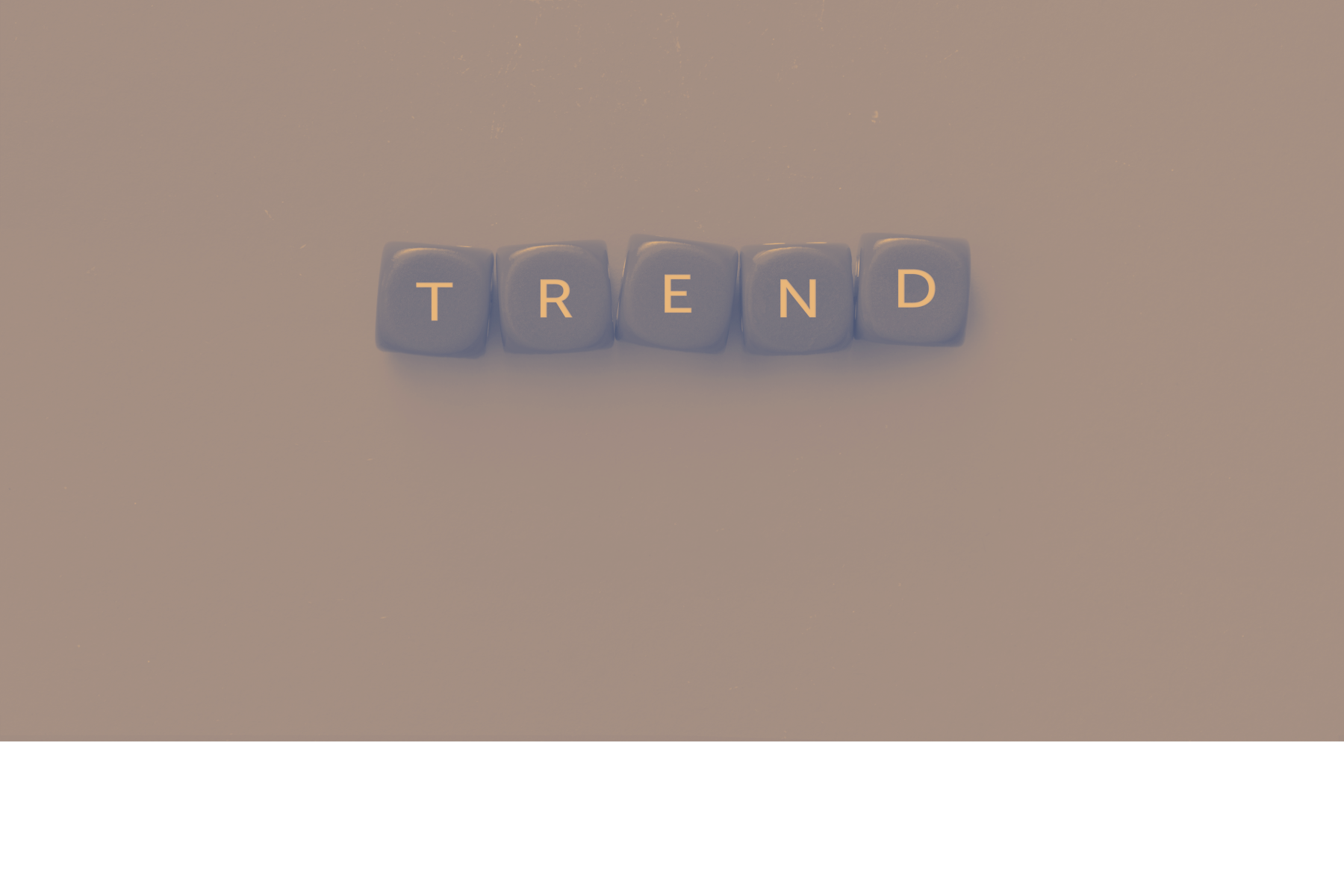}
\chapter{Introduction}

In the past couple of decades, trends calculations have become increasingly valuable across the scientific literature. Scientists, engineers, mathematicians, and teachers often find the derivative sign a good enough measurement of the local monotonicity. They knowingly and deliberately spare the complete information in the tangent slope and focus on its trend. Is the function ascending, descending, or constant? Humans and computer algorithms alike are increasingly asking this simple question at the core of several cutting-edge applications in multiple fields of study. 
For example, AI researchers are increasingly applying emerging “sign”
back propagation techniques that rely solely on the loss function's
derivative sign, mathematicians are extensively investigating locally
monotone operators, and chemical engineers are using Qualitative
Trend Analysis to classify process monotonicity across intervals.

Unfortunately, while it is a reasonable enough estimate in most real-life
and engineering applications, the derivative sign comes with inherent
differentiation caveats that sometimes render it suboptimal. In continuous
domains, for instance, differentiability is a must. We cannot apply
the derivative at singularity points like cusps and discontinuities.
Furthermore, the derivative sign often does not describe the local
trend coherently at a local extremum point, where it is zeroed, but
the function is not necessarily constant. Also, there is an inherent
redundancy in calculating the derivative sign of functions' quotient.

The acknowledgment of the importance of trends calculation motivates
us to explore what seems to be a trend of exploiting the trend - where
new theories emerge that inherently involve merely the information
about functions' trends. We take a close look at workarounds that
scientists have applied to differentiation caveats in multiple scenarios.
We show that researchers often spare the division operator and directly
calculate the sign of the one-sided difference when evaluating trends
based on discrete derivatives. We argue that avoiding the division
may spare up to 20\% of the runtime in this discrete implementation
while maintaining the trend value. Next, we will introduce a new Calculus
operator (“Detachment”) designated to measure trends, which we believe
is a natural step toward modeling the workarounds that engineers have
already been applying that would further improve trends estimates.
Finally, we will propose a mathematical theory emerging from the novel
trend operator: Semi-discrete Calculus.

\section{How to Read This Book}

The book is comprised of three main parts: Part \ref{trendland_part}
introduces "Trendland", where we review previous works on local
trends in the form of cherry-picked articles from the relevant scientific
literature. For each paper, we highlight the application of the trend
within the algorithmic or scientific framework. Part \ref{scientific_discussion_part}
discusses several scientific aspects of the papers reviewed in Trendland,
along with a review of known issues in the definition of trend as
the derivative sign and emerging workarounds. Part \ref{mathematical_discussion_part}
defines some of the recent workarounds as a standalone mathematical
operator and proposes a mathematical theory based on it.

Note that part \ref{trendland_part} (Trendland) is meant to convey
the message that trends calculations are becoming excessive in the
scientific literature. It is not meant to be read as-is but rather
skim through and focus on the reader's interest areas. For instance,
AI researchers will probably find most interest in subsection \ref{machine_learning_subsection},
which is part of the engineering Trendland, chemical physicists are
likely to find more valuable information in subsubsection \ref{chemical_physics_subsubsection},
which is part of the scientific Trendland, and teachers will likely
find interest in the educational Trendland portrayed in chapter \ref{educational_trendland}.
While part \ref{scientific_discussion_part} is intended for scientists,
engineers and mathematicians alike, it is likely that the mathematical
discussion in part \ref{mathematical_discussion_part} will be of
particular interest to mathematicians and scientists.

\nomenclature[N]{\(\mathbb{R}\)}{Real numbers} \nomenclature[N]{\(\mathbb{C}\)}{Complex numbers}
\nomenclature[N]{\(\mathbb{Z}\)}{Integers} \nomenclature[N]{\(\mathbb{Q}\)}{Rational numbers}
\nomenclature[N]{\(\mathbb{N}\)}{Natural numbers} \nomenclature[M]{\(f',\dot{f},\frac{df}{dx}\)}{The first derivative of $f$ (with respect to $x$)}
\nomenclature[M]{\(\frac{\partial f}{\partial x}\)}{The partial derivative of $f$ (with respect to $x$)}
\nomenclature[M]{\(f^{\left(n\right)},\frac{d^nf}{dx^n}\)}{The $n^{\text{th}}$ derivative of $f$}
\nomenclature[M]{\(f^;\)}{The detachment of $f$} \nomenclature[M]{\(\int_{a}^{b}f\left(x\right)dx\)}{The definite integral of $f$ in the interval $\left[a,b\right]$}
\nomenclature[M]{\(\int f\left(\overrightarrow{x}\right)\overrightarrow{dx}\)}{The multiple integral of $f$ in the domain $D$}
\nomenclature[M]{\(F\)}{An antiderivative of \ensuremath{f}} \nomenclature[M]{\(\text{sgn}\left(x\right)\)}{The signum of $x$: $-1$ if $x<0$, $0$ if $x=0$, $+1$ if $x>0$}
\nomenclature[M]{\(\boldsymbol{1}_{A}\)}{The indicator function of the set $A$}

\chapterimage{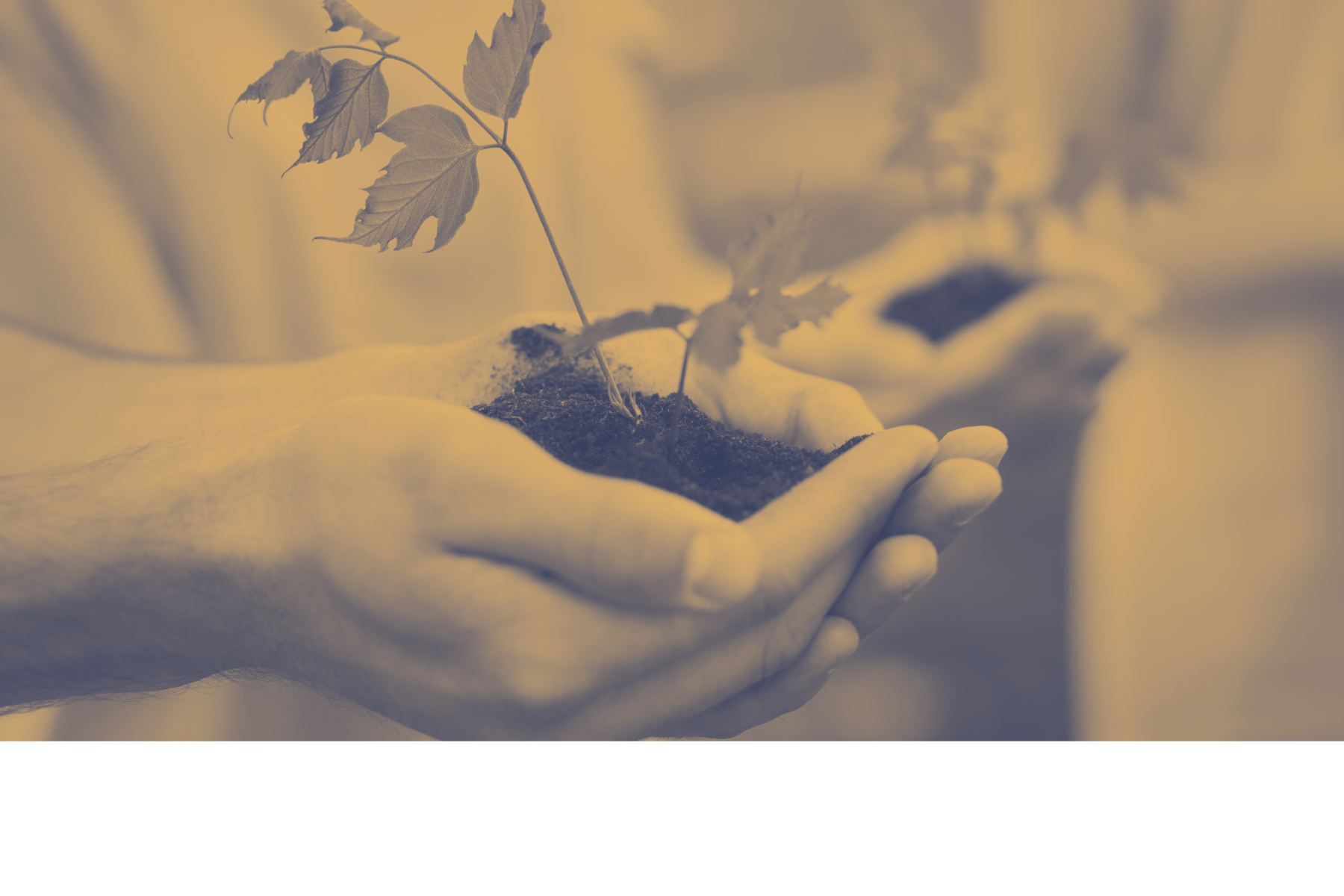} \printnomenclature{}

\part{Previous Work - Trendland}

\label{trendland_part}

\chapterimage{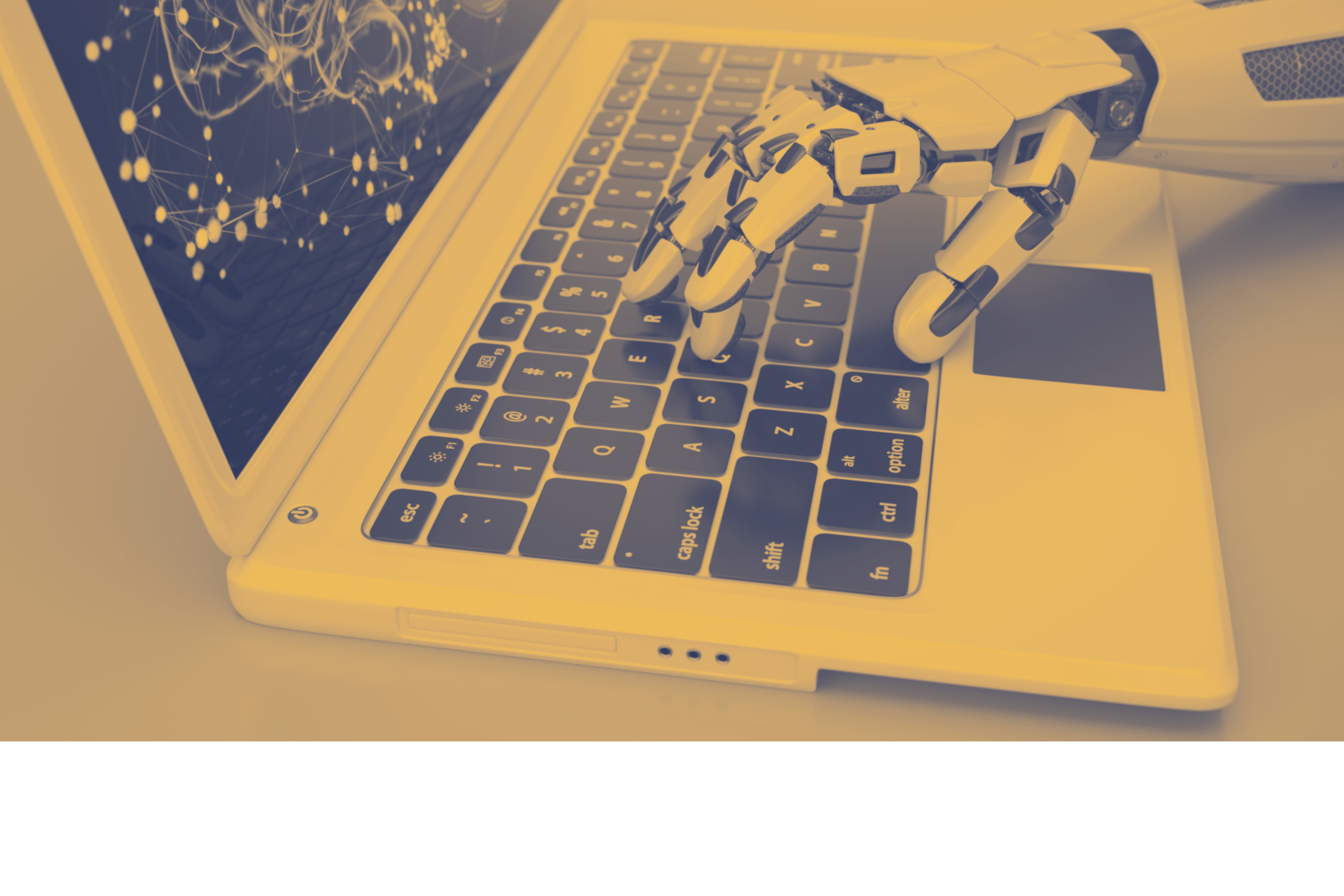} \index{Engineering Trendland}

\chapter{The Engineering Trendland}

In this chapter, we will see several examples of the use of the derivative sign in computer science and engineering, as well as electrical, systematic, mechanical, civil, agricultural, quantum, and aerospace engineering.

\index{Computer Science}

\section{Computer Science and Engineering}

\index{Machine Learning}

\subsection{Machine Learning}

\label{machine_learning_subsection}

\index{Machine Learning Optimization}

\subsubsection{Machine Learning Optimization}

While the gradient is the fundamental concept in many optimization
algorithms, it turns out that its sign is an essential stand-alone
component in some of them. Recently, several papers pointed out this
trend in the literature and surveyed several algorithms that incorporate
the derivative sign while analyzing their mathematical and convergence
properties (\cite{wang2019signadam++,moulay2019properties}). Let
us study examples of optimization algorithms in which the derivative's
sign was found lucrative to researchers due to its relative stability.

Researchers often treat the RProp (\cite{riedmiller1993direct}) algorithm
as a basis for other optimization approaches relying on the derivative
sign. Although originated in 1993, researchers still cite the approach
proposed by Riedmiller and Braun as one of the recommended optimization
frameworks, alongside its modern alternatives - even when it comes
to deep learning optimization. For example, \cite{behnke2003hierarchical}
recommends it, and \cite{wang2017origin} states that this unique
formalization allows the gradient method to overcome some cost curvatures
that we may not solve quickly with today's dominant methods.

The two-decade-old method is still worthy of our consideration. We
can formulate the algorithm as follows: 
\[
\theta^{t+1}=\theta^{t}-\eta_{+}I\left(sgn\left(\frac{\partial L}{\partial\theta}\left(\theta^{t}\right)\right)>0\right)+\eta_{-}I\left(sgn\left(\frac{\partial L}{\partial\theta}\left(\theta^{t}\right)\right)<0\right),
\]
Where $I(\cdot)$ is the indicator function. Based on the gradient
sign, we decide to proceed with a growing or shrinking step size in
each iteration. In other words, in case the optimizee trends upwards,
then to reach the minimum, our step should be of size $\eta^{+}$
to the left, and vice versa. The sizes of the steps vary exponentially,
allowing for efficient exploration and a binary search based exploitation.

Let us gain some intuition regarding this algorithm's efficiency.
We can think of optimization as a prediction task in which we impose
an intelligent guess of a lucrative step towards an optimum. Intuitively,
the magnitude of the gradient seems like a reasonable consideration.
However, often -- for example, in cases where the function's rate
of change plunges -- a simple binary search may land on the optimum
faster. Another way to think about it is that the gradient's magnitude
might 'overfit' the optimizer's prediction relative to the sign alone.
In other words, as a function increases, a step in the negative direction
is guaranteed to promote us towards a local minimum. However, one
can only speculates about a correct way to rely on the gradient's
magnitude for the step size, and speculations are often inaccurate
and give rise to challenging scientific debate.

In \cite{behnke2003hierarchical}, for instance, RProp mitigates the
challenge of vanishing gradients, as it overlooks their magnitude.
However, RProp entails a considerable challenge. It renders the optimization
method discontinuous upon conducting SGD. For example, imagine that
some mini-batches dictate a gradual deterioration of the gradient.
Suppose we proceed with a mini-batch whose gradient cancels the previous
ones. The researcher would expect the steps themselves to balance
out and that the optimization process would end up where it began,
which is the case in Stochastic Gradient Descent methods. In contrast,
the described case renders RProp proceed without canceling out the
accumulated gradients.

Other optimization algorithms leverage the gradient's sign only implicitly,
including AdaGrad (\cite{lydia2019adagrad}), AdaDelta (\cite{zeiler2012adadelta})
and RMSProp (\cite{hinton2012neural}). In such algorithms, we divide
by the weighted root of the mean squares of the previous gradients.
We can think of this standardization as a generalization of RProp
in that it considers the sign of the current gradient while doing
it smoothly relative to the process thus far, which is evident when
we restrict the window size to include only the most recent gradient.
The gradient's magnitude in the numerator then cancels out with its
absolute value in the denominator, resulting in the gradient's sign.

An alternative to this type of algorithms is an emerging family of
machine learning (ML) algorithms that use the derivative sign explicitly.
They are referred to in the literature as the ``Sign'' algorithms.
Thus, researchers often refer to RProp as the signed gradient descent;
SignSGD is a version of stochastic gradient descent that applies its
sign, SignAdam is the analog of Adam, and so on.

Several proposals on how to further improve the RProp algorithm have
been made, including the work of \cite{igel2003empirical}, which
focuses on the loss function's local trends and proposes a modifications of the original algorithm that improves its learning speed.\cite{anastasiadis2003efficient} introduces an efficient modification of the
Rprop algorithm for training neural networks
and proposes three conditions that any algorithm that employs information
of the sign of partial derivative should fulfill to update the weights.

Quickprop (\cite{roux2005optimization}) proposes looking  at the evolution of the sign of the gradient with respect to
one parameter for successive iterations; if it is the same, one should
follow the gradient descent direction; if it is different, a minimum
is likely to exist in between the preceding and current values and
one should expect to be in a situation where the second-order approximation
is reasonable.

\cite{cortes2005achieving} introduces the normalized and signed gradient
descent flows associated with a differential function. The flows characterize
their convergence properties via non-smooth stability analysis. They
also identify general conditions under which these flows attain the
set of critical points of the function in a finite time. To do this,
they extend the results on the stability and convergence properties
of general non smooth dynamical systems via locally Lipschitz and
regular Lyapunov functions.

\cite{zainuddin2005improving,prasad2013comparison,mushgil2015comparison}
discuss the pros and cons of RProp and compare several approaches
for backpropagation that leverage the derivative sign. Among them
are \textquotedbl Sign Changes,\textquotedbl{} \textquotedbl Delta-bar-delta,\textquotedbl{}
\textquotedbl RProp,\textquotedbl{} \textquotedbl SuperSAB,\textquotedbl{}
and \textquotedbl QuickProp.\textquotedbl{}

A key challenge in applying model-based Reinforcement Learning and
optimal control methods to complex dynamical systems, such as those
arising in many robotics tasks, is the difficulty of obtaining an
accurate system model. These algorithms perform very well when they
are given or can learn an accurate dynamics model. However, it is
often very challenging to build an accurate model by any means: effects
such as hidden or incomplete state, dynamic or unknown system elements,
and other effects, can render the modeling task very difficult.

\cite{kolter2010learning} presents methods for dealing with such
situations by proposing algorithms that can achieve good performance
on control tasks even using only inaccurate system models. In particular,
one of the algorithmic contributions that exploit inaccurate system
models is an approximate policy gradient method, based on an approximation
called the Signed Derivative, which can perform well when only the
sign of specific model derivative terms are known. 

In a study focusing on the associations between the fields of active learning and stochastic convex optimization, \cite{ramdas2013algorithmic} proposes an algorithm that solves stochastic
convex optimization using only noisy gradient signs by repeatedly
performing Active Learning, achieves optimal rates, and is adaptive
to all unknown convexity and smoothness parameters. 

\cite{zhai2013direct} proposes an algorithm which utilizes a greedy coordinate
ascent algorithm that maximizes the average margin over all training
examples, leveraging the derivative sign in an interval.

\cite{ganin2015unsupervised} introduces, Gradient reverse layer (GRL),
an identity transformation in forward propagation that changes the
sign of the gradient in backward propagation.

\cite{zamanidoost2015manhattan} proposes that we update the weights
by only using sign information of the classical backpropagation algorithm
in the Manhattan Rule training. Equations 3.4, 3.5 in \cite{wang2016study}
state that the partial derivatives of $y'$ equal the sign of the
derivatives of $w$ and $x$, respectively.

Parallel implementations of stochastic gradient descent (SGD) have
received significant research attention, thanks to its excellent scalability
properties. A fundamental barrier when parallelizing SGD is the high
bandwidth cost of communicating gradient updates between nodes; consequently,
researchers have proposed several lossy compression heuristics, by
which nodes only communicate quantized gradients. Although effective
in practice, these heuristics do not always converge.

An interesting example of parallel implementations of SGD is the work of \cite{alistarh2017qsgd}, who propose Quantized SGD (QSGD), a family of
compression schemes with convergence guarantees and good practical
performance. The \textquotedbl ternarize\textquotedbl{} operation in
Eq. 1 of \cite{wen2017terngrad} splits the gradient to its sign and
magnitude ingredients. The parameter in TernGrad is upgraded in turn
by Eq. 8, where the gradient sign is applied. Theorem 1 then proves
a convergence property of this gradient sign-based algorithm.

Backpropagation provides a method for telling each layer how to improve
the loss. Conversely, in hard-threshold networks, target propagation
offers a technique for telling each layer how to adjust its outputs
to enhance the following layer's loss. While gradients cannot propagate
through hard-threshold units, one may still compute the derivatives
within a layer. As illustrated in \cite{friesen2017deep}, an effective
and efficient heuristic for setting the target activation layer is
to use the (negative) sign of the partial derivative of the next layer's
loss.

Training large neural networks requires distributing learning across
multiple workers, where the cost of communicating gradients can be
a significant bottleneck. SignSGD alleviates this problem by transmitting
just the sign of each minibatch stochastic gradient. \cite{bernstein2018signsgd}
proves that it can get the best of both worlds: compressed gradients
and SGD-level convergence rate. The relative $\ell_{1}$ or $\ell_{2}$
geometry of gradients, noise, and curvature informs whether signSGD
or SGD is theoretically better suited to a particular problem. On
the practical side, the authors find that the momentum counterpart
of signSGD can match the accuracy and convergence speed of ADAM on
deep Imagenet models.

\cite{balles2018dissecting} interprets ADAM as a combination of two
aspects: for each weight, the update direction is determined by the
sign of stochastic gradients, whereas an estimate of their relative
variance determines the update magnitude; they disentangle these two
aspects and analyze them in isolation, gaining insight into the mechanisms
underlying ADAM. 

\cite{zhang2018sign} propose an algorithm based on the derivative
sign of the backpropagation error. \cite{bernstein2018convergence}
further studies the base theoretical properties of this simple yet
powerful SignSGD. The authors establish convergence rates for signSGD
on general non-convex functions under transparent conditions. They
show that the rate of signSGD to reach first-order critical points
matches that of SGD in terms of the number of stochastic gradient
calls but loses out by roughly a linear factor in the dimension for
general non-convex functions.

\cite{wang2019e2} explores the usage of a highly low-precision gradient
descent algorithm called SignSGD. The original algorithm still requires
complete gradient computation and therefore does not save energy.
The authors propose a novel \textquotedbl predictive\textquotedbl{}
variant to obtain the sign without computing the full gradient via
low-cost, bit-level prediction. Combined with a mixed-precision design,
this approach decreases both computation and data movement costs.

\cite{moulay2019properties} studied the properties of the sign gradient
descent algorithms involving the sign of the gradient instead of the
gradient itself. This article provides two convergence results for
local optimization, the first one for nominal systems without uncertainty
and a second one for uncertainties. New sign gradient descent algorithms,
including the dichotomy algorithm DICHO, are applied on several examples
to show their effectiveness in terms of speed of convergence. As a
novelty, the sign gradient descent algorithms can converge in practice
towards other minima than the closest minimum of the initial condition,
making this new metaheuristic method suitable for global optimization
as a new metaheuristic method.

\cite{wang2019signadam++} applies the sign operation of stochastic
gradients (as in sign-based methods such as signSGD) into ADAM, called
signADAM. This approach is easy to implement and can speed up the
training of various deep neural networks. From a computational point
of view, choosing the sign operation is appealing for the following
reasons:
\begin{itemize}
\item Tasks based on deep learning, such as image classification, usually
have a large amount of training data. \textquotedbl Training\textquotedbl{}
on such a data set may take a week, even a month, to shrink calculation
in every step, enhancing efficiency.
\item The algorithm can induce sparsity of gradients by not updating some
of the gradients. One of the ADAM's drawbacks is that it ignores these
gradients by using the exponential average algorithm. SignAdam++ can
prevent this issue by giving small gradients shallow confidence (e.g.,
$0$).
\item Nowadays, learning from samples in deep learning is always stochastic
due to the massive amount of data and models. The confidence of some
gradients produced by loss functions should be small. From the perspective
of the maximum entropy theory, each feature should have the equal
right to make efforts in a deep neural network.
\item The incorrect samples cause large gradients which may hurt the models'
generalization and learning ability. signADAM++ addresses this issue
by using moving averages after applying confidence for unprocessed
gradients and an adaptive confidence for some large gradients; hence,
improving the performance in some models.
\end{itemize}
\noindent As illustrated in \cite{kafka2019traversing}, sign changes
in the directional derivative along a search direction may appear
and disappear stochastically as the oracle updates the mini-batches.
In addition to the sign change of each mini-batch loss function, additional
sampling-induced sign changes may manifest along a search direction.
This occurs when the oracle switches between a negative and positive
directional derivative for essentially the same step along a search
direction. \cite{preechakul2019cprop} argues that only the signs
of the gradients would amplify small noisy gradients, making the conformity
score not useful. Instead, CProp measures the conformity by asking
the following:\textquotedbl{} Does the past gradients conform enough
to show a clear sign, positive or negative collectively?\textquotedbl{} 

\cite{chen2020just} argues in favor of the Gradient Sign Dropout
(GradDrop) method, noting that when multiple gradient values try to
update the same scalar within a deep network, conflicts arise through
differences in sign between the gradient values. Following these gradients
blindly leads to gradient tugs-of-war and to critical points where
constituent gradients can still be significant,making some tasks perform
poorly. \cite{chen2020just} demand that all gradient updates are
pure in sign at every update position to alleviate this issue. Given
a list of (possibly) conflicting gradient values, the method proposed
algorithmically selects one sign (positive or negative) based on the
distribution of gradient values and mask out all gradient values of
the opposite sign.\cite{joy2020fast} applies directional derivative
signs strategically placed in the hyperparameter search space to seek
a more complex model than the one obtained with small data.

\cite{agarwal2020role} explored whether one can generate an imperceptible
gradient noise to fool the deep neural networks. For this, the authors
analyzed the role of the sign function in the gradient attack and
the role of the direction of the gradient for image manipulation.
When one manipulates an image in the positive direction of the gradient,
they generate an adversarial image. On the other hand, if they utilize
the opposite direction of the gradient for image manipulation, one
observes a reduction in the classification error rate of the CNN model.

In \cite{ma2020qualitative}, the signSGD flow, which is the limit
of Adam when taking the learning rate to $0$ while keeping the momentum
parameters fixed, is used to explain the fast initial convergence.

Sign-based optimization methods have become popular in machine learning
also due to their favorable communication cost in distributed optimization
and their surprisingly good performance in neural network training.
\cite{balles2020geometry}, for instance, finds sign-based methods to be preferable
over gradient descent if the following conditions are met:
\begin{itemize}
\item The Hessian is to some degree concentrated on its diagonal
\item Its maximal eigenvalue is much larger than the average eigenvalue

Both properties are common in deep networks
\end{itemize}
\cite{safaryan2021stochastic} analyzed sign-based methods for non-convex
optimization in three key settings: Standard single node, parallel
with shared data, and distributed with partitioned data. Single machine
cases generalize the previous analysis of signSGD, relying on intuitive
bounds on success probabilities and allowing even biased estimators.
Furthermore, this work extended the analysis to parallel settings
within a parameter server framework, where exponentially fast noise
reduction is guaranteed with respect to the number of nodes, maintaining
1-bit compression in both directions and using small mini-batch sizes.
Next, they identified a fundamental issue with signSGD to converge
in a distributed environment. To resolve this issue, they propose
a new sign-based method, Stochastic Sign Descent with Momentum (SSDM),
which converges under standard bounded variance assumption with the
optimal asymptotic rate.

Due to its simplicity, the sign gradient descent is popular among
memristive neuromorphic systems. When implementing it with memristor
synapses, as in \cite{demirag2021online}, the LB sends a single UP
(or DOWN) pulse to instruct an increase (or decrease) of the synaptic
weights. Hence, a single SET pulse is applied to the PCM device, determined
by the gradient sign. However, the effective value of $\delta$ is
not constant due to the WRITE noise and is not symmetric because SET
operation in PCM is gradual, whereas RESET is abrupt.

\cite{li2021faster} investigated faster convergence for a variant
of sign-based gradient descent, called scaled signSGD, in three cases:
\begin{itemize}
\item The objective function is firmly convex
\item The objective function is non-convex but satisfies the Polyak-Łojasiewicz
(PL) inequality
\item The gradient is stochastic, called scaled signSGD in this case.

\cite{zou2021understanding} studied the optimization behavior of
signSGD and then extend it to Adam using their similarities; Adam
behaves similarly to sign gradient descent when using sufficiently
small step size or the moving average parameters $\beta_{1}$,$\beta_{2}$
are nearly zero.
\end{itemize}
\index{Adversarial Learning}

\subsubsection{Adversarial Learning}

One of the simplest methods to generate adversarial images (FGSM)
is motivated by linearizing the cost function and solving for the
perturbation that maximizes the cost subject to an $\ell_{\infty}$
constraint. Based on the gradient sign, one may accomplish this in
closed form for the cost of one call to back-propagation. In \cite{kurakin2016adversarial},
this approach is referred to as \textquotedbl fast\textquotedbl{} because
it does not require an iterative procedure to compute adversarial
examples and thus it is much faster than other considered methods.

Several other approaches have been proposed for fast adversial learning.\cite{kurakin2016adversarial},
for instance, introduce a straightforward way to extend the \textquotedbl fast\textquotedbl{}
method --- they apply it multiple times with small step size and
clip pixel values of intermediate results after each step to ensure
that they are in an $\varepsilon$-neighbourhood of the original image.

To achieve a rapid generation of adversarial examples in the \textquotedbl Fast
Gradient Sign\textquotedbl{} algorithm,\cite{goodfellow2014explaining}applies
the sign of loss function's derivative with respect to the input.

\cite{moosavi2016deepfool} add a perturbation to Neural networks
to fool them with the DeepFool algorithm. The authors used the sign
of the derivative in the iterative update rule with the supremum norm.
They also apply the fast gradient sign method, wherein the absence
of general rules to choose the parameter $\varepsilon$, they chose
the smallest $\varepsilon$ such that $90\%$ of the data are misclassified
after perturbation.

\cite{papernot2016crafting}crafts an adversarial sequence for the
Long Short-Term Memory (LSTM) model. The algorithm iteratively modifies
words in the input sentence to produce an adversarial sequence. The
LSTM architecture then misclassifies it. The optimization step applies
the sign of the gradient $J$.

\cite{tramer2017ensemble} visualizes the ``gradient-masking'' effect
by plotting the loss of $v_{adv}^{3}$ on examples $x^{\ast}=x+\varepsilon_{1}\cdot g+\varepsilon_{2}\cdot g^{\bot}$,
where $g$ is the signed gradient of model $v_{adv}^{3}$ and $g^{\bot}$
is the assigned vector orthogonal to $g$. In Section 4.1, they chose
$g^{\bot}$ to be the signed gradient of another Inception model,
from which adversarial examples transfer to $v_{adv}^{3}$. In appendix
E, properties regarding the gradient-aligned adversarial subspaces
for the $\ell_{\infty}$ norm utilize the signed gradient (lemmas
6, 7).

\cite{madry2017towards} lays the foundation of a widespread attack
method - the PGD (projected gradient descent) attack described in
section 2.1. This method uses just the sign of the gradient. Since
its discovery, the superiority of signed gradients to raw gradients
for producing adversarial examples has puzzled the robustness community.
Still, these strong gradient signal fluctuations could help the attack
escape suboptimal solutions with a low gradient.

The sign of the partial derivative of $J'$ in \cite{ma2017adversarial}
is defined as a standalone operator in equation 6 and applied to solve
the second problem using a bounded update approach. In turn, lemma
1 proves a bound on a number defined based on the derivative sign.

Selecting $\mu=0$ in Eq. 6 of \cite{dong2018boosting} yields, in
turn, the I-FGSM, with the gradient sign as the update direction.
Otherwise, the sign is applied to an approximation of the gradient.

The fast gradient sign method is tweaked in several aspects. One of
them is that the algorithm in \cite{sankaranarayanan2018regularizing},
which performs backward pass using the classification loss function.
Each gradient accumulation layer Gc stores the gradient signal backpropagated
to that layer via its sign (Eq. 5 there).

\cite{xie2019improving} surveys the performance of extensions of
the fast gradient sign method and proposes the method $M-DI^{2}-FGSM$,
whose special cases are $DI^{2}-FGSM$ and $MI-FGSM$ - all of which
leverage the gradient sign for some constellations of parameters.

Cheng et al. propose an approach for hard-label black-box attack which
models hard-label attack as an optimization problem where the objective
function can be evaluated by binary search with additional model queries.
Thereby, a zeroth-order optimization algorithm can be applied. In
\cite{cheng2019sign}, the authors adopt the same optimization formulation;
however, they suggest to directly estimate the sign of the gradient
in any direction instead of the gradient itself, which benefits a
single query. By using this single query oracle for retrieving the
sign of directional derivative, they develop a novel query-efficient
Sign-OPT approach for a hard-label black-box attack.

Combined with the fast sign gradient method, \cite{gao2020patch}
proposes the Patch-wise Iterative Fast Gradient Sign Method (PI-FGSM)
to generate strongly transferable adversarial examples. The authors
surveyed the development of the gradient sign-based attack method
in section 3.1 and their algorithm also applies the loss function's
gradient sign.

The adversarial bracketed exposure fusion-based attack in Eq. 6 of
\cite{cheng2020adversarial} is optimized with the sign gradient descent
(Eq. 9 there).

In \cite{lin2020black}, given a clean input image and a pre-trained
DNN, Differential Evolution first derives a gradient sign population,
children candidates competes with their parents using the corresponding
perturbed inputs and, finally, the authors perturbed the inputs with
the approximate gradient signs.

In the proof of the proposition in \cite{fowl2021preventing}, it
is only required to look at the sign of the derivative of \textquotedbl$\alpha$\textquotedbl{}
for some arbitrary perturbation $\Delta_{j}^{\ast}$. Then, because
the crafting algorithm uses signed gradient descent (either Adam or
SGD), the perturbations crafted using the online vs. non-online method
will be identical.

In \cite{geiping2021doesn}, it is possible to optimize the adversarial
perturbation through projected descent (PGD) for all attacks. The
authors found Signed Adam, with a step size of 0.1, a robust first-order
optimization tool to this end. While perturbation bounds were weakly
enforced in the original version by an additional penalty, the authors
optimized the objective directly by projected (signed) gradient descent
in line with other attacks. They find this approach to be at least
equally effective. The proposed method implements specific adversarial
training at training steps, starting from a randomly initialized perturbation
and maximizing cross-entropy for five steps via signed descent. Further,
it optimizes the surrogate attacks via signed Adam descent with the
same parameters described in the attack section.

The stronger the anti-adversary layer solver for Problem (2) in \cite{alfarra2021combating}
is, the more robust $g$ is against all attacks. To that end, the
anti-adversary layer solves Problem (2) with $K$ signed gradient
descent iterations, zero initialization, and $L$ is the cross-entropy
loss. Algorithm 1 summarizes the forward pass of $g$.

\cite{gao2021adversarial} suggests integrating the attack method
into any gradient-based additive-perturbation attack methods, e.g.,
FGSM, BIM, MIFGSM. The authors use the sign gradient descent optimization
with the step size $\lambda=\frac{\varepsilon}{T}$. $T$ denoting
the iteration number, and they fix it as ten as a typical setup in
adversarial attacks.

Algorithm 1 in \cite{pan2021explaining} defines the adversarial direction
in the Individual AGI based on the sign of the $f$ gradient.

The ``fast derivative sign'' method has become so prevalent, that
an analogous ``derivative magnitude'' method was introduced in \cite{rozsa2016adversarial},
emphasizing that it is the magnitude and not merely the sign that
is being exploited.

\index{Meta-Learning}

\subsubsection{Meta-Learning}

\cite{andrychowicz2016learning} presents LSTMs that can learn to
conduct Gradient Descent automatically. Their model learns how to
learn; it builds a favorable optimization method per each domain separately
and outperforms hand-crafted optimization algorithms such as Adam.
The authors mention a crucial challenge in training the optimizers:
different input coordinates can have different magnitudes, which can
make training an optimizer difficult because neural networks naturally
disregard small variations in input signals and concentrate on bigger
input values. To tackle it, the authors decided to replace the gradients
by two other features: $\left(\log\left|\nabla L\right|,sgn\left(\nabla L\right)\right)$.
In other words, to overcome the instability of the gradient operator,
they scaled it down exponentially to a positive number and added its
sign.

This approach is adopted by \cite{ravi2016optimization} in their
prominent MAML few-shot learning framework.

\paragraph{Neural Networks}

Section 6.3 of \cite{behnke2003hierarchical} discusses supervised
learning for recurrent networks. The difficulty of gradient computation
in recurrent networks makes it necessary to employ algorithms that
use only the gradient sign to update the weights. For example, the
magnitudes of the backpropagated errors may vary significantly. For
this reason, it is challenging to determine a constant learning rate
for gradient descent that allows for both stable learning and fast
convergence. Since the RProp algorithm does not use the magnitude
of the gradient, it is not affected by very small or substantial gradients.
Hence, it is advisable to combine this algorithm with backpropagation
through time. This training method for recurrent neural networks have
been experimentally shown to avoid the stability problems of fixed-rate
gradient descent while at the same time being one of the most efficient
optimization methods.

\cite{de2003qualitative} partitions the State-space into regions
with a unique derivative sign pattern. It is shown that a qualitative
abstraction yields a state transition graph that provides the discrete
picture of continuous dynamics.

\cite{salgado2003neuro} proposed an explicitation method of gradual
rules, which allows the extraction of rules starting from the ASP
units. To build a MGR, the sign of the output's derivative is analyzed,
among others.

Several backpropagation methods such as SSAB, RPROP and GNFAM are characterized
by using the derivative sign to increase or decrease the learning
rate (LR) exponentially. These methods are surveyed in \cite{allard2004neural}, who show that different problems require different
learning rate adaptive methods.

Failure modes 5 \& 6 in \cite{kurd2005artificial} refer to scenarios
where the SCANN function's derivative sign is positive or negative,
which prevents the opposite of what was intended in output changes
over input changes (for example, the output must increase if the input
increases). The safety argument for adhering to these bounds focuses
upon assuring that the derivative sign (rule function gradient) as
expressed by Eq. 50 is limited during generalization and learning.
The solution to this argument is that parameter $a_{i}\left(1\right)$
will always be positive to reflect increasing output and negative
to define decreasing output. In addition, the gradient, in any case,
can be zero as a result of the saturation performed by the rule output
bounds.

\cite{batt2008symbolic} introduces a hyperrectangular partition of the state space that forms the basis for a discrete abstraction preserving the sign of the derivatives of the state variables. The approach defines a fine-grained partition of the state
space which underlies a discrete abstraction preserving more substantial
properties of the qualitative dynamics of the system, i.e., the derivative
sign pattern.

Among the hand-chosen features in a work on fast classification and anomaly measurement, \cite{wulsin2011modeling}, the
Normalized Decay describes the chance-corrected fraction of data that
is decreasing or increasing based on the sign of the discrete derivative.

The lecture in \cite{hinton2012neural} introduces the RMSProp algorithm.
The gradient is divided by the (weighted) root of the mean squares
of the gradients in the proceeding steps. One may view this standardization
as a generalization of RProp in that it considers the sign of the
current gradient smoothly with respect to the process thus far. The
gradient's magnitude in the numerator cancels out with its absolute
value in the denominator, resulting in the gradient's sign.

At the deep neural net studied in \cite{poole2016exponential},the
number of times a single neuron has a sign change in the derivative
across the parameters serves as a crucial upper bound for exponential
expressivity.

To identify whether a given data point $x$ is a critical sample,
\cite{arpit2017closer} searches for an adversarial example within
a box of a given radius. To perform this search, the authors use Langevin
dynamics applied to the fast gradient sign method as shown in algorithm
14. They refer to this method as Langevin adversarial sample search
(LASS). While the FGSM search algorithm can get stuck at a point with
zero gradients, LASS explores the box more thoroughly. Specifically,
a problem with first-order gradient search methods (like FGSM) is
that there might exist training points where the gradient is $0$,
but with a large $2^{nd}$ derivative corresponding to a significant
change in prediction in the neighborhood. The noise added by the LASS
algorithm during the search enables escaping from such points.

\cite{wang2017origin} argues that RProp's (signed gradient descent)
unique formalization allows the gradient method to overcome some cost
curvatures that may not be easily solved with today's dominant methods,
depending on the problem's geometry.

\cite{hubara2017quantized} introduce a method to train Quantized
Neural Networks (QNNs) --- neural networks with extremely low precision
(e.g., 1-bit) weights and activations at run-time. At train-time,
the quantized weights and activations are used for computing the parameter
gradients. QNNs drastically reduce memory size and accesses during
the forward pass and replaces most arithmetic operations with bit-wise
operations. As a result, power consumption is expected to be drastically
reduced.

\cite{fan2017unifying} applies the sign of the loss function's gradient
to prove theorem 16. The results are reflected in the update in algorithm
3.

As explained in \cite{gad2018practical}, if the loss function's derivative
is positive, then an increase or decrease in the weight increases
or decreases the error, respectively; the opposite effects occur if
the derivative is negative.

In \cite{ghorbani2019interpretation}, the gradient sign of the dissimilarity
function that measures the change between the interpretations of two
images is applied to iterative feature importance attacks.

\cite{gramsch2020analysis} analyzes the sensitivity of the cause-and-effect
relations. The first partial derivative of the regression function
is computed with respect to each input parameter. A significant value
of the partial derivative indicates a considerable influence of the
corresponding input parameter, i.e., small changes in the input will
lead to substantial changes in the output. Furthermore, the sign of
the partial derivative is essential. A positive partial derivative
indicates that an increase in the input leads to a rise in output.
A negative sign means that an increase in the input leads to a decrease
in the output.

To overcome the vanishing gradient problem, \cite{abuqaddom2021oriented}
introduced a new anti-vanishing back-propagated learning algorithm
called oriented stochastic loss descent (OSLD). OSLD updates a random-initialized
parameter iteratively in the opposite direction of its partial derivative
sign by a small positive random number, which is scaled by a tuned
ratio of the model loss.

In \cite{hasan2021training}, the backward gain w.r.t w is $\beta=\frac{\partial y'}{\partial w}=\sigma x$
where $\sigma$ is the gain of the activation function $\sigma$.
If input $x$ is negative or $\sigma$ is negative, $\beta$ is negative,
and $A\beta$ becomes negative, which is unstable. To make the system
stable for negative backward gain, the forward gain $A$ needs to
be negative so that $A\beta$ becomes positive. The output of the
backward function for negative $\beta$ can be expressed as (10).
We can transfer the negative sign to the error $e$ and keep $A$
positive. Since the negative sign is there because of the negative
backward gain $\beta$, the sign can be replaced by the signum function
as in (12). For negative backward gain, we can continue using positive
forward gain with the error $e$ augmented by the sign of the backward
gain. For positive backward gain, no negative sign is necessary. Thus
(12) is consistent for both positive and negative values of $\beta$.

\paragraph{Federated Learning}

In the online learning algorithm presented by \cite{han2020adaptive},
we see an example of the use of an estimated sign of the derivative
of the objective function which gives a regret bound that is asymptotically
equal to the case where the exact derivative is available. Encrypting
and decrypting all elements of the model gradients have several shortcomings.
First, performing encryption on local clients is computationally extremely
expensive, causing a big barrier for real-world applications since
the distributed edge devices usually do not have abundant computational
resources. Second, uploading model gradients in terms of cipher text
incur high communication costs. Finally, the first two issues will
become computationally prohibitive when the model is large and complex,
e.g., DNNs.

To tackle the above challenges, \cite{zhu2021distributed} introduces
ternary gradient quantization (TernGrad) into their federated learning
framework. TernGrad compresses the original model gradients into ternary
precision gradients with values $\in\left\{ -1,0,1\right\} $ as described
in Eq. (13)..

In the signed gradient scenario of \cite{mahara2021multi}, computing
the gradient of the objective in (7) involves finding the sign of
the gradient with respect to $\left\{ w_{1},w_{2},\ldots,w_{N}\right\} $
instead of the full gradient, and the full gradient with respect to
$\alpha$; this provides a function of a relatively lesser dimensional
information. The gradient with respect to the importance coefficients
$\alpha$ does not depend on the neural network weights but on the
discrepancy and the losses. The method involves sending $O\left(N\right)$
parameters while the gradient with respect to neural network weights
$w$ can potentially involve millions $\left(>O\left(N\right)\right)$
of parameters. The signed gradient of the objective function (7) with
respect to the weight of the device $k$ includes the sign function
within the summation, as shown in (9).

The algorithm in \cite{phuong2021decentralized} provides decentralized
optimization in wireless device-to-device networks of pervasive devices
such as sensors or 5G handsets. The signs of the stochastic gradient
are used for descent steps.

Averaging masked gradients in \cite{tenison2021gradient} is equivalent
to averaging the gradients directly when all the client gradients
being considered are of the same sign.

\paragraph{Reinforcement Learning}

In a work on policy search for reinforcement learning \cite{kolter2009policy}, the approximation is based on the intuition
that it is often straightforward to guess the direction in which control
inputs affect future state variables, even if we do not have an accurate
system model.

The reward function, defined in Eq. 1 of \cite{kong2017collaborative},
can be thought of as the sign of the discrete numerical derivative
of the Intersection-over-Union (IoU) function.

The initial reward in \cite{tang2018deep} is defined as the sign
of the discrete derivative of the probability function that predicts
the video as belonging to a specific class across consecutive iterations.

While an optimal policy is deterministic (gradient ascent searches
among all threshold policies), the gradient is discontinuous at such
policies. Plain Gradient Ascent suffers from this problem. Therefore
\cite{massaro2019optimal} suggests taking integer steps to explore
just integer policies, calculate both one-sided gradients, and use
them in the update step. In turn, the gradient sign is then used within
this optimization method.

As shown in Algorithm 3 in \cite{zhang2020semantic}, when generating
adversarial samples, the adversary computes the perturbation with
$g_{i}=sgn\left(\frac{\partial J\left(G,y\right)}{\partial V}\right)$.
$V$ is the feature matrix, each row of which describes instructions
in a basic block $v_{i}$ of the graph $G$, and $y$ is the label
of the CFG. The attacker heuristically inserts a semantic Nop that
is closest to the gradient $g_{i}$ into the corresponding basic block
of the CFG. In each iteration, the attacker injects the closet semantic
nops to the sign gradient descent. The attacker then repeats this
procedure until a maximum number of iterations $T$.

\paragraph{Regression}

\cite{inoue2002identifying} derives a necessary and sufficient condition
for identifying the sign of the derivative of an unknown monotonic
function by the method of weighted average derivatives. While OLS
has a weighted average derivative representation, it does not necessarily
satisfies the condition except in restrictive cases.

As portrayed in \cite{schmidt2010graphical}, the orthant-wise learning
algorithm of Andrew et al. uses a direct approach. In particular,
they computed and set any value $d_{i}$ in $d$ to zero if its sign
does not agree with $-\nabla f(x_{k})$, as in Eq. 2.8 . However,
this algorithm does not satisfy the property of reducing to Newton's
method because of the PS sign projection. The problem with this projection
is that it may set elements of the Newton direction to zero for a
large portion of the zero and non-zero variables. However, to lie
in the correct orthant (to guarantee descent), the only requirement
is that the zero-valued variables in the search direction agree with
the negative pseudo-gradient sign. Thus, in the PSS sign projection
(PSSsp) variant, the authors apply the orthant-wise learning iteration
but use a less constrained version of the PSsign projection, denoted
by $P_{s}^{*}$.

Corollary 1 in \cite{pan2020implicit} claims that if we have two
roots for $f\left(x\right)=0$ on an interval and the tangent lines
on the two endpoints have the same derivative direction, there must
be another root between the two with a different derivative sign.

In the proof of part a of theorem 2 in \cite{xue2020regularized},
it is sufficient to prove sign properties regarding the partial derivatives
of $Q$ with respect to $\beta_{j}$.

\paragraph{Classification}

\cite{anastasiadis2005sign} introduces a new class of sign-based
schemes based on the composite nonlinear Jacobi process. An algorithm
of this class that applies the bisection method to locate subminimizer
approximations along each weight direction has been derived.

\cite{fernandez2011image} presents a conceptually simple and computationally
efficient family of texture descriptors. Three different methods have
been proposed, namely single-loop, double-loop and triple-loop binary
gradient contours, based on pairwise comparisons of pixel intensities
all along the periphery of a $3*3$ window. These models have been
comparatively analyzed from a theoretical standpoint. The Binary Gradient
Contours are based on the sign of the discrete derivative of the image
in different directions (Eq. 11).

Given that the traditional gradient descent algorithm suffers from
long-term dependency problems, a refined BP algorithm named Rprop,
which leverages the gradient's sign solely, is extended in \cite{chen2013multi}
to train multi-instance multi-label image classification (MIMLNN)
effectively.

\paragraph{Computer Vision}

There are many papers detailing the use of the discrete derivative
sign in computer vision. The circuit in \cite{higgins2000modular},
for instance, is nonlinear in the sense that it produces a fairly
narrow current pulse at the change of the derivative sign both for
sharp and smooth inputs due to the nonlinear feedback.

\cite{viola2001rapid,wang2007shape} apply the discrete derivative
sign implicitly. The authors distinguish between different rectangles'
corners in the plane in efficiently calculating the integral, assigning
different weights based on the corner type. Classification of corners
leverages the curve's derivative sign at its corners.

\cite{grau2002texprint} found a textured print by counting the number
of changes in the derivative sign in the gray level intensity function
by rows and columns.

The algorithm in \cite{navalpakkam2006optimal} optimizes the signal-to-noise
ratio of the saliency map by scrutinizing its derivative's sign at
each iteration (Eq. 11-13).

In a work on topological mapping using spectral clustering and classification \cite{brunskill2007topological}, one of the features used for the submap classifiers is the number of second derivative sign flips (a measure of bumpiness).

In a work on orientation-selective edge detection and enhancement using the irradiance transport equation, \cite{flores2010orientation} distinguishes
the sign of the derivative of the intensity pattern along an arbitrarily
selected direction.

In \cite{pham2010non}, the detection of the $\left(2n+1\right)$-neighborhood
maxima on a 1D scan-line is shown in detail in Figure 3a . If $g$
is the sign of the finite difference of $f,g$ is either $-1,0,$
or $1$ depending on the local slope of $f$. $g$'s finite difference
$h$, therefore, equals $-2$ at local peaks, $+2$ at local troughs,
and $0$ elsewhere. 1D peak and trough detection, therefore, requires
only one comparison per pixel. Next, each 1D peak is compared against
its $\left(2n+1\right)$-neighborhood with the knowledge of the extremum
detector $h$. Neighboring pixels that are on a consecutive downward
slope from the local peak are by definition more minor than the current
peak. Hence they don't need to be recompared and only pixels outside
the enclosing troughs of the current peak need an extra comparison.
The number of extra comparisons to obtain $\left(2n+1\right)$-neighborhood
maxima from an initial list of 3-neighborhood maxima is very small
for a smooth function $f$.

The method in \cite{volkov2011objects}, used for objects description and extraction by the use of straight line
segments in digital images, includes directional filtering
and searching for straight edge segments in every direction and scale,
taking into account edge gradient signs.

The algorithm in \cite{li2013locally} applies the second derivative
sign to distinguish between the requirement to use Newton's method
or a line search in the optimization process while matching images
observed in different camera views.

In contrast to existing approaches, the proposed cost-effective descriptors
in \cite{arrospide2013image} render fine orientation binning and
consideration of gradient sign affordable. Experiments revealed that
these descriptors achieve a significantly better trade-off between
cost and performance in the vehicle verification task than standard
HOG. The HOG technique was initially proposed for pedestrian detection,
for which the authors claimed that including the gradient sign results
in no performance gain. In effect, for humans, the wide range of clothing
and background colors renders the sign of contrast uninformative.
In contrast, when it comes to vehicles, the sign information might
well be significant.

\cite{redi20146} use the sign of the numeric discrete derivative
of the Euclidean distance function applied to frames to define a stop
motion measure.

The derivative sign is used explicitly in a global minimization algorithm
of \cite{tron2014quotient} that establishes a meaningful distance
between essential matrices during pose averaging.

\cite{hu2014deblurring} propose an iterative process to detect light
streaks in the input image automatically with the sign of the derivatives
of the latent image.

The computation of the second-order derivative in \cite{yang2016shape}
only considers a signed binary version of the first-order derivative.

\cite{huang2017robust} presents a computationally efficient yet powerful
binary framework for robust facial representation based on image gradients
(see Eq. 1,2 there)referred to as structural binary gradient patterns
(SBGP). To discover underlying local structures in the gradient domain,
image gradients are computed from multiple directions and simplified
into a set of binary strings. The SBGP is derived from certain types
of these binary strings that have meaningful local structures and
resembled essential textural information. It is shown that the SBGP
can detect micro-orientational edges and possess strong orientation
and locality capabilities, thus enabling significant discrimination.
The SBGP also benefits from the advantages of the gradient domain
and exhibits profound robustness against illumination variations.
The binary strategy realized by pixel correlations in a small neighborhood
substantially simplifies the computational complexity and achieves
extremely efficient processing.

In a study of study the problem of single-image depth estimation for images in the wild, \cite{chen2017surface} shows that adding normals improves ordinal error,
but only from the angle-based normal loss, not the depth-based normal
loss because depth-based normal loss emphasizes getting the same steep
slopes, but this does not make any difference to ordinal error as
long as the sign of the slope is correct.

In a proposed foreground-background separation \cite{lee2017depth} uses an optical phenomenon where bundles of rays from the background
are flipped on their conjugate planes. Using the Lambertian assumption
and gradient constraint, the foreground and background of a scene
can be converted to a binary map by voting the gradient signs in every
angular patch. Using light field reparameterization, the disparity
map can be obtained by accumulating the binary maps.

In \cite{kuznetsov2017new}, an image is binarized using Eq. 5 . One
can view it as an \textquotedbl or\textquotedbl{} between the signs
of its partial derivatives. The process repeats seven times before
the final image $g$ is calculated as a weighted average of the eight
images.

According to \cite{kudryavtsev2017autofocus}, the essential part
in the derivative approximation is its sign and not the amplitude
because the error in the amplitude can be compensated during optimization. The algorithm in \cite{silva2018weighted}, for example, detects a head-turning
movement in a video by assuming a frame within a short camera motion
interval if the sign of all curves' derivatives is constant at a point.

The LBDE component in \cite{xia2018novel} describes the amplitude
change among the neighborhood pixels. In Eq. 7 , the authors define
the local binary gradient orientation (LBGO) to characterize the orientation
information. The proposed LBGO is designed to extract gradient orientation
from center-symmetric pixel pairs. Compared to the original WLD, LBGO
uses the neighboring pixels and retains more local structure information.

The method proposed by \cite{dong2018accurate} is interested in the
general orientation of the gradient and not in the exact gradient
vector. Therefore, this approach uses only the sign of the gradient
as representative of the general orientation of the gradient. Each
edge point $p_{i}$ is assigned with either a positive or negative
direction by Eq. 1 . The function $X\left(p_{i}\right)$ becomes zero
when the Sobel derivatives $dx$ or $dy$ are zero, i.e., the gradients
are along the vertical or horizontal directions. Thus, after the classification
of curves, an edge pixel $p_{i}$ whose gradient sign function is
zero is removed from the edge image.

Canny edge detector with auto-thresholding is applied in \cite{dong2018fast}
to obtain the edge image from an input image, as shown in Fig. 2(B)
there. The coordinates and gradient at an edge pixel $p_{i}$ are
expressed as $\left\{ x_{i},y_{i},\eta_{i}\right\} $, respectively.
Since $\eta_{i}$ cannot be calculated accurately in digital images,
only the orientation of the gradient is used, denoted by its sign,
rather than the actual value of $\eta_{i}$. The gradient sign function
$X\left(p_{i}\right)$ at the pixel $p_{i}$ is defined as in Eq.
1. The edge pixels corresponding to horizontal and vertical gradients,
whose gradient signs are undefined, are discarded. The authors define
$\mathscr{Q}\left(e_{k}\right)$ as the direction of the arc that
lies in a quadrant. Consequently, the arcs of the positive gradient
direction rest on the first or third quadrants ($\mathscr{Q}(e_{k})\in\left\{ I\bigcup III\right\} $)
while the arcs of the negative gradient direction belong to the second
or fourth quadrants ($\mathscr{Q}(e_{k})\in\left\{ II\bigcup IV\right\} $)
as illustrated in Fig. 2(C, D). $I,II,III$ and $IV$ represent the
number of four quadrants. This property shall be used in section II(C)
again for arc classification.

In a work on computational modelling for human 3D shape perception from a single specular image, \cite{shimokawa2019computational}, finds a strong relationship
between vertical polarity and the surface second derivative signs.

\index{Image Processing}

\paragraph{Image Processing}

The signed differences $g_{p}-g_{c}$ in \cite{ojala2002multiresolution}
are not affected by changes in mean luminance. Hence the joint difference
distribution is invariant against grayscale shifts. The authors achieve
invariance with respect to the scaling of the grayscale by considering
just the signs of the differences instead of their exact values as
in Eq. 5 there.

A modified derivative sign binary method is proposed in \cite{zhang2002fringe}
to extract fringe skeletons from interferometric fringe patterns.

The velocity and acceleration signs are used as features in classifying
the signatures in \cite{herdaugdelen2004dynamic}. Unfortunately,
the sign of the derivative vector is sometimes unknown on an equiluminant
edge and is set arbitrarily in the current theory. However, choosing
the wrong sign can lead to unnatural contrast gradients (not apparent
in the original color). In \cite{drew2009improved}, this sign problem
is ameliorated using a generalized definition of luminance and a Markov
relaxation.

\cite{fraz2011retinal} reports an automated method for segmenting
blood vessels in retinal images using a unique combination of differential
filtering and morphological processing. The centerlines are extracted
by applying the first-order derivative of Gaussian in four orientations,
followed by the evaluation of derivative signs and average derivative
values . The shape and orientation map of the blood vessel is obtained
by applying a multidirectional morphological top-hat operator followed
by bit plane slicing of a vessel enhanced grayscale image.

After constructing the derivative sign binary image, \cite{el2012new}
efficiently extracted the fringe center lines.

In \cite{trujillo2013accurate}, both the parameters $m$ and $n$
are defined based on the signs of the partial derivatives of $f$.

Upon conducting simultaneous Super-Resolution of depth and images
using a single Camera in \cite{seok2013simultaneous}, the sign of
the gradient of the data cost function is used to evaluate the optimal
index during the optimization process.

In the fourth part of the algorithm of \cite{mccloskey2014improved},
the authors compute the average gradient sign vector, weighted by
the strength of edge points.

In \cite{tani2014derivative}, it is noted that the basins of attraction
of the true and alias solution are the positive and negative semiplanes,
respectively. Therefore, the sole knowledge of the first derivative
sign would provide a good enough initial condition for any iterative
solver to converge to the actual solution.

\cite{zhu2015modeling} describes a single-image super-resolution
method based on gradient reconstruction. In this approach, the sign
of the derivative of the input image's gradient magnitude is used
for Gradient Ridges and Mask Generation.

Equation (11) in \cite{byrne2015nested} is binarization. A nested
motion descriptor can be binarized by computing the sign of (10) .
A nested motion descriptor is constructed with binary entries. It
is an optional step that can be used to provide compact representation
and can be thought of as the discrete sign of the partial derivative
of $d$.

\cite{gallardo2015shape} discusses the problem of inferring the shape
of a deformable object as observed in an image using a shape template.
Proving that the sign of the derivative of a function remains constant
in an interval is an integral part of the algorithm's justification.
Further, step 4 of the proposed algorithm uses refinement to find
all solutions on each interval by forcing all different combinations
of signs for $\theta$ and $\theta'$.

The algorithm in \cite{pan2016robust} retains large image gradients
and removes tiny details from an intermediate image, where only the
gradient sign and its maximal absolute magnitude are relevant.

Three discomfort characterizations for depth jump cuts are defined
in \cite{delis2016automatic}, namely \textquotedbl mildly uncomfortable\textquotedbl ,
\textquotedbl uncomfortable\textquotedbl , \textquotedbl highly uncomfortable\textquotedbl .
According to the sign of positive and negative depth derivatives,
a characterization is given to a depth jump cut.

A partial differential equation, called SF, is used in \cite{huang2017multi}
for image sharpening and enhancement. The SF process can suppress
the edge diffusion, achieve image deblurring and deconvolution. Still,
it is susceptible to noise, and the noise is also amplified when the
image is enlarged. The SF is commonly generalized by Eq. 7, which
incorporates the sign of the second-order directional derivative of
the image gradient direction.

\cite{li2018a2} utilizes the sign of the discrete derivative of the
aesthetic score with respect to the number of steps the agent has
taken.

In a work on robustness via Curvature Regularization, \cite{moosavi2019robustness} opts to define the solution vector as
a function of the gradient sign rather than that of a gradient. It
does so to constrain its direction to belong to the hypercube of interest.

\index{Computer Music}

\paragraph{Computer Music}

In a work on motion-driven sound synthesis \cite{cardle2003sound} where it is required to determine which portions
of the source motion best match with those in the new target motion.
This is achieved by using a motion matching algorithm (Pullen and
Bregler, 2002). The algorithm is depicted in Figure 5, and the motion
curve is broken into segments where the sign of the first derivative
changes. For better results, a low-pass filter is applied to remove
noise beforehand. All fragments of the (smoothed and segmented) target
motion are considered one-by-one, and for each, the authors select
whichever chunk of source motion is most similar. By breaking up the
movement at first derivative sign changes, they enforce better audio
continuity over portions of motion that are constant. On the other
hand, segmentation based on second derivative changes, or inflection
points, gives better audio continuity at changes in the motion. Consequently,
the system generates two soundtracks, one for each segmentation strategy,
and the animator picks whichever best their expectations.

In \cite{urbano2012mirex}, working on hybrid sequence alignment with geometric representations, the authors follow a naive rationale for
the substitution score: if two spans have roughly the same shape,
they are considered the same, no matter how similar they are. The
authors only look at the direction of the splines at the beginning
and the end of the spans. If the two curves have the same derivative
signs at the end and the beginning of the span, the penalization is
the smallest. If the two curves have opposite derivative signs at
the end and the beginning of the span, the penalization is the largest.
If the two curves have the same derivative sign at one end of the
span but not at the other, the penalization is averaged.

Several authors have proposed cost functions using the derivative
information to avoid singularity points and reduce alignment bias.
Keogh and Pazzani also combine the Savitzky-Golay filter to estimate
the derivative and avoid problems with noise. Following these ideas,
\cite{schramm2014dynamic} proposes a novel cost function that handles
conducting gestures with a wide range of amplitudes better than the
simple derivative measures by decreasing the cost value when the signs
of the derivatives of both signals agree. Such addition prioritizes
the matching of monotonically coherent portions of both trajectories
(i.e., increasing and decreasing behaviors, which are characterized
by the sign of the derivative), which also implicitly helps to align
local extrema, even when the amplitude variation is considerable.
It is an essential feature in the paper's context since the amplitude
of the movement is highly user-dependent and may vary significantly
among different people (e.g., children are more minor in size and
consequently will make smaller movements than adults).

\index{Validation and Verification}

\subsection{Validation and Verification}

In a work on automated test suite generation for time-continuous Simulink models, \cite{matinnejad2016automated} proposes a model where the derivative feature is extended
into sign-derivative and extreme-derivative features. The sign-derivative
feature is parameterized by $\left(s,n\right)$ where $s$ is the
sign of the signal derivative and $n$ is the number of consecutive
time steps during which the sign of the signal derivative is $s$.

In \cite{deantoni2018towards}, working at an ultimate formally verified master algorithm, an entity often monitors a variable
and reacts when this variable crosses a specific value with one particular
derivative sign.

\index{Computer Simulations}

\subsection{Computer Simulations}

In \cite{sjoberg2003probabilistic}, to minimize the overshoot problem,
the Newmark time-stepping algorithm uses a variable time-step when
velocity sign changes are detected. The time-step is repeatedly bisected
until the absolute value of the response velocity at the end of the
reduced time-step is less than a preset fraction of the peak response
velocity.

In a paper on the development, modeling, and testing of skyhook and MiniMax
control strategies of semi-active suspension, \cite{zhang2009comparison} shows that the damper velocity $\nu_{D}$ or
the sign of the damper velocity $sign(\nu_{D})$ provides information
about the direction into which the damper is moving compression or
rebound.

In the work of \cite{golchi2015monotone} on monotone emulation of computer experiments, the experimenter knows
beforehand that the simulator response is monotone in some of the
inputs. However, they only know the sign of the derivatives, the magnitude
of the derivatives being unknown. Through the mechanism of a link
function, monotonicity information is encoded via virtual derivatives
(binary indicators of derivative sign) at points in a derivative input
set. An advantage of the approaches that use derivative sign information
at specific locations in the input space is that they offer flexibility
in incorporating monotonicity information. For example, by specifying
that derivatives are positive with respect to a particular variable
at particular locations, we have the flexibility to make predictions
of the response that has a monotone relationship with a predictor
in just a subset of the range of the predictor.

In \cite{ingraham2018learning}, the TM-score is the best possible
value of the preceding quantity for all possible superpositions of
two structures, where $D_{i}=\Vert x\left(\text{Model}\right)-x\left(\text{Data}\right)\Vert$.
This approach requires iterative optimization, which the authors implement
with a sign gradient descent with $100$ iterations to superimpose
the model and target structure optimally; backpropagating through
the unrolled optimization process and. the simulator.

\index{Computer Security}

\subsection{Computer Security}

In \cite{cristea2008statistical}, the cross-covariance function,
(24) and (25), depends not only of the density function, $f^{\ast}(x)$,
and the representation used for the chaotic map, $\sigma t(x)$, but
also of the chaotic map itself, $\tau\left(x\right)$, and the sign
of its derivative, $sign\left(\tau'\left(x\right)\right)$.

In lemma 7.1 of \cite{ghazi2018resource}, to bound $\lambda$, the
authors analyze the monotonicity of $f$ based on the sign of its
partial derivative. In lemma 7.2, we directly prove the monotonicity
properties of $Q$ based on its derivative sign.

\index{Software}

\subsection{Software}

Theorem 1 in \cite{okamura2013optimal} and its proof leverage monotonicity
information only, embodied by first derivative signs of the involved
functions.

\index{Search Engines}

\subsection{Search Engines}

In \cite{chang2005caption}, when two adjacent edges with opposite
gradient signs (i.e., one has a positive and the other a negative
gradient) are found within a specific distance, they form an edge
pair.

\cite{liu2017discretely} introduces the ADM (Adaptive Discrete Minimization)
algorithm, where the optimal hash code is optimized leveraging the
sign of the function's gradient at each iteration.

The main impact of utilizing the mutual relationship among adjacent
neighbors in \cite{banerjee2018local} is that it does not rely on
the sign of the intensity difference between the central pixel and
one of its neighbors only. Instead, the approach considers the sign
of difference values between its adjacent neighbors and the central
pixels and the same set of neighbors.

The iterative update of $\Delta$ in Eq. 16 in \cite{chen2018deep}
is based on the sign of the loss function at the previous step, where
the $sgn(\cdot)$ operator is applied on the matrix element-wise.

B-step: The B-subproblem in \cite{ying2020locality} is a binary optimization
problem and adopts the signed gradient descent optimization algorithm,
as Eq. 8.

\index{Computer Graphics}

\subsection{Computer Graphics}

In a work on motion capture assisted animation, \cite{pullen2002motion} matches the first derivative of the chosen
band of each of the angles. Including the first derivatives in the
matching helps determine fragments of accurate data more closely matched
in value and dynamics to the keyframed data. The derivative sign change
of only one of the angles is used to determine where to break all
of the data corresponding to the matching angles into fragments.

Different approximation techniques are discussed in \cite{fryazinov2010extending}
for the affine form of several functions: Optimal (Chebyshev), Min-range,
and Interval approximation. While optimal and min-range approximation
requires the function to be bounded, twice differentiable, and with
the same sign of the second derivative on the given argument range,
the interval approximation requires only the function to be framed.
However, the range of the function for interval approximation is wider
than for other approximations.

In \cite{namane2018fast}, algorithm 1, \textquotedbl points of $S_{x}$
generation along $O_{z}$\textquotedbl , the variable $\Delta x$
equals $\pm1$, and it can be determined by just testing the sign
of the partial derivative of $x$ with respect to $z$ based on the
implicit function theorem. In turn, the variables $\beta$ and $\gamma$
are based on a condition that tests whether the sign of the discrete
$x$ equals $\Delta x$.

\index{Computational Geometry}

\subsection{Computational Geometry}

The gradient-sign parameter in Eq. 3 in \cite{horng2002vehicle} accounts
for the ability of a vehicle type to cope with positive or negative
directional gradients. Different vehicle types may react with positive
or negative gradient differently. A bicycle may have a higher cost
associated with a positive gradient than a negative gradient. A car
may have equal costs for both positive and negative gradients while
a truck may have a higher cost associated with a negative gradient
than a positive gradient because the heavy load of a truck may cause
danger in going down the slope.

\cite{abrahamsen2021chasing} classifies all configurations into three
types, according to the sign of the partial derivative of distance
with respect to the puppy's position. The critical configurations
are further classified based on the second and third derivatives'
sign.

\index{Computational Modeling}

\subsection{Computational Modeling}

In \cite{florinsky2000relationships}, the flow convergence and deceleration
result in the accumulation of substances at soils caused by slowing
down or termination of overland and intra-soil transport. On different
scales, the intensity of these processes and the spatial distribution
of accumulated substances can depend on the spatial distribution of
the following land-form elements. Thus, the natural classification
of landform elements is formed by the signs of $k_{h}$ and $k_{v}$.

In a work on series of abstractions for hybrid automata, \cite{tiwari2002series} shows that whereas qualitative reasoning usually
uses the sign of only the first derivative, the deduction is conducted
based on the signs of first $n^{th}$ derivatives.

A monotone and cooperative study can also be performed using graph
theory. At the graph monotonicity analysis in \cite{de2013computation},
the species graph assigns a node for each model compartment. No edge
is drawn from node $x_{i}$ to node $x_{j}$ if the partial derivative
$\partial f_{j}/\partial x_{i}(x)$ equals zero, meaning that node
$x_{i}$ has no direct effect on node $x_{j}$. An activation arrow
($\longrightarrow$) represents that the derivative is strictly positive,
while an inhibition line ($\dashv$) denotes that it is strictly negative.
However, if the derivative sign changes depending on the particular
entries, both an activation arrow and an inhibition line are drawn
from node $x_{i}$ to node $x_{j}$.

 In a work on modeling length of hydraulic jump on sloping rough bed using gene expression programming,\cite{pasandideh2020modeling} proposes a Partial Derivative Sensitivity Analysis method that tests the relationship between the objective function and its parameters based on the signs of its partial derivatives.

\index{Electrical Engineering}

\section{Electrical engineering}

\index{Electricity}

\subsection{Electricity}

In figure 6 of \cite{abdelaziz2015influence}, two asymmetric applied-voltage
waveforms (positive- and negative- ramp) are applied to the reactor
to better understand the discharge activity's dependence on the slope
applied voltage slope and its sign. These waveforms are characterized
by combining the fast and slow slew rate of the applied voltage in
one waveform. The applied voltage of the positive ramp have a significant
negative slope and a smaller positive slope. In contrast, the applied
voltage of the negative ramp have a fast transition when negative
and a slow change when positive.

In \cite{kustanovich2018synchronverter}, because of the $V_{string}$
oscillations, the power $P_{string}$, which is extracted from the
string, has a component that oscillates at the same frequency. So
the correlation of the $V_{string}$ and $P_{string}$ (after a low
pass filter) is proportional to the sign of the slope $\frac{dP}{dV}$.
The point of the maximum power is a point where the derivative $\frac{dP}{dV}$
changes sign.

At the iterative adaption algorithm for slice management in radio access network, \cite{khodapanah2019slice}, define the matrix $J$ based on the sign of the derivative
of KPI with respect to $x$.

The sign of the spoke speed, defined in \cite{ortega2019influence}
by the slope of the bands observed in the $LS\left(\theta,t\right)$
contours. It is opposite to the slope sign in the examples of Figure
4 (spoke rotation in the channel). The rotation in the ionization
chamber is counterclockwise, while the rotation in the channel is
clockwise (for a frontal view of the thruster).

The voltage in \cite{sergeyev2019autonomous} starts on the coil,
its polarity depending on the derivative sign of $\frac{d\text{\ensuremath{\phi}}}{dt}$,
that is, whether a wheelset is approaching or leaving a permanent
magnet. Consequently, the running of each wheelset of the moving train
above the permanent magnet causes voltage impulses on the coil.

\index{Electrical circuits' Fault Analysis}

\subsection{Electrical circuits' Fault Analysis}

Turning on the switch $S$ in \cite{shahbazi2012open} increases the
inductor current $i_{L}$. Consequently, the sign of the slope of
$i_{L}$ remains positive during this time interval ($DT$s). Fig.
4 presents the general scheme of the proposed fault diagnosis. In
subsystem FD1, the inductor current ($i_{L}$) passes through a derivation
block and then through a sign block which computes $sgn\left(\frac{di}{dt}\right)$.
If $i_{L}$ increases, $sgn\left(\frac{di}{dt}\right)=1$ and if $i_{L}$
decreases $sgn\left(\frac{di}{dt}\right)=-1$. The calculated error
signal equals $1$ when the estimated and measured current slopes
are different. If there is no switch failure, the two signals $sgn\left(\frac{di}{dt}\right)$
and $S'_{q}$ have the same values, then the signal ``error'' is
$0$, as described in equation 1.

The fault diagnosis algorithm in \cite{jamshidpour2015photovoltaic}
needs only the sign and not the exact value of the inductor current
slope.

The behavior of $i_{L}$ as a fault indicating signature in \cite{bento2018comprehensive}
happens to be the same for all non-isolated single-ended DC-DC converters.
Switch OCF and SCF can be detected for all non-isolated single-ended
DC-DC converters using signed inductor current derivative for CCM.

The reliability of the proposed scheme in \cite{elgeziry2019non}
is enhanced remarkably owing to the non-communication operation. Moreover,
its security against mal-operation in transient cases is guaranteed
via two security conditions depending on the derivative sign of both
current and voltage signals.

In \cite{kasis2021primary}, a switch from off to on occurs, which
in turn causes the frequency to decrease - and vice versa. This change
in the derivative sign will cause an infinite number of switches within
some finite time, resulting in the chattering behavior.

\index{Control Systems}

\subsection{Control Systems}

The control law in \cite{huang2000control}, implemented to achieve
the desired energy, is a function of the sign of the angular velocity.
In \cite{valishevsky2002adaptive}, only the sign and not the value
of the derivative is used. Therefore, it's possible to derive formulae
to determine the derivatives' signs, which require less computational
power than those for calculating the value of the derivatives.

The system convergence in \cite{loizos2008adaptive} implies that
a correct phase for delay compensation has been selected in that the
gradient's estimated sign is correct. Having six available phases
to choose from means that at least one and at most two of the neighboring
phases will also give an accurate estimate, maintain convergence,
and exhibit limit cycles.

The parameters of the plant's transfer function with the asymmetric
dynamics in \cite{zlosnikas2008integral} change when the sign of
the output parameter's time derivative changes. Therefore, the classical
controllers with the constant parameters do not allow us to achieve
the excellent transient performance of the control of the mechatronic
system in such a case.

\cite{jarlebring2010invariance} demonstrates that the local behavior
of the root path $s\left(\tau\right)$ around any associated critical
delay $\tau\in T$ can be entirely characterized by the sign of the
imaginary ratio between two derivatives. Further, for simple imaginary
roots, the root tendency is invariant in the sense that the root tendency
for some delay determines the root everywhere.

The Sign of the first derivative of the active power signal of the
magnetic separator motor in \cite{alekseyev2014automated} determines
the direction of change in ore mill charge.

The calculation of $C_{0}$ in \cite{lumbreras2018stability} depends
on the derivative sign of $\varphi$, as stated in formula (13).

The velocity's sign is used throughout \cite{kikuuwe2018some} in
different contexts, e.g., in setting the domain of the parameter $\rho$
in Eq. 3.16b; in determining the value of the function G in Eq. 3.21
and its derivative in Eq. 3.38, 3.39; and in constructing the Lyapunov
function candidate in Eq. 3.40.

The equations of the system's motion in \cite{romacevych2019closed}
are linearized differential equations explicitly incorporating the
derivative sign.

\cite{hendrickx2019trajectory} considers trajectories where the derivative
sign is opposite to that of the corresponding entry in the gradient
of an energy function.

\index{Electronics}

\subsection{Electronics}

\index{Thermoelectricity}

\subsubsection{Thermoelectricity}

\cite{lee2013heat} shows that one can control heat dissipation by
changing the anchor groups from isocyano to amino. The authors further
prove that the slope sign of the transmission curve of a molecular
junction governs in which electrode the majority of the heat is dissipated.
Finally, they show that the sign could be changed from positive for
amino to negative for isocyano.

In \cite{mateos2021thermoelectric}, to get cooling, it is necessary
to apply a voltage bias with a positive or negative sign, depending
on the properties of $T\left(\epsilon\right)$, which define the behavior
of $L_{ij}$. Crucially, the sign of the off-diagonal coefficients,
$L_{12}=L_{21}$, depends on the sign of the slope of $T\left(\mu_{c}\right)$.
Thus, $\Delta\mu>0$ ($\Delta\mu<0$) for the positive (negative)
slope of $T\left(\varepsilon\right)$.

The ambiguous characteristics $IC\left(VCE\right)$ are obtained in
\cite{gorecki2021influence} at both types of cooling conditions.
The point of the electrothermal breakdown, in which these characteristics'
slope sign changes from positive to negative, is visible on each characteristic.
Such a shape of the considered characteristics also means that the
breakdown in the investigated transistor can appear at a value of
voltage $VCE$, which is considerably lower than the admissible catalog
value.

\index{Electrical Circuits}

\subsubsection{Electrical Circuits}

As illustrated in \cite{davydov1998pressure}, the derivative of the
high-frequency dielectric constant with respect to the pressure changes
sign as the compounds become more ionic. This is connected with the
sign change of the parameter $h$, which goes from positive to negative.

To simplify implementation and make (8) in \cite{roo2000cmos} causal,
the authors make a substitution for the slope term and add a delay
to obtain the equation used for the phase detector as in equation
(9) - which uses a delayed error and the slope sign calculated from
the output of the slicer.

\cite{holleman2008micro} applies a digital counter in conjunction
with the differentiators from the NEO to measure the width of the
spike. The first differentiator has an auxiliary sign output. A change
in the sign of the first derivative indicates a minimum or maximum
in the input signal. After a spike is detected, the following change
in the derivative sign starts the counter. The second change in the
sign output causes the counter value to be registered for readout
and count back down to zero. The additional delay allows time for
the extreme spike values to occur and be sampled by the peak detectors.
When the counter returns to $0$, the Ready signal is asserted to
initiate conversions of the maximum and minimum voltages. 

In the feature
extraction section of \cite{holleman2011ultra}, the peak detector
is based on the signal's derivative sign. A digital counter is used
with the differentiators from the NEO to measure the width of the
spike. The first differentiator has an auxiliary sign output. A change
in the sign of the first derivative indicates a minimum or maximum
in the input signal. After a spike is detected, the following change
in the derivative sign starts the counter. The second change in the
sign output causes the counter value to be registered for readout
and the counter to count back down to zero. The additional delay allows
time for the extreme values of the spike to occur and be sampled by
the peak detectors. When the counter returns to $0$, the Ready signal
is asserted to initiate conversions of the maximum and minimum voltages.
The counter is also intended to measure the spike width.

Based on equation (2.2) in \cite{long2015non}, the sign of the slope
is determined by the sign of the capacitor. For a negative capacitor,
the slope is always negative. In contrast, the reactance of the positive
capacitor monotonically increases with frequency. According to Foster's
theorem, it is clear that a negative capacitor is categorized as non-Foster
impedance, judging by the slope of the reactance curve. This unique
non-Foster characteristic is utilized for broadband application.

\cite{aridhi2015fast}, working at fast statistical analysis of nonlinear
analog circuits using model order reduction, proposes a method where currents are grouped, and voltages are
divided into groups based on their range and the derivative sign of
$x_{i}$.

Without the fast excitability elements, the fast $I-V$ curve in \cite{ribar2019neuromodulation}
is monotonic, the slow $I-V$ curve is ``N-shaped'', and the ultra-slow
$I-V$ curve is monotonic, so the system is slow excitable. The voltage
regions are now indicated with two signs so that the first sign corresponds
to the sign of the slope of the fast $I-V$ curve, and the second
sign corresponds to the sign of the slope of the slow $I-V$ curve.

The transition into MEP-lock initially involves a rapid MEP search
using sign-gradient descent, followed by fine-grained MEP tracking
across PVT and load variation (Fig. 19.1.3 in \cite{ur201919}). In
each search step, the system is briefly operated at the search voltage
to determine the Vdd-dependent variables, Cfly and Nclk, needed for
tEPC comparison.

\cite{rakitin2021functional} integrates the input current signal.
Its effect on the waveforms depends on the resistance change sign.
The long current pulses slow down or accelerate the transient process,
but short current pulses do not impact the resistance value. The phase
plane with axes $R_{1}$ and $R_{2}$ (Figure 6) can be exploited
to analyze different behavior versions of such a system. The analysis
is based on model Eq. (8). In this case, the trajectories of moving
the image points are straight lines, which pass at angles of $\pm\frac{\pi}{4}$
on phase plane. Four trajectories can pass through each point of phase
plane. The sign of $\frac{dR}{dt}$ defines one from them. The threshold
resistances specify the boundaries of the area of trajectories movement.
When the trajectory reaches the boundary, the sign of the derivative
$\frac{dR}{dt}$ changes and the trajectory is mirrored from the boundary.
The edges can shift themselves at this time point.

\index{Electrical Currents}

\paragraph{Electrical Currents}

The characteristic of a current flowing in a ring can be obtained
by the shift in Eq. 2 of \cite{dajka2004persistent}. The slope of
the current characteristic (2) (i.e., the sign of the derivative with
respect to $\Phi$) allows one to distinguish the parity. A current
that has a positive slope at $\Phi=0$ is called a paramagnetic current,
whereas a current with a negative slope at $\Phi=0$ is called a diamagnetic
current.

The observed photocurrents in \cite{priyadarshi2013all} vanish for
resonant excitation of excitons and reverse their direction with a
change of the sign of detuning. For non-resonant excitation, the phase
differences show a gradual, close-to-linear variation, with the slope
sign depending on the sign of detuning.

\index{Imaging}

\subsubsection{Imaging}

Figure 3 in \cite{geday2000images} illustrates how the calculation
of the order of $\delta$ is performed. The sign of the calculated
slope of $\left|\sin\delta_{0}\left(k\right)\right|$ determines the
interval to which $\delta_{1}$ belongs. In the case of a positive
slope, any $\delta_{1}$ is within a specific area and will be assigned
to the region $\left[m\pi;m\pi+\frac{\pi}{2}\right]$. Similarly,
if the slope is negative and $\delta_{1}$ is calculated to belong
to $B$ then it will be assigned to the region $\left[m\pi+\frac{\pi}{2};\left(m+1\right)\pi\right]$.

As discussed in \cite{yang2005approach}, the quality of the linear
approximation in a square depends on whether the second derivative
sign remains constant.

The possibilities illustrated in \cite{denny2007computing} are evaluated
using a heuristic approach for choosing the best estimates of the
principal values based on a measure of edge strength and the sign
of the third derivative of the broadened experimental spectrum.

At the suggested phase derivative (PD) method in \cite{guo2014phase},
the phase distribution of the tested object wave is firstly worked
out by a simple analytical formula; then, it is corrected to its proper
range according to the sign characteristics of its first-order derivative.

Phase-contrast MRI modulus images were difficult to segment in \cite{kachenoura2015right}
because of the flow-related contrast variations along time. Therefore,
process velocity images were preferred, presenting connected velocity
signs in tricuspid inflow regions.

The sign of the slope $\frac{dD}{dt}$ is used in \cite{wang2016scattering}
to determine to which branch the phase $\Phi\left(t_{i}\right)$ belongs.

Assuming a linearly chirped pulse with a chirp coefficient $\beta$,
the signal in \cite{baudisch2018time} is expected to change sign
when $\beta$ changes sign (equation 2.19). The more general Kovalenko
model also predicts this effect. A linearly chirped probe's instantaneous
frequency vs. time can be represented as a straight line with a slope
$\beta$. The coating strongly modulates the probe's GD, the sign
of the slope $\beta$ may change the sign for a finite wavelength
interval. If the slope was increased, the same modulation would no
longer be enough to reverse the sign of the slope at some point. It
follows that increasing the white light chirp should remove the artifact
flipping. The Lorenc model explicitly covers only moderate modulations,
where $\beta$ does not change sign.

In \cite{venegas2018development}, the step heating and cooling are
nearly identical except the derivative sign. Put differently, according
to figure 4e, the step cooling stimulation shows a very similar tendency
but with an opposite sign of the slope compared to the heating case.
The signal for pulsed stimulation has a behavior similar to the case
of cooling stimulation, where the signal experiences a rapid pulse
increase and then descends very quickly until reaching an asymptotic
value at zero.

In the context of the DOI discrimination with a double-threshold approach
in \cite{polyachenko2020lynden}, a time-amplitude correction was
performed to retrieve a monotonic evolution of $\Delta t\left(2-1\right)$.
A linear function was fitted through the scatter plot of $\Delta t\left(2-1\right)$
against the integrated charge of the first and last DOIs, giving a
negative slope. The resulting fit was then mirrored with a higher
slope (changing the sign of the slope and multiplying by a scaling
factor) for better separation of the different DOI regions by exploiting
the signal intensity information. Correcting each event with the parameters
of the correction curve yielded the distributions leading to a monotonic
increase of $\Delta t\left(2-1\right)$. The choice of the slope extension
was done by sweeping the scaling factor until there was minimal overlap
between the distributions. A higher slope gave no significant modification
of DOI resolution. Indeed, as soon as a monotonic behavior is achieved,
separating the regions is not helpful as intra-region separation of
the events co-occurs, so the overlap between the adjacent regions
remains of the same order. Even for the first shallowest DOIs, extending
the slope is less valuable since there is already a small monotonic
behavior.

Lemma 3.1 of \cite{bonettini2021variable} proves monotonicity and
concavity properties of the function h defined in equation 21. For
that, the authors leverage the sign of its partial derivative.

\index{Telecommunication}

\subsection{Telecommunication}

\cite{musa2003clock} selects the sampling phase based on the signs
of the error and slope, something which they illustrate in table 1..
A positive error/slope is denoted by $1$ and a negative error/slope
by $0$. The slope sign can be obtained by comparing the input's current
value with a sample of $u$ delayed by one symbol period.

Practical high-speed implementations of the LMS algorithm often use
only $1$-bit representations of the sign of the error and the slope.
\cite{musa2007modeling} applies this idea to the MMSE TR results
in the sign--sign MMSE (SSMMSE) rule, as detailed in equation 30.

The digital input to the multiplier in \cite{edwards200912} is the
sign of the slope. The difference between the following sample and
the previous sample is computed and quantified with a signed comparator.
The product of the amplitude error and the slope sign form the phase
error for one phase of the PRS. The outputs of both interleaves are
summed together in the transconductor to produce a continuous-time
differential phase error.

\cite{dong2018energy} applies the derivative sign of the energy efficiency
$\Lambda$ in step 6 of algorithm 1 and step 5 of algorithm 2.

\index{Electrical Energy}

\subsection{Electrical Energy}

\cite{pesty2005low} introduces a prediction for a sinusoidal excitation.
The phase shift of the oscillating intensity only depends on the intensity's
slope sign: $I$ oscillates in-phase with $dV$ when the slope is
positive, but out-of-phase (with a $\pi$ phase shift) when it is
negative. For a given diffraction pattern, the phase shift of each
spot is found to be correlated with the sign of its corresponding
measure $\frac{dI}{dE}$ derivative. The middle of Fig. 6 displays
the corresponding evolution in the $\left(1,0\right)$ spot intensity
energy. These values are extracted from the 'mean' images for the
same series of energies. The intensity displays peaks, maxima, and
minima in a wide range of energy. The $I$ vs. $E$ curve allows several
energy intervals where the $\frac{dI}{dE}$ derivative is positive
or negative. This series is not intended to provide values for the
derivative, purpose being only to study the correlation between the
slope sign and the value of the phase shift. An excellent correlation
is observed: the response is thoroughly found to be in-phase as the
slope is positive and out-of-phase (with a jump of $\pi$) as the
slope is negative.

In \cite{gudina2019effect}, there is a clear correlation between
theoretical $A$'s sign with the experimentally observed derivative
sign, $\frac{d\mu}{dT}$, in the actual temperature interval. It is
per the estimation of $N_{cross}$ for InGaAs: $\frac{d\mu}{dT}>0$
for $S$-subbands with large $n_{1}$, but $\frac{d\mu}{dT}<0$ for
$AS$- subbands with small $n_{2}$. The $\frac{d\mu}{dT}$ sign correlation
with the value of $n$ (and the theoretical $A$'s sign) corroborates
the EEI nature for the temperature dependence.

The estimated slope in \cite{figueroa2020leveraging} is more susceptible
to noise. The actual and the estimated state can be in different segments
of the SOC, which could cause divergence. The gain is set to zero
when the estimated slope has a distinct sign from the modeled slope.
Therefore, the system will run an open-loop for both voltage and force
when a slope mismatch occurs, which is done to avoid instability issues.

\index{Wind Power}

\subsubsection{Wind Power}

In \cite{nagy2013control}, it is shown that negative $V_{dr}$ can
change the positive sign of the slope into a negative one producing
an unstable region around slip $s=0$.

As discussed in \cite{stumpf2013study}, the slope $\frac{dT_{e}}{ds}$
at $s=0$ depends on the sign of $V_{dr}$, and it is always positive
when $V_{dr}>0$. The sign of the slope is crucial as it determines
the stability in the lack of feedback control. The slope can be negative
by negative $V_{dr}$, producing an unstable region around slip $s=0$.

\cite{martynowicz2021nonlinear} conducts a Hamiltonian derivative
sign analysis and presents an implementation of a nonlinear optimal-based wind turbine tower vibration control method. The obtained results prove the effectiveness and validity of the proposed approaches.

\cite{soulier2021low} applies detection methods used to validate
the sensor. They preferably use instantaneous criteria: an instantaneous
evaluation of the sign of the tangential velocity an immediate assessment
of the profile wake width.

\index{Signal Processing}

\subsection{Signal Processing}

In a work on temporal separation of the density fluctuation signal measured by light
scattering, \cite{antar1999temporal} shows that the phase derivative sign reflects the
groups with the highest intensity occurring the most frequently in
the analyzing volume.

A gradient sign algorithm for transmit antenna array adaptation has
been defined in \cite{banister2003simple}, and the algorithm's convergence
and tracking have been analyzed. The algorithm uses gradient sign
feedback from the receiver to generate a coarse gradient estimate
used by the transmitter to adjust the transmit weights recursively.

The $10Gb/s$ eye diagrams obtained with a standard single-drive-cut
modulator after modulation and $50km$ of propagation through a dispersive
fiber exhibit significant differences in the distortion of the signal
if the sign of the slope of the transmission response is not properly
chosen in \cite{courjal2004modeling}, whereas a modulator having
a domain inverted section can generate the same signal for both slopes.

The 'vector' in a work on characterization framework for epileptic signals (\cite{vazquez2012characterisation}) is split up into
$N$ intervals. Each interval ends in a change of sign of the slope
of $y_{k}$. In other words, the trends are constant in each interval.

The polarity in a work on automatic speech polarity detection via phase information from complex analytic signal representations (\cite{govind2014automatic}) is assigned to the frame
based on the analytic signal's slope sign.

The convenient indicator of the transfer function drop in the band
in \cite{vreznivcek2016amti} is the first derivative sign. If this
sign changes more than twice in the internal passband, it is evident
that the characteristics are bent.

The discussion in \cite{sapinski2017laboratory} demonstrates the
potentials of MRD action as a velocity-sign sensor and presents critical
issues which need to be addressed to enable its real-life applications.

Time instants in which the velocity sign is changed in \cite{rosol2019ability}
are determined with sufficient accuracy allowing the implementation
of switching algorithms. Moreover, $u_{sky}$ and $u_{grd}$ are determined
based on the derivatives signs (velocities) $sgn\left(x_{s}'\right)$
and $sgn\left(x_{e}'\right)$.

In a work on the applications of the three-point formulas, \cite{yin2020background} shows that the derivative sign might be opposite
to that of the actual value if three-point formulas are used. Therefore
the calibration direction changes with the normalized input frequency.

\index{Systems Engineering}

\section{Systems Engineering}

\index{Control Charts}

\subsection{Control Charts}

The MPPT algorithm in \cite{leyva2006mppt} measures the sign of $\frac{dy}{dt}$,
whereas the resulting dynamics are governed by $\frac{dy}{dt}$. Eq.
6 summarizes different cases discussed qualitatively.

The incremental conductance (IncCond) method in \cite{esram2007comparison}
is based on the fact that the slope of the PV array power curve is
zero at the MPP, positive on the left of the MPP, and negative on
the right, as given by the sign of the derivative $\frac{dP}{dV}$.

From the $P-V$ characteristics shown in Fig. 3 in \cite{killi2015modified},
it can be visualized that the slope is positive at the left of MPP
and negative at the right of MPP. Depending on the slope sign, the
duty cycle has to be perturbed to track the peak power, and the flowchart
of this conventional P\&O MPPT algorithm is shown in Fig. 4. The duty
cycle and the $PV$ voltage are inversely proportional to each other,
i.e., an increase in duty cycle causes the $V_{PV}$ to decrease and
vice versa. In the drift-free modified P\&O MPPT algorithm, $V$,
$I$, and $P$'s (discrete) derivative signs are all used.

\index{Energy Management}

\subsection{Energy Management}

The fringes of normal directions in \cite{wang2009method} are determined
by the directions in which the change of the gray distribution is
most significant. The positions of the light fringes' center lines
can be obtained in normal directions by using the 2D derivative-sign
binary map. In the process of implementing the method, the thinning
for the broad binary fringes is concerned. In some circumstances,
the thinning results are the geometric centers of the wide binary
fringes rather than the physical center lines. Hence, some errors
are brought. In the triangulation measurement system by a Bessel beam,
the most crucial characteristic of the ring-structured light fringe
pattern is that the derivative signs in the normal directions on both
sides of the fringes' center lines are opposite. In contrast, the
signs between contiguous black and white center lines are identical.
Based on the characteristics above, the extraction method of the ring-structured
light fringes' center lines based on the 2D derivative-sign binary
map is proposed. Unlike the traditional extraction method, the process
does not depend on the particular threshold and has performances of
solid applicability, high accuracy, and high automation degree.

The classification of the periodic orbits in \cite{moussi2015nonlinear}
is performed based on the curve's slope sign in the configuration
space $\left(x,u\right)$. When the amplitude of the second mass ($u$
component) increases, the amplitude of the first mass is limited by
the elastic stop. This behavior implies a change of the modal line's
slope sign (on the neighborhood of the origin) and the apparition
of a new oscillation.

Inspired by discontinuous control protocols, to get finite-time synchronization,
\cite{dong2015finite} proposes a modification to the signed gradient
method from the Kuramoto model, as in Eq. 1.5. and 2.7. In this paper,
they study the signed gradient type Kuramoto model with identical
oscillators in section 4.1.

The main problem of a preliminary digital signal processing is the
calculation of minima $\left(m_{1},m_{2}\right)$ and maxima $\left(M_{1},M_{2}\right)$
of rising and falling edges to calculate the real boundary of shadow
further. To solve this problem, \cite{chursin2015methods} applies
an algorithm shown in Fig. 10 designed to implement on a field programmable
logicdevice (FPLD). The derivative sign change detector 1 receives
serial data on voltage in CCD cells and clock pulses for cell counts.
When the derivative signs changes, the detector transmits a control
signal to FIFO buffers 4 and 5. FIFO buffers receive a cell number
and a control signal from the derivative sign change detector 1, and
then output the latter four cell numbers received.

For a global analysis using homogeneity, the Lyapunov function derivative
sign can be checked not in the whole state space but on the sphere
with the unit radius only (defined by the homogeneous norm), as illustrated
in \cite{efimov2016conditions}.

Based on the sign of the gradient estimation, a variable structure
controller in \cite{attallah2017histogram} generates the control
input for the nonlinear plant. Zero-mean white noise in performance
measurement is considered in the problem. The sliding mode observer
has dramatically improved the accuracy of gradient estimation by limiting
the rate of change of the estimate. Moreover, the variable structure
controller depends only on the sign of the forecast, not the magnitude,
adding valueto the robustness of the overall system.

It is clear that (2.1) in \cite{pandey2017chiellini} cannot describe
a system with a quadratic damping as the term involving $\left(x'\right)^{2}$
does not change sign and oppose the motion when the velocity reverses
its sign. The authors split (2.1) into two parts to remedy this feature
depending on the velocity sign. In turn, the Hamiltonians in 3.3 are
defined separately for positive and negative derivatives.

The signs of the surge, sway and angular velocities in \cite{karami2019adaptive}
are used in the differential equation (1). In turn, they are also
applied in equations (14), (15), and (20).

To avoid oscillations in \cite{nikolaidis2020enhanced}, the authors
distinguished the values of $\alpha$ according to the Lagrangian
function's derivative sign.

\index{System Dynamics}

\subsection{System Dynamics}

The feature vector in a work on the evaluation of alternative dynamic behavior representations for automated model output classification and clustering \cite{onsel2013evaluation} represents the original
behavior as a sequence of atomic behavior modes based on derivative
signs.

In a work on the stability domains of
the delay and PID coefficients for general time-delay systems, \cite{almodaresi2016stability} shows that the sign of the derivative of the generator function is shown to change alternately at the singular frequencies.

In the frequency response rating program used in \cite{shokhin2017use},
the signal amplitude is defined for positive values at changing signal
derivative sign from positive to negative.

\index{Mechanical Engineering}

\section{Mechanical engineering}

\index{Mechanical Friction}

\subsection{Mechanical Friction}

\cite{morel1996precise} proposes a model-free fine position control method using the base-sensor with application to a hydraulic manipulator. The base sensor estimated torque reproduces the input voltage sine wave with a disturbing torque whose sign is
changed when the velocity sign changes.

When using a broad purpose program, as in \cite{dimova2000numerical},
the output of the velocity sign history is required after each collision.
The size of the new time step then depends on the velocity sign. \cite{ryu2001nonlinear} shows that the velocity sign in  is an integral part of the formula and the decision flow.

If only static friction is considered, the modified LuGre model in
\cite{chen2014dual} is reduced to Eq. 8, indicating that, by fixing
the nominal micro stiffness, $\beta_{0}$, the adaptation of actual
micro stiffness, $\sigma_{0}$, effectively changes the level of static
parameters; for example, $F_{C}$ and $F_{S}$ in $g\left(v\right)$.
This approach takes into account the velocity sign and friction identification
is conducted to set initial values in the adaptive friction observer.
The model in Eq. 1-6 is reduced to Eq. 41., which incorporates the
sign of the derivative of the angular position.

When the system is in the slip state, based on Coulomb's law of friction,
the friction force in \cite{yuan2016new} can be expressed in Eq.
65, thus incorporating the sign of the tangential velocity during
the sliding.

In the stochastic analysis in \cite{lima2016stochastic}, the belt
velocity is modeled as a random process constant by parts. The number
of changes of the belt velocity sign is given by a random variable
with the Poisson distribution.

The static Stribeck friction model in \cite{ruderman2017break} and
other physical sizes explicitly leverages the velocity sign.

The Dahl friction in \cite{choi2017tension} is presented as a first-order
non-linear ordinary differential equation. The general form of the
Dahl friction model is given by Eq. 10 and 11, which incorporate the
derivative's sign. In turn, the modified form of the kinematics equation
that includes the Dahl model is given as in Eq. 12-15, also incorporating
the velocity sign.

A Coulomb friction model is used in \cite{ismail2017passive} to simulate
friction behavior in this paper can be formulated as in Eq. 1, incorporating
the velocity's sign. In turn, the sign of the difference velocity
$x'-u'$ is applied in Eq. 3 and 6. Other derivatives' signs are applied
in Eq. 5 and 7.

Kinetic friction formulas, such as Eq. 2 and 3 in \cite{chen2018global},
incorporate the derivative sign. They are in turn incorporated into
the global structures of the system (1) with the kinetic friction
force $F\left(v_{r}\right)$ having the form (3), stated in Eq. 4.

The dry friction forces between the tool, the workpiece, and the chip
in the three directions in \cite{wang2018bifurcation} are expressed
in Eq. 3 there, incorporating the signs of the velocities in the $x$,
$y$, and $z$ directions. These are applied in the governing equations
of the cutting tool vibration in the three directions (Eq. 6) and
the instantaneous thickness of the cut in Eq. 7.

The dimensionless state equations of the PD-controlled motion stage
without FI in \cite{dong2019friction} are given by Eq. 14-16. The
latter (representing the slipping equilibrium, $z=sgn\left(v\right)g\left(v\right)$)
incorporates also the sign of the derivative of $X_{p}$. It is also
evidenced in Eq. 24-25.

\index{Robotics}

\subsection{Robotics}

In \cite{tzafestas1999simple}, the control law dictates to maintain
the control action if and only if the sign of e does not agree with
the sign of its derivative. The control action can be either an increase
or a decrease of the control signal. The increase or decrease of the
control signal is realized via the use of fuzzy linguistic rules.

In \cite{pham2001identification}, the friction torques and other
physical sizes are explicitly modelized based on the sign of the joint
velocity.

Models based on Coulomb model in \cite{gervini2003new} indicate that
the friction is a function of the velocity sign. The tentative to
compensate the friction based on these models can generate limit cycles
around reference position or high-frequency vibration due to commutation
of the rotor velocity sign for velocities near zero (shattering).

New methods, ignoring the produced energy from the velocity sign change,
and holding the control force while the velocity is zero, are proposed
in \cite{ryu2005simulation} for removing the noisy behavior.

In \cite{walsh2006autonomous}, knee-on occurred at heel strike, and
the damper was programmed to exert a torque proportional to the rotational
velocity of the knee joint. Depending on the velocity sign, two different
gains were used to control knee rotation for knee flexion and extension.

A conventional friction model is utilized in \cite{chen2014experimental}
for the robotic hand DLR-HIT II joint. The friction model is expressed
as in Eq. 29 while leveraging the sign of the angle's derivative,
$\theta'$. In turn, the joint dynamics and its linearization also
leverage the derivative sign (Eq. 31), which then merges with the
derivative sign to form the derivative's magnitude.

The sign of the error derivative is applied in the proposed nonlinear
DED in \cite{jin2017robust} as part of the nonlinear desired error
dynamics. It is later used throughout the paper's formulas.

The dynamics of the robot in \cite{xie2018power} are modeled in Eq.
4-6. The matrix $C\left(\sigma\right)$ is comprised of the rotational
directions signs that are defined as positive according to the left-hand
rule along the $x_{r}$-axis. Further, the sliding and rolling frictional
forces acting on the wheel, respectively, which can be found by Eq.
8-9, are defined based on the sign of the vector projection of the
wheel velocity relative to the ground $V_{i}$ onto the unit vector
along the roller parallel direction.

When the external force $\tau_{e}$ in \cite{chen2018universal} is
equal to or greater than the maximum static friction force, the static
friction force will equal to the maximum static friction force, the
direction is different from the external force. Then, the friction
can be described as in Eq. 16, incorporating the sign of the motion
velocity. In turn, its sign is also applied in Eq. 17.

The Slope Sign Changes (SSC) feature, calculated in Eq. 4 in \cite{hassan2020teleoperated},
detects the changes in the slope sign of the $s_{EMG}$ signal and
counts them. It is represented based on the signs of the discrete
one-sided derivatives.

\index{Machinery}

\subsection{Machinery}

In a work on a model-free fine position
control method using the base-sensor with application to a hydraulic manipulator \cite{iagnemma1997model}, compensation for Coulomb friction at velocity sign changes
is accomplished much more rapidly than conventional methods.

In \cite{korzeniowski2008investigation}, working at the transient states of the hydraulic power unit cooperating with the servovalve, shows that when the velocity sign
is changing, rapid pressure increase occurs, caused by receiver inlet
flow decrease, reached as the effect of servo valve SV control.

In order to suppress the consideration of variants during optimization
in \cite{tsalicoglou2010design}, two additional criteria for the
curve shape are formulated. First, the radius must be monotonically
decreased from inlet to outlet. Second, the first derivative of the
radius must be monotonically increasing from negative values towards
zero. An example of a curve that does not satisfy these criteria is
shown in Fig. 8 there. The monotony criteria are checked by counting
the number of sign changes. The expected number of sign changes for
an acceptable curve shape is zero for both derivatives. The component
objective function applied on this criterion is a binary step, with
which curves with sign changes of the derivatives are penalized with
the value $1$.

As indicated in \cite{juuso2011intelligent}, typical reasoning systems
have three components: a language to represent the trends, a technique
to identify the trends, and mapping from trends to operational conditions.
The fundamental elements are modeled geometrically as triangles to
describe local temporal patterns in data (Figure 1). The parts are
defined by the signs of the first and second derivatives, respectively.
These elements, also known as triangular episodic representations,
have their origin in qualitative reasoning and simulation.

Although the bode plot is helpful in the design of suitable controller
gains for the proposed torque regulator, a further discussion on the
constraints of controller gains is held in \cite{zhang2015constant}
from the viewpoint of stability. The proposed torque regulator is
dependent on a triangular-wave carrier which is compared with the
controller output. In reality, the function of this carrier is to
periodically change the sign of torque slopes based on the result
of the comparison. The hysteresis logic in Fig. 3 illustrates it.
From Fig. 3, it is understandable that the absolute slope of $T_{c}$
should be smaller than the absolute slope of the triangular wave carrier.
Otherwise, the slope sign of $T_{c}$ can never be changed, and the
torque signal will eventually be out of control due to the unidirectional
increasing or decreasing force.

The simulations performed in \cite{lima2017construction} consider
just one value to the friction coefficient, $\mu$ and to the parameter
$\lambda$, which represents the expected value of the number of sign
changes of the base velocity per unit of time.

The jet angular velocity sign change phenomenon is studied in \cite{rassokha2017numerical}.
There are two groups of regions in the shell -- rotating counterclockwise
and clockwise, which is due to the presence of rifling on the shell's
outer surface.

The approximation function for non-elastic resistance in \cite{lazutkin2017non}
should have a part with the velocity sign.

The model introduced in \cite{fujii2018bouc} can be obtained by replacing
the symbol $A$ with the expression $A+A_{0}sgn\left(u\left[k\right]-u\left[k-1\right]\right)$,
meant to introduce the velocity sign sensitivity to the behavior of
the model which would lead to the asymmetric off-center hysteresis
loop formation.

Considering the friction force and the moment of inertia in \cite{zhao2019design},
the dynamic equation of the new accumulator is represented by Equation
(14), where the sign of the piston velocity in the fluid chamber is
taken into account. In turn, the fluid pressure also depends on this
velocity sign. Further, the dynamic equation of the inertial load
can be written as Equation (17), which incorporates the sign of the
derivative of the intertial load.

Both $P_{ch}$ and $F_{ch}$ in \cite{sahu2020static} can be calculated
based on the derivative sign of $\frac{dR}{dP}$, according to equations
(42) and (43).

\index{Vehicles Engineering}

\subsubsection{Vehicles Engineering}

In \cite{bang2003large}, for stability analysis by the controller
in Eq. (11) subject to nonlinearity in by Eq. (15), three different
cases are considered where the sign of the derivative $V'$ is studied.
For example, under the inequality condition by Eq. (18), $V'$ could
have either a positive or negative sign. Let us assume a case where
the initial \textbackslash omega is negative, then Eq. (21) is valid,
and Lyapunov stability holds. However, as the positive maximum control
input continues to act on, the angular velocity tends to increase
in the negative direction further with decreasing quaternion. If such
a situation continues, then the constraint equation (Eq. (18)) may
no longer be valid. Thus switching in the Lyapunov function derivative
sign is expected. Similarly, if the initial angular velocity is positive,
Eq. (20) may also produce a $V'>0$ result. Since the angular velocity
decreases, switching in the Lyapunov function is also expected when
$\omega$ crosses the zero line.

The sign of the azimuthal velocity in \cite{bourgeois2010unexpected}
is chosen with respect to the direction of rotation of ions due to
the Lorentz force.

The equations of the force terms for the Rudder model are given in
Eq. 2-10 in \cite{apri2011analysis}. Eq. 3 and 10 depend on the signs
of the surge and sway velocities $u$ and $v$, respectively. In turn,
the sign of the sway velocity is modeled per the sign of the longitudinal
force's derivative.

In \cite{cabecinhas2017hovercraft}, the drag coefficient multiplying
the velocities and the dry friction coefficient are unknown and must
be estimated. It leads to Eq. 9 and 10 for the force and torque. Both
leverage the signs of the different velocities in the systems. These
signs are used in the differential equations that enable finding the
speeds themselves (Eq. 11) and in the auxiliary matrix G, Eq. 14,
the second time derivative of the tracking error, Eq. 20, 22, 23,
and 25.

The vehicle Longitudinal Dynamic Model in \cite{fukuda2017basic}
models its velocity's derivative as its sign function (Eq. 2). The
velocity sign is also applied in calculating $N_{f}$ and $N_{r}$.

The derivative sign of the torque with respect to the percent of biodiesel
in \cite{bietresato2019use} dictates whether it's always or never
in the validity domain.

\index{Finite Elements}

\subsubsection{Finite Elements}

In \cite{kaya2015system}, for both sensor configurations, the linear
and spline interpolation methods fail to give a good estimation for
the amplitudes of the third and the fourth mode shapes, while the
MSBE method provides reasonably good assessment for both of the mode
shapes. This occurs mainly due to the change of the sign of the slope
of the mode shape and the height of the structure. The mode shapes
of a multi-story building can usually be divided into several linear
segments between points where the sign of slope changes. Therefore,
unless one middle sensor is placed at each of these locations, the
linear and cubic spline interpolation methods will always fail to
give a good estimation for the amplitude of the higher modes. To get
a good estimation from both interpolation methods, a sensor has to
be placed at each floor level where the mode shape's slope sign changes.

\index{Thermal Engineering}

\subsubsection{Thermal Engineering}

Since a Dirichlet condition in \cite{chalhub2013integral} is imposed
at the left boundary for the present test case, using $\Psi\left(\xi\right)$
values at negative $\xi$ does not change the sign of the numerical
derivative for a small enough $\delta$. If a Neumann condition was
imposed instead at the left boundary, the numerical derivative sign
would change for any value of $\delta$.

In a work on critical and optimal thicknesses of thermal insulation in radiative-convective heat transfer, \cite{zarubin2016critical} shows that the relation between physical
sizes, such as the heat flux and the thermal conductivity, is analyzed
based on their derivatives' signs.

In a work on heat transfer of power-law fluids in plane couette–poiseuille flows with viscous dissipation \cite{coelho2020heat}, the Couette-Poiseuille flow, even for Newtonian Fluids, Recrit, will depend upon the sign of the imposed pressure
gradient.

In \cite{yong2021categorization}, an inspection of phasor diagrams
that represent velocity fluctuations, pressure, heat release rate,
and characteristic wave amplitudes at the flame elicits characteristic
features of marginally stable, intrinsic thermoacoustic (ITA) modes.
The sign of velocity fluctuations and the sign of the gradient of
pressure fluctuations change across the flame. These sign changes
result from a reversal of direction of the velocity phasor across
the flame, affected by unsteady heat release exactly out-of-phase
with respect to upstream velocity fluctuations and of sufficient strength.

\index{Civil Engineering}

\section{Civil Engineering}

In a work on phase analysis of actuator response for
sub-optimal bang–bang and velocity cancellation control of base-isolated structures, \cite{austin2007phase} shows that only the sign of the displacement/velocity
matters (the magnitude of velocities and displacement is irrelevant).
Hence, the mean value of velocity is an aggregation of the local trends.

\cite{athanasiou2011modelling} proposes a model friction formulas 
and other physical sizes that explicitly leverage the sign of the velocity
and the acceleration in a work on modelling hybrid base isolation systems
for free vibration simulations.

The wbbl thickness in \cite{henriquez2014piv} increases linearly
during favorable horizontal pressure gradients (i.e., when the sign
of the pressure gradient is opposite to the sign of the fluid velocity).
The coefficient of determination $R^{2}$ between measurements and
a linear fit is usually above $0.9$. When the pressure gradient sign
switches, the boundary layer thickness shows a sudden increase. Highly
asymmetric waves have wbbl growth rates under the wavefront that are
roughly twice as large as under the corresponding wave back.

In \cite{giresini2015comparison}, comparing between rocking analysis and kinematic analysis for the dynamic out-of-plane behavior of masonry walls, the rebound effect offered by transverse walls can be numerically considered a change in sign of the velocity immediately
after impact and possibly as additional damping.

The change in the monotonic behavior (derivative sign) of the network
outflow function at a turning point in \cite{amirgholy2017modeling}
is reflected as a dramatic change in the arrival rate of the users,
which subsequently affects the variation pattern of system accumulation,
as well as the outflow.

In \cite{hazaveh2017experimental}, the defining control laws in Eqs.
(1)--(3) provide direction- and displacement-dependent forces based
upon the piston location within the device and the sign of piston
velocity. To control this semi-active device requires sensors across
the device for displacement and velocity. Depending on the signs of
displacement and velocity direction, the active orifices are closed
or opened. Fig. 2 there shows a step-by-step example of the control
mechanism and response for a $2-4$ semiactive viscous $D_{3}$ device
under sinusoidal displacement loading.

The dynamic behavior of a curved surface slider in \cite{saitta2018base}
is governed by three friction coefficients, relative to the onset
of motion, the dynamic phase, and the velocity sign's inversion.

In \cite{petikas2020calculation}, when the specific energy curve's
slope is positive, the flow is subcritical. A negative slope indicates
supercritical flow. Considering that the sign of the slope of the
particular energy curve changes at the critical depth, elevations
where the specific energy curve is not continuous or differentiable
need be checked, as it might indicate a respective critical depth.
More specifically, these elevations are the ending points of the previously
defined segments (Figure 8). Suppose the specific energy curve slope
at the end of one segment has a different sign than the slope at the
beginning of the next segment. In that case, this means that the flow
changes regime from subcritical to supercritical or vice versa, and
thus the common elevation of the two segments must be of a critical
depth.

\index{Agricultural Engineering}

\section{Agricultural Engineering}
Yield trends have many applications in agricultural engineering, which makes the derivative sign particularly important . In \cite{zymaroieva2020spatial}, for instance, when the model is linear, the yield trend may be classified based on the slope sign. If the slope is negative, yields always decrease; thus, they may be classified as ``yields collapsed.'' 

There wasn't any case with a trend that may be classified as ``yields collapsed'' in our data set. If the slope of the linear equation was positive, it meant that yields were increasing and thus were classified  as ``yields increasing.'' When the chosen model was quadratic, and if the quadratic term was positive, the trend was classified as ``yields increasing.'' When the quadratic term was negative, the trend was classified as ``yields stagnating.'' When the chosen model was quartic, the trend was classified as ``yields stagnating.'' 
	
The slope sign is also relevant to the shape of action, estimated as the number of turn changes in the sign of the envelope intensity slope (\cite{ngo2020research}).

\index{Quantum Engineering}

\section{Quantum Engineering}
In a work relevant to topological phase transitions, cite{santra2014local} shows that the monotonicity of the entire set $S_{\alpha}\left(\lambda\right)$ induces a characteristic of the phase unless the perturbation and the choice of bipartition are fine-tuned. The collective behaviour can be captured succinctly by the sign of the derivative $sign\left[\partial_{\lambda}S_{\alpha}\left(\lambda\right)\right]\forall\alpha$, which remains constant in the topologically disordered phase.

\cite{molignini2017sensing} finds that, at the phase boundaries between two phases with differing non-trivial topology, the slope of the heat current changes sign with respect to the tuning parameter. On the other hand, transitions between a zero and a nonzero Floquet Majorana Fermions (FMF) phase are tracked by changes in signs or discontinuities in the slope of the heat current. The quasi-energy spectra are completely gapped away from the transitions, and the heat transport is essentially mediated by FMFs. The high-frequency oscillations in the heat current are due to finite size effects and decrease with increasing $N$. 

The change in the slope sign of the heat current with respect to the control parameter effectively tracks the parity of the phase and is valid for any cut in the phase diagram. Since the actual sign of the heat current is determined by the bath parameters, it is not possible to assign a fixed parity to a phase. Instead, the heat current is sensitive only to changes in parity. Consequently, one cannot ascertain whether a given phase has an even or odd number of FMFs. For specific bath parameters, the heat current can indeed change sign within a given topological phase without a concomitant change in the sign of the slope of the current.

\index{Aerospace Engineering}

\section{Aerospace Engineering}
In a work concerning abrupt stalls, \cite{green2005f} shows that based on the Figures of Merit developed during the AWS program, abrupt stalls can occur when the slope of the coefficient of the wing root bending moment (WRBM) curve changes sign. 

The angle of attack (AoA) at which the slope changes sign is of particular importance. If, for any of the morphed configurations, the slope of the WRBM curve changes sign at a lower AoA than it does for the F/A-18C, this indicates that the particular wing parameter or parameters being modified may be contributing to an abrupt stall. The high entrainment in the near jet field, a unique feature of synthetic jets, is also observed as a change of velocity sign in \cite{hashiehbaf2014experimental}.

An interesting study concerning the relevance of derivative values for aircraft stability, \cite{allen2017aerodynamic} shows that the $Cm_{\alpha}$ term crosses zero as Mach increases for both designs. When the stability derivative values change in sign, this can have interesting implications on aircraft stability. The forces in the $y$ and $z$ directions during the contact with the wall in \cite{zhang2019effects} are modeled (Eq. 2-3) based on the signs of the respective velocities of the right heel.

\chapterimage{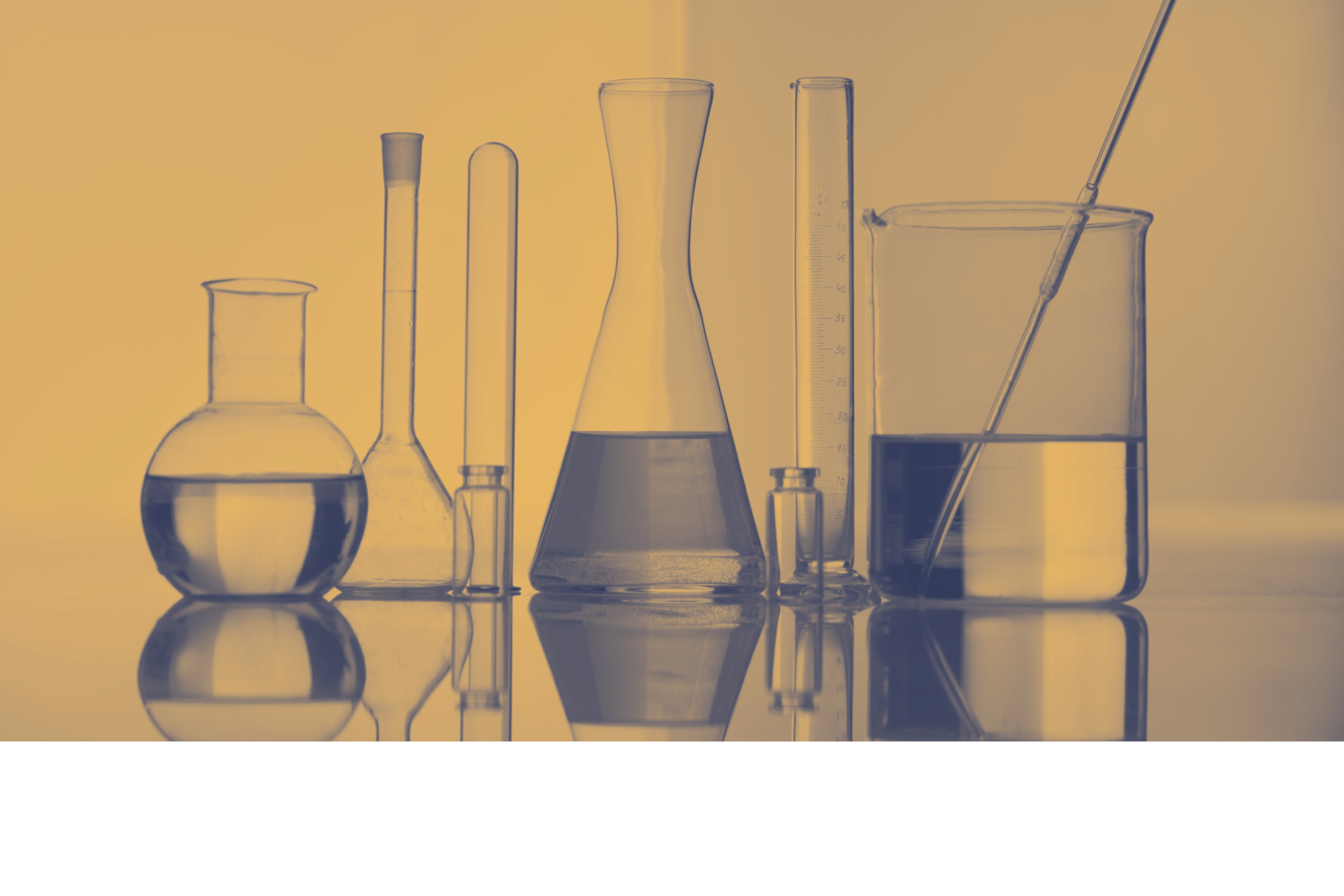}

\index{The Scientific Trendland}

\chapter{The Scientific Trendland}

Trends have a wide range of applications in data, material, management, earth, social, and natural sciences. Let us see some examples, first in data science and then in the other fields.

\index{Data Science}

\section{Data Science}

The run-length sequence studied in \cite{cammarota2011difference}, in a work on the difference-sign runs length distribution in testing for serial
independence,
is modeled based on local trends (events of a local maximum) rather
than rates.

An example of the importance of trends in time series and forecasting is the work of \cite{brockwell2016introduction}, who shows that one of the simplest methods for testing the estimated noise sequence is the difference-sign test. For this test we count the number of values of i such that $y\left(i\right)>y\left(i-1\right)$, or equivalently the number of times the differenced series $y\left(i\right)-y\left(i-1\right)$
is positive.

Another work on time series complexities with a focus on their relationship to forecasting performance \cite{ponce2020time}, the 2-regime symbolization is carried out
by employing the sign of the first difference.

Also in a work on time series complexities, \cite{novak2016linguistic} devises a definitive framework for trends
and an algorithm that finds intervals in which the time series trend
is monotonous in soft computing theories.

In a work on the local trend analysis method, \cite{tang2015evaluation} quantizes the derivative into three possible values indicating the local trend's
sign.

In \cite{kithinji2021adjusted}, working at adjusted extreme conditional quantile autoregression with application to risk measurement, an optimal threshold corresponds
to the start of approximate linearity of the mean excess plot with
the sign of the slope, indicating the specific family of the GPD.
A positive sign corresponds to the Frechet family, while a negative
sign implies the Weibull family.

\index{Materials Science}

\section{Materials Science}

One example of the use of the derivative sign in material science can be found in \cite{moreno2015stress}. In this work, the researchers explore the role of stress and its competition with purely erosive mechanisms (that emphasize the role of surface effects) to determine the sign of the velocity with which the ripple
pattern moves across the target plane. Based on their theory, the authors of the paper
discuss different situations and make specific testable predictions
for the sign change in that velocity.

In a work on sensorless control of low voltage permanent magnet
synchronous motor \cite{ccetin2017effect}, a negative or positive sign
of the slope of the resulting curves may
indicate whether a magnetic transition is either first order or second
order, respectively, according to the Banerjee's criterion. All samples in the paper show a change from FM state to PM
state of the second order because of the positive slopes of the Arrott
curves.

In a study on nonlinear friction compensation
method using adaptive control and its application to an in-parallel actuated
6-DOF manipulator (\cite{zhou2017model}), the expression of the modified KP operator
depends on the system's input derivative sign in several ways. First,
it serves to distinguish between two cases, as in Eq. 2. It also helps
to differentiate instances of the parameter $\xi$ in Eq. 2 (see Eq.
3). Another parameter from this equation, $q$, is the number of sign
changes of that derivative.

When considering Eq. 18 with non-zero
virgin levels in a work on perturbation analysis in finite LD-QBD
processes and applications to epidemic models (\cite{jin2019channeling}), the slope sign of the channeling spectrum depends on
two competing terms. The positive is very small in high-quality crystalline
samples. Thus, the slope is usually negative (with a declining yield
region between the peak and the end of damage) in most common non-metal
situations. However, in irradiated metals such as $N_{i}$, two reasons
might cause an overall positive slope:
\begin{itemize}
\item The small scattering factor value, $f$, while non-zero, can significantly
reduce the negative term
\item The irradiation-induced defect can reach much deeper depth than the
end of displacement profile predicted by SRIM, causing a gentle decrease
of defect density, meaning a small $N'\left(z_{m}\right)$.
\end{itemize}

In such cases, the negative term may be smaller than the positive one, causing
an overall positive slope.

In a work on improving laser interferometers, \cite{wei2020auxetic} shows that by examining the dominant spd-orbitals along with the phase factors of
the wave function, one can determine the
state's bonding and anti-bonding characteristics along a specific
direction. When strain is applied in that direction, the energy variation
with the strain of the state obeys the pattern schematically based
on the slope sign.

In a study of computationally enabled total energy minimization under
performance requirements for a voltage-regulated microprocessor in
65nm CMOS, \cite{wu2020influence} show that when the friction velocity is constant,
the expression of the steady friction force of the LuGre model can
be given by Eq. 4, thus incorporating the velocity sign. When the
cylinder moves with a uniform speed, the average elastic shape variable
in the LuGremodel remains unchanged. That is, when the model becomes
steady-state, $v$ is a constant ($\frac{dz\left(t,\zeta\right)}{dt}=0$),
then Equation (5) is set to $0$, and $z$ can be expressed by $g\left(v_{r}\right)$
up to the velocity sign, which can, in turn, be incorporated into Eq.
6 and 12 to obtain the steady friction force in Eq. 14, again leveraging
the velocity sign.

Using a simple RPA argument, \cite{pouget2021momentum} shows that
the experimental $q\left(T\right)$ can be understood if the electron-phonon
coupling (EPC) $g\left(q\right)$, necessary to set coupled electronic
and structural modulations, is momentum dependent. In this analysis,
the sense of $q\left(T\right)$ variation depends upon the sign of
$\frac{\partial g}{\partial q}\left(q\right)$. Further, using a detailed
analysis of the low-frequency phonon spectrum of the blue bronze,
a new scenario for the $q$ dependent EPC is proposed. In it, $g\left(q\right)$
is due to a momentum-dependent hybridization between the critical
phonon branch bearing the Kohn anomaly and other low-lying phonon
branches. This approach allows obtaining a sign of $\frac{\partial g}{\partial q}\left(q\right)$
in agreement with that deduced from the $q\left(T\right)$ analysis.

As reported by Banerjee, an inspection of the sign of the slope of
the straight line in the Arrott plots gives the nature of the magnetic
phase transition order. The magnetic transition is of second-order
if the slope is positive and first-order if negative. 

As shown in Figure 14 in an ECG signal analysis study with
temporary dynamic sequence alignment \cite{saidi2021effect}, the curves ($\mu_{0}H/Mvs.M^{2}$)
exhibit a positive slope for all the samples in the vicinity of $T_{C}$,
which indicates that the samples undergo a second-order ferromagnetic-paramagnetic
phase transition.

\index{Management Science}

\section{Management Science}

It is known that there is a relationship between the sign of the
first derivative of the utility function and a stochastic dominance
order, named the first-degree stochastic dominance (FSD) order. In
the context of \cite{beaud2016impact}, who discusses a new texture feature descriptor
for image retrieval, the FSD order allows the unambiguous
ranking of any two transport alternatives with the exact cost. Further,
Proposition 7 proves a link between the SSD order and the sign of
the first two successive derivatives of a travelers' preferences
function with respect to travel time. A similar analysis is also conducted
with respect to the sign of the third derivative.

An interesting qualitative discussion of the monotonic relationship between the signs of the partial derivatives of price and housing consumption
with respect to several parameters can be read in \cite{basso2017effects}. A different work relevant to this section is that of \cite{navarramarkup}, where one can infer from equation 4 the price
change, $\frac{dp}{dc_{f}}\left(c\right)$, implied by the efficiency
gains from the change in markup and the change in marginal
costs. Looking at the second term of the total derivative in (4),
foreign variable costs' partial derivative of variable expenses is
positive and equal to the reduction in costs due to cheaper foreign
inputs. The cost is larger if foreign inputs have spillover effects that
allow achieving efficiencies through insourced inputs' substitution.
Depending on the sign of the partial derivative of markup by costs,
the first term of the right-hand side can either be negative or null
(as firms are maximizing profits).

Finally, in a discussion on how the time evolution of the reporting COVID-19 rates controls the occurrence of the apparent epidemic peak, the derivative sign of $\rho\varepsilon$ is used inside the integration
process {chiarello2020non}, in lemmas 3.1-3.3. The sign of the difference velocity $u-v$ is applied throughout the
paper in various contexts.

\index{Earth Science}

\section{Earth Science}

In a study investigating the magnetic field effects on in-medium dissociation, \cite{lee2001slope} showed that if the boundary current initially has zero
relative vorticity and the bottom boundary layer is spatially uniform,
the torque is zero. However, if the water depth increases offshore, then the bottom stress has to act over increasingly thicker
water columns, producing a bottom torque. The outcome is a gain of
positive vorticity by the northward flow, even if the curl of the
bottom stress is negative. 

A simple scheme that portrays the slope-induced
bottom torque is illustrated in Fig. 1. The sign of the slope-induced
torque depends on the bottom slope; within a western boundary current
flowing over a steep continental slope, it opposes the bottom stress
curl term in the cyclonic side of the stream.

In a different line of work, \cite{sonneveld2003nonparametric} proposes a neural network for wind-guided compass navigation. From Eq. (4),  the authors derive
the probability of a wrong sign of the first derivative. A value of
$0.5$ indicates that, on average, the slope information is uninformative.
Above $0.5$ it has the wrong sign, and the more below $0.5$, the
more reliable the average slope. The nonparametric test calculates
the probability of a wrong sign of the first derivative of the individual
parameters, which measures the significance of the association between the independent and the dependent variables. Table 1 presents the probability of a
wrong sign of the first derivative for the USLE factors.

In a paper on the detection of degenerative change in lateral projection cervical spine
x-ray image (\cite{liu2005identifying}), one of the parameters among the $Q_{C}$ are the
along-beam perturbation velocity sign changes. The results show that VSC (the Velocity
Sign Change method) is more effective than VDC, while VDC is more
effective than MRF.

In a work discussing evidence of second-order nonlinear susceptibility sign reversal in
thermally poled samples, \cite{nevzorov2006glory} shows that the angle sequences of neighbor peaks
belonging to curves of the relative intensity have been related to
the derivative sign of the corresponding curve of the observation
angle.

In a paper on accelerating experimental design by incorporating experimenter hunches \cite{parrot2011statistical}, the data is smoothed, and a variation
is evaluated by comparison with the background level by using
the change of the derivative sign.

A paper on nested motion descriptors \cite{chen2013spectral} argues that change patterns should be analyzed in
a hybrid according to different spectral gradient signs (qualitatively)
and gradient values (quantitatively).

According to a study of gradient-only approximations
for line searches towards robust and consistent training of deep neural
networks \cite{corvaro2014fluid}, to evaluate the turbulence efficiency in influencing the motion of
\textquotedbl water particles\textquotedbl{},
a quadrant analysis is usually used. Suchan analysis investigates
the sign of the turbulence components ($u_{0};v_{0}$) with that of
the free-stream mean velocity $U_{0}$ according to four conditions
that are based on the signs of $u'$ and $U_{0}$ (where $u$ is the
velocity). In the classical analysis of the turbulent boundary layers,
the four cases almost directly assign each event to a specific dynamics:
sweep, ejection, high- and low-speed fluid motion. In the
problem presented in the study, the variation of the horizontal velocity with the wall distance
is not monotone. Hence, the presence of an extreme value of the horizontal
velocity close to the bed induces a change in the sign of $\frac{\partial u}{\partial Z}$
and then on the fluctuation components $u_{0};v_{0}$. This sign change
does not occur in the same place for horizontal and vertical components.
Hence the Reynolds stress may change its sign when the same phenomenon
(i.e., suction or injection) is observed at different levels, i.e.,
above or below the maximum values of the velocity components.

It is worth mentioning that trend analysis systems have three components: a language to represent
the trends, a technique to identify the trends, and mapping from trends
to operational conditions. The fundamental elements are modeled as
triangles to describe local temporal patterns. 
The elements in \cite{juuso2018intelligent}, a work on the interaction between the demand for saving and
the demand for risk reduction, are defined by the first and second derivative signs, respectively; they are also known as triangular episodic representations. In this work, the fuzzy
rule-based solution has been transformed to an equation-based solution
by the LE-based trend analysis.

\index{Social science}

\section{Social science}

\index{Sociology}

\subsection{Sociology}

The derivative sign has applications in many fields that can be broadly classified as social sciences. An example of this in sociology is the study of \cite{estrada2017effect}, who were interested in the sign of the partial derivative of
$g$ with respect to $d$ to sign the effect of an exogenous change
in demand for elite schools on stratification by family income. This paper showed that the sign of the change in the stratification is the same as that of the change in the share of high-income students
admitted to the elite school. For simplicity, let us call this share $\gamma$.

In \cite{ushchev2017technology}, the derivative sign is used to prove comparative
statics of equilibrium effort and utility \& welfare (see Propositions 10 and 11). More specifically, the proof is obtained based on a derivative sign analysis of the functions $x^{*}$ and $U$, respectively.

In a study on the use of information theory as a measure of sociodemographic changes, \cite{lancaster2018information} shows that the sign of the derivative indicates the current regime. While the influence of magnitude requires additional research, the
sign suggests a system gaining or losing in entropy. As such, the slope indicates how close the system is to regime change. If the slope becomes
less negative and if there is an inflection, the system will change to a growth
state, where complexity and diversity increase. Conversely, if the
slope moves from positive to negative across the inflection, the system
will move towards a regime of homogenization. This value can be used as a planning tool.

In a paper investigating the role of performance incentives in need-based grants for higher
education, \cite{montalban2018role} finds that the sign of the difference depends on
the cross-derivative sign. It is shown that the sign will be positive if
a given grant amount provides stronger performance when performance
incentives are higher.

In a study of distributive pensions in the developing world, \cite{kemmerling2019redistributive} derives $U$ according to $\tau$ and then devises (see Equation 13) a parameter that dictates the derivative
sign. The findings provide valuable data for a qualitative discussion regarding the
pension scheme preferences of young urban workers.

In the work of \cite{deuster2019climate}, we see an example of the use of linear regression analysis where the results align
with the model prediction on the sign of the partial derivative of
Equation (12) with respect to the climatic variation. Overall, the results of the linear regression analyses might only reflect correlations between different variables and are the results of a reduced form estimation. However, the linear regression analysis can be understood as the first step
in empirically validating the theoretical predictions of the model.

Lemma 1 in a study of social welfare and price of anarchy in preemptive priority queues (see \cite{chamberlain2020social}) suggests conditions to the
monotonicity of the function $C(\varphi)$. The data obtained is used to analyze equilibrium outcomes, equilibrium stability, and social welfare variables.

In a work on social and economic networks, the derivative sign is used for monotonicity analysis in the proofs
of propositions 1.3 and 1.4 (see \cite{cortes2020essays}).

Finally, Table 2 in a study of the social cost of carbon in a
non-cooperative world (see \cite{hambel2021social}), summarizes the model input
parameters' influence and comprises the respective partial derivatives
signs.

\index{Political Science}

\subsection{Political Science}

In one political science study focused on the political economy of power-sharing, \cite{tridimas2011political} reports the comparative static
properties of $k^{*}$ (see Table 1). The first row shows that when $A$ is more
likely to win the election and the fighting effectiveness in the war rises, more power is allocated to the election winner ($A$). On the
other hand, if $A$ is less likely to win the election, but fighting
effectiveness rises, less power is allocated to the election winner
(likely to be $B$) to give $A$ the incentive to participate in the
power-sharing arrangement, a result that is arguably intuitive. The derivative signs in the second row of Table 1 indicates when $A$ is the likely
election winner; the increase in electoral effectiveness further
reduces the probability that $B$ may win the election and is followed by an increase in the rents given to the election loser.
On the other hand, if $B$ is the likely election winner, an increase
in the electoral effectiveness is accompanied by more powers for the
election winner, as both groups now see the election offering them
a way to gain the rents of office. The third row of the same table presents the effect
of an increase in the destructive effect of war. It shows that irrespective
of whoever is more likely to win the civil war and the election, an
increase in the destructiveness of war induces both parties to award
more office rents to the election loser to avoid war.

In a competition effect analysis exploring voter turnout, \cite{herrera2016turnout} shows that
for any fixed cost distribution and cost mobilization function, turnout
depends only on the marginal benefit (henceforth $M_{B}$) and increases
with it. Hence, $M_{B}$ is studied as a proxy for turnout $T$. Fixing
the institutional setting $\gamma$, the authors focus only on the
sign of the derivative with respect to $Q$.

In \cite{acharyya2017asymmetric}, the sign below each argument in
(10), which indicates how the equilibrium wages change with respect
to change in that particular argument, follows from (7) and (8) (the
sign of the partial derivative). The authors of the study also provide a qualitative analysis
of these signs.

Also investigating voter turnout, \cite{justo2018liii} shows that the negative sign of unemployment reflects
that the withdrawal effect dominates the mobilization effect. This is
somewhat unexpected since vote-buying behaviors among vulnerable people was widespread
in some Argentinian provinces, implying a positive correlation between
unemployed and voter turnout. The negative sign
of the rate of growth of crime indicates that
withdrawal dominates mobilization. A priori, we expected that people
would try to use the electoral instrument in attempt to influence crime policy, as crimes where at a high level in most Argentinian districts at the time of the study.

\index{Economics}

\subsection{Economics}

\index{Trade}

\subsubsection{Trade}

It is likely no surprise that the derivative sign has been widely used in economics. In trade, we have several study examples, including one on the impact of exogenous asymmetry on trade and agglomeration
in core-periphery model (see \cite{sidorov2011impact}). In Proposition 6, the study proves that the derivative
$\partial/\partial\lambda\left(\frac{V_{H}}{V_{F}}\right)$ changes
its sign not more than twice for all $\lambda\in(0,1)$. Therefore
there is at most three values $\lambda\in(0,1)$ yielding welfare
equalization $V_{H}\left(\lambda\right)=V_{F}\left(\lambda\right)$.
As a consequence from corollary 2, $F\left(\varphi,\rho,\mu,\alpha\right)$
is strictly increasing with respect to $\mu$, i.e. $\frac{\partial F}{\partial\mu}>0$
for all admissible arguments. Thus, in accordance to the Implicit Function Theorem,
there are differentiable function $\mu_{B}\left(\rho,\alpha,\varphi\right)$
such that $\frac{\partial\mu}{\partial\alpha}<0,\frac{\partial\mu}{\partial\varphi}<0$,
because $F\left(\varphi,\rho,\mu,\alpha\right)$ increases with respect
to $\alpha$ and $\varphi$.

In a study of trade, firm selection and innovation, \cite{impullitti2018trade} shows that because of the positive sign of the
derivative in (21), reducing $\theta_{\tau}$ by increasing
$\tau$ and by reducing the share of exporters in the economy increases
the average profits for new firms $\pi$ in (22), thus increasing
firm entry and ultimately leading to a higher equilibrium in the number of firms.
Moreover, a trade-induced increase in competition reallocates resources
from the homogeneous good to all varieties (exporters and non-exporters)
in the differentiated sector. This outcome has has an additional positive effect
on the average profits and induces more entry. A larger n then reduces
the domestic markup $\frac{1}{\theta}$ and raises the domestic cutoff
$z^{*}$, hereby forcing the least productive domestic firms to exit.
Finally, a higher n also strengthens the reduction in the export markup
produced by trade liberalization, thus further increasing the export
cutoff $z_{x}^{*}$.

\cite{chattopadhyay2019nash}, investigating nash equilibrium in tariffs in a multicountry
trade model, shows that on the technical side, there are
two obvious candidates for further research. The first involves extending
the model to three or more goods. One can imagine that many of the
results in Section 5 will go through, provided that the best response
functions are increasing. To show the latter, one will have to grapple
with quasiconcavity and the sign of some cross partial derivatives.

\index{Economic Systems}

\subsubsection{Economic Systems}

To see the effects of the parameters on location in a work on the public sector and
core-periphery models, \cite{lanaspa2001public}
obtains the $V$ derivative for each of them; if the derivative sign
is positive (negative), then increases in the parameter favor dispersion
(agglomeration). It can be verified that $\frac{\partial V}{\partial t_{1m}}$
and $\frac{\partial V}{\partial t_{1a}}$ are both positive, showing
that a tax increase in region 1 favors the disappearance of the agglomeration
that is present there. $\frac{\partial V}{\partial t_{2m}},\frac{\partial V}{\partial t_{2a}}$
are both negative in such a way that an increase in taxes in region
2 helps to maintain the initial agglomeration in region 1. These results
are in line with the literature on tax abatement. The different regional
or local jurisdictions use this mechanism as an instrument to attract
firms to their territory.

In a paper on the choice
of quality assurance systems in the food industry, \cite{carriquiry2007reputations} tells us (see System (5)) that the
sign of the derivative of $y$ with respect to $\omega$ equals that
of the second partial derivative of $\pi$ with respect to $y$ and
$s$.

In \cite{kacperczyk2009rational}, a work on variables influencing investment choices, proposition 1  proves that the second
partial derivative of $U$ with respect to $K$ and $\sigma$ is positive,
based on the positive derivative sign of the trace term's cross partial
derivative. Similarly, proposition 4 proves that an increase in risk
aversion $\rho$ increases the dispersion of funds' portfolio returns.
Proposition 5 proves that an increase in aggregate shock variance
increases the difference between an informed investor's expected certainty
equivalent return and an uninformed investor's expected certainty
equivalent return.

In a related line of work (see \cite{jouvet2012irreversible}), proposition 1 proves that for a
given initial stock of capital investment, ambiguity aversion tends
to decrease the agent's optimal production levels. It does so by analyzing
the sign of covariance given by the derivative of $\Lambda$ with
respect to $\theta$.

In another related line of work (see \cite{laporte2015should}), section \textquotedbl Comparative statics relation between
$I$ and $H$\textquotedbl{}, an extensive
qualitative analysis is made regarding the derivative sign of the
investment and its qualitative consequences.

The importance of three-way interactions in a work on health behavior (see \cite{kucher2019health})
also follows from the comparative static effects of the weight environment
in equation 3.3. The authors find that, for a given effort level,
a marginal increase in the weight environment affects the transition
probabilities where the size and the sign of the effect depend on communication
efforts and weight environment $\Delta$. As the sign of the first
factor of equation (3.3) is always positive, the sign of the partial
derivative is determined by the second factor. Independent of the
parents' perception type, the directional effects of an increase in
$\Delta$ on the respective (conditional) perception probabilities
of children directly follow from our assumptions on $q_{U}$ and $q_{O}$.
More specifically, the likelihood of child over(under)-perception
falls (increases) with \textbackslash Delta. The effect on the probability
of correct weight perception cannot be signed.

\index{Environmental Economics}

\subsubsection{Environmental Economics}

In a paper investigating the relationship between energy and technological changes, \cite{khademvatani2009measuring} defines the ratios between the derivatives of different
functions to analyze the relationship between their change
directions. For example, the derivative of $SV$ with respect to technology
(equation 3.2.15) is positive if there is an energy augmenting technological
change where a lower level of energy is used to produce the same output
and profit. With technological change requiring more energy, the derivative
sign is negative, and in the case of energy-neutral technology, the
derivative is zero. This derivative helps us to find out the short-run
average change rate of the energy $SV$ with respect to technology
changes, as $\varepsilon_{SV_{t}}$ (equation 3.2.16), which is expected
to be positive when the derivative of $SV$ with respect to technology
is positive. Over time, the overall impact of technology changes is
likely to be in the same direction but greater in the short run.

The important of the sign of $\frac{\partial r}{\partial\rho}$ in the general model of the environmental economics presented in \cite{jeon2014three} is studied extensively in section 2.3.

An example of the monotonicity
of the function $\varphi^{*}$ by evaluating the sign of its derivative can be seen in the proof of proposition 5 in \cite{bell2018three}.

\index{Behavioral economics}

\subsubsection{Behavioral economics}

In order to measure the slope of the reaction function in a work on social preferences in voluntary
cooperation, \cite{thoni2015peer}
uses agent i's belief about agent $j$'s initial effort. In a subset
of the data ($n=110$), the authors elicited each agent's beliefs about
their co-agent's initial effort decision. For any belief, we can
observe two points on an agent's reaction function if the belief was
wrong, which provides a direct measure of the sign of the slope of a
monotonic reaction function by estimating the function $\Delta e_{i}$,
where $e'_{j}$ denotes the agent's $i$'s belief about $j$'s initial effort.

In a study of child labor, \cite{nwogwugwu2017impact} shows that the time spent
at work is negatively related to school enrolment, as indicated by
the sign of the slope coefficient. The finding implies that the higher the
time a child spends at work, the lower the enrolment rate.

In a paper on optimal term length for an overconfident
central banker (see \cite{farvaque2017optimal}) , Lemmas 1 and 2 state claims regarding
the monotonic relation between the degree of different types of bankers'
overconfidence and the duration of the optimal term. They do so by
analyzing the respective derivative sign.

Lemmas 5.2 and 5.3 in a study of opinion dynamics (see \cite{abebe2020opinion}) prove results on the
signs of the partial derivatives of $f$ and $z$ with respect to
$\alpha$. They are then used to prove monotonicity properties in
other theorems.

In the proof of propositions 5 and 6 in a work on social organization (see \cite{marcoasocial}), the
implications of different signs of the derivative $\Omega_{c}$ and
$\Omega_{s}$ are analyzed qualitatively.

In a study of overconfidence in games, \cite{santos2021overconfident} shows that given that $\frac{dP1}{da2}<0$,
the sign of the strategic effect is negative when $\frac{da_{2}}{d\lambda}>0$
and positive when $\frac{da_{2}}{d\lambda}<0$. Hence, an increase
in overconfidence makes the player worse off when it raises the effort
of the rational player.

\index{Risk}

\subsubsection{Risk}

Empirical evidence in a study of household income (see \cite{mariotti2016job}) shows that the sign
of the slope may change with the level of the wages. It is especially
true in a household contest (i.e., in a two-individual economy where
the two subjects strictly interact). What happens in such an environment
is that the sign of $\frac{\partial h_{i}}{\partial w_{i}}$ changes
both with the level of and with the level of $w_{i}$.

In \cite{augustin2018term}, the importance of global and country-specific
risk in explaining sovereign credit risk varies with the sign of the
slope of the term structure and the duration of its inversion. A model
shows that global uncertainty shocks determine spread changes when
the slope is positive and domestic shocks are more critical when
the slope is negative.

\cite{loffler2018pitfalls} illustrates the effect of
risk parameters on systemic risk measures (see see Table 1). Panel A presents possible
signs of the derivatives. A superscript $n$ marks cases where the
sign applies under normal conditions. Only very implausible parameters
constellations specified in the appendix would generate the opposite
sign. In the single instance of ``$+/-$'', the partial derivative
can be negative under normal conditions, but only when the partial
derivative to $\sigma(RS)$ is negative, too. Panel B reports the
partial derivatives' signs for the base case. Panel C reports signs
for a system with a dominant bank of high systematic risk.

In a study of meager performance, \cite{bardgett2019inferring} uses the slope of
the vehicle routing problem (VRP) term structure to show that as long as the realized
variance is low, i.e., outside the financial crisis, the strategy
yields a positive payoff equal to the difference between the three-month
and the six-month VRP. This difference becomes negative during the
financial crisis, causing losses for the portfolio holder and canceling
the previous gains. Therefore, we can interpret a switch in the sign
of the slope of the VRP term structure as a warning that the future
realized variance might increase, as a result of which the forward
variance risk premium is no longer positive. If one leaves the investment
on hold until the slope switches sign again, it is possible to avoid some of
the losses of the former strategy and generate a Sharpe ratio of $0.46$.
Changing position from selling future variance into buying future
variance whenever the slope of the VRP term structure is negative
further enhances the Sharpe ratio, which reaches $0.77$.

Propositions 10 and 11 in a work on incentive pay and systemic risk (see \cite{albuquerque2019incentive}) state
claims about functions' trends: the increase in leverage and systemic
risk is given by certain conditions.

A well-behaved $v\left(\cdot\right)$ is defined in \cite{de2019strategic}
as the scenario where the slope of $\alpha\left(R\right)$ changes
sign at most once. All results in Proposition 1 are now readily obtained
conditional on this scenario. Within the proof of this proposition,
further trend analyses are conducted based on the derivative sign
of $\alpha$.

In \cite{hartwig2020identifying}, working at identifying indicators of
systemic risk, if a candidate variable passes
both stages, the square's color is determined by the sign of the sum
of the slope coefficients in the first stage. This distinction turns
out to be crucial for the interpretation of the results.

In a related line of work, ( see \cite{hyndman2021dissolving}) Hypotheses 1 is about the monotonicity
of the risk with respect to the price, and hypothesis 2 is about the
monotonicity of the cost with respect to the buyer's and seller's
risk aversion. Their proofs are via the signs of the respective partial
derivatives.

In yet another paper on system risks ( see \cite{bouvardlabor}), Table 1 lists the derivatives $\frac{dy^{*}}{dx}$
of the column variables $y$ with respect to the row variables $x$
when all other row variables are held constant. A qualitative comparative
analysis follows.

\index{Insurance}

\subsubsection{Insurance}

In a study on long-term care (LTC) insurance,  \cite{zweifel2016long} shows that a positive trend of the variable $\alpha$,
denoted with $d\alpha>0$, represents an increased level of cost-sharing
and $d\alpha<0$, an increase in subsidization of LTC. The signs of
the first and second derivatives of the expected utility with respect
to the amount of private insurance are used to estimate $sgn\left(\frac{dI}{de}\right)$,
in section 2.

In a paper on the interaction between the demand for saving and
the demand for risk reduction, (see \cite{grainich2017interaction}), the first condition in definition 1 is necessary and sufficient for the $\left(n-1\right)$ first moments
of $G$ and $F$ to coincide. The second condition is sufficient (but
not necessary) for the nth moment of $F$ sign adjusted by $(-1)^{n}$
to exceed the $n^{th}$ moment of $G$ sign adjusted by $(-1)^{n}$.
In the expected utility model, preferences over $n^{\text{th}}$-degree
changes in risk in the sense that Ekern are identified by the signs
of subsequent derivatives of the utility function, which motivates
the definition of $n^{\text{th}}$ degree risk adversity and the following
theorem.

In a paper on subrogation (see \cite{polinsky2018subrogation}), the signs of the derivatives of $x$ and $p$ are used to prove proposition 3.

In a paper on natual disaster insurance (see \cite{wagner2019adaptation}),
the sign of the slope of the average cost curve is a test for the selection
on observable determinants of natural disaster risk. If the market
is adversely selected, homeowners' costs are positively correlated
with willingness to pay, so infra-marginal homeowners are costlier
to insure than marginal individuals. In this case, the derivative
is positive because lower-cost individuals cease to purchase insurance
at higher prices. Further, natural disaster insurance markets are
adversely selected if adapted houses are required to be elevated both
less costly and less likely to be insured, depending on prices.
In terms of the model, this is equivalent to testing for negative
derivatives of both $AC$ and $s$.

\index{Macroeconomics}

\subsubsection{Macroeconomics}

A key issue on the demand side in a work on embedding care and unpaid
work in macroeconomic modeling (see \cite{braunstein2011embedding})
is whether the $IS$ curve is upward or downward sloping -- whether
declines in profit share are associated with declines or increases
in output. The sign of equation (7) determines these relationships.
A qualitative analysis of the signs of the components of this equation
follows.

General functional forms cannot be used in the conditions of Legros
and Newman directly to verify whether assortative matching will occur
since it is hard to verify them. Instead, \cite{ghatak2011contractual}
uses the implicit function theorem and relies on the cross-derivative
sign to directly characterize the conditions for sub- or super-modularity
of joint payoffs. Further, in corollary 1 to proposition 3, the second
derivative sign proves that the join surplus is super-modular in the
skill levels.

The time-series of the logarithmic returns shown in Fig. 6a in \cite{da2012heavy}
must first be mapped in a series of events as shown in Fig. 6b. One
event is defined as successive instants in the original time series
having the same derivative sign, either positive or negative. Each
time the derivative changes sign, a new event starts. In the continuous
limit, events would correspond to the instants in the time series
with vanishing first-derivative.

In a paper on state dependent monetary
policy (see \cite{lippi2014state}), the phase diagram in figure 1 determines
three regions where the functions $u\left(z\right)$ and $p\left(z\right)$
change behavior according to their derivative sign. The only positive
solutions to the problem must stay in the third region, where both
functions decrease the entire $\left(0,1\right)$ interval. More precisely,
their path develops in the area that is dotted in the figure. Proposition
3 characterizes the marginal value of money when $\mu$ is constant,
based on the derivative signs of $p$ and $u$.

The signs of the reaction function slopes in a paper on exclusionary conduct of dominant firms, R\&D competition, and
innovation (see \cite{baker2016exclusionary})
are the same as the signs of $D$ and $W$, respectively. If $D$
and $W$ are both positive, each best response function is upward
sloping (strategic complements); if $D$ and $W$ are negative, each
best response function is downward sloping (strategic substitutes).
If $D$ and $W$ take on opposite signs, the best response functions
differ in the sign of their slopes.

In a study investigating the optimal ways of Kharkiv’s social-economic development by
component analysis (see \cite{mazurova2017defining}), the qualitative solution in  is reduced
only to the definition of the time derivative sign (increases, decreases,
or remains unchanged). This work is limited to the analysis of exclusively
qualitative problem-solving, as it is the most general.

In its proof, the formulation of Proposition 2 in a paper on tax and borrowing (see \cite{bengui2018macroprudential})
is translated to the signs of the partial derivatives of the tax with
respect to the different types of borrowing.

In equation (13) of a paper concerning medical trials, \cite{chemla2018subject} calculates the sign of
$\beta'$. It is later applied in propositions about the expected
treatment-control health quality difference given some monotonicity
assumptions.

The monotonicity and concavity of the BB-SSLD curve in a study of positive and normative implications of liability
dollarization for sudden stops models of macroprudential policy (see \cite{mendoza2019positive}) are studied extensively based on the signs of its first and second
derivatives.

In a paper on credit cards an recession, \cite{drozd2019credit} shows that the local monotonicity, as expressed
by partial derivatives' signs, is applied to prove several results. The sign portion of the gradient
is often the only necessary information for the debate, and the information
regarding the magnitude is discarded.

In \cite{ferret2021green}, $Y_{2}$ increases with $M_{1}$ only
if the positive influence of the middle term is stronger than the negative
influence of the last term.

\index{Taxation}

\subsubsection{Taxation}

Based on derivatives sign analysis, a work on tax evasion (see \cite{panades2004tax}) shows
that at the low-income report equilibrium, a tax rate increase results
in more income concealed from the tax authority.

In a paper addressing some of the effects of government enforcement and service
provision (see \cite{berman2013predation}), the derivatives $\frac{dy^{*}}{dx}$
of the column variables $y$ with respect to the row variables $x$
when all other row variables are held constant can be seen in Table 1. For example, the top
right positive sign indicates that $\frac{d_{i}^{*}}{dn}|v,m,g,I,<0$.
Further, proposition 5 states that government enforcement and service
provision increase rebel violence, ceteris paribus. Its proof is based
on the partial derivative signs analysis of $B$ and $EU$.

The theoretical model proposed by \cite{agrawal2015inter} in a work on Inter-federation competition provides the following
testable hypotheses regarding the sign on the coefficients from equation
18 there. The coefficient representing the horizontal interaction,
$t\left(-i\right)$, is expected to be positive on average based on
equation 10. Diagonal tax competition represented by $\tau\left(i,-j\right)$
should be positive on average because the diagonal interaction is
similar in sign to a horizontal interaction in the local region of
the border from equation 11. The diagonal reaction will be affected
by distance through $d\left(i\right)\tau\left(i,-j\right)$, and the
effect is expected to be negative as the lower branch of equation
11 is less than the upper branch.

In a study addressing foreign state policy, \cite{chirinko2017tax} shows that the sign of the slope of the reaction
function of the home state to foreign state tax policy depends on
the income elasticity of private goods relative to public goods. To
develop an intuition for this critical result, consider the case when
the capital tax rate for a neighboring state rises. In turn, mobile
capital (eventually) flows into the state, and the tax base increases.
They further show that the slope of the reaction function can be positive
(``racing to the bottom'') or negative (``riding on a seesaw'')
and that the sign of this slope depends on the sign of one critical
parameter: the income elasticity of private goods relative to public
goods.

Due to the complex structure of Equation (29) in a paper on the impact of taxes on competition for CEOs, \cite{krenn2017impact} shows that deriving necessary conditions with respect to the sign of the partial
derivative does not yield significant expressions. Nevertheless, it
is possible to observe that the wage tax rate positively affects the
offered fixed salary whenever the first two addends exceed the third
one. It can be summarized as corollary 2 there.

In a work addressing taxation and hidden income (see \cite{di2017productivity}) Proposition 2 analyzes the monotonicity
properties of the hidden income with respect to tax parameters, based
on its respective partial derivative signs.

Appendix D in \cite{niemann2019investment} is dedicated to proving
results on algebraic signs of selected partial derivatives.

Propositions 3, 4, and 6 in a study of profit-sharing rules and the taxation of multinational
internet platforms (see \cite{bloch2019profit}) state claims regarding
the monotonic relations between parameters such as the tax rate and
tax revenue. On top of that, an extensive qualitative monotonicity
discussion is provided in sections 3 and 4.

In a study of the impact of taxes and wasteful government spending on giving (see \cite{sheremeta2021impact}), Theorems 1 and 3 prove claims regarding
the monotonicity of the provision of the public goods with respect
to the tax rate and the degree of waste, respectively. Their proofs
are based on an analysis of the signs of the partial derivatives.

\index{Finance}

\subsubsection{Finance}

\paragraph{Public Finance}

In section 4 of a study on public debt, \cite{englmann2015can} holds an extensive analysis
regarding the sign of the steady-state rate of change of the present
value of public debt.

In a study of governmental structures, \cite{corhay2016government} shows that when the yield curve is downward-sloping shortening the maturity structure increases the government discount
rate, generating fiscal inflation and expanding output. The opposite
results are obtained when the yield curve is upward-sloping.

Lemma 6 and corollary 2 in a study of monetary authority (see \cite{bursian2018trust}) prove sufficient
conditions for the derivative sign of $U$ and $\Phi$ with respect
to $\theta$, respectively.

In a study of monetary and fiscal policy interactions (see \cite{haslag2020monetary}) the propositions do not
mention trends directly; nevertheless, their proofs apply the monotonicity of parameters
such as $\Gamma$ and $q^{u}$ by assuring their derivative signs
are constant.

Proposition 4 in a work on imperfect credibility versus no credibility of
optimal monetary policy (see \cite{chatelain2021imperfect}) proves results on the sign of the derivative of the inflation eigenvalue with respect
to $\varepsilon$. Further, once it is clear that the discretion equilibrium
is not a relevant theory for stabilization policy, the empirical issue
is the measurement of the sign of the slope of the the new-Keynesian
Phillips curve. If it turns out to be negative, the transmission mechanism
corresponds to an accelerationist Phillips curve instead of the new-Keynesian
Phillips curve. Unfortunately, nearly $50$ of the estimates have
a positive sign in a large number of estimations. Once this sign is
known, the optimal response of the policy instrument to inflation
will have the opposite sign under quasi-commitment.

In a paper of bond holdings, \cite{kocherlakota2021public} shows that the cross-sectional
average of agents' bond holdings is unbounded due to monotonicity
considerations via the sign of the partial derivative (see Corollary 1). Further, proposition
4 heavily relies on function's partial derivatives' signs when proving
that $W$ increasing in $\alpha$ leads to its decrease in $q$.

\paragraph{Corporate Finance}

The initial information of the illustrative model proposed by \cite{camara2014methodology}, a work meant to develop a methodology for the evaluation of interoperability improvements in inter-enterprises collaboration based on causal performance measurement models,
included the qualitative magnitudes of the variables in the initial
state determined using the values of the 'as-is' IEP (intermediate
landmarks). For example, the circle representing the $C_{t}$ variable
is positioned at the milestone of $183$ min. The initial information
also includes the derivative sign for the $C_{t}$, which decreases
according to the simulation results of the to-be business process.
The propagation of the derivative of the $C_{t}$ across the proportionalities
in the model enables us to deduce that $O_{u}$ and $O_{p}$ have
increasing derivative signs because they vary in the opposite direction
of $C_{t}$.

In a study of capital accumulation, \cite{ko2015corporate} assesses the derivative signs of several parameters. Equation
11 is quite similar to that of Lavoie. A higher interest rate does
not uniformly affect the rate of capital accumulation. The investment
function (equation (6)) shows that an increase in the interest rate
negatively affects the rate of capital accumulation. However, the
saving function (equation (5)) indicates that an increase in the interest
rate raises rentier capitalists' earnings and, in turn, their consumption
expenditures. An increase in consumption expenditures immediately
raises the rate of capacity utilization and, depending on the sign,
that is, provided $\left(\frac{\beta}{h}+\delta\right)\tau-\sigma\theta$
is positive, the rate of capital accumulation would rise. Lavoie called
this situation the \textquotedbl puzzling case.\textquotedbl{} Hein
called the opposite problem an increase in the interest rate leading
to a decrease in the rate of capital accumulation, the \textquotedbl normal
case.\textquotedbl{}

The second-order conditions in a paper that uses the derivative sign to study the relationships between different firms (see \cite{ho2021government}) state that
two of the terms in equation 16 are negative. Consequently, one may
establish the relationship between the slopes of these two firms based
on the other two. The LMF's R\&D reaction curve moves to the right
as the LMF's government R\&D subsidy increases if $\eta<1$ (i.e.,
$gx_{s}>0$) while it moves to the left if $\eta>1$ (i.e., $gx_{s}<0$).
Following, the authors analyze the different cases according to the
various relationships between the R\&D reaction curves of both firms
in the form of the derivatives' signs.

\paragraph{Financial markets}

The signs of the first-order partial derivatives of $RSC$, $BC$,
$x$, and $y$ are qualitatively analyzed in a paper investigating the link between margin loans and stock market bubbles (see in \cite{ricke2004link}). They are also summarized in tables 4-7, respectively.

Theorem 5 in a paper on geographical separation of oligopolists (see \cite{madden2006geographical}) proves that $\rho^{*}\left(n\right)$
approaches infinity by first establishing that the sign of $\frac{\partial F}{\partial n}$
is positive, rendering $F$ increasing with $n$.

Previous literature suggested that the key factor is the sign of the
first derivative of the utility function with respect to the perceived
type. This is contradicted by \cite{molnar2008revenue}'s findings,
a paper which focuses on the second derivative sign in a study of revenue maximizing auctions with market interaction
and signaling”.

Derivative signs analysis in \cite{haucap2012regulation}, who studies the effects of regulation on network and service quality, leads to
conclusions on monotonic relation between the different discussed
conditions. For example, a higher network charge decreases consumers'
willingness to pay for services leads to a reduction of investment
into service quality.

Proposition 1 in \cite{raykov2016share}, which studies the effect
of variation margin gain haircutting on trading, proves the location
of peaks via partial derivative signs analysis.

Proposition A.2 in a work that includes an analysis of market Shape ratio (see \cite{eisfeldt2017complex}) states that the equally
weighted market Sharpe ratio increases with total risk in general
equilibrium given monotonicity conditions on the function $\sigma$,
proved in proposition 3.4. The monotonicity discussion is introduced
and confirmed with partial derivatives signs analysis.

Equation 4.16 in a paper on “Newspapers’ political differentiation (see \cite{hansen2017newspapers}) shows that the demand
for each newspaper is increasing in the number of advertisers ($\frac{\partial D_{i}}{\partial k}>0$).
The authors find that the sign of the partial derivative with respect
to location depends on the size of the transportation cost $t$. To
interpret this result, they make use of the observation that demand
is decreasing in $t$ ($\left(\frac{\partial D_{i}}{\partial t}\right)<0$).
It is the effect Kim and Serfes identified as the aggregate demand
creation effect.

Studying random encounters and
information diffusion about markets, \cite{gabszewicz2018random} uses the sign of the second derivative of $p^{*}$ and that of the partial
derivative of $\theta^{*}$ with respect to $t$ in the proofs
of several propositions.

In a study of the effects of economic variables on Swedish stock market volatility, \cite{kejlberg2018effects}
dedicates an extensive analysis to the slope sign of many of the studied parameters.

Proposition 6 in \cite{siemroth2019informational} calculates the
slope sign of the RHS for the comparative statics via a detailed analysis
of their derivatives.

Proposition 1 in a study of the informational content of prices when policy makers react to
financial markets (see \cite{oehmke2019tragedy}) states the interpretation
of the different possible signs of the derivative of $t$. Lemma 1
states that for any given vector of product complexities, the derivative
signs of different variables agree.

Table 1 in a paper on financier search and boundaries of the angel and VC markets (see \cite{pandher2019financier}) displays the sign of the partial
derivatives for the entrepreneur's equity share, expected investment
return, entrepreneur effort, and financier effort (columns) with respect
to various model parameters (rows). When $\alpha_{F}>\alpha_{F}^{*}$
holds, the sign of the partial derivative as is given in the cell.

Corollaries 1 and 2 in a paper on platform competition and incumbency advantage
under heterogeneous switching cost (see \cite{siciliani2019platform}) prove claims
about the monotonicity of the Incumbent's market with respect to the
switching costs and the cross-group network benefits, respectively.

\cite{hennig2021labor} conducts a comparative statics analysis to
understand how parental bequests and future wages affect the thresholds
of educational frictions. Note that the threshold is continuous and
differentiable with respect to parental investment $x_{ij}\left(t\right)$,
and future wages. Consider $w_{ij}\left(t+1\right)$ the wage for
the employment type which requires a relatively lower level of educational
attainment (numerator), and $w_{i'j}\left(t+1\right)$ the wage for
the employment type which requires a higher level of educational attainment
(denominator). The comparative statics reveal the following: $\frac{\partial\tau_{ij}\left(t\right)}{\partial x_{ij}\left(t\right)}>0$,$\frac{\partial\tau_{ij}\left(t\right)}{\partial w_{ij}\left(t+1\right)}<0$
and $\frac{\partial\tau_{ij}\left(t\right)}{\partial w_{i'j}\left(t+1\right)}>0$.
In other words, these findings imply that higher bequests and a rising
wage in i\textbackslash prime raise the threshold of educational
frictions, implying that the marginal individual faces higher constraints.
It, in turn, means that upward mobility is more likely to occur as
more individuals will enter the sector requiring a higher level of
education. On the other hand, if wages in $i$ are rising, then the
marginal individual has a lower level of educational frictions, and
thus less upward mobility occurs.

In the section \textquotedbl explanatory variables\textquotedbl{} in a paper on Bayesian Markov switching model applied to corporate credit default swap spreads (see \cite{bulfone2021corporate}), the predicted signs of each regressor
are analyzed and summarized based on previous works.

\index{Employment}

\subsubsection{Employment}

In the empirical counterpart of equation (1) in \cite{cristini2007high},
the sign of the reference wage is a prior ambiguous. Generally, if
the role of future internal prospects is relevant and the latter are
permeable to the outside peer group, a non-negative effect of the
reference wage is more likely. On the contrary, where commitment is
less dependent on the expected rewards and/or these are somehow insulated
from the outside market, the usual negative coefficient on the reference
wage is likely to prevail. On the whole, we expect internal monetary
prospects to be particularly relevant for work attitudes related to
solid commitment and less so for loose commitment work attitudes.
Further, in section 3.5 there, productivity, rents, and amenities
are analyzed based on a classification of workplace practices by total
derivative sign, summarized in table 1 there.

Proposition 4 in \cite{visschers2007employment} proves that both
uncertainties on the selling price and the emerging BTL technology
development decrease the capacity choice of the decision-maker in
the pre-treatment process. It does so also based on a partial derivative
sign analysis of $\Delta\left(\theta\right)$ and $\Lambda\left(\theta\right)$
with respect to the variable $\theta$.

\cite{schutz2009endogeneity} investigate
the sign of $\frac{dz}{d\omega}$ based on a qualitative analysis
of the relation between the wages and the profit share. After calculating $\frac{dz}{d\omega}$, continues to discuss the stability of the system.

In a paper investigating the effects of daily work time length on productivity and costs, \cite{delmez2018long} shows that, mathematically, the sign of the slope (or derivative) of the average
productivity is determined by the difference/ratio
between the average and marginal productivity. If $r\left(H\right)<1$,
we necessarily have a negative slope for the average productivity,
meaning that we are beyond its maximum. And marginal productivity
of hours is declining.

To understand the formula of the sign of the slope of the transition
path, equation (13) in a paper on unemployment and vacancy
dynamics with imperfect financial market (se \cite{carrillo2018unemployment}) implies that
aggregate firm profits and the number of vacancies must move in the
same direction along the equilibrium path. However, aggregate firms'
profits depend positively on the number of jobs and negatively on
the wage paid to workers. Both are inversely related to the unemployment
rate. The slope of the transition path then depends on how responsive
wages are to changes in the unemployment rate.

All the results \cite{di2019envelope} regard the monotonic behavior
of $h$, $F$, and $B$ with respect to their parameters; they are
proved based on the respective partial derivatives signs.

\index{Economic Growth}

\subsubsection{Economic Growth}

A total differentiation of equation (3) in a work on the systemic fragility of
finance-led growth (se \cite{bhaduri2015systemic})
yields the derivative of the output with respect to the capital gain,
and qualitative analysis of its sign follows.

\cite{sukharev2016structural}, investigating structural modelling of economic growth, defined the total by the law of velocities
change and their correlation and by the sign of change - velocity
increase or velocity decrease.

The effect of a change in $z$ on equilibrium employment and distribution
in \cite{tavani2017endogenous} depends on the partial derivative
sign. If it is positive, that is, if wages are more responsive than
productivity to the employment rate, an increase in labor market protection
lowers employment and productivity growth while raising the wage share.
Vice versa, if the sign is negative, an increase in z has a positive
effect on equilibrium employment but an adverse effect on the labor
share. Either way, workers face a trade-off between jobs and productivity
on one end and the wage share on the other. Such exchange contrasts
with the steady-state implications of the Goodwin model, where an
increase in employment protection would reduce employment but would
have no impact on income distribution.

In formula 20 in \cite{palley2017inequality}, if the Keynesian multiplier
stability condition holds, the numerator is positive, and the sign
of the slope of the IS depends exclusively on the sign of the denominator.
The authors then discuss the sign of the numerator and the denominator
due to workers' ownership and wage share. Furthermore, a discussion
is held regarding the sign of the rate of other parameters such as
the locus, capitalist managers' share of the wage bill, and the normal
capacity utilization.

\cite{carvalho2017debt}, in a work on debt-financed knowledge
capital accumulation, capacity utilization and economic growth, applies the signs of the derivatives in the
Jacobian matrix for a qualitative monotonicity analysis in the long-run
equilibrium discussion.

Throughout the discussion on the balanced-growth equilibrium with creative class
competition in a paper on Schumpeterian creative class competition,
innovation policy, and regional economic growth, \cite{batabyal2018schumpeterian} analyzes the monotonicity
of the R\&D expenditures with respect to other parameters extensively
based on the signs of its partial derivatives.

To analyse the behaviour in the long term, and therefore the stability
of their model model, \cite{perez2019economic} calculates its derivative
with respect to time $\frac{dG}{dt}$. The sign of the derivative
depends solely on the substitution parameter $\alpha$. Thus it is
obtained that if $\alpha\rightarrow0$ (that is when the elasticity
of the substitution is $1$), the rates of variation of capital and
production are constant over time. If there is substitution in production
($\sigma>1;-1<\alpha<0$), the rates of variation of capital and product
are increasing over time. If there is no substitution in production
($0<\sigma<1;\alpha>0$), the rates of variation of capital and product
are decreasing over time and tend to be zero in the long term.

The sign of the partial derivative of net exports with respect to
gross exports in \cite{dasgupta2020explains} depends on the precise
value of import propensity ($m$) compared to the critical value of
import propensity ($m_{c}$). Thus, the sign of the partial derivative
would be ambiguous as there is no guarantee that the import propensity
would be lower than this critical value.

The effect of integration and scale on the sign of the partial effects
is studied in \cite{ott2021institutional}.

In a work on the
Janus face of economic integration, \cite{foreman2021fertility} shows that the sign of the partial derivative
of child demand with respect to human capital is negative so long
as $m_{2}$ is less than $50\%$, which must be valid outside the
fourteenth century. The rising elasticity of substitution means the
effect of human capital reducing child demand increases with economic
development. This human capital effect is one contributor to the fertility
transition. As $m_{2}$ falls, there is a greater effect in absolute
value on the demand for children from a rise in human capital.

\index{Economic Inequality}

\subsubsection{Economic Inequality}

Proposition 4 in a study of inequality aversion in long-term contracts (see \cite{cato2014inequality}) proves monotonicity properties
of the function $\omega_{2}$ based on the agent risk. It is established
based on its partial derivative signs analysis.

\cite{nahuis2000rising} shows that the tension between non-rivalness
and appropriability of R\&D output is crucial for the sign of the
slope of the skill-demand curve. A necessary condition for an upward
sloping demand curve is the ability of firms to appropriate the intertemporal
returns from non-production activities.

In the theoretical discussion about fairness, mobility and position and their relevance to inequality aversion (see \cite{bergolo2019we}, the signs of the partial derivatives of
$\gamma$ with respect to $e$, $M$, and $P$, respectively, help
establish qualitative conclusions.

\index{Investments}

\subsubsection{Investments}

In a paper on stochastic skew in currency options, \cite{carr2007stochastic} finds valuable patterns from the implied volatility quotes. One of them is that the curvature
of the implied volatility smile is relatively stable, but the slope
of the smile varies significantly over time. The slope sign switches
several times in the sample. Therefore, although the risk-neutral
distribution of the currency return exhibits persistent fat-tail behavior,
the risk-neutral skewness of the distribution experiences substantial
time variation. It can be positive or negative on any given date.

In \cite{thwaites2015real}, a familiar expression shows that the
sign of the slope of the savings schedule in $\left\{ s,r\right\} $
space depends on the intertemporal elasticity of substitution $\frac{1}{\theta}$.
When this substitution elasticity is high (i.e., above unity), a fall
in interest rates causes a falling saving, as agents substitute away
from relatively expensive retirement consumption. Infinite-horizon
households pin the interest rate down at $r=\frac{1}{\beta}-1$ and
are thus equivalent to OLG households with linear period utility functions.
When the elasticity is below unity, retirement saving is akin to a
Giffen good: lower interest rates raise the savings rate out of wages,
as the desire to offset the negative effect of lower interest rates
on retirement consumption outweighs the higher its price. When the
elasticity is precisely one, these two effects cancel, and the savings
schedule is vertical.

As illustrated in \cite{leung2016speculative}, S\&P-GSCI carries
out rolling of the underlying futures contracts once each month, from
the fifth to the ninth business day. On each day, $20\%$ of the current
portfolio is rolled over in a process commonly known as the Goldman
roll. The S\&P-GSCI roll yield for each commodity is defined as the
difference between the average purchasing price of the new futures
contracts and the average selling price of the old futures contracts.
In essence, it is an indicator of the sign of the slope of the futures
term structure.

In \cite{skott2017weaknesses}, who investigates the weaknesses of wage-led growth, the derivative calculation
is often a step towards evaluating functions' trends, which in turn
help establish a qualitative discussion on the relationship between
parameters.

From a practical point of view, the asymptotic slope is a useful ingredient
for model calibration in \cite{pinter2017small}: E.g., if the market
slope is negative, then a simple constraint on the model parameters
forces the (asymptotic) model slope to be negative, too. The authors'
numerical tests show that the slope sign is reliably identified by
a first-order asymptotic approximation, even if the maturity is not
short. With the authors' formulas, the model parameters determine
the asymptotic slope (and its sign). For instance, the slope of the
NIG (Normal Inverse Gaussian) model is positive if and only if the
skewness parameter satisfies $\beta>-12$.

Annex 1.3 in \cite{de2017debt} presents the sign of the derivative
of equity value with regard to $z$. It is negative. We know that
$z$ increases with the additional growth in asset value $\eta$.
This growth in asset value first benefits the debt value. The authors
have shown that it induces a mechanism of transfer of value to creditors.
The equity value is also indirectly harmed. The mechanism is the following:
an increase in the asset growth rate will also increase the default
threshold value, $A_{b}$. From Equation (10) there, it is easy to
show that $\frac{dE}{dA_{b}}$ is negative. The economic sense is
straightforward: A rise in the threshold triggering a zero equity
value will result in a lower equity value.

The standard Keynesian stability condition, as stated in \cite{basu2018does},
is that savings are more responsive to changes in incapacity utilization
than investment, which translates into $I_{u}<s_{h}$. This assumption
ensures that the denominator in the expression for the local slope
of the $IS$ curve is positive. Hence, the sign of the slope of the
$IS$ curve depends on the sign of the numerator. When the numerator
is negative, i.e., $I_{h}-s_{u}<0$, the $IS$ curve is negatively
sloped, and the economy is in a stagnationist (wage-led) regime. An
increase in the profit share is associated with a decline in the capacity
utilization rate (because the local slope of the $IS$ curve is negative).
When the opposite happens, i.e., $I_{h}-s_{u}>0$, the $IS$ curve
is positively sloped, and the economy is in an exhilaration (profit-led)
regime. An increase in the profit share is associated with an increase
in the capacity utilization rate.

Proposition 2.2.5 in \cite{olsen2018optimal} states that the optimal
threshold of the option to invest in the second project stage, $P_{2}^{\ast}$,
is increasing in $\sigma$. It also proves that the first price threshold,
$P_{1}^{\ast}$, is increasing in $\sigma$ if $P_{1}^{\ast}\geq P_{2}^{\ast}$.
If $P_{1}^{\ast}<P_{2}^{\ast}$, then $P_{1}^{\ast}$ is increasing
in $\sigma$ if and only if a provided condition is met. It is proved
with the sign of its partial derivative.

In \cite{kraft2019consumption}, the relation between the terminal
condition $\pi^{\ast}\left(T\right)$ and the optimal demand $\pi_{\infty}$
in an infinite horizon setting is crucial for the question of whether
the stock demand is increasing or decreasing over the life-cycle.
Formally, the reason is that the form of the ODE (3.20) induces a
monotonic behavior of the demand since the slope is either decreasing
or increasing but cannot change its sign.

In \cite{shahzad2020clean}, the slope $b_{S_{j}}$ discriminates
as if the trend of $S_{j}$ is negative or positive. The sign of the
slope is used to ascertain that $x\left(t\right)$ has a positive
(negative) trend in $S_{j}$. Thus, the average fluctuation function
assesses asymmetric cross-correlation scaling properties when $x\left(t\right)$
exhibits piecewise trends. The trend-wise directional $q$-order average
fluctuation functions are calculated in equations 5 and 6.

Throughout \cite{cicerothree}, qualitative analyses are held where
the monotonicity of one parameter with respect to another is analyzed
based on the sign of the partial derivative. For example, following
equation 1.43, analyzing the sign of each term, there's an ambiguous
impact of variations on capital stock composition on the proportionate
rate of change of the same variable. The sign of this term depends
on the sign of $\frac{\partial u^{*}}{\partial k}$.

In proposition 1 in \cite{chang2021bonds}, the monotonic relation
between credit spread and asset volatility is analyzed in terms of
the partial derivative sign. Holding asset volatility constant, the
partial derivative of investment with respect to credit spread is
negative, and holding credit spread constant, the partial derivative
of investment with respect to volatility is positive. We thus provide
the elasticities of investment when observing asset volatility and
credit spread. Given Assumptions 1-5, the sign of these partial derivatives
matches the empirical results. The first term on the right-hand side
of equation (2) is negative due to the concavity of $k\left(\iota\right)$
in the denominator. All other terms are positive, and thus the sign
of the elasticity of investment to credit spreads is always negative.

\index{Economic Value}

\subsubsection{Economic Value}

\paragraph{Goods}

Throughout \cite{van2017economic}, and especially in the discussion
where the authors' analysis of the single batch fixed time problem,
lemmas and corollaries prove monotonicity properties regarding the
amount of the good the seller decides to supply with respect to other
parameters, based on the signs of its partial derivatives.

The vast majority of the propositions in a study of expanding distribution channels (see \cite{matsushima2017expanding})
have to do with the monotonicity of $w^{*}$, $q\left(w^{*}\right)$
or $\pi$. They are proved by inspecting the signs of the partial
derivatives of these variables.

In a study of the demand for caffeinated beverages (see\cite{liao2021essays}), qualitative monotonicity based
analyses are conducted and trend related theroems are being proved.
For example, lemma 1 states that the marginal effect of a change in
the probability of choosing size category $y$ due to a change in
the expected utility from size category $x$ is positive, i.e. $\frac{\partial Pr(y)}{\partial\omega\left(x\right)}>0$,
when the ratio of partials, $\frac{\frac{\partial Q_{2}}{\partial\omega\left(x\right)}}{\frac{\partial Q_{1}}{\partial\omega\left(x\right)}}$,
is sufficiently large, otherwise $\frac{\partial Pr(y)}{\partial\omega\left(x\right)}<0$.
Furthermore, $\frac{\partial Pr(y)}{\partial\omega\left(x\right)}>0$
is only possible when $\delta>0$, i.e. when consumers are forward-looking
in their decision-making.

\paragraph{Services}

\cite{alizadeh2017capacity} calculates the partial derivatives of
the expected duration of lay-up with respect to underlying parameters.
The signs of these derivatives indicate that everything else being
constant, the expected time of the lay-up period decreases as long-run
freight level and speed of mean increase. Still, the expected duration
of the lay-up period also increases as freight market volatility increases.
More precisely, the negative sign of the partial derivative of the
expected time to recovery with respect to the long-run mean of freight
rates implies that the lower the long-run mean of the freight rate,
the faster freight rates recover, and the expected duration of lay-up
will be shorter. Similarly, the negative sign of the partial derivative
of the expected time to recovery with respect to the mean reversion
rate of freight rates implies that the faster the speed of mean reversion,
the faster freight rates recover, and the expected lay-up duration
will be shorter. Finally, the positive sign for the partial derivative
of the expected time to recovery with respect to the freight rate
volatility implies that the higher the freight rate volatility, the
longer the adjustment to the long-run mean of freight rates and hence
the expected duration of lay-up.

Theorem 7 in a study of stability regions for a delay Cobweb model (see \cite{matsumoto2016stability}) suggests a formula for
calculating the sign of the partial derivative of $\gamma$ with respect
to $\beta$.

In a paper discussing a theory of social finance (see \cite{cornee2018theory}), propositions 1, 3, and 4 and lemma 2
are proved based on a sign analysis of partial derivatives, for example,
the derivative of SCFI relative to $e$.

\paragraph{Preference and Demand}

The necessary and sufficient condition for prices (gross equity premium)
to increase (decrease) with supply is determined by the sign of the
slope of the asset Engel curve in \cite{kubler2011theory}. This observation
allows the authors to derive:
\begin{itemize}
\item Sufficient conditions directly in terms of the representative agent's
risk aversion properties for general utility functions
\item Necessary and sufficient conditions for the widely used HARA (hyperbolic
absolute risk aversion) class.
\end{itemize}

In section 2.2 there, \cite{kim2012endogenous} discusses incorporating
vehicle choice in the monocentric model with an extensive comparative
analysis primarily based on the partial derivative signs of $p$ with
respect to different parameters such as $y$ and $\alpha$. In turn,
the derivative signs of $q$, $x$, and $u$ are analyzed throughout
the appendix sections.

Proposition 1 and 2 in a study of income pollution path (see \cite{figueroa2015beyond}) suggest sufficient
conditions for the monotonicity of the flow of the pollution with
respect to the capital in the additive and general case, respectively.

\paragraph{Pricing}

A comparative analysis of the buyers' distribution among contract
types is conducted in \cite{noll2004optimal} based on partial derivatives
signs analysis, in equations 40-45. The proportion of both buyer groups
decreases with a more considerable maximum potential loss. The number
of buyers opting for low-price contracts falls with a higher probability
of failure for low-quality goods. Still, the percentage rises with
increasing failure probabilities for high-quality goods, since then
ceteris paribus buying the high-price contract becomes less attractive.
As expected, the opposite holds for the high-price group. Alleviating
the assumption of socially optimal high-quality production (i.e. $\beta>c_{H}$)
renders the last derivative sign positive for some special cases (e.g.,
$\beta<c_{H}\leq\beta+2\pi HF$). In these cases, high-quality production
would not be socially optimal but still profitable for the seller.
Higher failure probabilities would lead to higher demand for contracts
with penalty clauses.

\cite{gaudin2016pass} shows that comparing both pass-through rates depends on three factors: demand
curvature, its derivative, and firms' bargaining power. In the canonical
case where $\theta=1$, however, only the demand curvature's derivative
sign matters, as indicated by corollary 1.

The very specific shape of $P_{3}^{*}\left(\cdot\right)$ stated in
lemma 5 in a paper on the unimodality of the price-setting
newsvendor problem with additive demand under risk considerations (see \cite{rubio2018unimodality}) makes the analysis easier,
especially if the sign of the slope at the point where this function
and $P_{2}^{*}\left(\cdot\right)$ intersect is known. For example,
if the slope at the joint is negative, then the mode of $P_{3}^{*}\left(\cdot\right)$
has already occurred when this function becomes part of $P^{*}\left(\cdot\right)$,
and therefore the maximum of $P^{*}\left(\cdot\right)$ will take
place at the interval where $P^{*}\left(z\right)=P_{2}^{*}\left(\cdot\right)$.
Conversely, if the slope at the joint is positive, the maximum of $P_{3}^{*}\left(\cdot\right)$
will occur in the section of $P^{*}\left(\cdot\right)$ where $P^{*}\left(\cdot\right)=P_{3}^{*}\left(\cdot\right)$.
This idea is illustrated in Figure 7 there.

\cite{diep2021cross} shows that when a positive prepayment shock
is value-increasing for the overall MBS portfolio, securities that
load positively on prepayment risk earn the highest returns in the
cross-section. On the other hand, when a positive prepayment shock
is value-decreasing for the overall MBS portfolio, securities that
load negatively on prepayment risk earn the highest returns. In other
words, the sign of the change in wealth of a specialized MBS investor
with respect to a positive prepayment shock changes over time. It
thereby changes whether MBS investors require additional compensation
to bear the risk that prepayment is too high or, conversely, too low.
The authors provide additional support for segmented markets for MBS
by demonstrating that the price of aggregate stock market risk is
negative for MBS, meaning that securities that load more positively
on systematic equity market risk earn lower returns on average.

\index{Natural Science}

\section{Natural Science}

Trends calculations are prevalent where one may expect the direction
to be of importance. For this reason, trends calculation are also used in Physics, Chemistry, and Biology.

\index{Physics}

\subsection{Physics}

\index{Chemical Physics}

\subsubsection{Chemical Physics}

\label{chemical_physics_subsubsection}

In \cite{martini2010gold}, a careful analysis of the anisotropy spectrum
as a function of the emission wavelength for excitation at $480nm$
shows that the $r$-value depends on the emission wavelength. Each
curve can be divided into three central regions characterized by a
change in their derivative sign: the blue edge, the mean range, and
the red edge. In the blue edge (for fluorescein roughly from $500$
to $507nm$), the $r$-value decreases rapidly as the wavelength increases.
In the middle range (between $507$ and $535nm$), the $r$-value
remains approximately constant. Finally, for emission wavelengths
superior to $535nm$ (red-edge), the anisotropy value re-increases
slowly.

In a work on Mooij rule and weak localization (see \cite{gantmakher2011mooij}), the derivative sign of the resistivity of alloys is shown to correlate
with the resistivity. 

The model predictions for benzene + cyclohexane in \cite{eslamian2012thermodiffusion}
were the best due to the less complex structure of this binary hydrocarbon
mixture. It shows that the proposed model, which assumes that the
sign of thermodiffusion coefficient is determined by the sign of the
derivative of the mixture viscosity with respect to concentration,
has the potential for further refinement and development for more
complex mixtures.

In \cite{batignani2015energy}, a work on energy flow between spectral components in 2D broadband stimulated
Raman spectroscopy, the associated time-dependent phase
induces a blue or red frequency shift, depending on the derivative
sign.

The slope sign at the tipping point in \cite{bodai2015global}, a paper on global instability in the Ghil–Sellers model, determines
whether the `vertical' order of the stable and unstable branches
of the diagram of $\Delta T$ with respect to that of $\left[T\right]$
flips in the vicinity of the tipping point. When it does flip, $\delta W\left(\mu\right)\left(\delta U\left(\mu\right)\right)$
becomes a convex (concave) function, while $fW\left(\mu\right)\left(fU\left(\mu\right)\right)$
remains concave (convex). The bifurcation diagram prompts that $\frac{d\Delta T}{d[T]}$
is negative. However, it should be relatively small because of the
sharp tipping point.

\cite{cina2016ultrafast} briefly explores an alternative strategy
for bringing quantum beats to light in transient-transmission data.
Even small-amplitude beats are expected to produce sign changes in
the derivatives of a spectrum with respect to inter-pulse delay. For
instance, the second derivative should be negative in the vicinity
of a peak or downward-curving shoulder in the signal and positive
near a trough or upward-curving shoulder. Figure 14 there shows a
plot of the sign of the second derivative of the calculated SE signal
at each probed frequency with respect to $t_{d}$ --- without background
subtraction. Figure 15 shows similar plots of the second derivative
sign for experimental transient-absorption signals from PC577 and
methylene blue.

In a study of universality of the critical point
mapping between Ising model and QCD at small quark mass, \cite{pradeep2019universality} shows that the relative orientation of the
slopes, i.e., the slope sign difference, is determined by the sign
of the Jacobian of the $\left(a,b\right)\rightarrow\left(\mu,T\right)$
mapping. It is positive in the case of the mapping without reflection
and negative otherwise. In that sense, it is topological. The authors
show how to determine the sign on Fig. 2 thereby comparing the phase
diagram in the vicinity of the tricritical point in $\left(a,b\right)$
coordinates with the standard scenario of the QCD phase diagram in
$\left(\mu,T\right)$ coordinates. The two graphs are topologically
the same: the first-order transition is to the right of the tricritical
point, and the broken (order) phase is below the tricritical point.
This means that the Jacobian of the $\left(a,b\right)$ to $\left(\mu,T\right)$
is positive (no reflection is involved). It means that, since $h=0$
slope is negative, the $r=0$ slope must be less steep, or if $\alpha_{1}$
itself is small, $\alpha_{2}$ could be slightly negative. In the
random matrix model, both slopes are negative and small (i.e., $\alpha_{1}>\alpha_{2}>0$
in the model).

The inversion curve in \cite{abdusattar2021joule} acts as a boundary
between the cooling and heating regions, and cooling (heating) does
not occur on the inversion curve. Therefore, we can distinguish between
the cooling and heating region by checking the sign of the slope of
the isenthalpic curves. The positive sign of slope stands for the
cooling region and the minus for the heating region. The authors conclude
that the temperature and pressure are different for different values
of $\omega$ and $M$. The inversion point moves from the positive
pressure to the direction of negative pressure when the $\omega$
changes from $-1$ to $\frac{1}{3}$. The temperature rises when the
pressure decreases in the heating region. By contrast, the temperature
is reduced as the decreasing of pressure in the cooling region. Moreover,
the cooling-heating region shrinks as $M$ grows.

\index{Optics}

\subsubsection{Optics}

The light rays' rotation around the focuses in a study of wave and ray spatial dynamics of the light field in the generation,
evolution, and annihilation of phase dislocations (see \cite{aksenov2002wave})
implies the presence of a slope near the focuses. The slope does not
change its sign (the rotation has the same direction); i.e., the derivative
with respect to the azimuthal angle does not change the sign in tracing
around the focus. When we return to the start point, the peak's height
differs from the initial value because of the constancy of the derivative
sign.

\cite{giuliani2005laser} shows that the interferometric waveform can be inverted
for either its increasing or decreasing branch. It requires that the
derivative sign of the actual displacement is known.

The finite recovery time of the bleached absorber in \cite{li2006maximum}
is presented as one of the possible mechanisms accounting for the
increase--maximum--decrease in pulse energy with the pumping rate
in $c_{w}$-pumped $Q$-switched solid-state lasers passively by analytically
evaluating the sign of the derivative of the energy with respect to
the pumping rate.

In a study of polar decomposition of the Mueller matrix, \cite{sanz2011polar} calculates, for each measured scattering region
(for normal incidence, and in $1{^{\circ}}$ steps), the difference
between the number of measurements with positive and negative local
derivative. A new parameter, $\Upsilon R-G$, is defined accordingly.

Figure 4b in \cite{roxworthy2018electrically} shows the motion phase
near the second-order flexural mode resonance. Dispersive coupling
produces the opposite phase at different signs of detuning, indicating
a change in the sign of the reflectance derivative $sign(\partial_{x}R)$
with detuning. Reactively coupled devices produce the same phase ($sign\left(\partial_{x}R\right)$)
regardless of detuning.

In a paper on the optics and optimal control theory interpretation of
the parametric resonance, \cite{schitov2019optics} shows that only the equality to zero of the ``conjugated''
function and its derivative sign initial value are important.

\index{String Theory}

\subsubsection{String Theory}

In a study of scattering-type scanning near-field optical
microscopy with low-repetition-rate pulsed light source through phase-domain sampling (see \cite{xiao2015phase}),there are two possible shapes of $bGC\left(x\right)$.
They are similar to the cases in the $GG$ ensemble. The only difference
is that the left endpoint of the $bGC\left(x\right)$ is at $x=q$
here. The difference between the two patterns is the sign of the slope
of the curves at $x_{max}$. So one can conclude that there must be
a transition line on the $\Phi-q$ plane on which the $bGC\left(x\right)$
curve has $\frac{dbGC\left(x\right)}{dx}=0$ at $x_{max}$.

From equation 4.5 in a study of inverse anisotropic catalysis in holographic QCD (see \cite{gursoy2019inverse}), it is possible to
relate to the sign of the slope to the jumps of the derivatives of
$\chi_{a}$ and $s$. For $x=0$ and $0<\frac{a}{\Lambda}<1$, the
jump $\Delta\chi_{a}$ is positive which agrees with $T_{c}$ decreasing
with $\frac{a}{\Lambda}$ in this region. At large values of $\frac{a}{\Lambda}$,
where $T_{c}$ increases with $a$, $\Delta\chi_{a}$ also has the
opposite sign.

In \cite{attali2019averaged}, a paper on averaged null energy condition and the black
hole interior in string theory, only one of the solutions, (3.1) and
(3.4), survive an arbitrarily small slope of the dilaton. The sign
of the slope determines which one survives. It suggests that if setting
$Q=0$, while not specifying which of the limits are taken, $Q\rightarrow0^{+}$
or $Q\rightarrow0^{-}$, there are no folded string solutions at all.
A simple way to see this is the following. Suppose that setting $Q=0$,
the Virasoro constraints are (3.2). Taking the derivative of (3.2),
we get that the derivatives' product is zeroed. However, if we glue,
say, $\partial_{+}X_{0}=1$ with $\partial_{+}X_{0}=-1$ at a certain
point, then at that point the second derivative blows up, which is
not consistent with (3.5) and the fact that at that point we have
either $\partial_{+}X_{0}=1$ or $\partial_{+}X_{0}=-1$.

\index{Astronomy}

\subsubsection{Astronomy}

In a study of emission beam geometry of selected pulsars derived from
average pulse polarization data, (see \cite{everett2001emission}) the inner and outer line of sight trajectories are defined based on the sign of the slopes of the first and second
derivatives of $\psi$ with respect to $\varphi$.

The solution in \cite{dabrowski2003phantom}, a paper on phantom cosmologies, is either a monotonic
expansion or a monotonic collapse, as a function of the derivative
sign.

In a study of entropy of static spacetimes and microscopic density of states, \cite{padmanabhan2004entropy} shows that the sign of $\beta$ depends on
the sign of the derivative of $N$ near the horizon. For example,
it is positive for the Schwarzschild black hole horizon while negative
for the de Sitter horizon.

In \cite{eiroa2004cylindrical}, a study of thin-shell wormholes associated
with global cosmic strings”, the throat collapses to zero radii,
remains static, or expands forever, depending only on the sign of
its initial velocity.

\cite{stuchlik2005aschenbach} shows that the possibility to have three changes of the sign of $\frac{\partial V\left(\phi\right)}{\partial r}$
in constant specific angular momentum tori is limited from bellow for black holes, and from above for naked singularities.

Should a rotating motion have been considered, \cite{xia2005time}
would have had an interesting result that the transverse velocity
changes its sign during the decaying phase of macro-spicules.

In a study of fast magnetoacoustic waves in curved
coronal loops-I. Trapped and leaky modes (see \cite{verwichte2006fast}), the sign of the third term in Eq. (7)
dictates the propagatory nature of the solution. Where $V_{ph}\left(r\right)>V_{A}\left(r\right)$,
this term is positive and the solution is oscillatory. On the other
hand, where $V_{ph}\left(r\right)<V_{A}\left(r\right)$, this term
is negative and the solution is evanescent. It is equivalent to considering
the slope of the AlfvÃ©n frequency profile, $\omega_{A}=V_{A}\frac{m}{r}$.
For an AlfvÃ©n speed profile of the form (4), there are three possibles scenarios:
\begin{itemize}
\item $\alpha<-4,\frac{d\omega_{A}}{dr}>0$: the solution is always oscillatory
for small $r$ and evanescent for large $r$.

\item$\alpha=-4,\frac{d\omega_{A}}{dr}=0$: the solution is either oscillatory
or evanescent, depending whether the phase speed is above or below
the AlfvÃ©n speed profile.

\item $\alpha>-4,\frac{d\omega_{A}}{dr}<0$: the solution is always evanescent
for small and oscillatory for larger.
\end{itemize}

The temporal evolution of the wormhole throat in \cite{bejarano2007thin}
is determined by the sign of its initial velocity. If it is positive,
the wormhole throat expands monotonically; when it is negative, the
wormhole throat collapses to the core radius in finite time; and,
if it is null, the wormhole throat remains at rest.

The derivative of the bulk field in \cite{setare2009braneworld} maintains
its sign during the cosmological evolution if and only if the coupling
function always lies on one side of the phantom divide.

In a study of solar granulation from photosphere to low
chromosphere observed in Ba II 4554 Å line (see \cite{kostik2009solar}), the contrast sign reversal of granulation occurs at heights around 200-300km. At the same heights, on average,
the velocity sign reversal also occurs.

In \cite{sheminova2009origin}, the gradients of matter velocity,
magnetic field strength, and inclination, which are the main cause
of usual classical asymmetry, remain important factors in forming
unusual profiles. The number of profile lobes is proportional to the
frequency of changes in the magnetic field strength gradient sign
along the line of sight.

The descending elements in a work on the properties of convective motions in facular regions (see \cite{kostik2012properties}) can first
change their contrast and then the direction of motion at a lower
height.

The simplified analysis in section 4 in \cite{triana2015internal}
is helpful because it provides a first idea of the sign of the slope
of an unknown rotation profile just by plotting the splittings and
the period spacings and allows to see if they vary in phase. In this
case, it is not very clear if they vary in phase or not, and so instead,
the authors make use of the linear trends of the splittings as indicated
with the linear fits (dotted lines) in Figure 6 there. When period
spacings and splittings vary in phase, the linear trend is downward
(negative slope). Conversely, the linear trend is upward when period
spacings and splittings are in anti-phase (positive slope). The observed
splittings of KIC 10526294 have an increasing trend, as shown by the
linear fit (dotted red line) in Figure 5. It is thus associated with
a decreasing rotation rate.

The extension to a positive slope has been worked out by Furlanetto
\& Piran. Note that for the halo barrier, ellipsoidal collapse predicts
a positive slope. However, section 4 in a paper on spherical evolution
for modelling void abundances (see \cite{achitouv2015testing})
shows that for the void threshold, it seems that a negative slope
is in better agreement with the Lagrangian barrier.

The circular orbits in \cite{tursunov2016circular} are analyzed using
the so-called force formalism by separating the circular orbits into
four qualitatively different classes according to the sign of the
canonical angular momentum of the motion and the orientation of the
Lorentz force.

In \cite{mishin201727}, a work on preonset phenomena, two main onsets,
field-aligned current systems, and plasma flow channels in the ionosphere and in the
magnetosphere, a change in the derivative sign provides
the timing of EO1 and EO2.

In addition to its amplitude, a crucial feature of the result (4.23)-(4.24)
in \cite{garcia2018primordial} is that the sign of the entropic mass
squared $m^{2}$ can be positive or negative --- with significant
observational consequences --- depending on whether the slope of
the potential is positive or negative respectively. It is unusual
in inflationary models to find a physical quantity that depends on
the sign of the slope of the potential. In standard single-field inflationary
models, one can arbitrarily change the definition of $\varphi$ into
$-\varphi$, and hence the sign of $V$, without physical consequences.

The non-Keplerian velocity along the planetary wakes in \cite{casassus2019kinematic}
undergoes an abrupt sign reversal across the protoplanet. Also, the
morphology of the flip in HD 100546 is similar to that predicted for
disk--planet interactions. Especially in its azimuthal extension
and in the sign of the velocity deviations.

In \cite{steinwachs2020higgs}, the value $t_{0}$ is related to a
value $t_{crit}$ at which the $RG$ improved effective $EF$ potential
has an inflection point. There are three qualitatively different scenarios,
depending (also) on the sign of the slope at that point: Universal,
Critical, and Hilltop.

\cite{arguelles2021formation}, in a study of the formation and stability of fermionic dark matter haloes
in a cosmological framework, demonstrates that changes of stability
for any individual perturbative mode will occur, at a given point
in the series, if and only if two specific conditions are met:
\begin{itemize}
\item The slope of the $\partial_{x}f$ vs. $x$ curve is infinite (i.e.
the tangent is avertical line)
\item The sign of said slope shifts at that point.
\end{itemize}

\paragraph{Gravity}

From Theorem III.1 in \cite{maciel2015cosmological}, the cases where
gMcVittie contains only a white hole are restricted to cases where
$\dot{\xi}\rightarrow0^{+}$, that is, when the derivative goes to
zero from positive values. Symmetrically, the only black hole case
corresponds to $\dot{\xi}\rightarrow0^{-}$. It is because the sign
of $\dot{\xi}$ is the sign of the slope of $r^{-}\left(t\right)$
for large $t$ since the denominator is positive at the $r$\textminus horizon.
The authors assume that the denominator is not degenerate. That is,
$r^{+}$ and $r^{-}$ do not coincide. When its slope is negative,
it is easier for null rays to reach the horizon from above the apparent
horizon, which corresponds to the regular region, which leads to a
black hole. In the same manner, when the slope of $r^{-}$ is positive,
it is easier for null rays to traverse from below the apparent horizon,
laying in the anti-trapped region, characterizing the limit of a white
hole region. In both cases, if the absolute value of function $\xi\left(t\right)$
decreases fast enough, we have the case in which the limit surface
corresponds to a pair of white-hole/black-hole horizons, separated
by a bifurcating two-sphere. The cases in which the limit surface
has only one character are those in which the $\xi$ function does
not decrease faster than the exponential that modulates it in Eq.
(41). Upon building models, one first needs to choose if the slope
of $r^{-}$ will be positive or negative for large times, which by
Eq. (42) means choosing the sign of $\dot{\xi}$ for large times.
Inspecting Eq. (24), one observes that the term proportional to $M$
is positive by our initial assumptions. The sign of $\dot{H}$ can
be either plus or minus, but physically realistic models usually correspond
to $\dot{H}<0$. It means that we can tune the functions $m\left(t\right)$
and $H\left(t\right)$ so that the leading term for large $t$ is
positive or negative.

\cite{chabab2018joule} notes that when expanding a thermal system
with a temperature $T$, the pressure always decreases, yielding a
negative sign to $\partial P$. In this context, we can consider two
different regimes with respect to the so-called inversion temperature,
defined as the temperature $T_{i}$ at which the Joule-Thomson coefficient
vanishes $\mu_{JT}(T_{i})=0$: If $T<T_{i}$ ($T>T_{i}$), then the
Joule-Thomson processus cools (warms) the system with $\partial T<0$
and $\mu_{JT}>0$ ($\partial T>0$ and $\mu_{JT}<0$) respectively.
When the system temperature tends to $T_{i}$, its pressure is referred
to as the inversion pressure $P_{i}$, so defining a special point
called the inversion point ($T_{i},P_{i}$) at which the cooling-heating
transition occurs.

The expression of $CV$ in \cite{nakarachinda2021effective} is very
long and difficult to consider. It is not worth showing. However,
some of its behavior is consequently known from the sign of the slope
of the temperature. The divergent points of the heat capacity are
directly obtained from the points at which slopes of the temperature
vanish. Hence, there is no divergence in $CV$ for $\eta>\eta_{0}$
while two divergent points appear when $\eta<\eta_{0}$. It is easy
to check that $M$ is a monotonically increasing function in $u$
for fixed $V$. The sign of $CV$ is thus the same as that of the
slope $\frac{\partial T}{\partial u}$. One can conclude that there
is no locally stable range of $u$ for $\eta\geq\eta_{0}$, but there
exists the locally stable range of $u$ for $\eta<\eta_{0}$ as shown
in Fig. 13 there.

\index{Acoustics}

\subsubsection{Acoustics}

\cite{lei2014novel} investigated the effect of the sensor substrate
size on the sensitivity through analyzing the correspondence between
the regression function monotonicity and its partial derivative sign.

Two types of thermal loads are investigated in {[}417{]}, namely $q=Bv$
and $q=Qsgn\left(v\right)$. This type of surface heating represents
an abrupt change of heat flux in space as the beam moves through the
upper half-plane towards the lower half-plane. The Euler-Bernoulli
theory assumes that the flexural displacement $v$ is homogeneous
across the cross-section, while the stress and the temperature in
the presence of thermoelastic coupling are not. It follows that, when
considering these two types of heating, once the geometric center
of the cross-section is positively (negatively) displaced, i.e., $v>0$
($v<0$), the whole cross-section experiences surface heating (cooling),
as shown in Fig. 1(d) and (e) there.

\index{Geophysics}

\subsubsection{Geophysics}

In a study of correlated
grace monthly harmonic coefficients using pattern recognition and neural networks, \cite{piretzidis2016identification} shows that the number of derivative sign changes is one of the extracted features. 

The required input values in a paper proposing automatic boundary layer algorithm (see \cite{poltera2017pathfinderturb}) are
the profiles of $\Theta$ and the wind. The stability conditions,
essential for choosing the correct threshold value, are derived from
the sign of the slope of the linear fit of $\Theta$ in the first
$30m$.

The relationship between the time and parallel position x on an orbit
in \cite{hutchinson2018prediction} explicitly relies on the velocity
sign.

In \cite{iverson2019basal}, equations that differ slightly from (32)
apply for cases with adverse surface slopes (i.e., slopes with $\beta<0$).
If $\beta<0$ and $tan\beta>\left(1-\frac{1}{\kappa}\right)tan\theta$
each apply, then basal shear stresses on a smoothly sloping bed and
staircase treads have opposite signs, such that $\tau>0$ and $\tau<0$
each apply. In this case a derivation that parallels the derivation
of (32) yields the result $tan\varphi_{tread}=-\left[\kappa\tan\theta+\tan\left(\varphi-\theta\right)\right]$.
This result and (32) can be consolidated into equation 33, where $sgn\left(\beta\right)$
denotes the sign of $\beta$ (Table 4). This equation implies that
$\tan\varphi_{tread}\geq0$ is always satisfied, because the sign
of $\beta$ is the same as that of $\kappa\tan\theta+\tan(\varphi-\theta)$
for scenarios that satisfy a limiting equilibrium force balance.

According to a paper comparing estimators of the conditional
mean under non-stationary conditions (see \cite{vogel2020comparison}), the sign of the bias
associated with FM will depend on the sign of the slope coefficient
$\beta$. For positive (negative) trends, $E\left[FM\right]$ will
generally be smaller (larger) than its true value, with that bias
increasing as both $f$ and $n$ increase.

\index{Climate}

\subsubsection{Climate}

The effect of initial THC strength on the sea ice effect can be seen
in Fig. 9b in \cite{levermann2007role}: the temperature effect is
also reduced for potent initial THC. In this case, however, it has
the inverse influence on the THC weakening because the sea ice effect
tends to strengthen the THC (Fig. 6). Thus the modulation of the THC
weakening through the sea ice effect has the correct qualitative behavior
to explain the slope sign in Fig. 2. Further, the authors suggest
separating permanently ice-free areas ($20^{\circ}-60{}^{\circ}N$;
curve with stars) from regions affected by changes in sea ice cover
($60^{\circ}-80{}^{\circ}N$; curve with solid circles). It allows
identifying the effects dominating the sign and the slope of the curve
with diamonds in Fig. 11. The change in heat loss is negative for
the more southern region, which determines the sign for the entire
North Atlantic and is consistent with the decrease in $T_{o}-T_{a}$
(Fig. 5). Finally, the authors hold an extended qualitative discussion
of the sign of the slope in section 5.

\cite{mcewing2015environmental} shows that as the depth of the water table
increases, the relationship between $CH_{4}$ emission and soil temperature
switched from negative to positive, with the sign of the slope of
the relationship changing near the point where the water table is
just above the soil surface.

In \cite{ross2015sea} the sign of the slope $\beta$ is positive,
consistent with the hypothesis that a northward displacement of the
Gulf Stream will increase oceanic salinity offshore on the Mid-Atlantic
Bight and drive more saline water into the estuary. The terms are
also positive at all of the remaining USGS locations except Chester.

As stated in \cite{song2018seasonally}, during AMJ, the Northern
Hemisphere is rapidly warming up ($\frac{dT}{dt}>0$), and the Southern
Hemisphere is cooling down ($\frac{dT}{dt}<0$). In contrast, JAS
is the warmest and coldest season in the Northern Hemisphere and Southern
Hemisphere, respectively, with a weak temperature tendency.

In subsection 2.2 in \cite{setiawan2021climate}, the climate's trend
is analyzed using a linear equation to get its slope. The slope sign
indicates whether the trend is increasing (positive) or decreasing
(negative). Furthermore, the Mann-Kendall Test was used to judge whether
the trend was significant or not, subject to a confidence limit.

\index{Nuclear Physics}

\subsubsection{Nuclear Physics}

The low-energy behavior of the $S$-factor in a paper on low energy behavior of the astrophysical S-factor
in radiative captures to loosely bound final states (see \cite{mukhamedzhanov2002low})
is defined by the sign of the derivative of $F\left(E\right)$.

The Extremum Seeking Control Algorithm in \cite{agarwal2016integration}
measures the sign of $\frac{dy}{dt}$, whereas the resulting dynamics
are governed by $\frac{dy}{dx}$ (formula 29).

For MPG-6 graphite with the close filler and binder crystallite sizes,
irradiation under the same conditions does not change derivative sign
on the dose dependences of elastic modulus. Its absolute value grows
monotonically. This behavior of the \textquotedbl single-phase\textquotedbl{}
MPG-6 graphite under irradiation is observed in \cite{gurovich2019radiation}
at much more significant volume changes (approximately four times)
compared to GR-280 graphite. It contradicts the model according to
which after the exhaustion of the accommodative ability of graphite
to compensate for dimensional radiation changes, a decrease in density
and strength characteristics should be observed (accommodation mechanism).
A derivative sign change on the dose dependence of elastic modulus
for GR-280 graphite at irradiation temperature is observed upon reaching
a neutron fluence.

In \cite{cooper2021dual}, for counter-propagating beams, the same
effect is present, but the slope sign is reversed. Since both co-propagating
and counter-propagating beams are present in the vapour cell, the
dual-frequency optical pumping effects produce diagonal line features
with positive and negative slopes, creating the observed grid patterns.

\index{Electromagnetism}

\subsubsection{Electromagnetism}

The results in \cite{jung2009experimental}, a paper on sensing capability of an electromagnetic
induction system for an MR fluid damper-based control system, show that the electromagnetic interference system
can be used as a ``velocity-sign'' sensor because the sign change
of the emf signal agrees well with that of the velocity signal.

Based on the derivative sign, cite{mende2012new} shows that the resistance can induce either receiving energy from the outside or omitting it
to external chains.

The binary barcodes in \cite{wang2013raman} were generated based
on the second derivative spectra. A binary value ($0$ or $1$) was
assigned to each calculated second derivative spectral data point
primarily based on the sign of the second derivative, i.e., $1$ for
positive second derivatives (upward curvature), and $0$ for negative
second derivatives (downward curvature).

The electrons in \cite{kuzikov2014flying} wiggle in the $r_{f}$
field of the first spatial harmonic with the phase velocity directed
in the opposite direction in respect to the bunch velocity so that
particles can irradiate high-frequency Compton's photons.

The sign of the slope can be used to assign the neutral state of the
molecule in the stability plots even if no ZFS is detected in the
Coulomb blockade regime or the SET, as illustrated in \cite{burzuri2015observing}.
Eight samples show a positive slope ($\Delta S=-1/2$), and four offer
a negative slope ($\Delta S=1/2$).

In \cite{konoplev2020design}, a study of of frequency selective conductive materials
for electromagnetic fields control, the second derivative sign changes
following the first foil becoming partially or fully transparent.

\paragraph{Magnetism}

\cite{trukhanov2011peculiarities} applies the Banerjee criterion and shows that in the ferromagnetic state, the positive value,
throughout the all concerned interval, of the slope of the tangent
to the Arrott isotherms means the magnetic phase transition of the
second order, while the change of the tangent slope sign from positive
to negative means the phase transition of the first order. A positive
value of the slope of the tangent corresponds to the increase of the
Arrott isotherm or, equivalently, a positive value of the derivative
$\frac{d\left(M_{2}\right)}{d\left(H/M\right)}$ (Fig. 4b).

The transition to the paramagnetic state in \cite{kourov2015specific}
is accompanied by a change in the primary scattering mechanism of
conduction electrons. It explains the experimentally observed change
in slope sign of the dependence $S\left(T\right)$ at temperatures
$T>T_{C}$ in the studied half-metallic ferromagnets. At low temperatures
$T<T_{C}$, the decisive factors are the mechanisms of elastic scattering
of conduction electrons and the specific features of the band structure.
In the paramagnetic state, the temperature dependence of the thermoelectric
power is determined by the mechanism of the inelastic scattering of
charge carriers. It leads to a change in the slope of the curve $S\left(T\right)$
near the Curie point.

According to Banerjee's criteria, the slope sign of the $\frac{H}{M}$
versus $M^{2}$ curve in \cite{ho2015magnetic} gives information
related to the nature of the FM-PM transition. If all the $\frac{H}{M}$
versus $M^{2}$ curves have a positive slope, the magnetic transition
is of second-order; if some of the $\frac{H}{M}$ versus $M^{2}$
curves show a negative slope at some point, the transition is of the
first order. Clearly, at some temperatures, the $\frac{H}{M}$ versus
$M^{2}$ curves of $La_{0.6}Ca_{0.4}MnO_{3}$, show negative slopes
at low-magnetic fields ($H<10kOe$) but positive slopes at high-magnetic
fields ($H>10kOe$). Particularly around the TC, the slopes are positive
over the entire field range. It indicates the existence of the tricritical
point sets a boundary between FOMT and SOMT in the undoped sample.

The mode number in \cite{mink2016toroidal} is given by the slope
of the best fitting line, calculated from a least square regression.
In this context, the slope sign determines the direction of the propagation
in the laboratory frame. A positive (negative) slope is connected
to a propagation in the ion(electron) - diamagnetic or co(counter)-current
direction. In the presented case of two core modes, a positive slope
reproduces the data better than a negative one.

Although the negative sign of the slope in an Arrott plot in \cite{bustingorry2016second}
can be considered a clear signature of an underlying first-order transition,
Banerjee made an important observation: the theory predicts the value
of the (negative) slope to increase with increasing temperature. In
fact, it follows from Eq. (35) that the slope of the Arrott, when
$m^{2}\rightarrow0$, changes sign at $\tau=\tau_{2}>\tau_{c}>\tau^{\ast}$.
More importantly, most of the subsequent works using the Banerjee
criterion to identify first-order transitions show that the (negative)
slope of the Arrott plot, $S=\frac{\partial\left(H/M\right)}{\partial\left(M_{2}\right)}$,
decreases when increasing the temperature.

The nature of the magnetic phase transition in \cite{hussain2017magnetocaloric}
is determined by using Banerjee's criterion, according to which the
slope sign of $\frac{H}{M}$ vs. $M_{2}$ (Arrott plot) gives information
about the order of FM-PM transition. In this work, the standard Arrott
plots were derived from the M-H curves and shown in Fig. 6. Positive
slopes can be seen in all $\frac{H}{M}$ vs. $M_{2}$ curves over
the whole field interval, implying that the magnetic phase transition
in all the investigated samples is of second order.

The compounds in \cite{edstrom2017magnetocrystalline} differ in the
sign of the variation of the orbital magnetic moment with $\theta$,
although they both have the same sign of $K_{1}+K_{2}$. In a plot
of the energy as a function of $\theta$ vs. the anisotropy, this
appears as a difference in the curves' slope sign.

\cite{inghirami2020magnetic} tries to settle the disagreement between
theory and experiment on the sign of $\Delta v_{1}\left(y\right)$.
It is possible that in the authors' formalism, there is a delicate
interplay between the properties of the magnetic field in the medium,
related to the rate of expansion in comparison to the decrease in
the magnetic field with time, which the inclusion of temperature-dependent
conductivity and viscosity might drastically alter. Another possible
source of error might be the prescription to determine charge-dependent
spectra, which assumes a chemical potential to modify the particle
species abundances without considering any modification of the momentum
distribution due to the electromagnetic field.

The contribution from the interstitial region in \cite{tran2020shortcomings}
is one order of magnitude smaller and has an opposite sign (negative),
which is due to the reverse polarization of the $4s$ electrons. The
difference between the two functionals is not uniformly positive or
negative; some lobes (which differentiate orbitals) have opposite
signs. It is visible for the Co atom in FeCo, for instance.

\index{Condensed Matter Physics}

\subsubsection{Condensed Matter Physics}

The solution in equation 16 of a paper on longitudinal resistance of a quantum Hall system
with a density gradient (see \cite{ilan2006longitudinal}) describes
a current concentrated within a distance of order $\left|\ell x\right|$
of one edge of the sample. Which edge this is depends on the sign
of $\ell_{x}$ (i.e. the sign of $\frac{\partial\rho_{xy}}{\partial x}$
since $\rho_{xx}$ is positive). For $\ell_{x}>0$, the current is
concentrated close to $y=0$; for $\ell_{x}<0$, the current is concentrated
close to $y=w$. Further, for $w\gg\ell_{x}$ the side along which
the current flows is determined by the sign of $\ell_{x}$.

\cite{fan2020gap}, in a paper on gap and embedded solitons in microwave-coupled binary condensates,  discusses the Vakhitov-Kolokolov (VK) or anti-VK
criteria, which relates the slope sign to the necessary stability
condition for solitons supported by the self-attractive or repulsive
nonlinearities. It is valid in the system presented when both the
nonlocal and local nonlinearities are self-repulsive. Indeed, the
families satisfy the anti-VK criterion, $\frac{d\mu}{dN}>0$, and
are certainly completely stable. On the other hand, when the solitons
are supported by the combination of the nonlocal repulsion and contact
attraction, the VK/anti-VK criterion does not hold. The reason is
that it is not possible to identify the dominant nonlinear term: the
change of the sign from $\frac{d\mu}{dN}>0$ to $\frac{d\mu}{dN}<0$
does not lead to destabilization of the solitons (non-compliance with
the VK criterion occurs in other models too).

For smooth functions, the order of accuracy for the forward divided difference (FDD) ($O\left(\Delta t\right)$)
in a paper on quantifying the dynamics of topological defects in active nematics (see \cite{morgan2020quantifying}), is less than the centered divided
differences, but there is no guarantee that our function is smooth.
Then the authors suggest recalculating with the forward divided differences
and removing the outliers (Fig. A.24 there). The behavior is an increased
slope with the FDD. After applying the outlier removal scheme, the
forward divided differences with outlier removal (FFDOR) still have
an overall \% increase from the initial results reported (Table A.4
there), but the slope sign is still unaffected. If we analyze every
other frame due to concerns of oversampling, both derivative approximation
schemes continue to have positive slopes.

\index{Dynamics}

\subsubsection{Dynamics}

\paragraph{Classical Mechanics}

The direction of the peaks in the contact mode in \cite{bhushan2004surface}, a work on surface topography-independent friction measurement
technique using torsional resonance mode in an atomic force
microscope (AFM), depends on the slope sign. However, in the case of the amplitude of the tip motion (TR) mode, it
is always downward, and the surface slope's pattern does not correlate
with that of the TR amplitude. Further, there is a sign reversal in
the surface slope in the Trace and Retrace scan, and the sign reversal
in the friction force only occurs in the contact mode.

As illustrated in \cite{ramdas2013algorithmic}, a study of algorithmic connections between active learning and
stochastic convex optimization, there may be irregularities
with a positive slope in a real pile of grains due to erosion. In
this case, the velocity must depend on the slope sign, or else we
will have avalanches climbing up the pile at the points with positive
slope, with the same velocity as in the negative slope side. To correct
this defect, we considered the equation for v as stated in equation
15.

In the sensitivity analysis method in \cite{skalna2008systems}, a work on systems of fuzzy equations in
structural mechanics, the
knowledge of the derivative sign of the function allows us to calculate
the upper and lower bound of the solution by using the endpoints of
the interval.

$Pos\left(\omega^{2}D\right)$ and $Pos\left(\omega^{2}G\right)$
are introduced as a function of distance from the edge in a work on stress transfer mechanisms at the submicron level
for graphene/polymer systems (see \cite{anagnostopoulos2015stress}).
There is a gradual change of the slope sign over a distance of $1.5\mu m$
from both edges. It indicates that most of the compression is gradually
relaxed, and the flake in that region is subjected to tension. At
higher strain levels, the region from $2$ to $4.5\mu m$ appears
to be free of residual strain and shows the highest rate of tensile
stress takeup. The region on the right-hand side of the flake is already
in compression. Therefore, it lags behind the rest of the flake.

The backaction force in \cite{okamoto2015cavity} is dependent on
strain and the displacement, $z$. In this red-detuned case, the sign
of force gradient is negative. This negative $\nabla F_{p}$ and the
corresponding time delay result in the efficient amplification effect
around the mechanical resonance frequency while reducing the damping
factor. In contrast, when the photon energy is blue detuned, the upward
bending decreases the backaction force, whereas the downward bending
increases. Therefore, $\nabla F_{p}$ is positive in this blue-detuned
case, and it leads to the efficient damping effect. The above detuning
dependence is the opposite of sideband amplification/damping. In this
excitonic optomechanics, the feedback is caused by strain-induced
modulation of the number of $e-h$ pairs. Thus, the slope sign in
the absorption (PLE) spectrum and the sign of the piezoelectric coefficient
determine the polarity.

Theorem 1 in \cite{de2021telegraph} states that the conditional probability
law of $\left(X\left(t\right),V\left(t\right)\right),t\geq0$, depends
on (also) the sign of the velocity. Theorem 2 further states that
the bounded linear operator also depends on the velocity's sign, which
reflects in the following expected value calculations, e.g., in Eq.
3.13.

\paragraph{Quantum Mechanics}

Without loss of generality, one can set $t^{\ast}=0$ and restrict
the attention to factorized pure initial states in \cite{benatti2003environment}.
In other words, the two subsystems, initially prepared in a state
$\rho(0)=\tilde{\rho}(0)$, will become entangled by the noisy dynamics
induced by their independent interaction with the bath if $E(0)=0$
and $\partial_{t}E(0)<0$, for a suitable vector $|\psi>$.

\cite{godoy2005effects}, investigating the effects of nonparabolic bands in quantum wires, shows that if the condition on the derivative sign is not met, the energy is increased or decreased. The process iterates
until the necessary conditions are fulfilled.

\cite{achuthan2011synaptic}, working at the synaptic and intrinsic
determinants of the phase resetting curve for weak coupling, shows that the result that the sign of
the slope of phase resetting curve $H'\left(\varphi\right)$ at zero
is sufficient to give the stability of synchrony for identically coupled
identical oscillators, which turns out to help develop an intuition
for when synchrony can occur.

In \cite{budich2016dynamical}, the change in the dynamical topological order
parameter (DTOP) $\Delta\nu D\left(t_{c}\right)$
in the vicinity of a critical time $t_{c}$ can be directly related
to the sign of the slope $s_{kc}$ at the critical momentum. This
result affords an intuitive geometric interpretation: critical momenta
are located on the equator of the relative Bloch sphere. $\Delta\nu D\left(t_{c}\right)$
is then directly related to whether $df\left(k\right)$ traverses
the equator of the relative Bloch sphere from the northern to the
southern hemisphere ($sgn\left(s_{kc}\right)=-1$) or from the southern
to the northern hemisphere ($sgn\left(s_{kc}\right)=+1$) at the critical
momentum.

The derivative sign of the band structure in \cite{baireuther2016scattering}, a work on the scattering theory of the chiral magnetic effect in a Weyl semimetal:
interplay of bulk Weyl cones and surface Fermi arcs,
is explicitly used in several formulas.

The sign of the Seebeck coefficient in \cite{gehring2019single} is
given by the slope sign of $T\left(\varepsilon\right)$ at the Fermi
energy. It can be used to determine if the highest-occupied molecular orbital or the lowest-unoccupied molecular orbital dominates
transport.

\cite{hollestelle2021some} assumes $w\left(k\right)$ changes during
$\Delta t$, i.e. $w\left(k\right)=w\left(k\right)\left(t\right)$
does not remain the same function, and the sign of the slope $w'=w'\left(t\right)$
changes during $\Delta t$. It implies center a changes within the
$k$-domain and average momentum $k_{changes}$ accordingly during
$\Delta t$. A change of $w'$ can only be assured by measurement.
For instance, a change from $a$ to the opposite $-a$, when reflection
occurs of the original wave during $\Delta t$ and momenta $k$ and
$k_{change}$ to their opposites. It depends on the sign of $w'$
whether $\left|\Delta k_{-}\right|.\left|\Delta q\right|\approx c_{-}$
is more or less than $1$. When a sign change occurs during $\Delta t$
and one assumes the $\Delta t$ time interval average $\left\langle c\text{\_}\right\rangle =1$,
the description is within average similar to the time-independent
situation with $w'$ equal to zero.

The transition time in \cite{vashistha2021transition} is proportional
to the slope of the curve in Fig. 3(b). Thus the sign of the transition
time is solely determined by the slope sign. Pointedly the slope of
the curve is positive when $\varepsilon\geq0.5$ and in this region,
the phase is leading with the energy, giving a positive value of $\tau$.
The slope is negative when $\varepsilon\leq0.5$ and here the phase
is lagging with the energy, giving a negative value of $\tau$.

\paragraph{Fluid dynamics}

\cite{fistul2002symmetry}, investigating symmetry broken motion of a periodically driven brownian particle, goes beyond the adiabatic limit ($\omega=0$)
and explains the peculiar reversal of the velocity sign found previously
in the numerical analysis.

The parameter $\sigma=sgn\left(u\right)$ from the mass conservation
equation (eq. 24) in \cite{perazzo2003thin} is the opposite of the
sign of the slope of the free surface in the $x$-direction, measured
with respect to the horizontal, not with respect to the plane.

Lemma 3.1 in a study of Steady Euler–Poisson systems, \cite{jerome2009steady} proves that $\varepsilon$ achieves
a unique positive minumum at $s_{0}$ based on an analysis of the
derivative sign properties.

In \cite{pessoa2010experimental}, the surge motion increases a little
before the discussed period, and right after this period, there is
a drop in the surge motion. It is due to the derivative sign shift
of the phase in pitch motion in this particular period. It means that
the body will not encounter the incident wave crest when the pitch
motion is reaching its maximum amplitude but a little after, thus
creating a little smaller pitch force and a smaller surge force as
well.

Equation 5 in \cite{duprat2011wall} applies the pressure gradient
sign and that of the streamwise velocity to calculate the nondimensional
velocity's gradient. In turn, the velocity itself also depends on
these gradients' signs, as evidenced in eq. 7.

In \cite{holmen2012methods}, a point belongs to a vortex center based
on conditions involving the velocity signs near it.

The examination of the proper orthogonal decomposition modes in \cite{watanabe2015three} shows
that the longitudinal structure of the vertical velocity fluctuation
is generated along the jet axis, having the opposite sign of velocity
fluctuation on both sides of the jet axis.

The definition of oscillatiory and non-oscillatory sequences in \cite{wheeler2015solitary}
is based on sign changes and lack thereof. The authors also prove
several facts about the relationship between the sign of $w$ (which
determines the sign of the slope $\eta'$ of the free surface) and
the pressure disturbance $R$.

A change of sign of the pressure derivatives in \cite{alencar2017pressure}
was observed for all $G$-band components of the double-wall CNTs
at $\sim1GPa$. It is not possible to establish if there is a corresponding
change in the RBLM peaks. At $\sim2GPa$ the $G$-band peaks energy
pressure slope starts to evolve to become positive again associated
with the loss of the $R_{1},R_{2},R_{3},R_{4}$, and $R_{6}$ RBLM
peaks. These two simultaneous observations can be assigned to the
onset of the collapse in these CNTs. Furthermore, the four $G$-band
components tend to evolve towards a monotonic behavior that is reached
at about $\sim5GPa$, which a graphite-like response can then explain.
Further, the authors assigned the change of the $G$-band's slope
sign to the onset of the collapse and the graphitic behavior to the
fully collapsed geometry. In contrast, the change of sign was previously
assumed as marking the complete collapse of the tube.

In fig. 2 of \cite{abiev2017analysis}, illustrates the scheme of
the Taylor gas-liquid flow in the microchannel and the calculated
pressure profile in the liquid phase. Axial pressure distribution
has a positive slope in the film, whereas the pressure distribution
slope is negative in the liquid slugs. The velocity field in the fixed
coordinate system in the nose and tail areas of bubbles in the liquid
film velocity has an exceptionally positive sign, the same as in the
continuous medium: in liquid slugs moving between bubbles (Fig. 3b).
Thus, there is a phenomenon consisting of a change of sign of the
velocity. In the bulk liquid and areas near the nose and tail of the
bubble, the velocity is positive, whereas in the area with constant
film thickness, the liquid velocity is negative.

The formula for very early-time dynamics with concise time intervals
(Eq. 32 in \cite{hill2019group}) depends on the signs of $v$ and
that of $v_{0}$.

A method for manipulation of microparticles in volatile liquid layers
hundreds of microns thick is proposed in \cite{al2021transport}.
It relies on the control of Marangoni flows by changing a sign of
the temperature gradient in the liquid by the local action of the
heat source and/or the heat sink. The method's applicability to perform
a wide range of manipulations with the particle ensembles is demonstrated
partly by creating ring-like patterns by changing the temperature
gradient sign during the particle assembling process.

The normalization and sign of the velocity at the center in \cite{faran2021non}
are determined only by the time dependence of the pressure at fixed
$\xi$. The relation between the velocity sign and $\lambda_{p}$
is explained as follows. When $\beta>0$, the fluid expands, and the
pressure decreases. Since the pressure profile is constant at $r\rightarrow0$,
$\lambda p$ must be negative. When $\beta<0$, the fluid is compressed,
and the pressure increases, which requires $\lambda p>0$. According
to this result, negative velocities near the origin, corresponding
to $\lambda p>0$, are achieved for $k\sim0.92$.

A preliminary leading-order analysis of a Couette flow DNS in \cite{monkewitz2021late}
yields an increase of logarithmic slope (decrease of $\kappa$) at
a $y+break\approx400$. The correlation between the sign of the slope
change and the flow symmetry motivates the hypothesis that the breakpoint
between the possibly universal short inner logarithmic region and
the actual overlap log-law corresponds to the penetration depth of
large-scale turbulent structures originating from the opposite wall.
More specifically, according to hypothesis (2.1), the sign of this
slope change depends on the flow symmetry, with a slope decrease in
channel and pipe flows and an increase in Couette flow.

\paragraph{Field Theory}

A dynamical study of the generalized scalar-tensor theory in the empty
Bianchi type I model is made in \cite{fay2000dynamical}. The authors
use a method to derive the sign of the first and second derivatives
of the metric functions.

In a paper on road signs for UV-completion, the sign of the derivative couplings for which there is no consistent
Wilsonian UV-completion is the one that allows for consistent classicalons
in \cite{dvali2012road}. The information about the chosen road is
encoded in the couplings' derivative sign, such as the quartic coupling
for a Goldstone-type particle.

The change in slope sign in \cite{kiefer2017numerical}, a study of numerical convergence study regarding homogenization
assumptions in phase field modeling, indicates
that phase 1 experiences radial stretching, whereas phase 2 is radially
compressed.

\paragraph{Solition}

Theorem III.1 in \cite{sivan2008qualitative}, a qualitative and quantitative analysis of stability and instability dynamics
of positive lattice solitons, applies the Vakhitov-Kolokolov
condition, stating that the optical power function decreases in $\mu$.

In \cite{duque2019generation}, the change in the slope sign at $w_{m}=2.1624$
for the plot of the width coefficient $A_{2}$, when approaching a
general nonlocal nonlinearity (GNN) regime. A second transition in
the slope sign is then observed at $w_{m}=3.0846$ if the characteristic
length keeps increasing. Thus, the NVA approach naturally suggests
three regions defined in Fig.1(b) there. In region I, the slope is
positive, and it stands for a suprarange localization, where the weak
nonlocality (WNN) is defined as the limit where the width of the beam
is well outside the range of the nonlocal response, particularly at
$w_{m}\rightarrow0$ the authors recover a local Kerr theory. Region
II stands for critical-range localization. It can be associated with
a GNN regime where the range of the nonlocal response is close to
that of the beam's width. In this case of an NLGR, it is characterized
with a negative slope. Region III has a positive slope again, which
stands for a subrange localization. The strong nonlocality (SNN) is
defined as the limit where the width of the beam is well within the
range of the nonlocal response.

\index{Chemistry}

\subsection{Chemistry}

\index{Analytical Chemistry}

\subsubsection{Analytical Chemistry}

In \cite{de2000infrared},
four distinct groups of polar tensor results are seen, one for each
possible derivative sign alternative.

The points' brightness in \cite{el2002multiple} is based on the derivative
signs.

The sign of the first and second derivatives of the common-mode input
impedance in \cite{spinelli2005two}, a paper on two-electrode biopotential measurements, is analyzed to classify its extrema
points.

The basic idea of Derivatives Sign Differences (DSD) is to count the
points where either the monotony or the concavity (first and second
derivative signs, respectively) of the spectra differ. Therefore,
the lower value returned the lesser spectral difference. The proposed
measure in \cite{gutierrez2010new} does not compare absorbance values
but the signs of first and second derivatives tuples. Thus, DSD correctly
matches different spectra from the same substance because these spectra
do not differ in monotony and concavity.

Alternating the polarity of the gradient pulses every other scan in
\cite{aguilar2016robust}, a paper on robust NMR water signal suppression for
demanding analytical applications, seems to improve results slightly.

The system presented in \cite{sakah2016measuring} cannot resolve
flow direction because of the symmetry of Bessel beams about their
axis. For laser doppler vibrometers, the acousto-optic shifting of one of the beams, causing
the fringes to move in one direction, is used to resolve the velocity
sign.

For $f$, as in example 3.10 in a paper on optimal measures for p-frame energies on spheres (see \cite{bilyk2019optimal}), the icosahedron
minimizes energy integral over symmetric measures on the sphere $S_{2}$.
Note that the constant term can be ignored, so it suffices only to
consider the sign of derivatives. In particular, if $b>0$ and $d$
becomes sufficiently small in magnitude, the example's inequalities
will hold.

The frequency-domain method in \cite{sun2020optimization}, a paper on optimization of velocity and displacement measurement with optical encoder
and laser self-mixing interferometry, employs
the non-linearity of signal to recover the speed sign in the frequency
domain directly. The target speed is measured by the signal frequency,
whereas the signal phase evaluates the speed sign. Another method
to improve the resolution is to perform offline signal processing
to invert the function and reconstruct the target displacement accurately.
It requires knowing the derivative sign of the actual displacement.

\index{Chemical Engineering}

\subsubsection{Chemical Engineering}

\paragraph{Qualitative Trend Analysis}

In a study of fuzzy-logic
based trend classification for fault diagnosis of chemical processes, \cite{dash2003fuzzy} shows that the fuzziness of trends is defined based
on the primitives that are classified by the sign of the derivatives.

\cite{dash2004novel} proposes a novel interval-halving framework for automated identification of
process trends and introduces an interval-halving algorithm for
trend extraction that leverages derivative signs of different orders.

In \cite{maurya2007signed}, a study of signed directed graph
and qualitative trend analysis-based framework for incipient fault diagnosis, episodes are defined as time segments
in which the sign of one or more derivatives does not change.

In a similar line of work also conducted by \cite{maurya2010framework}, the flowchart for online trend-extraction is based on primitives classified by derivative signs.

In a study of generalized
shape constrained spline fitting for qualitative analysis of trends, the branch-and-bound algorithm (\cite{villez2013generalized}) searches
for optimal argument values in which the sign of the fitted function
and/or one or more of its derivatives change.

In a later study conducted by the same authors \cite{villez2014qualitative,thurlimannqualitative,villez2016shape,thurlimann2018soft},
episodes are defined as time segments in which the primitives do not
change.

\paragraph{Chemical Thermodynamics}

In \cite{poole2005density}, all along the locus $\Delta$ (the union
of $\Delta_{max}$ and $\Delta_{min}$) the condition $\frac{\partial\rho}{\partial T}P=0$
is satisfied. Formal thermodynamic analysis shows that changes of
sign of the slope of $\Delta$ in the T-P plane are associated with
intersections with certain response function extrema. The point A
in Fig. 2(a) there, where $\Delta$ has an infinite slope, is coincident
with a point on a locus $\Lambda$ along which $\frac{\partial K_{T}}{\partial T}=0$,
where $K_{T}$ is the isothermal compressibility.

The sign of the derivative $\frac{d\gamma}{dp}$ in \cite{baidakov2012surface}
is determined by that of the concentration factor. In particular,
if the composition of the surface layer is intermediate between the
compositions of the coexistent phases, this factor is negative, and
the surface tension decreases with increasing pressure.

Many of the GST materials in \cite{volker2015low} display a prominent
feature. Upon crystallization, their electrical resistivity starts
high and can be decreased tremendously upon annealing. This effect
is accompanied by a continuous change in the temperature coefficient
of resistivity (TCR), which eventually changes its sign from negative
to positive. The high resistivity and the negative TCR have been attributed
to the disorder-induced localization of carriers in the vicinity of
vacancy clusters due to the random occupation of the Ge/Sb/vacancy
lattice sites. These localization effects dominate the electrical
transport even at room temperature, as evidenced by high resistivity,
a negative TCR, and a small mean-free path. Grain boundaries, on the
contrary, do not provide the dominant contribution to scattering,
as can also be seen from data on single crystalline GeTe nano-wires,
which also reveal disorder-induced localization. Further, the last
column of Table 1 indicates the metallic or insulating nature of the
samples by listing the sign of the slope of $w\left(T\right)$ as
defined in Equation (7) there at the lowest accessible temperature
(LAT). Lastly, the proposed method does not reduce the requirements
in terms of low-temperature data. If in the low-temperature limit
this quantity is positive and its slope $\frac{dw\left(T\right)}{d\left(T\right)}$
is negative, the corresponding sample must be insulating.

A negative or positive sign of the slope of Arrott plots in \cite{estemirova2018structural}
corresponds to a first-order or second-order magnetic phase transition,
respectively. The results obtained for $S_{1}$, $S_{2}$ and $S_{3}$
show clearly the positive slope in the entire range, indicating a
second-order magnetic phase transition.

In figure 9 of \cite{caine2019experiment}, the structure of the typical
skeleton of the primary benzene sulfonamide series AP-BSA (where substituent
variation occurs on the Ph ring), with bonds labeled in red ($+$)
or blue ($-$) depending on the sign of the slope when regressed against
pKa. In figure 11 there, the signs of the slopes of bond length vs.
pKa for $n$-butylsulfonylureas substituted at the phenyl group, where
red denotes a positive slope, and blue denotes a negative slope.

The sign of the second derivative of $P$ with respect to $V$ in
\cite{fijan2019interactions} is a parameter that plays an essential
part throughout this work. For example, the second derivative of the
TMD line is positive but negative in the VT projection. The only mechanism
by which this can happen for a TMD is if the second derivative of
$P$ with respect to $V$ is negative. However, this means that such
a point cannot intercept the liquid-vapor spin-odal as, at that point,
the second derivative is positive. It means that the avoidance of
a collision between the TMD and liquid-vapor spinodal lines is necessary
if the TMD line passes through the infinite gradient and changes gradient
sign in the case where the signs of second derivatives of the density
anomaly lines are always opposite.

The sign of Seebeck coefficient in \cite{wang2020thermal} can be
positive or negative, depending on the sign of the slope of the transmission
function at the Fermi energy EF. The sign of the Seebeck coefficient
is related to the nature of charge carriers: The Seebeck coefficient
is positive for hole-dominated transport and negative for electron-dominated
transport. Therefore, measurements of the Seebeck coefficient of MJs
are of great importance in determining the dominant transport mechanism
and the location of frontier molecular orbitals in MJs. In addition
to the above-described phenomena, bithermo-electricity effects in
MJs. coexistence or sign switching of positive and negative Seebeck
coefficients of the same MJ, have also been reported. More importantly,
the positive sign of the Seebeck coefficient unambiguously indicates
hole ($p$-type or HOMO) conduction in these MJs, which was not accessible
with other electrical measurements. The Seebeck coefficient of molecules
can also switch its sign when molecular length increases, indicating
an alteration of dominant charge carriers.

As seen in the high-pressure phase diagram in \cite{hull2020liquid},
due to the existence of the inflection point, the graphite melt line's
slope sign changes. As the entropy change upon melting should be positive,
the change in sign is then due to a change in volume attending the
phase transition. At low pressure, the melt line slope is positive,
indicating a liquid that is less dense than graphite, while at high
pressure, the liquid is denser than graphite, evidenced by the negative
slope of the melt line. Several investigators have interpreted this
change in the slope of the melting line as evidence that the liquid
may undergo a first-order liquid-liquid phase transition (LLPT) from
a low-density liquid to a higher-density liquid.

The curve scale model in \cite{barz2021paraffins} uses the sign of
the rate of change of the temperature, $sgn\left(\frac{dT}{dt}\right)$,
to distinguish between different (sub-)models for heating and cooling.

The mixing enthalpy and the permutation enthalpy in \cite{berthier2021effective}
provide the same information but differently. Whereas the sign of
the mixing enthalpy indicates that the alloy tends to phase separation
or form ordered structures, the sign of the slope of the permutation
enthalpy gives the tendency of the alloys since the permutation enthalpy
is the derivative of the mixing enthalpy. A positive (respectively
negative) slope characterizes a tendency to form ordered structures
(respectively to phase separation). The permutation enthalpy is also
determined for each configuration. It corresponds to the change in
energy when an A atom replaces a random B atom of a given configuration.
Here, $\Delta H\left(c\right)$ has a nonmonotonic behavior. For $c<0.8$,
when $\Delta H_{mix}$ is negative, the slope of $\Delta H_{perm}\left(c\right)$
is positive; it becomes negative for $c>0.8$. The slope of the triplet
contribution is close to $0$; the slope of the chemical contribution
is thus given by the slope of the pairs, and it is positive.

\paragraph{Quantum Chemistry}

The classification of critical points in \cite{de2019quantifying}
is based on the second derivative sign.

In \cite{nara2019sensitivity}, the surprising oscillating behavior,
a double change of sign of the $v_{1}$ slope, points to the appearance
of a hitherto unknown first-order phase transition in excited QCD
matter at high baryon densities in mid-central Au + Au collisions.

\paragraph{Molecular Structure}

In \cite{shashkin2001metal}, a work on metal-insulator transition in a 2D electron
gas the critical electron density for the
metal-insulator transition in a two-dimensional electron gas, can be
determined by a sign change of the temperature derivative of the resistance.

According to Eq.(2) of \cite{jadzyn2007interactions}, the behavior
of the field-induced entropy increment is determined by the permittivity
derivative temperature dependence. In particular, the sign of the
increment $\Delta S$ depends straightly on the sign of the permittivity
derivative. In the case of a less polar 7CHBT, the permittivity derivative
attains zero as its final value at the I--N transition. For strongly
polar 7CB, the $\frac{d\varepsilon}{dT}\left(T\right)$ dependence
shows a critical-like behavior, and at about $10$ degrees before
the phase transition, one observes a change of the permittivity derivative
sign (Fig. 1b). The result seems to be necessary because the $\frac{d\varepsilon}{dT}\left(T\right)$
dependence reflects temperature behavior of the electric field-induced
entropy increment directly. A negative value of the derivative $\frac{d\varepsilon_{s}}{dT}$,
i.e., $\Delta S<0$, means that the entropy decreases due to an applying
of the electric field to the dielectric material. That decrease is
apparent: forced by the field, an orientation of the dipoles causes
an increase of the molecular order. It is a normal behavior of dipolar
liquids for which the static permittivity increases when the temperature
decreases ($\frac{d\varepsilon}{dT}<0$). A change of the permittivity
derivative sign to the positive, observed in the prenematic region
of 7CB, means that the entropy increment is also positive, $\Delta S>0$.
So, in that region, an electric field applied to the isotropic dipolar
liquid increases a disorder on the molecu-60J.

The relations between interaction energies and substituent constants
for pyridine complexes with $p$-substituted iodotetrafluorobenzenes
and for the complexes of $p$-substituted pyridines with iodotetrafluorobenzene
are illustrated in \cite{szatylowicz2015substituent}. The slopes
of both regression plots are similar in magnitude but obviously of
opposite sign. Therefore, the influence of the substituents on the
interaction strength is identical in either the halogen-bond donor
or acceptor aromatic molecules. Furthermore, it was shown that the
sign of the slope for correlations between the chemical shifts and
the substituents constants depends on the position of the carbon atom
in the ring, indicating their different sensitivities to the substituent
effect.

The $CP\left(r\right)$ function defined in equation 7 of \cite{de2018faldi}
returns the slope of the tot-ED. However, it is with an adjusted sign
depending on the sign of the slope of the nonbonding-ED contribution.
Since the sign of the directional derivative depends on the direction
in which it is measured, the derivative sign factor is used to enforce
the $CP\left(r\right)$ function to be negative throughout. The exceptions
are regions where the sum of the bonding and antibonding gradients'
sign is opposite to the nonbonding gradient.

\cite{feliu2019sign}, in a paper on sign-sensitivities for reaction networks, studies the sign of the derivative of the concentrations
of the species in the network at a steady state with respect to a
small perturbation on the parameter vector.
\paragraph{Crystallography}

The strength of the interactions in a paper on Hirshfeld surface analysis and density functional calculations
of a new steroid derivative, (see \cite{ruiz2014unusual}) is classified
based on the second derivative sign.

The graphs of lattice enthalpies vs. molar volumes of $LnPO_{4}$
(with CSE of formation of $LnVO_{4}$ from oxides) are presented in
Figure 1 (monazite structure) and Figure 2 (xenotime structure) of
\cite{petrov2014lattice}, a study of lattice enthalpies, polarizabilities and shear moduli of lanthanide orthophosphates
LnPO 4. The slopes are negative and the negative
sign of the slope accounts for the trend of changes of lattice enthalpies
vs. molar volumes within the light and heavy lanthanide orthophosphates.
Hence, lower approximate limits have resulted for the shear moduli
of $LnPO_{4}\left(m\right)$, $G\approx61GPa$, and $G\approx49GPa$
for $LnPO_{4}\left(x\right)$.

In a study of designer topological insulator with enhanced gap and suppressed bulk
conduction in Bi2Se3/Sb2Te3 ultrashort-period superlattices (see \cite{levy2020designer}), the Hall resistance plots of the SL samples show that
the slope sign indicates the conductivity type: $n$-type or $p$-type
character.

The current sign in \cite{prywer2020first}, a study providing the first experimental evidences of the ferroelectric nature of struvite, depends on the derivative
sign of the triangle voltage pulse.

\index{State of Matter}

\subsubsection{State of Matter}

Exponential growing and damping in \cite{soshnikov2007collisionless}, a study of collisionless damping of electron waves in non-Maxwellian plasma,
appear only in cases where the derivative sign is constant.
For the central fields in \cite{sinha2013hypernuclear}, the magnetized
hypernuclear matter shows instability, signaled by the negative sign
of the derivative of the pressure parallel to the field with respect
to the density, leading to vanishing parallel pressure at the critical
value. It limits the range of admissible homogeneously distributed
fields in magnetars to fields below the critical value.

\cite{adam2020net} calculates the probability that at least one derivative
has a different sign from the remaining ones. Further, the derivative
of the polynomial function changes sign with the squared root of $sNN$,
thereby indicating a non-monotonic variation of the measurement with
the collision energy.

In \cite{filippov2020electrostatic}, the sign of the derivative of
the chemical potential with respect to the total number of dust particles,
the positiveness of which is the third condition for the thermodynamic
stability, is shown to coincide with the sign of the isothermal compressibility
of the dust subsystem. Therefore, it is concluded that the dusty equilibrium
plasma is thermodynamically unstable.

The order of the magnetic phase transition in \cite{jeddi2020improvement}
can be ascertained from the sign of the slope of Arrott curves. The
positive slope observed for all studied temperatures implies that
the magnetic phase transition between the FM and PM state is of the
second order.

\cite{correyero2020physics} applies a partial differential equation
called SF for image sharpening and enhancement. The SF process can
suppress the edge diffusion, achieve image deblurring and deconvolution.
Still, it is susceptible to noise, and the noise is also amplified
when the image is enlarged. The SF is commonly generalized by Eq.
7 there, which incorporates the second-order directional derivative
signs.

\index{Chemical Solutions}

\subsubsection{Chemical Solutions}

In \cite{rebelo2002double}, while for polymer blends (thus large
size-large size systems, but with components of ``similar'' size)
one finds that the golden rule is $\frac{dT}{dp}>0\left(v_{E}>0\right)$,
commonly, in polymer solutions (large size-small size systems) $\frac{dT}{dp}<0\left(v_{E}<0\right)$,
which may evolve to a change of sign at high pressures. Similar trends
have been observed for long-chain oligomers + small chain oligomers.
This behavior was discussed in terms of the ($T-p$) minimum location,
which locates the p-DCP. And they conclude that, most probably, for
those systems where $\frac{dT}{dp}>0$ at atmospheric pressure, there
is a pressure-hypercritical region lying in the ``hidden'' mechanically
metastable domain of $p<0$.

From Fig. 9a of \cite{thapa2020studies}, a paper on on aggregation and
counterion binding nature of didodecyldimethylammonium bromide in presence of added
salts, it is clear that the deviation
occurs in the presence of added Sodium chloride (NaCl). However, the authors found
that the deviation from the benzene (CH) relation still has a negative slope
value, which is consistent with the sign of the slope for the C-H
equation. Unlike in the case of added NaCl, the deviation in the CH
plot (Fig. 9c) for didodecyldimethylammonium bromid at a low concentration of added NaBz
is more drastic with a reversal in the slope sign from negative to
positive, which highlights the limitation of CH to mixed counterion
solutions.

The slope in \cite{orlovapolarimetry} is proportional to the second
virial coefficient $A_{2}$, the sign of which indicates the thermodynamic
``quality'' of solvent is. Thus, in this case, at concentrations
of solutions lower than $0.05$ and greater than $1mol\cdot L^{-1}$,
the solvent is suitable ($A_{2}>0$), while at the intermediate concentrations
the solvent poor ($A_{2}<0$).

\index{Biochemistry}

\subsubsection{Biochemistry}

In \cite{quiros2017nucleophile}, the Hammett plots obtained for all
products using both nucleophiles showed a change of the slope sign,
with a concave shape for all the compounds in both reactions. It suggests
that different mechanisms operate, depending on the tethers' electronic
properties in bisallenes with both nucleophiles. It is remarkable
given the inherent electron-withdrawing nature of the sulfonamide
group.

The hydrophobic behaviour in \cite{niether2018unravelling} is reflected
in a sign reversal of the temperature-dependent slope of the Soret
coefficient, which is observed in experiments and non-equilibrium
computer simulations at $\sim5M$ concentration of urea in water.
A positive Soret coefficient indicates that the solute accumulates
on the cold side (thermophobic), while a negative sign denotes drift
towards the warm side (thermophilic).

A plot of the $\ln\left(Ka\right)$ vs. $\ln\left(\left[NaCl\right]\right)$
in \cite{swenson2021evaluating} fit to a simple linear regression
provides a magnitude of the slope. The slope sign relates to the change
in the number of ions involved in the duplex formation. A negative
slope suggests ions are ejected into the bulk solution, and a positive
slope suggests that ions are incorporated into the duplex.

\index{Physical Chemistry}

\subsubsection{Physical Chemistry}

The sign of shifting from the s level in \cite{shpatakovskaya2001quasiclassical}
depends on the sign of the velocity's second derivative. The energy
decreases with increasing orbital for the positive sign, and for the
negative sign, it increases with the orbital. The latter situation
occurs in atoms, and both variants may occur in clusters.

Lemma 1 of \cite{shevkunov2011effect}, a study of the effects of chlorine ions on the stability of nucleation cores in condensing
water vapors, states that given some monotonicity
conditions, a particular order is preserved. Neither the derivative
nor its sign is required to prove the claim. Further, in section 10.3.3
there, the sign of the derivative is examined to show the monotonicity
of the ratio. It, in turn, offers the ordering relation given in lemma
1 is valid for this family.

The phases of the signals in \cite{tarasov2011observation}, a paper on the observation of electric quadrupole spin resonance of
Ho 3+ impurity ions in synthetic forsterite, depend
on the sign of the derivative of the resonant frequency with respect
to the magnetic field.

In a study of femtosecond time-resolved Faraday rotation in thin magnetic
films and magnetophotonic crystals, \cite{chetvertukhin2012femtosecond} shows that the time derivative sign is
different for maxima and minima of spectral interference oscillations.

The charge, charge flux and dipole flux in \cite{silva2014atomic}, a paper showing that the atomic charge transfer-counter polarization effects determine
infrared CH intensities of hydrocarbons,
can be positive when both derivative contributions are of the same
sign, reinforcing the total intensity, or negative when the contributions
have opposite signs, decreasing the total power.

In a study of nonmonotonic energy dependence of net-proton number Fluctuations, \cite{adam2021nonmonotonic} calculates the derivative sign across
different sets and calculates the probability of having two sets with
other derivative signs.

\paragraph{Particle Physics}

Based on an analysis of the sign of the slope, it was found in \cite{orek2016ab}
that the electric dipole moments for the $Ar^{-}NO^{+}$ and $Kr^{-}NO^{+}$
and $Xe^{-}NO^{+}$ systems are positive for the considerable distance
but suddenly change the sign for a shorter distance.

The factor $g$ given by Temkin isotherm in \cite{abidar2016orthophosphate}
is positive, as the sign of the slope of the isotherm (logarithmic
form) is positive. The interactions involved are so repulsive and
weak. It confirms the excellent correlation of Langmuir that neglects
interactions between adsorbed species.

In \cite{melnikov2020metallic}, a study of metallic state in a strongly interacting spinless two-valley electron
system in two dimensions, the sign-changes of the derivative
of the resistivity yield critical electron densities for the MIT.

\paragraph{Electrochemistry}

Plotting $\frac{E}{I}$ vs. $I^{-1}$ in \cite{ponce2007strategies}
produces a curve with two sections corresponding to reactions (2)
and (3), respectively. Each section consists of three zones separated
by turning points where the slope of the curve changes sign or direction.
Figure 6 there shows the $\frac{E}{I}$ vs. $I^{-1}$ curve for a
rotation rate of $100rads^{-1}$ where the zones and turning points
for reactions (2) and (3) are indicated. In the first zone, the current
is small, both terms $\frac{E}{I}$ and $I^{-1}$ are large, and the
curve is steep. Zone two occurs when the curve approaches the limiting
current region, the current becomes constant, and the slope of the
curve changes sign after the first turning point. A peak is observed
if the limiting current region is completely horizontal. It is not
always the case since a secondary reaction often accompanies the main
reaction, and often other complications such as IR drop and charge
transfer effects exist. As shown in the figure for the $Cu\left(I\right)\rightarrow Cu\left(0\right)$
process, zone two is more commonly found. Zone three arises when the
potential and the current increase beyond the limiting current region;
the sign or direction of the slope changes again at the second turning
point, and both terms $\frac{E}{I}$ and $I^{-1}$ become smaller,
making the curve very steep again. The changes in the sign of the
slope in this zone depend on whether the $\frac{E}{I}$ vs. $I^{-1}$
curve includes data of the secondary reaction or not.

In \cite{mcpherson2017electrochemical}, a study of electrochemical carbon monoxide (CO) oxidation at platinum on carbon studied
through analysis of anomalous in situ IR spectra, the center initially increases
with potential. Deviation from this linear trend is observed at the
pre-peak current onset, and the slope sign is completely reversed
by $0.3V$. Finally, by $0.47V$, the center stabilizes and remains
constant until the end of the pre-peak. The initial positive $\frac{\partial\Delta v}{\partial E}$
slope is similar to the slopes reported previously for different sized
$\frac{P_{t}}{C}$ catalysts and is consistent with the electrochemical
Stark effect. The subsequent negative slope is also compatible with
results, which showed a pre-peak at $0.3V$ that correlated with a
reversal in the sign of $\frac{\partial\Delta v}{\partial E}$. The
negative slope region has been attributed to a decrease in dipole
coupling strength as the CO coverage decreases. The negative slope
has further been interpreted as evidence for high mobility of CO on
the surface on the basis that diffusion would be necessary to enable
the whole adlayer to equilibrate with the lower coverage and show
an overall redshift.

\cite{kebede2018red} demonstrates, using a simple electrostatic dipolar
model, that not only can the surface-induced frequency shift for OsH
and OHfbe described by the same model but also OHw. The model is expressed
by Eqn (5) there. All three OH groups are part of the same general
scheme where the main ingredients are the external electric field
from the surroundings as well as the permanent and induced dipole
moment derivatives along the OH stretching coordinate. The authors
concluded that it is the sign of $\frac{d\mu}{dr}$ which is the ultimate
origin behind the different frequency shifting behavior of the water
molecules and the OH- groups.

\paragraph{Diffusion}

Fig. 7 of \cite{polyakov2006study} presents predictions from the
lattice model for the Soret coefficients of equimo-larn-alkane/benzene
mixtures as a function of temperature. In agreement with the experimental
data shown in Fig. 3, the $ST$ values calculated from Eq. (11) increase
with increasing chain length. As in the case of the experimental data,
the slope of the Soret coefficients as a function of temperature decreases
with increasing chain length. However, at this composition, the predicted
slope of $ST$ versus $T$ is positive for all chain lengths, whereas
the experiments show a negative slope for the longest chains. The
sign of the slope is composition-dependent. For low alkane concentrations,
both theory and experiment show positive slopes for all chain lengths.
As the alkane content increases, the slope decreases and becomes negative
for the longest chains at high alkane concentrations. For tridececane,
for example, the experimental data presented in Fig. 4 show the Soret
coefficient to increase with temperature for $x=0.25$, to be almost
independent of temperature for $x=0.5$, and to decrease with temperature
for $x=0.75$. The calculated $ST$ values for tridecane change from
increasing with temperature to reducing with the temperature at a
higher alkane content ($x=0.92$) and only after the estimated Soret
coefficients have become positive. For heptadecane, the change in
behavior in the experimental data occurs for a concentration smaller
than $x=0.5$, while the calculated values change behavior near $x=0.78$.
Fig. 8 shows Soret coefficients as a function of chain length N of
the alkanes at a fixed temperature of $30^{\circ}C$ for the same
mixtures as in Fig. 7. A comparison between theory (open symbols)
and experiment (filled symbols) shows that the model describes well
the trend in the chain length dependence but that the calculated $ST$
values are always between $0.5$ and $1.3\times10^{-3}K^{-1}$ higher
than the experimental values at this composition.

The drift coefficient in \cite{ho2020fractional} is related to the
negative slope of $U_{\ell}\left(x\right)$ by $D_{2}\left(x\right)=-U'_{\ell}\left(x\right)$.
As such, one can gain some qualitative understanding of how the peak of
the probability density moves just from the sign of the slope of $U_{\ell}\left(x\right)$.
The peak of $p\left(x,t\right)$ tends to move to the right (left)
when it is at a position $x$ such that $U'_{\ell}\left(x\right)$
is negative (positive), until the stationary distribution is reached.
Fig. 1 there depicts $U_{\ell}\left(x\right)$ for $g=0.5$ and $\ell=0$
(the original Rayleigh process), $1$ and $5$. There can be a sign
change of the slope of the drift potential in a certain region near
the left wall. In such a region, the peak of the probability density
function will move in different directions for different $\ell$.
Particularly, at $x_{0}=1.2$, the sign of the slope of $U_{5}\left(x\right)$
is different from those of $U_{0}\left(x\right)$ and $U_{1}\left(x\right)$.
Thus one expects that for the initial profile $P\left(x,0\right)=\delta\left(x-x_{0}\right)$
with the peak initially located at $x_{0}=1.2$, the peak will move
to the right for $\ell=5$ system, while for the other two values
of $\ell$, the peak will move to the left.

\index{Biology}

\subsection{Biology}

\index{Zoology}

\subsubsection{Zoology}

The performance of two chemotaxis strategies were contrasted in \cite{itskovits2018concerted}:
The first obeys the sign of the first derivative only and follows
the classical biased-random walk strategy. The second strategy implements
the ability to adapt to the first derivative of the gradient. These
simulations were intended to examine the possible benefits of adapting
to the magnitude of the experienced first derivative rather than simulating
a fully-detailed model to fit the experimental observations.

\index{Ecology}

\subsubsection{Ecology}

Ordered linguistic variables can be said to be increasing, steady
or decreasing with respect to the quantity against which support set
elements are ordered. Given an ecologically meaningful interpretation,
it may be helpful to define relationships between the direction of
change expressed as $\left\{ +,0,-,zero\right\} $ and influencing
variable values, as illustrated in \cite{mcintosh2003qualitative}.

In equation 3 of \cite{dambacher2007understanding}, the community
matrix incorporates the influence of one or more species on a pairwise
interaction. It does so by generating terms that either modify the
intensity of the pairwise interaction or establish what can be formally
considered as direct effects emanating from $N_{j}$ to species $N_{i}$
and $N_{k}$ involved in the pairwise interaction. Its sign structure
(which is also influenced by the signs of the partial derivatives)
tells which species directly affects other species and suppresses
other interactions.

The equilibrium stability in Eq. 12 of \cite{prasad2014dynamics}, a study of dynamics of dissolved oxygen in relation to saturation and health of
an aquatic body,
depends on the sign of the derivative of $-\frac{ax}{1+x}-bx+c$ evaluated
at the equilibrium. This equilibrium will be stable if this sign is
nonpositive and unstable otherwise. Further, this equilibrium is asymptotically
stable if the derivative sign is negative.

In the comparative dynamic analysis of \cite{lafforgue2019dynamic},
a sensitivity analysis of the key variables of the model is conducted
with respect to the set of parameters. The results are described in
Table 1 there. Each box indicates the sign of the partial derivative
of the variable mentioned inline with respect to the parameter given
in the column. This sign can be positive (\textquotedbl$+$\textquotedbl )
or negative (\textquotedbl$-$\textquotedbl ). An empty box means
no relation between the variable and the parameter, whereas \textquotedbl$?$\textquotedbl{}
indicates an ambiguous sign.

\cite{rigal2020method} applies the sign of the first and second derivatives
upon fitting a quadratic polynomial that captures the trend of a time
series.

\cite{holen2021coping} explores the effect on acceptance rates of
small changes in ecological parameters that affect one (or two) of
the composite parameters a to f (appendix E). The findings are summarized
in table 2 there; further, the discussion is given under the various
applications.

\index{Biotechnology}

\subsubsection{Biotechnology}

As it is possible to observe in Fig. 4 of \cite{ivorra2014continuous},
an example of two doughs, in which the evolution of the ratio $\frac{\Delta A}{\Delta H}$
and $R^{2}$ of Pearson are drawn, peaks and valleys could be identified
(between broken lines). Peaks were considered when data from $R^{2}$
change their derivative sign from positive to negative (derivative
zero value), and the function value is equal or higher than the previous
peak. $R^{2}$ of Pearson and $\frac{\Delta A}{\Delta H}$ had an
inverse tendency with time. When $\frac{\Delta A}{\Delta H}$ decreased,
because the higher velocity of $H$ changed, $R^{2}$ increased, evolving
the shape of the dough surface to a theoretical arc. Inverse behavior
could be obtained when the A velocity was higher. The recount of the
number of peaks at $100min$ (NP100, Table 3 there) showed how this
number is related to dough evolution and could be used to discern
the final behavior of doughs (Fig. 6). In doughs that did not substantially
vary in their transversal area between $100min$ and their last fermentation
time (first doughs), the number of peaks did not increase, reaching
their highest number. On the other hand, doughs which increased their
transversal area, also increased their number of peaks (Table 3).

\index{Physiology}

\subsubsection{Physiology}

In a study of stance and swing detection based on the angular velocity of lower
limb segments during walking, cite{grimmer2019stance} shows that it is required to detect the events of the sign changes (heel-strike
and toe-off) to separate the stance and swing phase, on top of the velocity sign changes.

In a machine learning study of breaststroke, \cite{zanchimachine} shows that due to fluctuations in the speed measuring, the used algorithm
could not recognize the periodic shape. Another method to filter the
swimmer's profile has been to apply the change of the sign of derivative
as a counter to isolate periods.

In a related line of work, the forces in the $y$ and $z$ directions during the contact with
the wall are modeled (see Eq. 2-3 in \cite{mori2020simulation}); they
are based on the signs of the respective velocities of the right heel.

Also in a related study, \cite{tahmasian2020dynamic} neglects the symmetric drag force acting
on the main body and only considers the fin's drag force in the form
of Eq. 14, which incorporates the velocity sign. In turn, it allows
us to write Eq. 13 in the form that also depends on the velocity sign.
It also plays a part in other equations, such as 18, 19, 25, and 26.

\index{Medicine}

\subsubsection{Medicine}

\paragraph{Medical Experiments}

In a study of medical randomized controlled trials, \cite{chemla2019controls} shows that if the conditions of Proposition 4  are
satisfied, the experimenter cannot rely upon the sign of the treatment-control
difference to distinguish between the two efficacy states. In this
situation, the experimenter would need to rely upon magnitudes of
the treatment-control difference to determine the state. However,
interpreting treatment-control volumes is more difficult since magnitudes
depend upon unobservables such as mental effect functions. However,
with the more effective control, the treatment-control difference
is shifted downward so that the sign of the difference suffices to
infer the state.

\paragraph{Biomedicine}

Let us classify the technologies that leverage trends based on the
medical field to which they are related. 

In a study of bottom-up approach to uniform feature extraction
in time and frequency domains for single lead ECG signal (see \cite{srikanth2002bottom}), Step 6 of identifying the fiducial points
starts with the $P$ wave detection. The process is similar to $T$ wave
peak detection, except that the search is in the other direction of
QRS complex. A peak is defined as a local maximum where the sign of
the derivative changes and smooth descent occurs on either side. With
this step, the first stage of wave detection gets completed for the
normal beat. The absence of a $P$ or $T$ peak is also noted. Cross-checks
are introduced for negative wave detection. 

In another EEG study (see in \cite{molina2013ecg}), the \textquotedbl Score\textquotedbl{} measures the similarity between the derivative
sign of the samples of the pattern of the encoded signal.

In a related study, \cite{moghari2014three} proposes a signed gradient descent algorithm with a constant step size was
developed to register the reference 3D-LOC
to the other 3D-LOC images acquired at different cardiac cycles using
a 3D translational parameter ($t_{SI}$, $t_{AP}$, and $t_{RL}$)
which estimates the bulk translational displacement of the heart.

The signs of the one-sided numerical derivatives calculated
in Eq. 1 in \cite{soudani2018atrial}, a study of atrial fibrillation detection based on ECG features
extraction, are applied in the local peak
counter mechanism (figure 4 ).

In a related line of work (see \cite{meddah2019fpga}), the (discrete) derivative
sign is applied to detect maximum peaks (equations 7 and 8).

\cite{seo2010monitoring} proposes  signal-processing methods to investigate
the feasibility of monitoring ablative therapy for the myocardium
by identifying the point at which the slope of the thermal strain
curve changes sign caused by the speed of sound and thermal expansion
variations with temperature. 

In a study of sector-wise golden-angle phase contrast with high temporal
resolution for evaluation of left ventricular diastolic dysfunction (see \cite{fyrdahl2020sector}), the final voxel location is chosen based on two criteria, with one of them
being the peak velocity (sign change).

\cite{de2012prediction} is an example of a technological application
in the diabetes domain. Cooperative systems form a class of monotone
dynamical systems in which the partial derivatives are positive. Graph
theory also allows analyzing monotone and cooperative systems by using
a species graph, in which a node is assigned for each compartment
of the model. If the node $x_{i}$ has no direct effect on node $x_{j}$,
the partial derivative $\frac{\partial f_{j}}{\partial x_{i}}\left(x\right)$
equals zero; thus no edge is drawn from node $x_{i}$ to node $x_{j}$.
If the effect of the node $x_{i}$ on node $x_{j}$ is positive, the
derivative is strictly positive, and an activation arrow ($\rightarrow$)
is drawn. Finally, if the effect is negative, an inhibition line is
drawn. However, if the derivative sign changes depending on the particular
entries, both an activation arrow and an inhibition line are drawn
from node $x_{i}$ to node $x_{j}$. A spin assignment is an allocation
in which each node has a sign, such that nodes connected by an activation
arrow ($\rightarrow$) have the same sign, while nodes connected by
an inhibition line have different signs. If at least one consistent
assignment exists, the dynamical system is monotone. Furthermore,
the system is cooperative if all nodes are connected by activation
arrows ($\rightarrow$).

In a study of minimum spanning forest-based method for noninvasive cancer
detection with hyperspectral imaging, \cite{pike2015minimum} applies the derivative
sign difference (DSD) to calculate the number of times the pixels'
spectral derivatives are of opposite signs.

Applications of the derivative sign to muscle tracking studies include \cite{luppescuclassification,waris2018effect,wahid2018subject,cengiz2020detection,fajardo2021emg},
where SSC, the number of times the slope sign changes, is extracted
as a feature of the EMG signal in . The number of peaks was measured
based on the product of the signs of the one-sided numeric derivatives
at each datapoint.

Technological applications in the skeleton imaging includes an algorithm that has been used in \cite{rodriguez2004plane} for
the detection of the fringes generates a binary image with the sign
of the angle of the derivative vector at one of its steps. 

In \cite{jebri2015detection}, a study of the detection of degenerative change in lateral projection cervical spine
x-ray images,
the intensity change may be gradual at a transition point instead
of a step function. Indeed, one can see a gradual decrease in brightness.
To better address gradual brightness changes, the sign of the derivative
is used. 

In a similar line of work, \cite{shi2018bionic} placed electrodes on
the flexor digitorum superficialis (FDS) and extensor digitorum (ED)
muscles, and performed feature extraction by picking Hudgins' features.
One of them is SSC. It indicates the frequency information of the
EMG signal, as the number of times the slope changed from positive
to negative or vice versa. 

The friction torque in \cite{zhang2018lower}, a study of industrial handling
augmentation used to provide spinal support,
can be modeled as in Eq. 4 there. It incorporates the sign of the
derivative of the motor-side angle. It is further developed in Eq.
5, 7, 9, 11.

Examples of technological applications involving the human eyes include \cite{mendonca2006segmentation}, where each one of the four directional images resulting from
the DoOG filters is searched for
specific combinations of signs on the expected direction of the vessel
cross-section. The search is performed on one-pixel-wide lines with
orientation corresponding to the vessel cross profile, which means
that the scanning direction is distinct for each of the four images
under analysis. As actual vessels do not have the ideal profile presented
in Fig. 3(a), the authors empirically assessed several combinations
of filter responses that can characterize a vessel. The result was
the set of four combinations indicated in Fig. 3(c). In this figure,
plus and minus signs correspond to positive and negative derivative
responses, respectively, 0 is associated with a null output, and X
is a do not care condition meaning that the derivative sign is not
evaluated (conditions 2 and 3). However, in these two cases, the average
value of the derivative magnitudes (ADV) for the intensity profile
under analysis must be positive for condition 2, and negative for
condition 3. The ADV value gives a good indication if the vessel is
located in a region with a slowly varying baseline on the vessel cross
profile direction. It can make the values of the derivatives dominantly
positive (condition 2) or negative (condition 3). To illustrate this
process, consider the simple example for vessels with a predominantly
vertical orientation. We need to analyze the derivative signs in the
direction of the vessel cross-section. 

\cite{fraz2012approach} proposes a method to localize the retinal blood vessels using
bit planes and centerline detection. The centerlines are extracted using first-order derivative of a Gaussian
filter in four orientations, and then evaluation of derivative signs
and average derivative values is performed.  

The iteration-dependent weighting function in \cite{de2017automated} depends on the sign
of the derivative of the reflectivity recorded in a discrete voxel
position.

In a study proposing a method for quantitative assessment of retinal vessel tortuosity
in optical coherence tomography angiography applied to sickle cell retinopathy (see \cite{khansari2017method}), the number of critical points at which the first
derivative of centerlines vanishes was quantified
for each centerline based on frequency of changes in sign of the slope
of the tangent lines. 

Finally, the initial attempts of \cite{chakravarty2018supervised}
to segment the layers in retinal OCT images employed simple image
processing techniques and focussed on the segmentation of only a few
prominent layers. Each $A$-scan in the OCT slice was segmented individually
based on peak, valley, and/or signed gradient analysis of the intensity
profile. This approach had some problems, to which this work suggests
workarounds. The interaction between each pair of adjacent points
on the lth boundary is modeled as a linear combination of a shape
prior and an appearance term. The shape prior between $\left(x\left(n\right),x\left(n+1\right)\right)$
is a soft constraint that penalizes large deviations of the signed
gradient of the height values $\left(x\left(n\right),x\left(n+1\right)\right)$
to preserve the local smoothness of the $l^{th}$ boundary. A Gaussian
function models the deviation. The mean and the standard deviation
of the signed gradient are pre-computed for each layer and column
using the ground truth layer markings of the training images.

Attempts to trace the human brain that leverage the trends are also
abundant. For example, consistent with the view that AD patients are
more impaired in semantic fluency tasks than letter fluency tasks,
AD patients were found to have a negative slope for category type.
It indicates that they recalled more items from letter categories
than semantic categories, as illustrated in \cite{diaz2004category}.
On the other hand, normals had a positive slope for category type,
which indicates that they recalled more items from semantic categories
than letter categories. This difference in the slope sign was actual
even when log-transformed fluency was the dependent measure. Thus,
AD patients recalled a smaller proportion of exemplars from semantic
categories than from letter categories compared to controls. 

In \cite{shen2011use}, the apparent diffusion coefficient (ADC) values and the relative ADC values (rADC) in hyperacute
and acute lesions had gradient signs that these lesions increased
from the center to the periphery. The ADC values and the rADC values
in subacute lesions had adverse gradient signs that these lesions
decreased from the center to the periphery. 

In \cite{godino2017towards}, a study of the identification of Idiopathic Parkinson’s Disease
from the speech,
PE quantifies the probability that, within a signal, a segment will
resemble the next. Equally, changes in the direction of the signal
(slope sign) result in complexity increases, while a steadily positive
or negative slope would be associated with less complexity. Thus,
a signal with only one phase per cycle would have lower complexity
than others with various phases. 

In a study of background connectivity integrating categorical and spatial attention, \cite{tompary2018attending} shows that because of the mean-centering
of both regions, the best-fit line
passes through the origin. Thus points in these quadrants support
a positive slope. If the one-time course has a positive value and
the other has a negative value, the time point falls in the second
or fourth quadrant, which supports a negative slope. The relative
balance of points in quadrants $1/3$ versus $2/4$ thus determines
the sign of the slope. Because of the variance normalization, the
value of the slope is the Pearson correlation coefficient. 

One of the extracted features in \cite{karuppiah2018towards}, a study of adaptive seizure prediction framework using active learning heuristics,
is ``peak amplitude,'' describing the base-10 logarithm of the mean-squared
amplitude of the peaks, where a peak is defined as a change from negative
to positive in the signal derivative sign. 

In \cite{schwartz2020pupil},
a change in the sign of pupil coefficients across the breakpoint indicates
a nonmonotonic relationship between pupil and firing rate. The direction
of the sign change shows if a cell is a $U$ or inverted $U$. A difference
in the sign of the slope between segments indicated a nonmonotonic
relationship between pupil size and spiking activity. The accuracy
of this model was compared to a similar model in which slope could
vary between segments but where both line segments were constrained
to have the same sign. The effect of the pupil-associated state usually
had the same sign for both spontaneous and sound-evoked activity within
a single neuron, in contrast to earlier results suggesting that intermediate
pupil sizes were associated with opposite changes in spontaneous and
evoked activity. 

A significant difference between the HFB
activity and the low-frequency bands in \cite{piantoni2021size} is
the sign of the slope of the correlation between  intracranial electrocorticography (ECoG) and blood oxygenation level dependent (BOLD) activity.
For the high-frequency broadband (HFB), higher ECoG activity correlated with higher BOLD activity,
while the opposite was true for alpha.

In a study of respiratory systems, the algorithm in \cite{benetazzo2014respiratory}
automatically analyses the derivative sign, detects when it changes,
and checks if that sign is kept for at least three samples. In detail,
if the sign changes from negative into positive, a new breath is detected.
If the sign changes from positive to negative, the breath passes from
inhalation to exhalation. The algorithm checks that the change of
the sign persists for at least three samples to avoid disturbances
overlaid on the signal, which may cause an incorrect count. This way,
the algorithm automatically counts the number of breaths.

Trends are also useful when analyzing blood vessels. For instance, the problem of
detecting the ridge points, is reduced in \cite{croasmun2012fast}
to the problem of detecting the sign changes of the gradient vectors
projected onto two scanlines with sufficiently different orientations.
It suggests a simple scanline algorithm using the horizontal and vertical
scanlines, $S_{x}$ and $S_{y}$, respectively. The authors define
a set of gradient sign changes that can be used to identify ridge
points present in grayscale images. The ridge points are extracted
from the topological surfaces by detecting the gradient sign changes
on two orthogonal scanlines.

\paragraph{Drugs}

To spot potential errors, \cite{dykstra2010visualization} graphs
the data split by the number of gradient changes within the data to
spot potential errors. A change in the sign of the gradient of three
successive measurements is given if an increase or vice versa follows
a decrease. To further highlight changes in gradient, we indicate
increases with green upward-pointing triangles, declines with red
downward-pointing triangles, and no movement with gray squares. Figure
1 depicts observed concentration versus time after the dose, split
by the number of changes in the gradient sign during the period studied.
The majority of the profiles have at most one change in the sign of
the gradient. Selecting subjects for whom there are unexpected changes
in the gradient sign will often expose data groups that may require
further analysis or query.

\cite{brown2021general}, a study of General Purpose Structure-Based drug discovery neural network
score functions with human-interpretable pharmacophore maps, features for which models in the ensemble
agree on the derivative sign most routinely are interpreted as those
of most importance to the ensemble's performance. Consistency is
thus insensitive to the magnitude of a feature's influence.

\paragraph{Health Physics}

The trend of the function's rate (the sign of the second derivative)
is applied in \cite{shende2019determination} to classify the inflection
point for a dosimetric analysis of unflattened beam using the first principle of derivatives
by python code programming

\paragraph{Neurology}

The sign of the slope of the E-PG phase equation in \cite{okubo2020neural}
indicates the direction of the bump's circular movement. In most
cases, the trend is positive, meaning that clockwise wind shifts produce
clockwise bump rotations, as viewed from the posterior side of the
head.

\paragraph{Epidemiology}

In \cite{arino2010effect}, a study of the effects of a sharp change of the incidence
function on the dynamics of a simple disease, the sign patterns of the signs of the Jacobian
matrix aid in classifying the local stabilities at the equilibria.

In a study of perturbation analysis in finite LD-QBD
processes and applications to epidemic models, \cite{gomez2018perturbation} is interested in analyzing the impact
that small perturbations in the parameters of $\left(\beta_{1},\beta_{2},c,\beta\right)$
have on the summary statistics, whence Table 2 there lists values
of elasticities (i.e., $\left(\theta^{-1}D\right)^{-1}\frac{\partial D}{\partial\theta}$)
for summary statistics $D$ and parameter $\theta$. Further, the
authors hold an extensive qualitative discussions with insights that
follow from the sign of elasticities (which is identical to the sign
of the partial derivative $\frac{\partial D}{\partial\theta}$).

The apparent epidemic peak in \cite{carletti2020covid} occurs when
$I'=0$, whereas we know from the fact that SIRD-like dynamics govern
the epidemics that the actual peak happens when $D''=0$, i.e., when
the number of deaths/day reaches a maximum. Thus, Eq. (4) can be interpreted
as follows. Whether the apparent peak is observed before or after
the true peak depends on the sign of the rate of reporting, $\alpha'\left(t\right)$.
More precisely, if the testing activity is steadily ramping up ($\alpha'>0$),
the true peak will occur earlier than the apparent one. The reported
infected will have a maximum for $D''<0$ i.e. past the maximum of
the $D'$. Conversely, if the testing rate decreases, this will anticipate
the apparent peak, giving a false impression that the worst might
be over. At the same time, the actual number of infected is, in fact,
still increasing. We find that this analysis applies to all countries
considered in this paper (see also supplementary material), whereby
either the former or the latter scenarios are invariably observed.

The derivative of the basic reproduction number with respect to the
commuter ratio in \cite{seno2020sis}, a paper proposing an SIS model for the epidemic dynamics with two phases of the human
day-to-day activity, is analyzed extensively. In
lemma 1 the authors prove that the reproduction number is monotonically
decreasing, according to the limit of its derivative sign.

In a study of deterministic and stochastic non-local
SIR Models, \cite{ascione2020construction} mentions that they do not have any
sufficient condition for monotonicity of functions by knowing the
sign of their Caputo-type derivative. The lack of such satisfactory
condition can also be seen from the phase portrait in Figure 3, as
the maximum is reached in a region in which $\frac{dI}{dt}$ is still
strictly positive. Recalling the Fermat theorem on extremal points
becomes an inequality in the non-local context, justifying that, after
the function reaches a maximum and starts decreasing, the non-local
derivative could still be non-negative.

In a study of non-pharmaceutical interventions in a generalized
model of interactive dynamics between COVID-19 and the economy (see cite{datta2021non}), proving monotonicity properties regarding
the function $g$ in lemma 4 based on its derivative sign, then leveraging
the sign consistency while applying the Dulac criterion based on equation
16.

\paragraph{Well-being}

In \cite{awalgaonkar2019learning}, the response to a preference query
gives information about the sign of derivative of utility function
at the indoor air temperature where we ask these queries. Essentially,
the experimental data $y$ is noisy observations of the sign of derivatives
$u'$. The likelihood function is a model of the measurement process,
and it establishes the connection between $y$ and $u$. The authors
define $p$ as a function of $v$, $y$, and $u'$. The proposed likelihood
encodes the following intuitive characteristics. First, the possibility
is high when $y$ and $u'$ have the same sign. Second, it is low
when $y$ and $u'$ have opposite signs.

\index{Systems Biology}

\subsubsection{Systems Biology}

The species graph has $n$ nodes (or ``vertices''), which \cite{sontag2006monotone}, in a study of monotone and near-monotone network structure,
denotes by $v_{1},\ldots,v_{n}$: One node for each species. No edge
is drawn from node $v_{j}$ to node $v_{i}$ if the partial derivative
$\frac{\partial f_{i}}{\partial x_{j}}\left(x\right)$ vanishes identically,
meaning that there is no direct effect of the $j^{\text{th}}$ species
upon the ith species. If this derivative is not identically zero,
then there are three possibilities: (1) it is $\geq0$ for all $x$,
(2) it is $\leq0$ for all $x$, or (3) it changes sign depending
on the particular entries of the concentration vector $x$. In the
first case (activation), we draw an edge labeled $+$, $+1$, or just
an arrow $\rightarrow$. In the second case (repression or inhibition),
we draw an edge labeled $-$, $-1$, or use the symbol $\dashv$.
In the third case, when the sign is ambiguous, the authors draw both
an activating and an inhibiting edge from node $v_{j}$ to node $v_{i}$.

Part D of Fig. 2 in \cite{baldazzi2012importance} shows the Jacobian
matrix with the sign of the derived regulatory interactions between
the slow variables of the system. Further, Tables 1 and 2 show the
relative expression changes for the genes that are included in the
models $M_{glyco}$ and $M_{neo}$. As shown in Table 1, the sign
of the changes in expression is consistent between the two data sets,
bearing in mind the experimental uncertainty. Finally, Fig. 6 shows
an example of a qualitative simulation of the glucose--acetate diauxie
(Section 2.3). The vertical axis shows the symbolic values of concentration
variables, and the horizontal indicates the qualitative states of
the system. The selected pathway illustrates the typical dynamics
of protein concentrations following the shift to acetate. The glycolytic
enzymes are the first to respond. Global regulators respond later
once a sufficiently high level of Pps $A$ is reached. The derivative
sign of concentration variables in each qualitative state is explicitly
indicated.

\index{Genome Biology}

\subsubsection{Genome Biology}

Each qualitative state in \cite{batt2004model}, a study of checking genetic regulatory networks using GNA and CADP, corresponds to a self-transition
(loop) state in the Lts. The label of this loop encodes all the properties
of the corresponding qualitative state: its name, the range and derivative
sign of protein concentrations, and additional properties.

In \cite{balasubramaniyan2005clustering}, a paper on clustering of gene expression data using a local
shape-based similarity measure, SRC is compared to a frequently
used qualitative measure that compares the sign of the first-order
differences (i.e., the ups and downs) of two series.

As illustrated in \cite{kim2008linear}, linear time-varying models can reveal non-linear interactions of
biomolecular regulatory networks using multiple time-series data. By inspecting the sign of
the slope for two given time points, it can be decided whether the
slope is tilted to the right or left. The average value is then calculated
from multiple points on the curve. Depending on the sign of the slope
index, it can hence be deduced whether $x_{i}\left(t\right)$ activates
or inhibits $x_{j}\left(t\right)$. In this article, the authors adopt
this approach in the case of time-varying models. The slope index
is now defined using $x_{j}\left(t\right)$ and $a_{ij}\left(t\right)x_{j}\left(t\right)$,
since the authors consider the direct effect of $x_{j}\left(t\right)$
on the time derivative of $x_{i}\left(t\right)$. The slope index,
$SI$, is therefore defined in equation 7.

In a study of propagation of genetic variation in
gene regulatory networks conducted by \cite{plahte2013propagation}, a simple sign rule relates the sign
of the derivative of the feedback function of any locus to the feedback
loops involving that particular locus.

\cite{chu2016single} classifies genes into upregulated
and downregulated groups. It was done by their expression trend along
the recovered order of these $172$ cells. The two groups were defined
by the sign of the slope coefficient in the gene-specific linear fitting.
The genes with positive (negative) slope coefficient were defined
as up- (down-) regulated from early-$36$ hcells to late-$36$ h cells.

The peak detection algorithm in \cite{sedlyarova2017natural}, a study of natural RNA polymerase aptamers regulate transcription in E.
coli, is based
on changes in the sign of the numerical derivative.

\index{Cellular Biology}

\subsubsection{Cellular Biology}

In \cite{sarkar2019field}, a study of field induced cell proliferation and
death in a model epithelium, the zeros of $f\left(h\right)$ correspond
to steady states, which can be stable or unstable depending on the
sign of the slope $\frac{df}{dh}$. Negative slopes correspond to
stable and positive slopes to unstable situations. Depending on the
$\alpha$ and $\beta$ values, one, two, or zero steady states may
exist.

\index{Microbiology}

\subsubsection{Microbiology}

The sign of the derivative of $S_{n}$ in \cite{beretta2003effect}, a study of the effect of time delay on stability in
a bacteria-bacteriophage model,
is that of the real part of $\lambda$ at $i\omega$.

The signs of the partial derivatives of the total mass of drug released
with respect to various other parameters in \cite{bernardes2019fighting}
indicate the previewed behavior.

\index{Evolution}

\subsubsection{Evolution}

The sign of the gain function derivative in \cite{paenke2009influence}
is computed in equation 13 there. It assumes a monotonic and continuously
differentiable fitness landscape. The influence of constant learning
on evolution solely depends on the second derivative of logarithmic
fitness: Positive (negative) $\left[\ln f\left(x\right)\right]''$
implies learning-induced acceleration (deceleration) for this type
of learning. Further, given some evolutionary data (in the absence
and the presence of learning), we can deduce the sign of the gain
function. In other words, we learn something about the effect of learning
on fitness.

In figure 1 of \cite{osmond2017predators} the authors illustrate
the derivative signs of several parameters and how they combine to
affect the rate of evolution eventually. Pathways by which predation
intensity, $k$, can affect the rate of prey evolution, $\frac{dz}{dt}$.
Boldface lines show our two examples: the evolutionary hydra effect
(top) and the selective push (bottom). Positive and negative symbols
give the sign of the partial derivative of the right variable with
respect to the left variable in our examples (e.g., the negative symbol
betweenkandNindicates that prey density, $N$, declines with increasing
predation intensity, $\frac{\partial N}{\partial k}<0)$. Increasing
predation intensity increases the rate of prey evolution (toward larger
trait values) via a specific pathway when the product of the signs
along that pathway is positive.

Following equation 6 in \cite{day2018role}, by definition of conflict
occurring at the evolutionarily stable strategy (ESS), we have $\frac{\partial H\left(c^{*},v^{*}\right)}{\partial v}<0$.
As a result, host fitness will increase - and thus, conflict from
the perspective of the host will decrease --- only if $v<v^{*}$.
From above, we see that this requires that the slope of the reaction
norm be positive. Otherwise, the fitness of the host will decrease,
and thus it will experience more conflict. Last, we note that the
sign of the slope of the optimal reaction norm is determined by the
sign of the mixed partial derivative. This measures how selection
on the clearance rate of the host changes as parasite virulence increases.

Two of the nine vocal parameters in \cite{ju2019four} are new and
frequency-time excursion length (Ju and Podos et al.). They potentially
reflect phonological complexity: frequency-time excursion length and
changes in concavity. Treating a tonal song trace in a spectrogram
as a function in frequency-time space, the value for changes in concavity
is the number of times the slope of that function goes from positive
to negative or vice versa (i.e., the number of critical points in
the derivative of the trace). In other words, changes in concavity
measure the number of times a song trace changes direction (up vs.
down) in frequency. These two measures are distinct; for instance,
the same high frequency-time excursion length can be accomplished
either by traversing a high-frequency bandwidth with a few changes
in concavity or a low-frequency bandwidth with many modifications
in concavity (high convolution).

\chapterimage{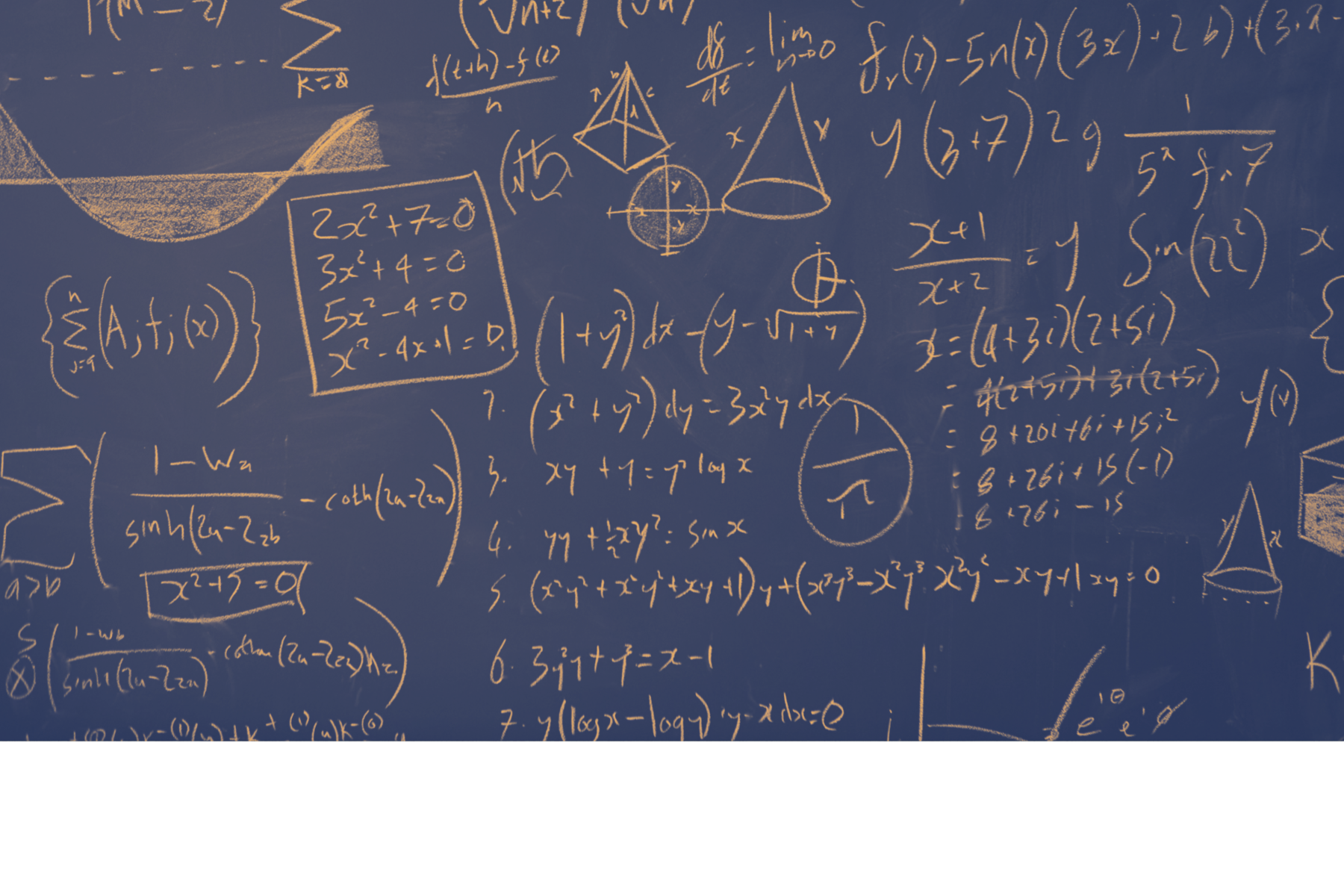} \index{The Mathematical Trendland}

\chapter{The Mathematical Trendland}

In mathematics, trends are used in analysis, algebra operations, mathematical optimization, dynamical systems, set theory, geometry, probability, statistics, game theory, and decision theory. Let us review some examples.

\index{Mathematical Analysis}

\section{Mathematical Analysis}

\index{Differential Equations}

\subsection{Differential Equations}

For some crucial classes of partial differential equations, which have the common feature of
a single-value $\lambda\left(A\right)$ function, the criteria to
understand and anticipate the dynamics of nonlinear systems by the
analysis of steady-state solutions are based on the sign of the derivative
$\lambda'\left(A\right)$. \cite{politi2007modified} addresses this topic in a paper on stability of
stationary solutions and the consequent dynamics.

Lemma 3.7 in \cite{bernetti2008exact}, a paper proposing an exact solution of the Riemann
problem for the shallow water equations with discontinuous bottom geometry, studies the monotonicity of
the function $\sigma$.

In \cite{makarenkov2011asymptotic}, classical conditions for asymptotic
stability of periodic solutions bifurcating from a limit cycle rely
on the sign of the derivative of the associated bifurcation function
at a zero. The paper shows that, for analytic systems, this result
is topological. This means that it is enough to impose a sign change
at the zero without any assumption on the successive derivatives.

Some oscillatory phenomena in physics, biomedicine and
biochemistry are described in \cite{pavsic2015sign} by positive functions
having sign-changing first derivatives. Here, these phenomena are studied for all
positive, not necessarily periodic, solutions of a large second-order
non-linear differential equations class, based on a new reciprocal
principle. The classic oscillations of the corresponding reciprocal
linear equation cause the sign-changing first derivative of every
positive solution of the main equation. The first main result (theorem
3.1) shows the oscillating solutions $y\left(t\right)$ of
the reciprocal equation (2.2) with the positive solutions $x\left(t\right)$
of Eq. (1.1) having sign-changing $x'\left(t\right)$. The result is then
used to derive various criteria for the sign-changing $x'\left(t\right)$
of every positive solution.

The encircled numbers in \cite{zegeling2018nonstandard}, a study of nonstandard finite differences for a truncated
Bratu–Picard model, indicate
the period for each case. More specifically, the sign indicates whether the first extremum
in the solution is a maximum ($+$) or a minimum ($-$), respectively.

In a study discussing the applications of Lyapunov functions
to Caputo fractional differential equations (see \cite{agarwal2018applications}), one of the sufficient conditions for stability
is connected with the derivative sign of the Lyapunov function.

In \cite{fiedler2020coexistence}, a study of coexistence of infinitely many large, stable, rapidly oscillating periodic
solutions in time-delayed Duffing oscillators, the slope of $\log\left|H\left(t\right)-H_{n}\right|$,
asymptotically with respect to time $t\rightarrow\pm\infty$,
coincides with the Floquet exponent and determines the instability
or stability of the periodic solution, depending on the positive or
negative sign of the slope.

In a paper on the florin problem for quasilinear diffusion
equation taking into account nonlinear convection (see \cite{turaev2020florin}), the main results in theorems 1 and 2 require conditions and prove claims regarding the derivative
sign of $s$.

\cite{almenar2020sign} provides results on the sign of the Green
function (and its partial derivatives) of an $n$-th order boundary
value problem subject to a wide set of homogeneous two-point boundary
conditions.

In a study of oscillatory behaviour
of higher order neutral differential equations with several delays and with a super
linear term (see \cite{panda2020oscillatory}), condition 3.10 (assumed in
several results) states that the first $n-1$ derivatives are monotonic
and of a constant sign.

\index{Real Analysis}

\subsection{Real Analysis}

\cite{gkioulekas2014generalized}, in a study of generalized local test for local extrema in single-variable functions, formulates generalizations to the
derivative tests based on the signs of derivatives of different orders
and the mean value theorem.

\index{Complex Analysis}

\subsection{Complex Analysis}

The sufficient hypothesis in \cite{gimeno2008euclidean} is the monotony
of the function between any two grid points. The function might
be globally non-monotonous by being allowed to change derivative sign
at the grid points themselves. If the function is strictly decreasing,
then it is bound. To arbitrarily shrink the error in our knowledge
of the initial condition, one needs to arbitrarily shrink the grid
spacing to constrain the function in every subinterval.

In \cite{berdellima2019note}, a study discussing a conjecture proposed by Khabibullin, the signs of the first and higher-order
derivatives of $\varphi$ (Eq. 2.3, lemma A.3) and other terms (Eq.
A.1, A.2) are applied in the analysis of monotonicity properties.

\index{Numerical Analysis}

\subsection{Numerical Analysis}

In a study of convergence of a generalized fast-marching method for an eikonal
equation with a velocity-changing sign, \cite{carlini2008convergence} uses the numerical computation
of dislocations dynamics where the velocity of the front can change
sign.

\cite{weynans2013consistency} extends the construction of the the total variation diminishing
particles remeshing schemes to nonlinear conservation laws with a
possible change of velocity sign. The results have applications to Burgers and
Euler equations.

\index{Algebra}

\section{Algebra}

According to \cite{kashiwagi2014derivative}, when the sign of the derivative of the determinant changes, we may use techniques such as the bisection method to narrow
the interval within which the sign changes and, thus, pinpoint singular
values.

In \cite{ghys2015signatures}, a paper discussing signatures in algebra, topology and dynamics, if $F$ has $n$ distinct real roots,
then the total signature is defined as the weighted sum of its derivative
signs.

\cite{maignan2016fleshing} applies Hardy's notion of \textquotedbl False
derivative\textquotedbl{} of a function whose sign agrees with the
function's derivative sign at the zeros of the original function.

\index{Mathematical Optimization}

\section{Mathematical Optimization}

In a study of methods for nonlinear optimization (see \cite{svanberg2007mma}), the signs of the function's discrete derivatives (the finite difference)
are applied in some of the formulas.

According to a paper on optimal abstraction on real-valued programs (see \cite{monniaux2007optimal}), if a polynomial and its derivative are co-prime, then the sign diagram
of its derivative is used to compute its sign diagram.

In \cite{lafond2015online}, a paper on the online Frank-Wolfe algorithms for
convex and non-convex optimizations, $C$ is an $\ell_{1}$ ball constraint and
the linear optimization in Line 4 of Algorithm 1 or (3) in Algorithm
2 can be evaluated based on the gradient sign as $a\left(t\right)=-r\cdot sign\left(\left[\nabla F_{t}(\theta_{t})\right]_{i}\right)\cdot e_{i}$,
where $i=argmax\left(\left|\left[\nabla F_{t}\left(\theta_{t}\right)\right]j\right|\right)$
subject to $j\in\left[n\right]$.

The first step in proving the main result of \cite{chorwadwala2015eigenvalue}, a study proposing an eigenvalue optimization problem
for the p-Laplacian,
is reducing the problem to analyzing the sign of the derivative of
a function defined on the real line.

In order to prove theorem 1 in \cite{holzhauser2017fptas}, a study of the parametric knapsack
problem, it suffices
to show that the sign of the first derivative of each function changes
at most twice while $\lambda$ increases. Since the denominator
in the derivative's expression is always positive, the authors bound
the number of times the sign of the numerator changes.

In a paper discussing the applicaitons of RProp, \cite{kotsialos2019constrained} praises several aspects of it, including
the simplicity of implementation, the relatively low computation effort,
as only one function and one gradient evaluation are required per
iteration, and the excellent convergence properties that make RProp a highly
efficient algorithm for large-scale problems. More so, RProp can tolerate errors
in the gradient evaluation since it is based on the partial derivatives'
sign rather than their values, allowing its application to non-smooth
Lipschitz continuous objective functions.

In an analysis of opportunities of analytical method
of optimization in chemical technology, \cite{prishchenko2020analysis} classifies optimums based on the signs
of the first and higher-order derivatives and also relies on the local
trend near the point.

As an alternative active set estimation scheme, in addition to the
$\varepsilon$ margin, \cite{kan2021pnkh}, proposing a projected Newton–Krylov
method for large-scale bound-constrained optimization, considers the sign of the
partial derivative so that curvature information is used for those
constraints predicted to become inactive. Thus $A_{aug}$ leverages
the signs of the derivatives of $f$.

The parameter $s_{i}$ in equations 18 and 19 of \cite{lahmdani2021smoothing}, a paper proposing a smoothing sequential convex programming
method,
is defined based on the product of the signs of the (discrete) one-sided
derivatives of $x_{i}$ at $k-1$.

In a study of flexibility index of black-box models with parameter uncertainty through
derivative-free optimization (see \cite{zhao2021flexibility}), the first and second derivative signs are used to define the vertices' directions.

\index{Bayesian optimization}

\subsection{Bayesian optimization}

\cite{li2017bayesian} presents an algorithm to detect the monotonic detection of the underlying function. The novel Bayesian Optimization algorithm proposed incorporates the monotonicity of the underlying function to optimize
towards a target value.

In a paper on correcting boundary over-exploration deficiencies in Bayesian optimization
with virtual derivative sign observations (see \cite{siivola2018correcting}), the 'virtual derivative sign' is leveraged
for correcting boundary over-exploration deficiencies in Bayesian
optimization.

The function in \cite{li2019accelerating}, a study of accelerating experimental design by incorporating experimenter hunches, is modeled using monotonic
GP by placing the consistent derivative signs across the search space.

\index{Optimal Control}

\subsection{Optimal Control}

\cite{schorsch2018identification}, in a study of optimal control of fructo-oligosaccharide production, observes that the Hamiltonian is affine in the control input. In general, no maximum exists
in this context. However, because of the linear inequality constraints
on the control variable, corresponding to the hardware constraints,
a solution exists by resorting to the bang-bang method with singular
arcs. This method evaluates the sign of the partial derivative of
the Hamiltonian with respect to $Q$, i.e., the value of $\psi$.

Lemma 4.2 in \cite{cheng2018reaching}, a paper on GPS-denied or costly areas, relates the sign of the partial
derivative in (4.29) to a geometric property of the optimal trajectory
at the terminal time. Lemma 4.4 connects the geometric property of
the optimal trajectory at the terminal time with the sign of the derivative
in (4.43).

\index{Stochastic Optimization}

\subsection{Stochastic Optimization}

The number of derivative sign changes is an essential part of Lemma
1 in a paper on robust mean-covariance solutions for stochastic optimization (see \cite{popescu2007robust}).

In section 3.2 of \cite{fateen2014note}, the authors modify the original
Cuckoo Search algorithm to incorporate information about the gradient of the
objective function. Any modification to the algorithm should not change
its stochastic nature in order to avoid affecting its performance. A
modification is made to the local random walk in which a fraction
($1-pa$) of the nests are replaced. In the original algorithm, when
new nests are generated from the replaced nests via a random step,
the step's magnitude and direction are both random. In the modified
algorithm, the randomness of the magnitude of the step is reserved.
However, the direction is determined based on the sign of the gradient
of the function. If the gradient is negative, the step direction is
made positive. If the gradient is positive, the step direction is
made negative.

In \cite{wang2019spi}, the historical gradients lag the update of
weights in the period $\left[t_{1},t_{2}\right]$ when the gradient
direction gets reversed, and lead to severe oscillation about the
optimal point. To ease the fluctuation, the proposed SPI-Optimizer
isolates the integral component of the controller when the inconsistency
of current and historical gradient direction occurs, as shown in Eq. 4.
The SPI-Optimizer is described by Eq. 5. The key insight here
is that the historical gradients will lag the update of the weights if
the weights do not keep the previous direction, i.e., $sgn\left(\nabla L\left(\theta_{t}^{\left(t\right)}\right)\right)$
does not agree with $sgn\left(v_{t}^{\left(i\right)}\right)$, leading
to oscillation of gradients about the optimal point until the gradients
compensate the momentum in the reversed direction. In this way, SPI-Optimizer
can converge as fast as MOM and NAG, leading to a much smaller maximum
overshoot.

Although the approach in a paper proposing a stochastic three points (STP)
method for unconstrained smooth minimization (see \cite{bergou2020stochastic}) is designed
not to use explicitly derivatives, it covers some first-order methods.
For instance, if the probability law is chosen to be the Dirac distribution
concentrated at the gradient sign, then STP recovers the Signed Gradient
Descent method, as proved in appendix B.

\index{Dynamical Systems}

\section{Dynamical Systems}

The sign of the derivative of $V$ in a paper discussing a multistate friction model with
switching parameters (see \cite{capone2005instability})
is linked through a simple relation to the eigenvalues; it allows
localizing the sources of instability, i.e., the points at which the
instability begins.

In \cite{solis2005chaos}, Chaos in the one-dimensional wave equation is due to the changes of sign of
the derivative of the energy function, the so-called 'self-excited
oscillations.' The discussion here is about a particular case where
this is not a necessary condition.

In \cite{yilmaz2015reliability}, the different distributions such
as the Farlie Gumbel Morgenstern Distribution with Identical Marginals,
Marshall-Olkin Trivariate, and Gumbel Type I Trivariate Exponential
Distributions are analyzed in terms of their monotonicity based
on their derivative sign (see part 10.3).

The sign of the slope coefficients in \cite{butner2017modeling} captures
the type (e.g., attractor, repeller, limit cycle) and strength of
attraction for the dynamic implied by the equations. The
fixed effects can be interpreted as indicating the likelihood
of being in a pattern. A positive sign means that declines
in a food pile corresponded to reductions in the pattern. A negative
sign suggested that decreases in a food pile corresponded to increases
in the pattern.

\cite{kalabuvsic2018global}, in a paper on global dynamics of
certain mix monotone difference equation,
prove global attractivity results, noticing that the
sign of the partial derivative with respect to the first variable
at the equilibrium point depends on the sign of $b-a\beta$. They
use this insight to prove lemma 3 and theorem 7.

According to \cite{shah2018reprogramming}, it has become experimentally possible to \textquotedbl reprogram\textquotedbl{}
a cell's fate by externally imposed input stimulations. In
several of these reprogramming instances, the underlying regulatory
network has a known structure and often it falls in the class of
cooperative monotone dynamical systems. Their monotonicity is reflected
in the constant signs of their partial derivatives. This paper introduces a new monotonic
property - sign-symmetry, which is the
equality of the signs of the partial derivatives.

In \cite{aguilera2021particular}, a paper on the physics of the Free Energy Principle, the sign of the slope of $f$ can
change, but the slope of the derivative is always negative given
the presence of a global attractor. This study illustrates how, even in straightforward
examples, these quantities can have radically different behaviors
and the conditional average flow does not necessarily capture
the actual behavior.

\index{Set Theory}

\section{Set Theory}

In a paper proposing a direct approach for determining the switch points in the Karnik–
Mendel algorithm, \cite{chen2017direct} calculates
the partial derivatives and finds a switch point $k\in\left[1,N\right]$
for which $\frac{\partial c}{\partial u}(k)\leq0$ and $\frac{\partial c}{\partial u}\left(k+1\right)\geq0$.
The paper shows that it is possible to find the switch point directly by locating the value
of $k$ where the sign of the partial derivative changes.

\index{Geometry}

\section{Geometry}

The proof of lemma 9.8 in \cite{kazez2015approximating}, a study of approximating C1, 0–foliations, states that
the orientation of $\Theta$ restricted to the binding is determined
by the sign of the slope of $\gamma_{i}$ as expressed in $\left(\lambda_{i},\nu_{i}\right)$
coordinates.

In a study of polar tangential angles and free elasticae (see \cite{miura2020polar}), the sign of the polar tangential angle function dictates properties.

\cite{bellet2020symmetry}, in a work on symmetry group of the equiangular cubed sphere, proves several lemmas by applying a monotonicity
analysis to the function via its derivative sign.

\index{Probability}

\section{Probability}

\cite{kofanov2003kolmogorov} obtains a new exact Kolmogorov-type
inequality, which considers the number of changes in the sign of the
derivatives over periods.

\index{Statistics}

\section{Statistics}

\index{Stochastic Processes}

\subsection{Stochastic Processes}

\cite{liu2010spde} extends the usual framework of
Stochastic Partial Differential Equations with monotone coefficients to include a large class of cases
with merely locally monotone coefficients.

In a study of organized criticality in a network of economic
agents with finite consumption, \cite{da2012self} defines an event as a (typically small)
set of successive instants in the original time-series having the
same derivative sign.

In a study of short-term asymptotics for the implied
volatility skew under a stochastic volatility model with Lévy jumps, \cite{figueroa2016short} shows that the order of convergence and the sign of the ATM (At the Money) implied volatility
slope can be easily recovered from the
model parameters.

The \textquotedbl Proportion of slope sign changes\textquotedbl{} is
proposed in \cite{duchen2017inference} as another way of assessing
convergence by taking the last $n$ values of the expectation–maximization algorithm and
counting the number of times there is a change in the sign of the
slope between consecutive values. If convergence is reached, the number
of slopes with a positive sign is expected to be similar to those
with a negative sign.

\cite{gess2020random} proves the existence of random dynamical systems
and random attractors for a large class of locally monotone stochastic
partial differential equations perturbed by additive Lévy noise.

In \cite{willers2020adaptive}, a study of adaptive stochastic continuation with a modified lifting procedure
applied to complex systems, a straight line can be parametrized
into two different directions. The relation between $G_{\lambda}$'s
slope sign and the stability of the steady-state depends on the direction
of the parametrization. For one direction, there is stability for a
positive slope, while for the other direction there is stability for the negative slope.

In a work on implied stochastic volatility models (see \cite{ait2021implied}), the sign of the leverage effect coefficient $\rho\left(v_{t}\right)$
is determined by the sign of the slope $\sigma$.

\index{Gaussian Processes}

\subsubsection{Gaussian Processes}

A method for using monotonicity information in multivariate Gaussian
process regression and classification is proposed in \cite{riihimaki2010gaussian}.
Monotonicity information is introduced with virtual derivative observations,
and the resulting posterior is approximated with expectation propagation.
The behavior of the method is illustrated with artificial regression examples.
The method is used in a real-world health care classification problem
that includes monotonicity information with respect to one of the covariates.

As stated in \cite{blum2013optimization}, Gaussian processes are
a powerful tool for nonparametric regression. Rprop, a fast and accurate
gradient sign-based optimization technique initially designed for
neural network learning, can outperform more elaborate unconstrained
optimization methods on real-world data sets, converging more
quickly and reliably to the optimal solution.

\cite{riutort2019gaussian} proposes a spatio-temporal model that
considers the derivative information by jointly modeling the regular
process and its derivative process using the Gaussian approach. Derivative observations
of both the sign and the values of partial derivatives are used to
induce monotonicity (non-decreasing) and long-term saturation as a
function of time. Furthermore, to force the functions to be zero at
the starting timepoints ($t=0$), noise-free pseudo-observations are
used at these points. It is shown that constraining the model using derivative sign
observations is beneficial in predictive performance and application-specific
interpretability.

In \cite{riutort2020correlated}, all of the judgments about $f$
that the facilitator has elicited from the expert have followed normal
distributions. This fact has allowed properties of multivariate normal
distributions to be utilized. However, if a condition is placed on
the derivative sign of f at a point, the Gaussian process model leads
to a truncated normal distribution.

\index{Statistical Theory}

\subsection{Statistical Theory}

In \cite{gosling2007nonparametric}, a property of Gaussian processes
can be manipulated to include judgments about the derivatives of the
density, which allows the facilitator to incorporate mode judgments
and judgments on the sign of the density at any given point.

\cite{carobbi2008absolute}, in a work on the absolute maximum of the likelihood function of
the rice distribution, analyzes the loss function's derivative
sign to prove it has a lower bound.

The proofs of the main results in \cite{bebbington2012discrete}, a study of the discrete additive Weibull distribution,
are based on analyzing functions' trends (derivative sign).

Theorem 2.1 in \cite{qian2012fisher}, a study of the Fisher information matrix for a three-parameter exponentiated Weibull
distribution under type II censoring, is illustrated through visualizing
the sign of the derivative of the hazard function.

In \cite{chetverikov2017nonparametric}, the ill-posedness of the
inverse problem of recovering a regression function in a nonparametric
instrumental variable model leads to estimators that may suffer from
a prolonged, logarithmic rate of convergence. In this paper, the authors
show that restricting the problem to models with monotone regression
functions and monotone instruments significantly weakens the ill-posedness
of the problem. In stark contrast to the existing literature, in this study the
presence of a monotone instrument implies boundedness of their measure
of ill-posedness when restricted to the space of monotone functions.
Based on this result, the study proposes a novel non-asymptotic error bound
for the constrained estimator that imposes monotonicity of the regression
function.

In a paper on consistent parameter estimation
for Lasso and approximate message passing (see \cite{mousavi2018consistent}), the sign change of the derivative is leveraged to prove an essential theorem within the algorithm aiming
to consistently estimate parameters for Lasso.

Table 8 in \cite{galvao2018testing} reveals a small but statistically
significant effect of the market portfolio return on risky asset risk
premium. Interestingly, the sign of the slope parameter for this market
factor becomes negative for the highest quantiles of the distribution
of excess asset returns and suggests the decoupling between the market
portfolio return and firms' asset returns for substantial firm returns.

According to \cite{qasim2020biased}, multicollinearity maximum likelihood (MLE) scenarios come with
the wrong sign of the slope parameters. However, biased estimation
methods may change the sign of the slope parameters. For instance,
theoretically, pinnacle away win odds and maximum market away win
odds negatively affect the number of full-time away-team goals, while
the MLE shows a negative effect. The method proposed in this study, on the other hand, shows
a positive effect, and it is considered a good approach to tackle
the problem of multicollinearity.

\index{Descriptive Statistics}

\subsection{Descriptive Statistics}

Functions' monotonicity is a crucial assumption throughout the formulation
and proofs of several lemmas in \cite{nock2015conformal}.

In the proof of lemma 2.5 of \cite{yang2017converses}, a paper on conformal divergences and
their population minimizers, the monotonicity
of $V$ is proved by analyzing its derivative sign. 

Lemma 2 in \cite{aghamolaei2020approximating} formulates a sufficient
condition for a segment in the normalized FDS to correspond to a segment
in the original space based on the monotonicity of the edges of the
input curve. Algorithm 1 outputs a monotone path from $\left(P\left[0\right],Q\left[0\right]\right)$
to $\left(P\left[n-1\right],Q\left[n-1\right]\right)$.

\index{Game Theory}

\section{Game Theory}

An alternative way of viewing the problem in \cite{bayramoglu2018climate}
is by noticing that the sign of the second derivative of payoff function
Eq. (2) depends on the sign of term $A$ with respect to the other players'
mitigation levels. Thus, if $A>0$, the payoff function is not concave
but convex in the other players' mitigation levels. Upward sloping reaction
functions could lead to more optimistic outcomes in a coalition formation
game (i.e., larger coalitions). The intuition is that, if mitigation
levels are strategic substitutes, any additional increase in signatories'
mitigation efforts is countervailed by a decrease of non-signatories
mitigation efforts. In climate change, this has been called (carbon)
leakage, which makes it less attractive to join an agreement. Thus,
upward sloping reaction functions may be viewed as a form of anti-leakage
or matching, which may be conducive to forming large stable coalitions.

In \cite{corchon2018handbook}, the sign of the slope of the best
response function at a point in the strategy space is solely determined
by the cross-effect on the marginal payoff function since the denominator
of Eq. (7.5) is unambiguously negative.

\cite{hoffmann2020endogenous} studies the presence of a first-mover
advantage or a second-mover incentive, which also depends on the sign
of the cross partial derivatives of the payoff functions at the NE. Plain and strategic complements and substitutes are defined
based on the payoff trend relative to the equilibrium strategy.

\index{Decision Theory}

\section{Decision Theory}

The sign of the slope of the consumption stream in \cite{drouhin2001lifetime}
depends on comparing the subjective factor of discount and the economic
aspect of discount. In the model, the subjective factor of discount
is a function of an objective parameter, the probability of surviving.
It is not necessary to know the particular shape of the utility function
to make some predictions about the slope of the consumption function.
Agents with a low probability of survival will consume more in the
first period than in the second (positive slope of the consumption
stream). In contrast, those with a high probability of survival will
consume less in the first period than in the second (negative slope).
When the income of the
second period is zero, the sign is the same as the difference between
the coefficient of relative intertemporal substitution resistance
and one.

Proposition 2 in \cite{blomberg2008have} proves that an increase
in $n$ decreases $p$. The importance of Proposition 2 is that one can use this theory to fit one of our three empirical
regularities - namely, smaller campaigns, i.e., those with lower
$n$, should have a higher fraction of heroes. An increase
in the group size reduces the probability that at least one person
will take action; this is based on numeric simulations of the derivative
sign of $p^{*}$.

In \cite{bertola2015hidden}, as long as the first and second-order
conditions of optimality for individual choices both hold at the equilibrium,
it is also possible to show that the equilibrium effort is quite intuitively
lower when more insurance is available. Result 4 proves that the individually
optimal equilibrium effort is negatively related to equilibrium insurance
volume based on a partial derivative sign analysis.

Proposition 2 in \cite{zheng2018willingness}, a study of willingness to pay for reductions in morbidity risks under anticipated
regret, proves three results
on the upward monotonicity of willingness to pay with respect to its different parameters
$\lambda,p$, and $w$. A comparative analysis based on the derivative
sign-based proof is provided.

Before closing this section, it is worth mentioning that the study of monotonicity properties and of locally monotone operators are important areas of research in Analysis.

\chapterimage{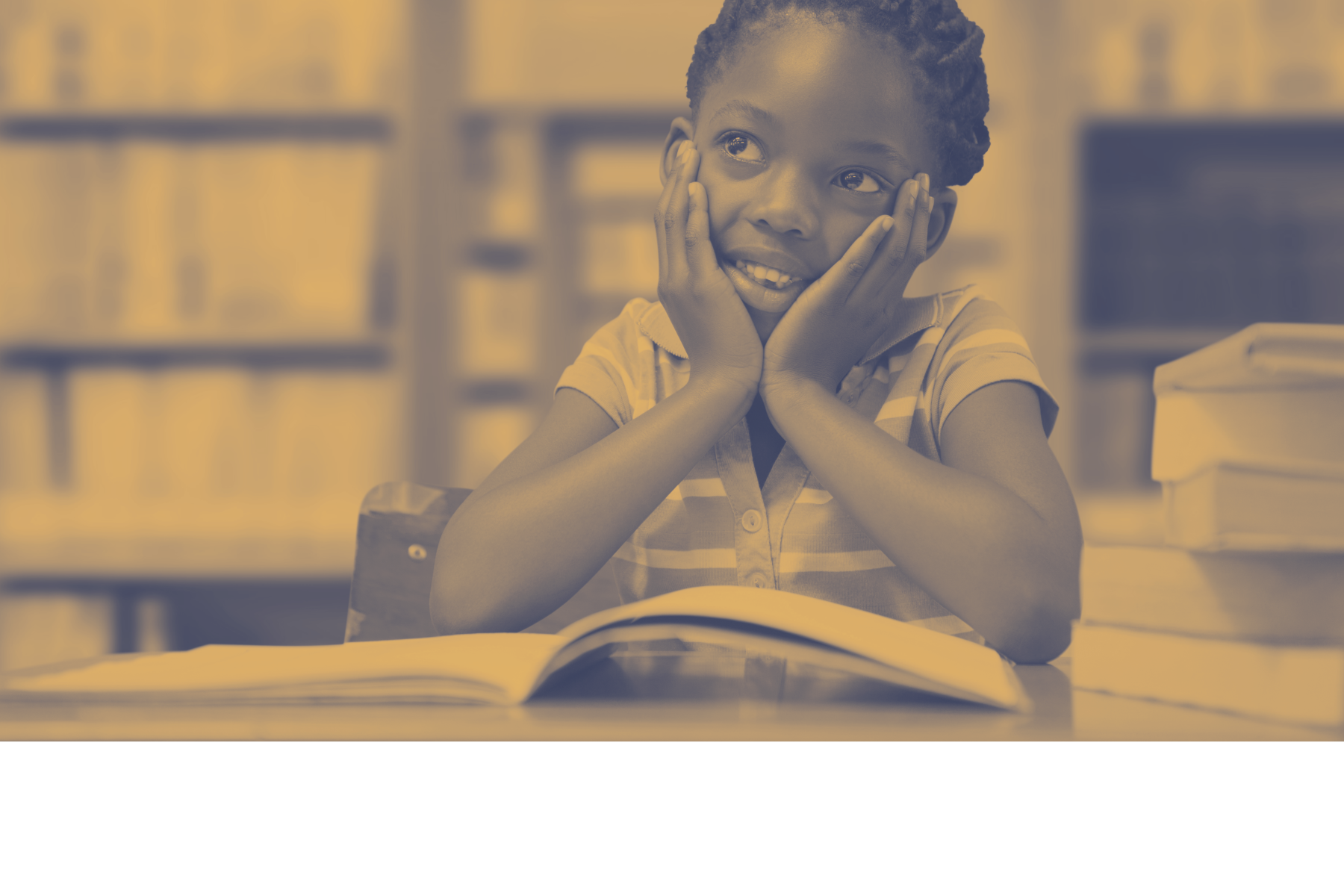}
\index{Science Education}
\chapter{The Educational Trendland}
\label{educational_trendland}
Science education researchers have been dedicating their attention to the way teachers illustrate trends
at schools and have pointed out that trends are an essential part of any student's toolbox. However, in education, trends are often defined in terms of the derivative sign. There are several caveats and disadvantages of this approach, as exemplified by many research studies.

One example of how trends are used in educational settings is the paper of \cite{nemirovsky1994ways}, where the velocity sign is introduced as a
helpful educational tool, more specifically, as a way to learn how to symbolize situations of physical
change.

Interpreting the derivative sign and the monotonicity in \cite{asiala1997development}, a study of the development of students’ graphical understanding of the derivative, is crucial in plotting the function.

\cite{da2002analyzing}, in a study of functions' behavior in a computational environment, recommends dedicating many questions to the
relation between the derivative sign and the function's monotonicity
as visible in its graph.

\cite{hahkioniemi2004perceptual} showed that while the vast majority of the
students also determined the trend (velocity sign), some students
had difficulties with functions and could not draw a tangent.

In \cite{hahkioniemi2007derivative}, a study investigating how the derivative becomes visible, the student seemed to consider
the derivative as an object with some separate properties, such as
sign and magnitude.

In a similar line of work, \cite{hahkioniemi2008durability} found that although the student's intuition
of the monotonicity classification is \textquotedbl Positive when
the line is ascending and negative when it is descending,\textquotedbl{}
she expressed it as the \textquotedbl sign of the slope of the tangent,
pencil as a tangent,\textquotedbl{}.

\cite{christensen2010investigating} discusses the role of the derivative
sign in Thermodynamics education, specifically in physics and calculus concepts.

\cite{garcia2011characterizing} found that students understand the functions
monotonicity as the function's derivative sign and they see concavity
as the second derivative's sign.

The derivative sign is mentioned in \cite{eun2011calculus} as one
of the first building blocks towards internalizing the concept of
the derivative.

In \cite{planinic2012comparison}, many learners identify the slope with
the angle between the straight line and the x-axis, or they evaluate
the sign of the slope according to the quadrant in which the line
is drawn. The same problem appears in other studies, as slope/height
confusion is common in Physics and
Mathematics.

\cite{christensen2012investigating} discuss the role of the derivative
sign in a study of graphical representations
of slope and derivative without a physics context.

According to \cite{hempel2018noncongruence}, studies have identified
two misconceptions that students have about the graphical representations of
the derivative. One assumption is that the graphs of a function and
its derivative resemble each other in terms of the trends. The other assumption is that the derivative
of an increasing (decreasing) function is always positive (negative),
in ignorance of the option that the derivative can be zero.

\cite{park2012transition} uses the monotonicity in an interval and
the derivative sign to ease the transition from the pointwise derivative
to the derivative function. Switching from a single
value of the derivative at a point makes it easier to think about
a single value (its sign) in the interval.

In \cite{criscuolo2013study}, students are shown the derivative of
the original function, which is also a rational function.
The roots of the numerator of the derivative are plotted as $M_{1}$
and $M_{2}$. These are locations at which the derivative is zero.
A graph of the sign function of the derivative is also provided (see Figure 8). The instructor encourages
students to interpret the location of $M_{1}$ and $M_{2}$, explicitly
asking how their location will impact the shape of the graph. The
sgn graph of the derivative (labeled as $h\left(x\right)$ in Figure
8) helps students answer the questions. For instance, the sign of
the derivative to the left of $M_{1}$ is negative; to the right,
positive. The information provided suggests that the function has a relative minimum (i.e.,
a \textquotedbl turning point\textquotedbl ) at $M_{1}$ and a relative maximum at $M_{2}$. Students use
this information to revise their sketch, as shown in Figure
9.

\cite{kostic2014extreme} mentions that the function's first derivative
sign and monotonicity are traditionally addressed after the derivative,
and its zeros, are found.

When interpreting the slopes in \cite{sokolowski2014constructivist},
the interpretation of the units of the slope (subsection 3.1) is
done separately from the interpretation of its sign (subsection
3.2); however, they are both inferred from the derivative.

\cite{martinez2015students} describes a study where many students made an effort
to decide the sign of a directional derivative without representing
the direction vector in a three-dimensional space. One student's
notion of partial derivative was constrained as she could not do actions
to form different quotients when considering a tabular representation; the only things she was able to do were the actions
associated with determining the sign of a partial derivative given
the surface graph. The student was able to eventually determine the sign
of the directional derivative on problem 3. To do this,
she coordinated used the function of two variables to determine the
base point, the schema of vectors to represent the vector's direction,
and the derivative of a function of one variable to determine the
sign of the derivative. Another student got confused when
asked about the partial derivative signs. The interviewer tried to
help him and gave a series of hints but they were not helpful.

According to \cite{bollen2016generalizing}, most students can
correctly identify the direction of motion in a linear kinematics
graph. However, some students appear to struggle with the reference
point implicit in distance-time graphs. Furthermore, the authors found
that a qualitative understanding of kinematics graphs is necessary
but insufficient for students in algebra-based courses to determine
instantaneous speeds correctly; for students in calculus-based classes,
it is neither necessary (though highly desirable) nor sufficient.
These findings imply that both the qualitative and quantitative aspects of linear
kinematics graphs should be taught.

On set $S_{3}$ in \cite{ivanjek2016student}, the main observed difficulty
was that students sometimes identified the sign of the slope with
the sign of the y coordinate. This situation may be regarded as a special
case or a consequence of slope-height confusion. One student's
explanation illustrates this difficulty: \textquotedbl The gross domestic product (GDP) growth
rate is negative between 2006 and 2010, because GDP is negative in
that period\textquotedbl .

\cite{borji2018application} claims that according to the Action–Process–Object–Schema theory,
the sign of the derivative is the first recommended action in the
process of sketching the function's derivative.

In \cite{fuentealba2018understanding}, the decomposition of the logic relationship of double implication
between the positive sign of the first derivative in an interval and
the strict growth of the function in the said interval allows to
generate the variables $V_{11}$ and $V_{12}$.

According to \cite{seethaler2018analyzing}, grasping the meaning of the rate of change sign is one of the main difficulties that
students are facing.
Studies have shown that undergraduate students struggle with
negative rates of change in various contexts, including kinematics
(the meaning of negative velocity and negative acceleration), light
intensity over distance from a point source, and discharge of a capacitor
in a simple circuit. For example, when determining whether something
is slowing down or speeding up, students may base their responses
on the sign associated with the slope of the position versus the time
graph rather than the change in the magnitude of the slope. Students struggle
to attend to the magnitude or absolute value and to the sign of the rate
of change simultaneously and find it especially confusing when rates
are negative but are increasing in magnitude. When solving equations or
interpreting graphs, students commonly confuse the slope sign with
the sign of the $y$-coordinate or carelessly drop the negative sign.
These findings have implications for the curricular treatment of the negative
sign associated with the consumption of reactants.

In \cite{ceuppens20199th}, the slope sign is applied to distinguish between validation results (see table 1). The items with the highest
average scores are \textquotedbl compare given a graph\textquotedbl{}
questions, three out of four have a positive slope, and three are
kinematics questions. The study highlight the important effect
of the slope sign on the average accuracy and the difficulty students
have with negative slopes in kinematics. These results are significantly influenced by the inclusion of the possible minus sign for the
one-dimensional velocity as a criterion for a correct answer. The
results show that most students do not include the minus sign in kinematics,
i.e., they consider the magnitude and omit the direction. This pattern
is not present in the mathematics questions, so students are far more
likely to include the minus sign when isomorphic equations and graphs
are used. 

Also according to \cite{ceuppens20199th}, the most preferred method in kinematics is, by far, calculating
the ratio of differences (mainly in questions with a graph). The method
is not always applied correctly, though. After calculating the ratio
of differences, students often omit the minus sign (when present)
from the result; from the students' explanations it is clear that this
is triggered by the use of the formula $v=\frac{\Delta x}{\Delta t}$
in kinematics and $a=\frac{y_{2}-y_{1}}{x_{2}-x_{1}}$ in mathematics.
Although this is essentially the same formula, there are a few differences
in students' use. In mathematics, the formula is usually used correctly.
In physics students usually write down the correct formula but often
calculate $v=\left|\frac{\Delta x}{\Delta t}\right|$. Illustration
of an algebraic interval or point confusion is given by the answers
to $K_{10}$ in Fig. 10. Interestingly, in the comparable question
$M_{12}$ - in which only the sign of the slope or velocity is different---the
students apply the (assumed) expert-like strategy $S_{1}$ in which
the slope is identified through its location in the equation. In questions
with a negative slope, and mainly in kinematics questions, students
sometimes make statements related to the motion of the cyclist, such
as: \textquotedbl the cyclist is returning\textquotedbl{} or \textquotedbl is
riding backward\textquotedbl{} or \textquotedbl is slowing down.\textquotedbl{}
The first two statements indicate that these students include a sense
of direction in their interpretation, but some have difficulties understanding
and/or expressing movements in opposite direction to the direction of the position
axis. A respondent repeatedly wrote \textquotedbl riding backward
or back,\textquotedbl{}, showing doubts about the correct
interpretation. The ``slowing down'' statement illustrates the difficulty
students have with interpreting $x\left(t\right)$ graphs with constant
velocity by confusing them with $v\left(t\right)$ or $v\left(x\right)$
graphs. The results show that the sign of the slope in mathematics
is not a problem for most students, but when confronted with a negative
sign for the velocity in kinematics, students frequently omit the minus
sign. Moreover, understanding the sign of the velocity in kinematics may be particularly challenging in specific linguistic contexts; for example, there is no distinction between the words speed and velocity in Dutch
and Hebrew.

According to
\cite{cabanas2019didactic}, one of the critical steps used to determine the sense of functions'
variation is calculating the sign of the derivative. It is in this step where variations are
observed, according to at least two procedures. A solution is given by
algebraic methods using inequalities and the \textquotedbl probative
number\textquotedbl{} approach: choosing one smaller value and one
greater value than the root (probative value) for each stationary
point. These values are substituted in the analytical expression of
the derivative and the signs of these values are observed.

In \cite{nagari14constrained}, sketches that represent incorrect
values but with the correct sign (negative or positive) suggest that
students connect between the increasing (decreasing) domains of the
function and the positive (negative) domain of the derivative, but
might ignore concavity upward and downward of the function domain.
Sketches that follow correct increasing and decreasing domains but
contain mistakes in values suggest that students connect the concavity
upward and downward of the function domain and the increasing and
decreasing domains of the derivative.

\cite{kartal2019mathematical} claims that students sometimes understand
the derivative intuitively as the \textquotedbl instantaneous change\textquotedbl{}
of the function or \textquotedbl the change with respect to $x$.\textquotedbl{}
Put differently, the changing trend assimilates in the rate of change,
and the derivative is presumed to capture the function's momentary
change as a whole. Then, the derivative sign tests contradict this
intuition.

Item 1M in \cite{carli2020testing} contains an algebraic expression
for the first derivative of a function. The students had to relate
the coefficients in this expression to the tangent's slope sign. In
the parallel physics item (1P), an algebraic expression for the time
derivative of position was given, and the students were asked to determine
when the object's velocity was negative. The students performed much
better on item 1P than on item 1M. For item 1M, the most common incorrect
answer was distractor D, consisting of using the wrong coefficient
to determine the slope sign. Items 4M and 4P concerned the relationship
between the first derivative and the function's maxima. The two similar
items were formulated a little differently. In item 4M (context of
mathematics), the students were given information about the sign of
a function's derivative, and they had to decide where the function
had its maximum value within the given interval. In item 4M (context
of physics), information was given about the sign of an object's acceleration
and the object's velocity at a point. The students were required to
choose the correct option describing the object's velocity at another
point. A similar percentage of students answered these two items correctly
in the two contexts, but the $\Phi$ coefficient was low, suggesting
that the two performances are weakly correlated. In fact, by checking
the students' answers in more detail, it turns out that the number
of students who answered only one item (either $M$ or $P$) correctly
is comparable to the number of students who answered both items correctly
or both items incorrectly.

In \cite{ikram2020mathematical}, students were able to establish reversible
reasoning if they recognized the relationship between the sign of
the first derivative with the increase/decrease of the function.

Most of the students' answers to the first question about the function's
monotonicity in \cite{ishibashi2020effect} were based on the first
derivative's sign. However, many answers were provided based on \textquotedbl the
difference between the y values\textquotedbl and \textquotedbl looking
at the graph.\textquotedbl, even though they were not taught
to think about it this way. In turn, when students were asked to code
this during the lab class, many struggled to come up with a way to
represent their understanding visually.

In the immediate post-test $Q_{1}$ of \cite{hyland2021introducing}, a study about introducing direction fields to
students learning ordinary differential equations,
a common strategy was to compare the sign of the slope (which is positive
except at the origin where it is zero) with the slopes indicated by
the direction fields.\newpage{}

\part{Scientific Discussion}

\label{scientific_discussion_part} \chapterimage{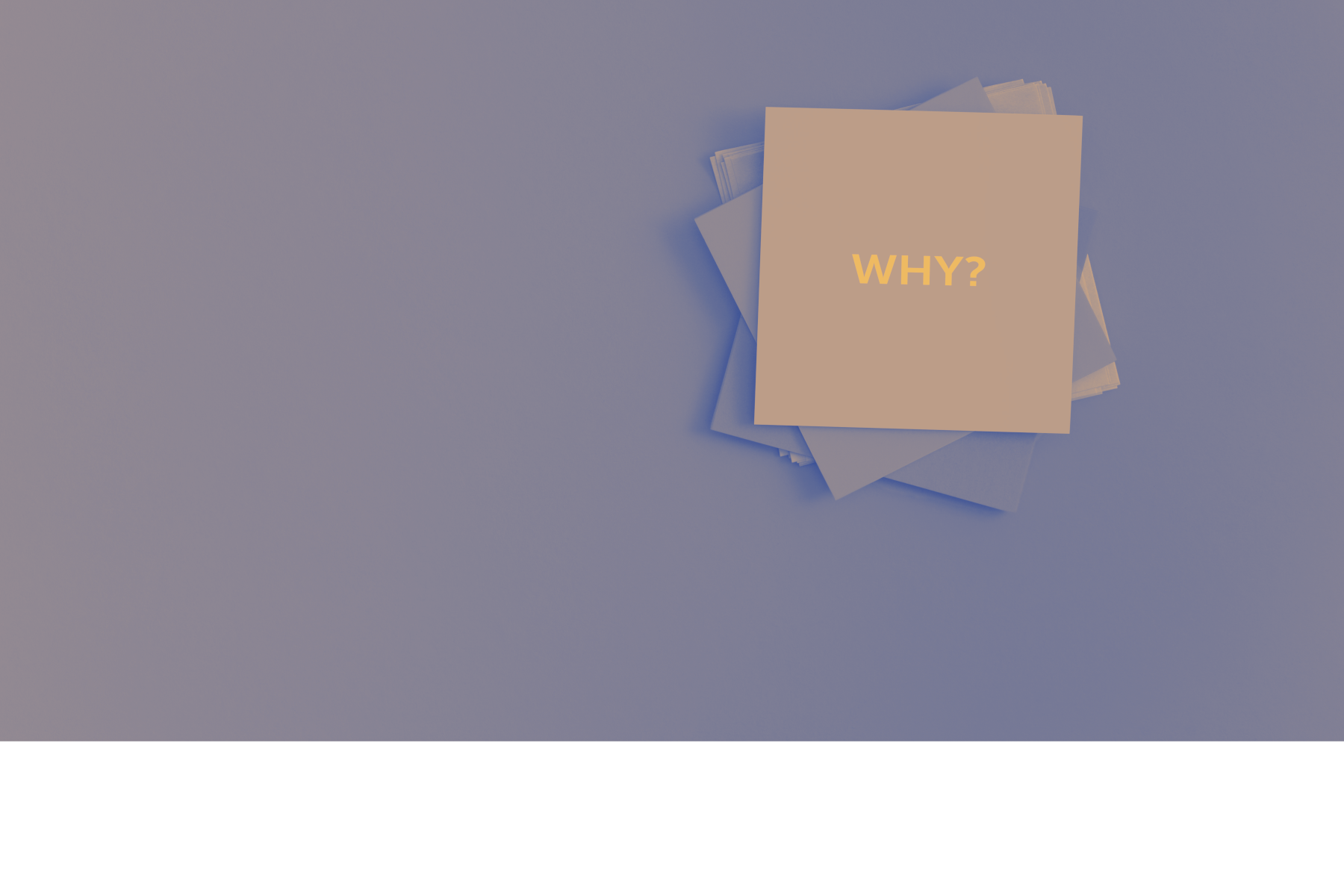}

\chapter{Reasons Why Trendland Is Trending}

There are several reasons to exploit the trend rather than the complete
rate information that the derivative embodies. In continuous domains, for instance, trends are often used to classify a function's monotony or curvature to prove a theorem
(see \cite{bebbington2012discrete,berman2013predation,kazez2015approximating,bengui2018macroprudential,gabszewicz2018random,de2019strategic,fowl2021preventing}).

Researchers are often interested in a mere portion of the available
data for a qualitative analysis of their findings, something which is
prevalent in Qualitative Trend Analysis (see \cite{dash2003fuzzy,dash2004novel,maurya2007signed,maurya2010framework,villez2013generalized,villez2014qualitative,thurlimannqualitative,villez2016shape,thurlimann2018soft}),
Qualitative Reasoning (see \cite{cerbah1992integrating,tiwari2002series,carriquiry2007reputations,ivanjek2016student,ikram2020mathematical}),
and Static Analyses (see \cite{lanaspa2001public,carr2007stochastic,carriquiry2007reputations,kacperczyk2009rational,haucap2012regulation,thoni2015peer,thwaites2015real,chirinko2017tax,tavani2017endogenous,skott2017weaknesses,impullitti2018trade,augustin2018term,loffler2018pitfalls,bardgett2019inferring}).

Trends are also a natural tool for separating data into cases and classifying
different results (see \cite{de2000infrared,dash2003fuzzy,herdaugdelen2004dynamic,hahkioniemi2008durability,arino2010effect,ruiz2014unusual,moussi2015nonlinear,chu2016single,dong2018accurate,dong2018fast}),
including in analytical formulas (see \cite{morel1996precise,dimova2000numerical,ryu2001nonlinear,chen2014dual,yuan2016new,lima2016stochastic,ruderman2017break,choi2017tension,ismail2017passive,chen2018global,wang2018bifurcation,li2018bifurcation,dong2019friction,alaci2020proposed,capace2021modelling,hanisch2021rolling,odabacs2021adaptive}).

In discrete domains, the derivative sign helps tackle several issues, including:
\begin{itemize}
\item The vanishing and exploding gradients issues, as in \cite{abuqaddom2021oriented}
and \cite{ravi2016optimization}, respectively.
\item Noisy data, leading to slow convergence, as in the RProp algorithm
(\cite{riedmiller1993direct}).
\item Fluctuations around the minima point, as in \cite{wang2019spi}.
\item Energy and computing resources sparing, as in \cite{hubara2017quantized}.
\item Linearizing the cost function and solving for the perturbation that
maximizes the cost subject to an $\ell_{\infty}$ constraint, as in
\cite{goodfellow2014explaining}. This method only uses the sign of
the gradient. Since its discovery, the superiority of signed gradients
to raw gradients for producing adversarial examples has puzzled the
robustness community. Still, these strong fluctuations in the gradient
signal possibly help the attack escape suboptimal solutions with a
low gradient.
\item Generally improving the training process's stability and ease of convergence,
as in \cite{li2018a2}.
\item Reducing overfitting during the training process by using trend-based
features, as in \cite{wulsin2011modeling}.
\item Achieve invariance for the grey shade in Object Detection. For
example, suppose the objects are uniform (such as vehicles). In such
case, it is better to discretize the gradient, as in \cite{arrospide2013image},
where the original Histogram of Oriented Gradients does not generally work well with the derivative
sign, but it does in the particular case of vehicles. \cite{ojala2002multiresolution}
illustrates a similar example.
\item When it is impossible to calculate the gradient precisely, as in Equation 1 in \cite{dong2018fast}.
\item When there is a diversification in the behavior of specific entities,
and we require a feature that would capture them all reasonably, as
in \cite{schramm2014dynamic}.
\item When clipping the gradient forces the solutions in a specific domain, as
in \cite{moosavi2019robustness}.
\item Improving the runtime performance compared to the complete derivative
calculation. For example, if we calculate the quotient between two
numerical derivatives, settling with their signs will spare the division,
as in eq. 2 in \cite{huang2017robust}.
\end{itemize}
Note that engineers often apply the discrete derivative sign as the
sign of the function's change from a particular point on the grid
to the following one, such as in the Scipy implementation of peak detection
(\cite{scipy2018}).

\chapterimage{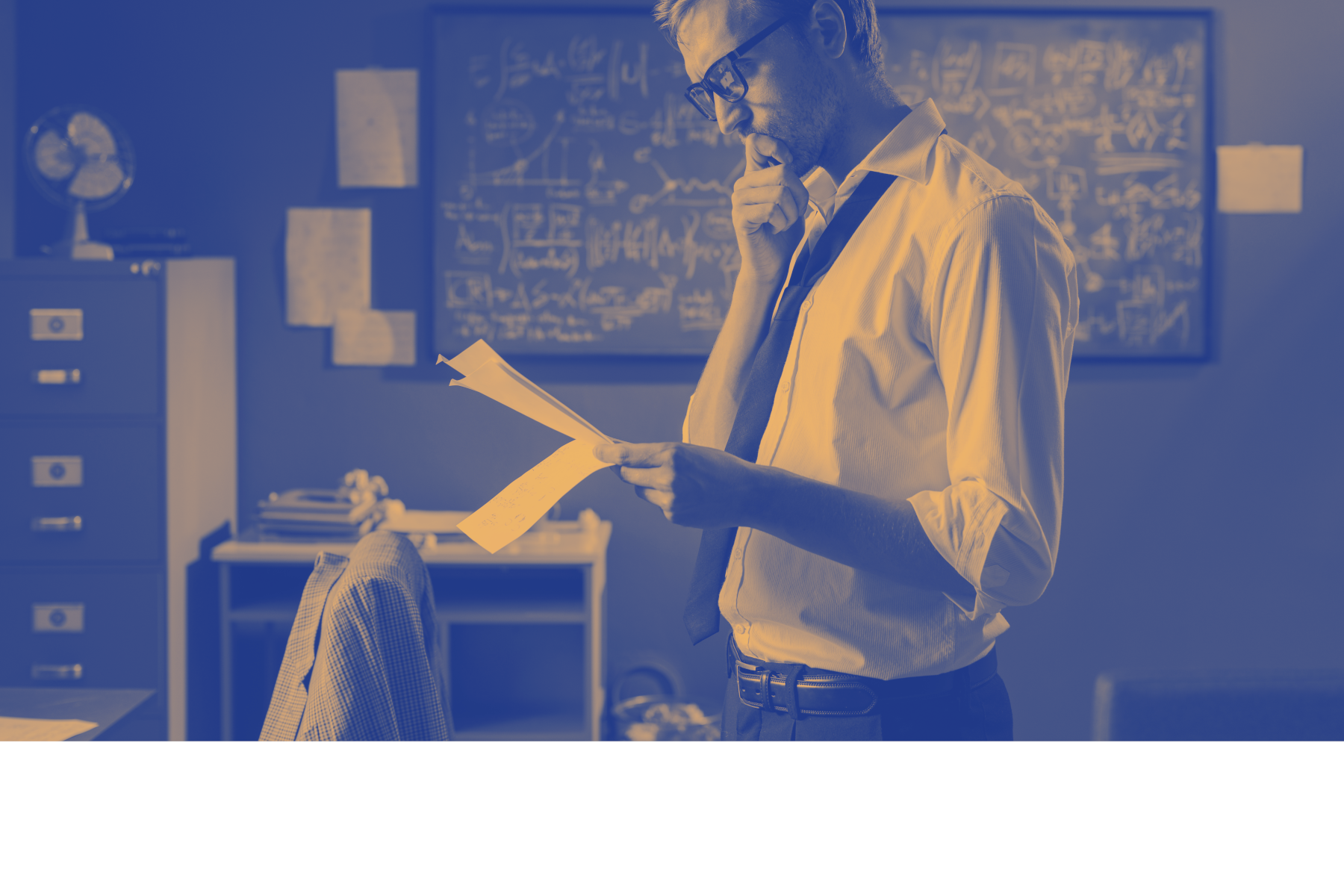}

\chapter{Reviewing Previous Discussions}

There are several mathematical studies of convergence properties of sign-based
optimization techniques in Machine Learning.

In \cite{cortes2005achieving}, the authors introduce the normalized
and signed gradient descent flows associated with a differentiable
function. They characterize convergence properties via nonsmooth
stability analysis and identify general conditions under which
these flows attain the set of critical points of the function in a
finite time. To do this, the researchers extend the results on the stability and
convergence properties of general nonsmooth dynamical systems via
locally Lipschitz and regular Lyapunov functions.

In appendix C of \cite{karimi2016linear}, the authors analyze the
convergence characteristic of the signed gradient descent (RProp).

\cite{bernstein2018convergence} provides an analysis of the convergence
rate of the sign stochastic gradient descent (signSGD).

\cite{balles2018dissecting} suggests that we can expect the sign
direction (as applied in Adam) to be beneficial for noisy, ill-conditioned
problems with diagonally dominant Hessians.

\cite{moulay2019properties} provides two convergence results for
local optimization, one for nominal systems without uncertainty and
one for systems with uncertainties. Sign gradient descent algorithms,
including the dichotomy algorithm DICHO, are applied to several examples
to show their effectiveness in terms of speed of convergence. The
sign gradient descent algorithms allow converging in practice
towards other minima than the closest minimum of the initial condition,
making these algorithms suitable for global optimization as the proposed
metaheuristic method.

According to \cite{balles2020geometry}, sign-based optimization methods
have become popular in machine learning due to their favorable communication
cost in distributed optimization and their surprisingly good performance
in neural network training. The authors find sign-based methods preferable
over gradient descent if the Hessian is to some degree concentrated
on its diagonal and its maximal eigenvalue is much larger than the
average eigenvalue. Both properties are common in deep networks.

\cite{li2021faster} investigates faster convergence for a variant
of sign-based gradient descent, called scaled signGD in three cases:
the objective function is firmly convex, the objective function is
non-convex but satisfies the Polyak-Łojasiewicz (PL) inequality, and
the gradient is stochastic, called scaled signSGD.

The proof outline of the main results for Adam in \cite{zou2021understanding}
is based on the fact that Adam behaves similarly to sign gradient
descent when using a sufficiently small step size or the moving average
parameters $\beta_{1},\beta_{2}$ are nearly zero. These findings motivate the
author to study the optimization behavior of signGD and then extend
it to Adam using their similarities.

\cite{safaryan2021stochastic} analyzes sign-based methods for non-convex
optimization in three key settings: standard single node, parallel
with shared data, and distributed with partitioned data. Single machine
cases generalize the previous analysis of signSGD, relying on intuitive
bounds on success probabilities and even allowing biased estimators.
Furthermore, they extend the analysis to parallel settings within
a parameter server framework, where exponentially fast noise reduction
is guaranteed for the number of nodes, maintaining 1-bit compression
in both directions and using small mini-batch sizes. Next, the researchers identify
a fundamental issue with signSGD to converge in a distributed environment.
To solve this issue, they propose a new sign-based method, Stochastic
Sign Descent with Momentum (SSDM), which converges under standard
bounded variance assumption with the optimal asymptotic rate.

In Adversarial Learning, \cite{faghri2021bridging} studies the impact
of optimization methods such as sign gradient descent and proximal
methods on adversarial robustness.

\begin{table}[H]
\centering %
\begin{tabular}{ccc}
\toprule 
\textbf{Algorithm} & \color[HTML]{FFA006} {\textbf{Rate Version}} & \color[HTML]{0039BD}{\textbf{Trend Version}}\tabularnewline
\midrule 
Gradient Descent & \color[HTML]{FFA006}{$x_{t+1}=x_{t}-\alpha_{t}\nabla f_{t}$} & \color[HTML]{0039BD}{$x_{t+1}=x_{t}-\alpha_{t}\text{sgn}\left(\nabla f_{t}\right)$}\tabularnewline
Stochastic Gradient Descent & \color[HTML]{FFA006}{$x_{t+1}=x_{t}-\alpha_{t}g_{t}$} & \color[HTML]{0039BD}{$x_{t+1}=x_{t}-\alpha_{t}\text{sgn}\left(g_{t}\right)$}\tabularnewline
\bottomrule
\end{tabular}\caption{The cannonical gradient descent algorithms and their signed analogues.
The term $g_{t}$ refers to stochastic gradients calculated at iteration
$t$.}
\end{table}

\chapterimage{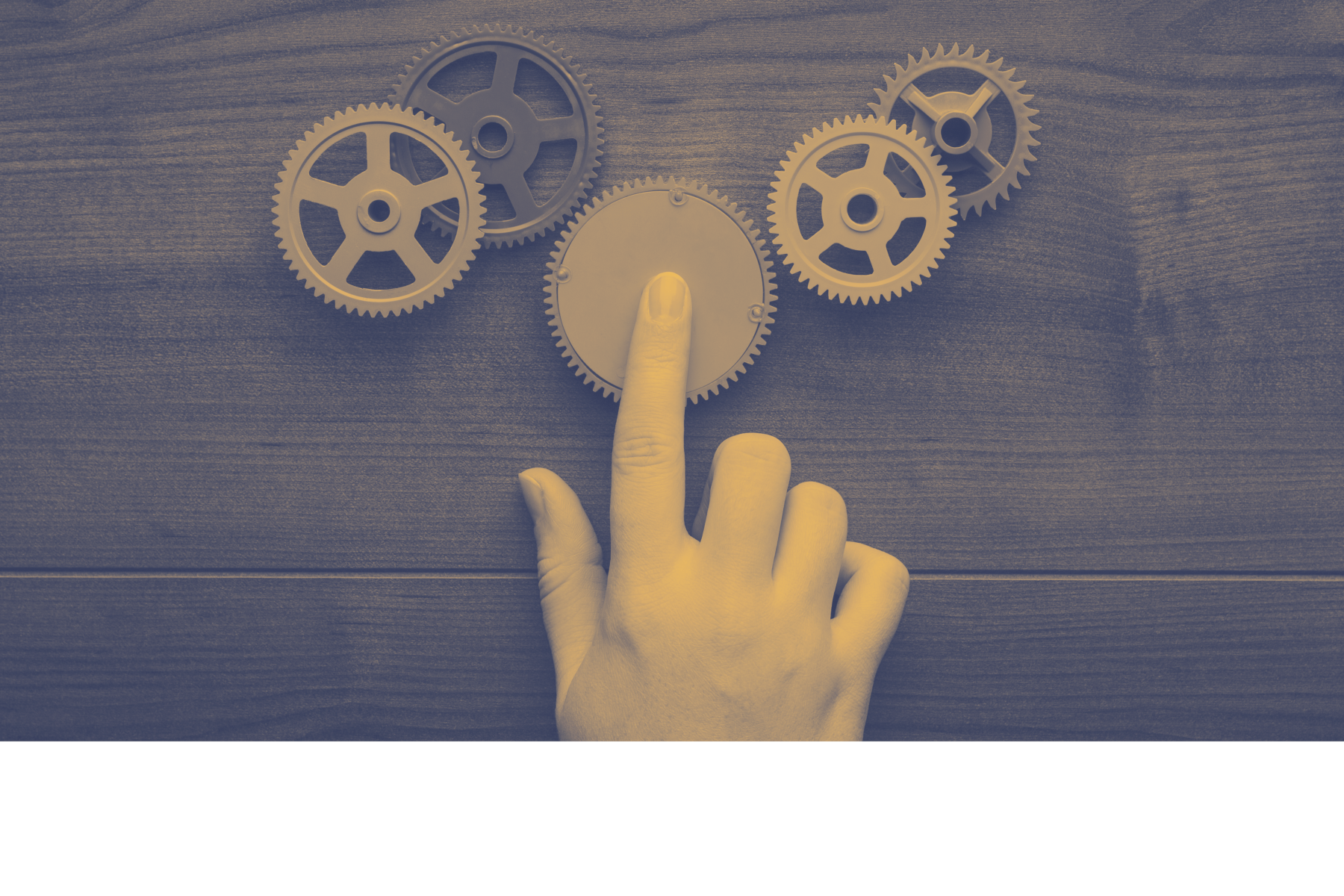}

\chapter{Trending Workarounds for Trends Calculations}

To evaluate local trends, scientists usually calculate the derivative
sign. Examples include equations 6,7,8 in \cite{meddah2019fpga} and
the code in the appendix of \cite{alfarra2021combating}. In these example, the authors first
calculate the derivative and then deduce its sign.

However, researchers are increasingly applying simple workarounds to calculate
the derivative sign or its approximation without going through the
derivative. In technological applications, for instance, this approach saves computational
time (in case there are runtime constraints). In theoretical applications,
scenarios where the derivative is not computable or its sign does
not reflect the trend are abundant. In Physics, studies of nowhere differentiable functions are increasingly common, particularly
when describing phenomena such as Quantum Fluctuations. The Baire category theorem, for example, implies that almost all the continuous
functions are nowhere differentiable. This means that one can calculate
the local rates of a ``negligible'' set of functions but local
trends may be well defined even if rates are not. The workarounds this issue are discussed in the following paragraphs.

\index{Trends in Discrete Domains}

\section{Workarounds In Discrete Domains}

In discrete domains, researchers and engineers often spare the division
operator upon calculating the derivative sign. \cite{schwartzlecture}, for instance, in a lecture of phase transition, omits the division by $T_{s}$ in Eq. 3, mentioning
that the value is not essential but the sign.

According to \cite{valishevsky2002adaptive}, since only
the sign and not the value of the derivative of the loss function
is used in RProp, it is possible to derive
formulae to determine the sign of the derivatives, which require much
less computational power than the one required for calculating their value.

\cite{guan2012detection}, in a work on the detection of the period of voice based on wavelet, introduces an $\epsilon-\delta$ definition
of a \textquotedbl trend operator,\textquotedbl{}, referred to as the
maximum definition.

In \cite{cheng2019sign}, the authors propose to directly estimate
the gradient sign at any direction instead of the gradient itself,
which enjoys the benefit of a single query. Using this single query
oracle for retrieving the sign of directional derivative, they develop
a novel query-efficient Sign-OPT approach for a hard-label black-box
attack.

In a section about the physics of solar cells in \cite{de2016analysis},
the authors calculate the directional derivative sign while sparing
the division operator.

In Eq. 3 of \cite{costanzo2019novel}, a paper proposing a novel MPPT technique for single stage gridconnected
PV systems, the sign of the discrete derivative
spares the division by $T_{P\&O}$.

If the conditions of Proposition 4 in \cite{chemla2019controls}, a paper on controls, belief updating, and bias in
medical randomized controlled trials are
satisfied, the experimenter cannot rely upon the sign of the treatment-control
difference to distinguish between two efficacy states. In this
situation, the experimenter would need to rely upon magnitudes of
the treatment-control difference to determine the state. However,
interpreting treatment-control magnitudes is more difficult since
magnitudes depend upon unobservables. Indeed, in many cases, the sign
of the difference suffices to infer the state.

\cite{li2020field}, in a paper on field-mediated locomotor dynamics on highly deformable surfaces, relies directly on the sign of the difference
of $B$ and $\varphi_{\text{prec}}$, without going through their
derivatives.

\cite{cheng2018peak}, in a study of gold nanoparticles translocations in nanopipette detection, calculates the sign of the discrete numeric
derivative in Eq. 3.2 based on Matlab's find peaks method, while sparing
the division.

 Algorithm 1 in \cite{huang2017robust}, a work on face recognition with structural binary gradient
patterns, calculates the sign of the
numerator of the discrete derivative to deduce the local trend.

In an efficient implementation of (5) in \cite{chen2020split}, to
obtain $\Delta y_{S_{n}}/y_{S_{n}'}$, a divider is needed.
However, the realization of the divider is very complex, and there
is a situation where the denominator $y_{S_{n}'}$ equals zero. To
avoid these disadvantages, the authors use the sign but not the value
of $y_{S_{n}'}$, as shown in (6) there.

Using other numerical tricks to calculate the trend without going
through the derivative's calculation to save energy is also common.
For example, \cite{wang2019e2} only calculates the most significant
bit of the derivative. This method allows to bound the error of calculating
its sign.

\index{Trends in Discrete Domains}

\section{Workarounds In Continuous Domains}

It is prevalent to spare the evaluation of the denominator when calculating
the sign of the derivative of a quotient. Since the quotient rule
squares the denominator, its sign does not affect that of the quotient.
Examples are abundant: the derivative of Eq. 3.82 in \cite{xiao2017passivity};
the analysis of the sigmoid fitness function case in \cite{gajer2009examining}, where
the sign of the first derivative of the gain function depends only
on the sign of its numerator; in \cite{shevkunov2011effect}, when
calculating the derivative of the quotient $r_{45}$; in the analysis
following Eq. 38 in \cite{palley2019unemployment}; in the analysis
of Eq. 17 in \cite{offiong2016determining}; in the proof of theorem
6.1 in \cite{van2017economic}; when calculating the sign of Equation
32 in \cite{seno2020sis}; in the proof of Lemma 3.1 in \cite{bonettini2021variable}; at Eq. 14 and 15 in \cite{chen2017direct}; in the analysis following Eq. 4 in \cite{sheremeta2021impact}; at the proof of claim 5 in \cite{bell2018three};
in the analysis following Eq. 13 in \cite{silvar}; in the study following
Eq. 6 in \cite{morel2005wood}; at the final step of proving Theorem
2 in \cite[p.~38]{valavi2020time}.

Recently, researchers have started to define the derivative
sign as an operator or parameter of its own, in several cases: to apply it recursively,
as in Eq. 5, 6 in \cite{ma2017adversarial}; for abbreviation,
by defining the parameter $\gamma$ as the derivative sign in Eq. 3 in \cite{horng2002vehicle}; for defining the
parameter $s\equiv sgn\left(y'\right)$ in \cite{hasan2021training};
and for defining the parameter $\epsilon_{i}$ that measures the velocity
sign in \cite{hasan2021training}.

Additional workarounds apply in scenarios where the one-sided derivatives
do not capture trends correctly, such as in one of the following
scenarios.

\index{Zeroed Derivative}
\subsection{Zeroed Derivative}

In the case of zeroed derivatives, the function's derivative exists and equals zero, while
the function is not necessarily constant; this happens at extrema points.
As a workaround, some researchers define the trend there as the limit
of the derivative sign. For example, \cite{maciel2015cosmological}
summarizes the possible asymptotic structures of gMcVittie spacetime
based on the sign of the one-sided derivative of $\xi$. Other examples include Lemma
1 in \cite{seno2020sis} and Definition 2 in \cite{ishitsubo2009experimental}.
\index{Cusps}

\subsection{Cusps}

A cusp of a curve defined in the plane by a pair of functions

\begin{align}
&\begin{aligned}
x & =f\left(t\right)\\
y & =g\left(t\right)
\end{aligned}
\end{align}

is a point where both derivatives of $f$ and $g$ are zero, and the directional derivative, in the direction of the tangent, changes sign (the direction of the tangent is the direction of the slope. Intuitively speaking, a single variable function incurs a cusp where it is continuous and non-differentiable at a point because its derivative approaches $\infty$ on the left to $- \infty$ on the right (or vice versa). 
For example, in Kato's cusp, quantum physicists are interested in the functions' trend (see \cite{mumtaz2019deformation}). \cite{arguelles2021formation}, on the other hand, applied cusps in Cosmology.

Several natural phenomena
can be described by the power laws involving fractional exponents. In Astronomy, we may naturally rewrite Kepler's third law with a fractional exponent
if we isolate $T$ or $r$; the initial mass function of stars involves
different fractional exponents, and so is the $M-\sigma$ relation.
In Physics, examples include the Angstrom exponent in Aerosol Optics,
the frequency-dependency of acoustic attenuation in complex media,
Stevens's power law of Psychophysics, behavior near second-order phase
transitions involving critical exponents, and the Curie--von Schweidler
law. In Biology, Kleiber's and Taylor's laws also involve fractional
exponents; these rules incur cusp points where they are
not differentiable; as a trend calculation workaround, one resorts
to a tangent-free approach.

\index{Discontinuities}
\subsection{Discontinuities}

Here we refer to the case where the function is discontinuous. This case
is prevalent in phenomena such as Sorption and Phase Transitions. Examples include the works in Cosmology of \cite{kumar2020phase} (see Figure 2) and 
\cite{schwartzlecture}, who provides an account of the struggle to analyze the monotonicity
of the heat capacity at a singularity (see Figure 12).

\index{Nowhere differentiability}

\subsection{Nowhere differentiability}

\label{nowhere_differentiability} Modern Physics is abundant with examples of
phenomena described by everywhere continuous and nowhere differentiable
functions. A prominent example is fluctuations, which are random invisible movements
of objects in their seemingly steady-state. Fluctuations are studied in the
fields of Thermodynamics and Quantum Mechanics, among others.
Nowhere differentiability can also be found in Electrical Engineering, specifically in a phenomenon
called Chattering (\cite{levant2010chattering}). Another prominent
example is the Brownian motion, whose statistical model - Wiener Process
- resembles fluctuations. Researchers devised the following methodologies
to describe the trends of such phenomena:
\begin{enumerate}
\item The theory of Detrended Fluctuation Analysis (see \cite{peng1994mosaic})
helps estimate the trend across an \textbf{interval}.
\item To describe \textbf{pointwise} trends, scientists devised the following
qualitative description of Wienner processes:
\begin{enumerate}
\item For every $\epsilon>0$, the function $w$ takes both (strictly) positive
and (strictly) negative values on $\left(0,\epsilon\right)$.
\item The function $w$ is continuous everywhere but differentiable nowhere.
\item Points of local maximum of the function $w$ are a dense countable
set.
\item The function $w$ has no points of local increase, that is, no $t>0$
satisfies the following for some $\epsilon$ in $\left(0,t\right)$:
first, $w\left(s\right)\leq w\left(t\right)$ for all $s$ in $\left(t-\epsilon,t\right)$,
and, second, $w\left(s\right)\geq w\left(t\right)$ for all $s$ in
$\left(t,t+\epsilon\right)$. (Local increase is a weaker condition
than $w$ is increasing on $\left(t-\epsilon,t+\epsilon\right)$).
The same holds for local decrease.
\end{enumerate}
\end{enumerate}

\section{Simplifying Trends as the Next Natural Step}

Since we showcased several workarounds that omit the division operator
upon calculating trends via the derivative sign in discrete domains,
one may ask, what if we could also do the same in continuous ones?
Perhaps we could establish a “trend operator” that also captures trends
in cases where the derivative sign is undefined or does not reveal
the correct trend information?

For example, we could apply such an operator to reformulate the conditions
(a)-(d) of Wienner process in subsection \ref{nowhere_differentiability}
in terms of the trend operator with a simple sentence: \textbf{The
function \ensuremath{\boldsymbol{w}} is continuous everywhere but
locally trending only at its dense countable local optima.} This description
is more concise and elegant than the former. It is also more positive;
it states a property (the existence of a local trend) at a dense subset
(local optima), as opposed to stating the absence of differentiability
everywhere.

Given these motivations in applications and theory, we dedicate the
following parts to propose a new trend operator and to explore its mathematical
properties.

\newpage{}

\part{Mathematical Discussion}
\label{mathematical_discussion_part}

\chapterimage{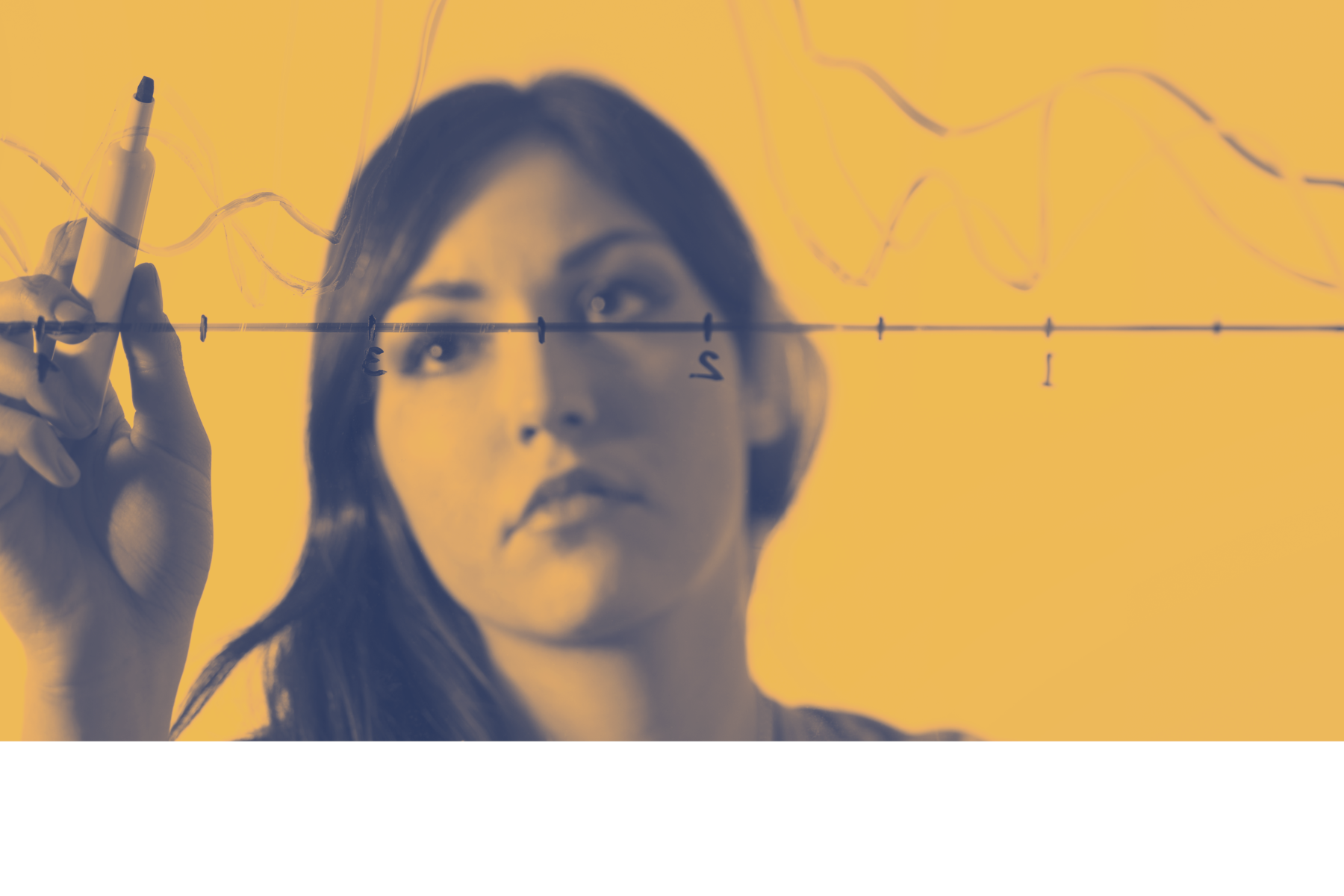}
\chapter{Single Variable Semi-discrete Calculus}
\section{Preliminary Notes}

\subsection{Motivation}
What if you could improve your AI algorithms performance by up to twenty percent? Describe natural phenomena better? And enhance classical math theorems? What if you could do all that with a new Calculus operator – on top of the derivative and the integral?

Here we present a new and simple operator in basic Calculus which renders prominent deep learning frameworks and various AI applications with a higher level of numerical robustness. In addition, we show that this operator provides advantages in continuous domains, where it classifies trends accurately, regardless of the functions’ differentiability or continuity.

\index{Discrete Domains}
\subsection{A Trends Calculation Workaround in Continuous Domains}
Upon analyzing custom loss functions' trends numerically, one may find value in treating momentary directions concisely and efficiently.

Consider the case where we approximate the derivative sign numerically with the difference quotient:

\begin{align}
&\begin{aligned}
\text{sgn}\left[\frac{dL}{d\theta}\left(\theta\right)\right]\approx \text{sgn}\left[\frac{L\left(\theta+h\right)-L\left(\theta-h\right)}{2h}\right]
\end{aligned}
\end{align}

This is useful for debugging your gradient computations, or in case your custom loss function, tailored by domain considerations, is not differentiable. The numerical issue with the gradient sign is embodied in the redundant division by a small number. This approach does not affect the final result, the numerator, $\text{sgn}\left[L\left(\theta+h\right)-L\left(\theta-h\right)\right]$; however, it amounts to a logarithmic or linearithmic computation time in the number of digits and occasionally results in an overflow. We’d better avoid it altogether.

In the above example, if the derivative exists in the extended sense, then $\text{sgn}\left(\pm \infty\right) = \pm 1$ represents the function’s trend altogether. However, infinite derivatives are often considered undefined, and we should pay attention to that convention. Moreover, there are cases where the derivative doesn’t exist in the extended sense, yet the function’s trend is clear. For example,

\begin{align}
&\begin{aligned} f\left(x\right)=\begin{cases}
x+x\left|\sin\left(\frac{1}{x}\right)\right|,  &x\neq0\\
0, &x=0
\end{cases}
\end{aligned}
\end{align}

at $x=0$.
There are several examples of various discontinuities types where the trend is clear (see below). To define the instantaneous trend of such functions, we can use the sign of their (different) Dini derivatives, if we are keen to evaluate partial limits. Otherwise, we can to introduce a more concise way to define trends.

Given this analysis, we would find it convenient to have a simple operator that is more numerically stable than the derivative sign; one that defines trends concisely and coherently whenever they are clear, including at critical points such as discontinuities, cusps, extrema, and inflections.

Let's use an example. Whenever you scrutinize your car's dashboard, you notice Calculus. The mileage is equivalent to the definite integral of the way you did so far, and the speedometer reflects the derivative of your way with respect to time. Both physical instruments merely approximate abstract notions.

The gear stick shows your travel direction. Often, its matching mathematical concept is the derivative sign. If the car moves forward, in reverse, or freezes, then the derivative is positive, negative, or zero, respectively. However, calculating the derivative to evaluate its sign is occasionally superfluous. As Aristotle and Newton famously argued, nature does nothing in vain. Following their approach, we probably need not go through rates calculation to define the instantaneous trend of change. If the trend of change is an important term in processes analysis, shouldn't we reflect it concisely rather than as a by-product of the derivative?

This occasional superfluousness of the derivative causes the aforementioned issues in numeric and analytic trend classification tasks. To tackle them, we will attempt to simplify the derivative sign as follows:
\begin{align}
\index{Detachment}
&\begin{aligned}
\text{sgn}\left[f_{\pm}’\left(x\right)\right] & =\text{sgn}\underset{{\scriptscriptstyle h\rightarrow0^{\pm}}}{\lim}\left[\frac{f\left(x+h\right)-f\left(x\right)}{h}\right]\\
&\highlight{\neq} \underset{{\scriptscriptstyle h\rightarrow0^{\pm}}}{\lim}\text{sgn}\left[\frac{f\left(x+h\right)-f\left(x\right)}{h}\right]\\
&=\pm\underset{{\scriptscriptstyle h\rightarrow0^{\pm}}}{\lim}\text{sgn}\left[f\left(x+h\right)-f\left(x\right)\right]
\end{aligned}
\end{align}
Note the deliberate erroneous transition in the second line. Switching the limit and the sign operators is wrong because the sign function is discontinuous at zero. Therefore, the resulting operator, which is the limit of the sign of $\Delta y$, does not always agree with the derivative sign. Furthermore, the multiplicative nature of the sign function allows us to cancel out the division operation. These facts may turn out in our favor, considering the issues we saw earlier with the derivative sign. Perhaps it is worth scrutinizing the limit of the change's sign in trend classification tasks.

This novel trend definition methodology is similar to that of the derivative. In the later, the slope of a secant turns into a tangent as the points approach each other. In contrast, the former calculates the trend of change in an interval surrounding the point at stake, and from it deduces the momentary trend of change, by applying the limit process; you can gain an intuitive understanding of this by looking at the following diagram:

Clearly, the numerical approximations of the (right-) derivative sign and that of: $$\underset{{\scriptscriptstyle h\rightarrow0^{+}}}{\lim}\text{sgn}\left[f\left(x+h\right)-f\left(x\right)\right]$$both equal the sign of the finite difference operator, $\text{sgn}\left[f\left(x+h\right)-f\left(x\right)\right]$, for some small value of $h$. However, the sign of the difference quotient goes through a redundant division by $h$. This roundtrip amounts to an extra logarithmic- or linearithmic-time division computation (depending on the precision) and might result in an overflow since $h$ is small. In that sense, we find it lucrative to think of trends approximations as $\underset{{\scriptscriptstyle h\rightarrow0^{\pm}}}{\lim}\text{sgn}\left[f\left(x+h\right)-f\left(x\right)\right]$, rather than the derivative sign. On average, it spares 30\% runtime in the calculation of the trend itself, relative to the numerical derivative sign. Accounting for the overhead of other atomic operations of the algorithm itself, the percent of time spared can still be up to 20\%, depending on the algorithm.

Similar considerations lead us to apply the quantization directly to $\Delta y$ when we are after a generic derivative quantization rather than its sign. That is, instead of quantizing the derivative, we calculate $Q\left[f\left(x+h\right)-f\left(x\right)\right]$ where $Q$ is a custom quantization function. In contrast with the trend operator, this quantization does not preserve the derivative value upon skipping the division operation. However, this technique is good enough in algorithms such as gradient descent. If the algorithmic framework entails multiplying the gradient by a constant (e.g., the learning rate), we may spare the division by $h$ in each iteration and embody it in the pre-calculated constant itself.

We can introduce a coarse estimation of the percentage of computational time spared by considering the computations other than the trend analysis operator itself. For example, suppose we can spare a single division operation in each iteration of gradient descent. Assume that the function is defined numerically at a finite discrete set of points. Then avoiding the division in the optimization iterations is significant, as that is the most time-consuming operation in each optimization iteration. In contrast, if the optimized function is computationally involved, for example, one that includes logical operators, then the merit of sparing a division operator, while it still exists, would be humbler.

Another numerical advantage of this operator relative to the derivative sign is the bound on its estimation error; that's prevalent in case the first derivative vanishes. While the estimation error of the derivative is $\mathcal{O}\left(\Delta x\right)$, that of higher-order derivatives is by orders of magnitude smaller, $\mathcal{O}\left(\Delta x^k\right)$. As we show below at the Taylor series expansion of this trend operator, it equals the sign of a higher-order derivative up to a sign coefficient (more precisely, the first one that does not vanish). It, thus, turns out that when estimating the trend operator with the first sign of the first non zeroed derivative, the error estimation has a tighter bound than that of the derivative sign.

The exploding gradient issue hinders the derivative sign as well, and we can mitigate it with this trend operator. Other numerical issues with the derivative are naturally transferred to the derivative sign and often tackled with the trend operator.

\begin{figure}[H]
\includegraphics[scale=1.3]{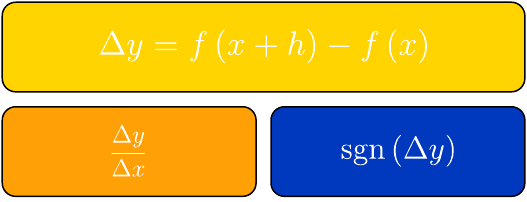}
\caption{The idea behind the definition of the local trend operator is as follows. Let us observe the term: $\Delta y=f(x+\Delta x)-f(x)$. If $f$ is continuous then $\underset{\Delta x \to 0}{\lim}\Delta y = 0$. Thus, we refrain from applying the limit process directly to $\Delta y$. The derivative, however, is rendered informative by comparing $\Delta y$ to $\Delta x$, via a fraction, prior to applying the limit process. The trend operator leverages less information, and quantizes $\Delta y$ via the function $\text{sgn}\left(\cdot\right)$. A function’s local trend does not reveal the information regarding its rate of change. In return, it is defined for a broad set of non-differentiable functions, as we show in the discussion below.}
\end{figure}

\index{Continuous Domains}
\subsection{A Trends Calculation Workaround in Continuous Domains}
Given this operator's practical merit in discrete domains, we proceed with theoretical aspects in continuous domains.

\begin{figure}
\begin{tabular}{ccc}
  \includegraphics[width=45mm]{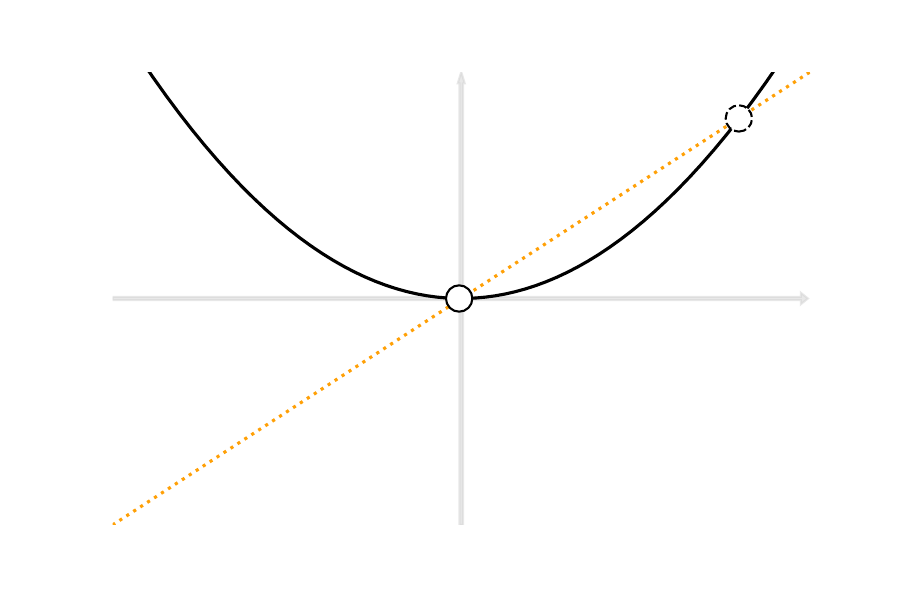} &   \includegraphics[width=45mm]{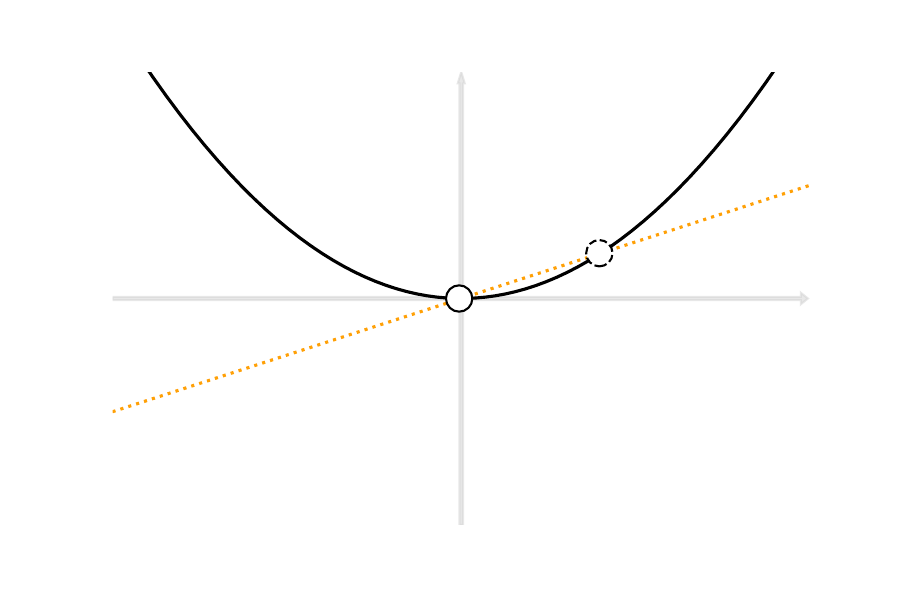} & \includegraphics[width=45mm]{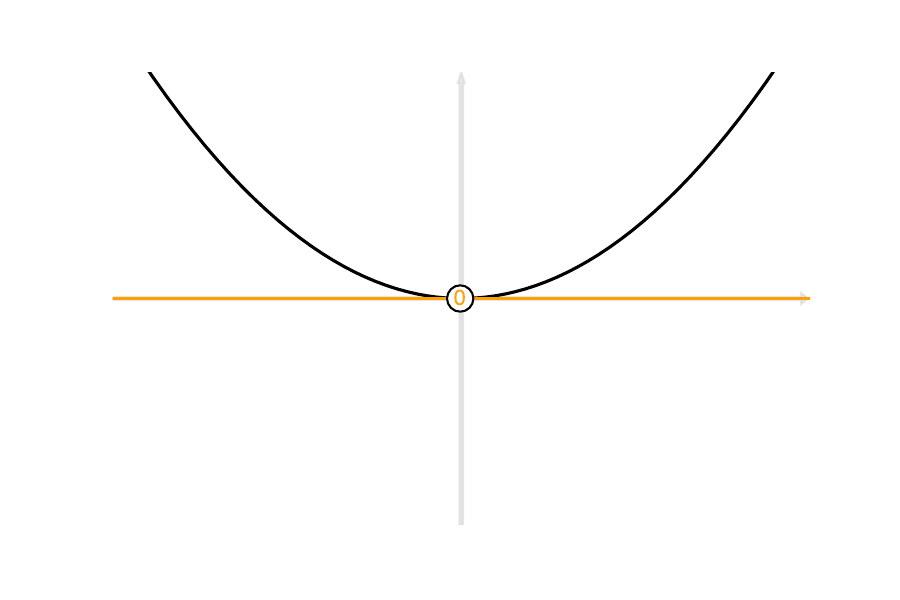} \\
 \includegraphics[width=45mm]{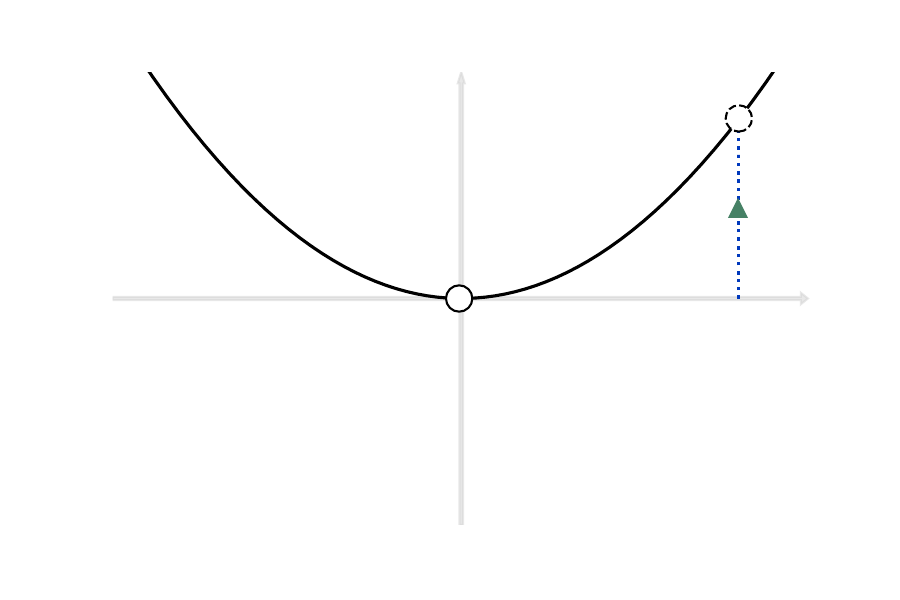} &   \includegraphics[width=45mm]{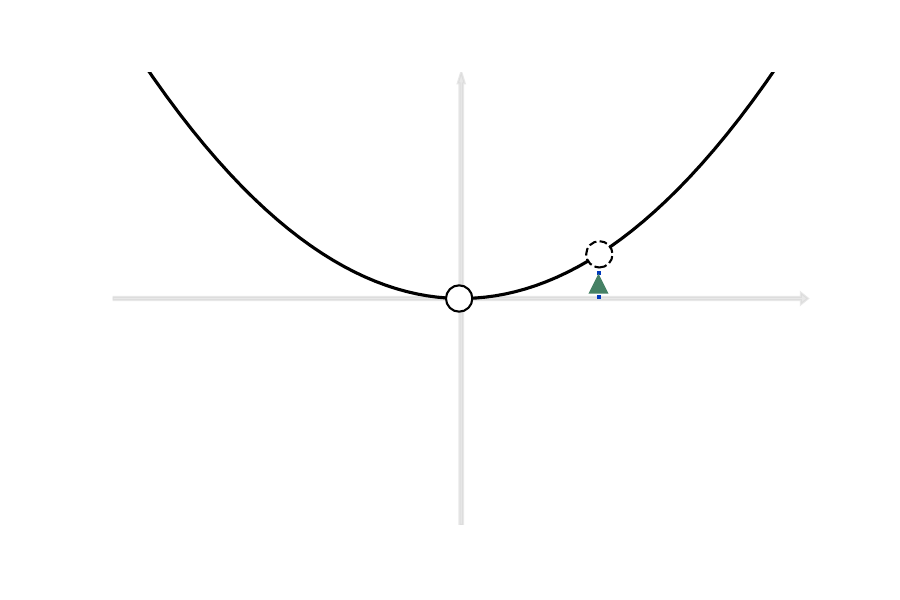} & \includegraphics[width=45mm]{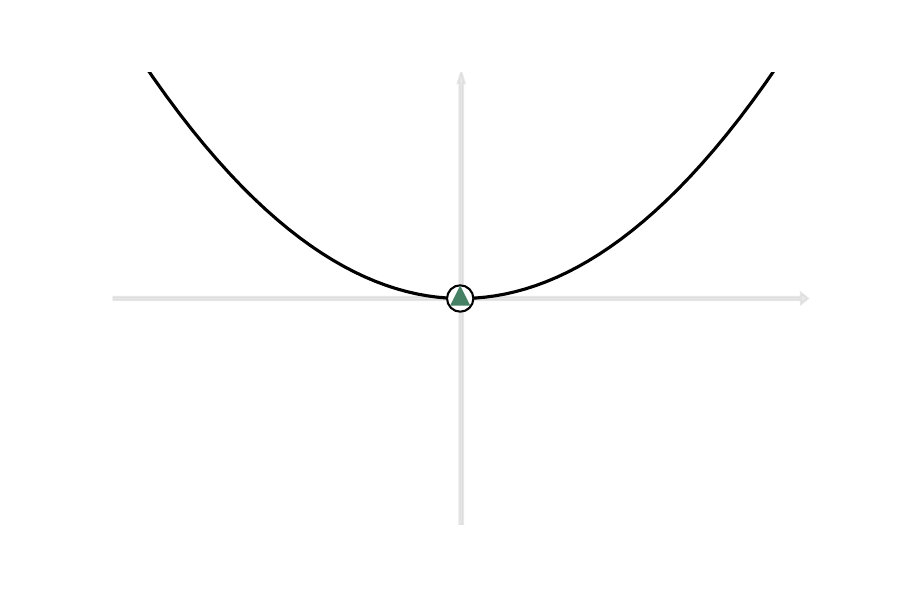}\\
\end{tabular}
\caption{The secant line's slope is positive throughout the finite steps of the limit process. However, the tanget's slope (calculated at the limit) is zeroed and does not reflect the positive local trend at the local minimum. In contrast, the difference sign between the functions' values and their value at the point remains positive at the limit.}
\label{limit_process_derivative_vs_detachment}
\end{figure}

Let us compare the way this operator describes local trends relative to the derivative. We define a family of monomials and attempt describing their local trend as follows: $f\left(x ; a,r \right)=a x^r$, where $f\left(0\right)=\begin{cases}
\underset{x\to0}{\lim}f\left(x\right), & \left|\underset{x\to0}{\lim}f\left(x\right)\right|<\infty\\
0, & \text{otherwise.}
\end{cases}$. To gain intuition, let us scrutinize the one-sided limit $\underset{\Delta x\rightarrow0^{+}}{\lim}\text{sgn}\left(\Delta y\right)$ and compare it to the right-derivative for cherry-picked cases. Let $r \in \left\{-1, 0, 0.5, 1, 2, 3 \right\}$, capturing discontinuity, constancy, cusp, linearity, extremum, and inflection, respectively. We allow opposite values of $a$ to cover all types of trends. An animation illustrating the calculation of these functions' trends via the derivative sign vs. the suggested trend operator is available in \cite{shachar_2020}.
A static and partial version is available in figure \ref{limit_process_derivative_vs_detachment}.

In table \ref{derivative_vs_detachment_summarizing_limits_table}, we summarize the calculations results and generalize it to all possible values of $r\in\mathbb{R}$, where $a>0$.

\begin{table}[h!]
\centering
        \begin{tabular}{ccc}
        \toprule
        \textbf{Scenario} & \color[HTML]{FFA006} $\text{sgn}\left(\lim_{x\to0^{+}}\frac{\Delta y}{\Delta x}\right)$  & \color[HTML]{0039BD} $\lim_{x\to0^{+}}\text{sgn}\left(\Delta y\right)$ \\
        \midrule
        $r\in\left(-\infty,0\right) \bigcup \left(0,1\right)$ & \color[HTML]{FFA006} $+\infty$ & \color[HTML]{0039BD} $1$ \\
        $r=0$ & \color[HTML]{FFA006} $0$ & \color[HTML]{0039BD} $0$ \\
        $r=1$ & \color[HTML]{FFA006} $1$ & \color[HTML]{0039BD} $1$ \\
        $r\in\left(1,\infty \right)$ & \color[HTML]{FFA006} $0$ & \color[HTML]{0039BD} $1$ \\
        \bottomrule
        \end{tabular}
\color{black}
\caption{A summary of the trend values as calculated with the derivative sign and the suggested trend operator. Note that, except for two cases (where $k\in\left\{0,1\right\}$, accounting for constancy and linearity) where both operators yield identical results, the concise trend operator represents local trends more coherently than does the derivative sign.}
\label{derivative_vs_detachment_summarizing_limits_table}
\end{table}

As we have seen, in all these cases, $\underset{\Delta x\rightarrow0^{+}}{\lim}\text{sgn}\left(\Delta y\right)$ reflects the way we think about the trend; that is, it always equals $a$ except for the constancy case, where it is zeroed, as expected. From table \ref{derivative_vs_detachment_summarizing_limits_table}, we conclude that the derivative sign does not capture momentary trends except for $k\in \left\{0,1\right\}$. \\
Thus, this operator does a better job in capturing trends at critical points. Let us calculate these limits rigorously with $\epsilon-\delta$ calculus in the following discussion and establish a first result that explains why this operator outperforms the derivative sign.

\section{Intuition}

\subsection{Why Does it Work}

Let us recall Fermat's stationary point theorem from differential Calculus.

\label{Fermat's Theorem}
\index{Fermat's theorem}
\begin{theorem}[Fermat's stationary point theorem]
Let $f:\left(a,b\right)\rightarrow\mathbb{R}$ be a function, and let $x\in \left(a,b\right).$ If $f$ has a local extremum at $x$ and is differentiable there then:
$$\underset{h\rightarrow0}{\lim}\frac{f\left(x+h\right)-f\left(x\right)}{h}=0.$$
\label{fermat_for_derivative}
\end{theorem}

We can establish a more rigorous justification by noticing how the definition of local extrema points coalesces with that of the operator at stake. In contrast with its basic Calculus analog, the following claim provides both a sufficient and necessary condition for stationary points:

\begin{theorem}[A semi-discrete Fermat's stationary point theorem]Let $f:\left(a,b\right)\rightarrow\mathbb{R}$ be a function and let $x\in \left(a,b\right).$ The following condition is necessary and sufficient for $x$ to be a strict local extremum of $f$:

$$\exists \underset{h\rightarrow0}{\lim}\text{sgn}\left[f\left(x+h\right)-f\left(x\right)\right]\neq0.$$
\label{fermat_semi_discrete}
\end{theorem}

The proof of this semi-discrete analogue of Fermat's theorem immediately follows from the Cauchy limit definition.

\begin{proof}Without loss of generality, we will prove the theorem for maxima points. We show that the following definitions of local extrema are equivalent: $$\begin{array}{ccc}
& \exists\delta>0:\left|x-\bar{x}\right|<\delta\Longrightarrow f\left(x\right)>f\left(\bar{x}\right)\\
& \Updownarrow\\
& \underset{{\scriptscriptstyle h\rightarrow0^{\pm}}}{\lim}\text{sgn}\left[f\left(x+h\right)-f\left(x\right)\right]=-1
\end{array}$$

First direction. Assume $\underset{{\scriptscriptstyle h\rightarrow0^{\pm}}}{\lim}\text{sgn}\left[f \left(x+h\right)-f\left(x\right)\right]=-1.$ Then according to Cauchy limit definition,
$$\forall\epsilon,\exists\delta:\left|x-\bar{x}\right| < \delta\Longrightarrow\left|\text{sgn}\left[f\left(\bar{x}\right)-f\left(x\right)\right]-\left(-1\right)\right| < \epsilon.$$
In particular, for $\epsilon_{0}=\frac{1}{2}$, $$\exists\delta:\left|x-\bar{x}\right|<\delta\Longrightarrow\left|\text{sgn}\left[f\left(\bar{x}\right)-f\left(x\right)\right]+1\right|<\frac{1}{2}.$$
The only value in the sign function’s image, $\left\{ 0,\pm1\right\}$, that satisfies the above inequality, is $-1$. Therefore:
$$\exists\delta:\left|x-\bar{x}\right|<\delta\Longrightarrow \text{sgn}\left[f\left(\bar{x}\right)-f\left(x\right)\right]=-1,$$
which can be written as:
$$\exists\delta:\left|x-\bar{x}\right|<\delta\Longrightarrow f\left(x\right)>f\left(\bar{x}\right).$$

Second direction. Let $\epsilon>0$. We know that there exists $\delta$ such that $\left|x-\bar{x}\right| < \delta$ implies that $f\left(x\right)>f\left(\bar{x}\right)$, which can be rewritten as $$\text{sgn}\left[f\left(\bar{x}\right)-f\left(x\right)\right]=-1.$$ Thus $\text{sgn}\left[f\left(\bar{x}\right)-f\left(x\right)\right]-\left(-1\right)=0,$ and in particular $$\left|\text{sgn}\left[f\left(\bar{x}\right)-f\left(x\right)\right]-\left(-1\right)\right|<\epsilon,$$hence the limit definition holds.

\end{proof}

\subsection{Where Does it Work}

Next, let us check in which scenarios is this operator well defined. We will cherry-pick functions with different characteristics around $x=0$. For each such function, we ask which of the properties (continuity, differentiability, and the existence of the operator at stake, that is, the existence of a local trend from both sides), take place at $x=0$.

\begin{figure}[h!]
\includegraphics[scale=0.7,trim={0 9cm 0 7cm},]{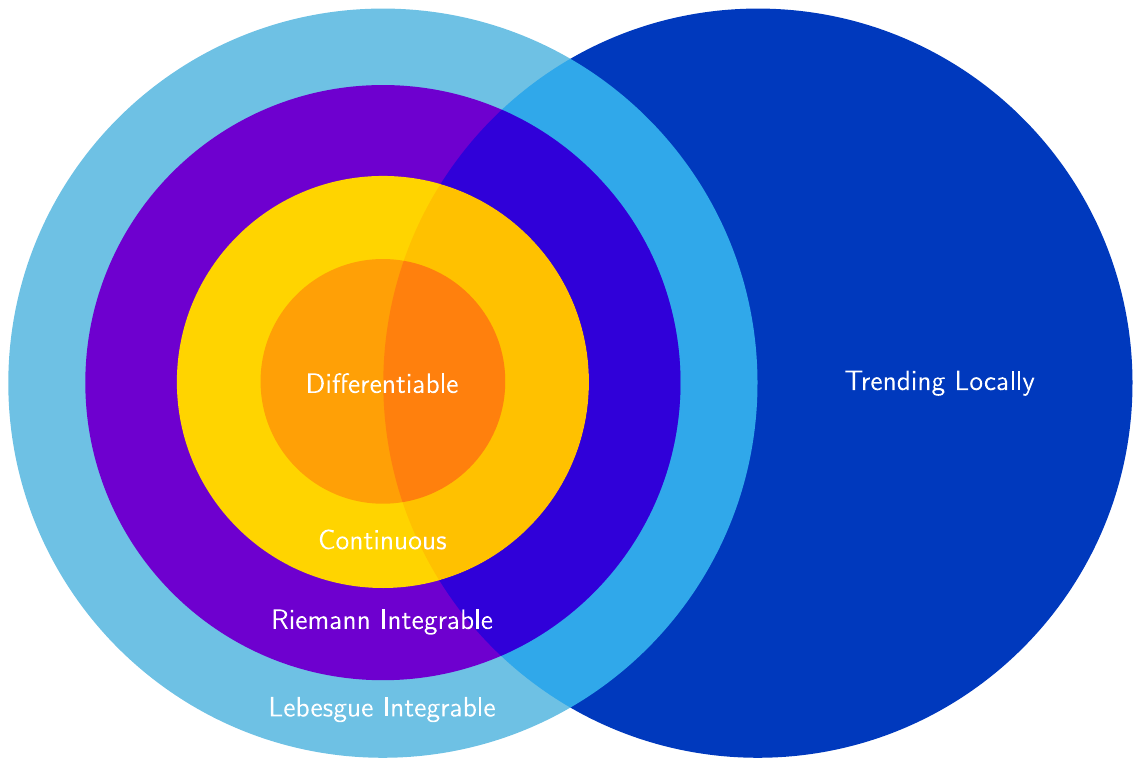}
\caption{Let us illustrate the set of detachable functions in the space of real functions. Examples of functions adhering to these properties in a pointwise fashion or an interval are available in tables \ref{pointwise_properties_examples}, \ref{interval_properties_examples}, respectively.}
\label{calculus_structure_interval}
\index{Function's Properties in Calculus}
\end{figure}

We would also like to find out which properties hold across an entire interval (for example, $[-1,1]$). To that end, we add two interval-related properties: Lebesgue and Riemann integrability. Feel free to explore these properties in \ref{calculus_structure_interval}.

We may extend the discussion to properties that hold in intervals almost everywhere, rather than throughout the interval. This is out of this chapter's scope, but as an example, we will mention the function $$f(x)=\sum\limits_{n=1}^{\infty}f_n(x),$$ where $f_n(x)=2^{-n}\phi\left(\frac{x-a_n}{b_n-a_n}\right)
, \phi$ is the Cantor-Vitali function and $\{(a_n,b_n):n\in\mathbb{N}\}$ is the set of all intervals in $\left[0,1\right]$ with rational endpoints. It's possible to show that this function has trend everywhere, it's strictly monotonic, but its derivative is zeroed almost everywhere. In this example, the notion of instantaneous trend is coherent with the function's monotonic behavior in the interval, in contrast with the vanishing derivative. Note that according to Baire categorization theorem, almost all the continuous functions are nowhere differentiable. Therefore, we could use an extension to the set of functions whose monotony can be analyzed concisely. Finally, we mention the function $$g\left(x\right)=x^{1+\boldsymbol{1}_{\mathbb{Q}}}$$ defined over $\left(0,1\right)$. In its definition domain, $g$ is discontinuous everywhere, but detachable from left almost everywhere.

\section{Definitions}
\index{Detachment}
\subsection{The Instantaneous Trend of Change}

Let us summarize our discussion thus far. The momentary trend, a basic analytical concept, has been embodied by the derivative sign for engineering purposes. It's applied to constitutive numeric algorithms across AI, optimization, and other computerized applications. More often than not, it does not capture the momentary trend of change at critical points. In contrast, $\underset{\Delta x\rightarrow0^{\pm}}{\lim}\text{sgn}\left(\Delta y\right)$ is more numerically robust, in terms of finite differences. It defines trends coherently wherever they exist, including at critical points. 

Given these merits, we dedicate a definition to this operator. As it "detaches" functions, turning them into step functions with discontinuities at extrema points, let us define the one-sided detachments of a function $f$ as follows.

\begin{definition}[Detachment of a real function]
Let $f:\mathbb{R}\rightarrow\mathbb{R}$ be a function and $x\in\mathbb{R}$ a point in its definition domain. Then we define its one-sided right- and left- detachments at $x$ as follows:
 $$\begin{array}{ccc}& f_{\pm}^{;}:\;\mathbb{R}\rightarrow\left\{ -1,0,+1\right\} \\
& f_{\pm}^{;}\left(x\right)\equiv \pm \underset{{\scriptscriptstyle h\rightarrow0^{\pm}}}{\lim}\text{sgn}\left[f\left(x+h\right)-f\left(x\right)\right].
\end{array}$$
\end{definition}

We say that a function is detachable if both those one-sided limits exist. We add the $\pm$ coefficient for consistency with the derivative sign. For convenience and brevity, from now on we denote by $f^;$ either one of the one-sided detachments separately ($f^;_+$ or $f^;_-$), without assuming that they necessarily agree.

Geometrically speaking, for a function's (right-) detachment to equal $+1$, for example, its value at the point needs to strictly bound the function's values in a right-neighborhood of $x$ from below. This is in contrast with the derivative's sign, where we make the assumption of the existence of an ascending tangent.

\begin{table}[h!]
\centering
        \begin{tabular}{cccc}
        \toprule
        \textbf{Function} & \color[HTML]{FFD400}{\textbf{Continuous}} & \color[HTML]{FFA006}{\textbf{Differentiable}} & \color[HTML]{0039BD}{\textbf{Trending Locally}} \\
        \midrule
        $f_1\left(x\right)=x^2$ & \color[HTML]{FFD400}{V} & \color[HTML]{FFA006}{V} & \color[HTML]{0039BD}{V} \\
        $f_2\left(x\right)=\left(\frac{1}{2}-\boldsymbol{1}_{\mathbb{Q}}\right)x^{2}$ 
        & \color[HTML]{FFD400}{V} & \color[HTML]{FFA006}{V} & \\
        $f_3\left(x\right)=x^{\frac{1}{3}}$ & \color[HTML]{FFD400}{V} & & \color[HTML]{0039BD}{V} \\        
        $f_4\left(x\right)=\text{sgn}\left(x\right)$ &  & & \color[HTML]{0039BD}{V} \\
        $f_5\left(x\right)=\begin{cases}
\sin\left(\frac{1}{x}\right), & x\neq0\\
0, & x=0
\end{cases}$ & & & \\
        $f_6\left(x\right)= x^{1+2\cdot\boldsymbol{1}_{\mathbb{Q}}}$ & \color[HTML]{FFD400}{V} & & \color[HTML]{0039BD}{V} \\        
        $f_{7}\left(x\right)=\begin{cases}
\sin\left(\frac{1}{x}\right), & x\neq0\\
2, & x=0
\end{cases}$ & & & \color[HTML]{0039BD}{V} \\
        $f_{8}\left(x\right)=\begin{cases}
x^{2}\sin\left(\frac{1}{x}\right), & x\neq0\\
0, & x=0
\end{cases}$ & \color[HTML]{FFD400}{V} & \color[HTML]{FFA006}{V} & \\
        $f_{9}\left(x\right)=\sqrt{\left|x\right|}$ & \color[HTML]{FFD400}{V} & & \color[HTML]{0039BD}{V} \\
        $f_{10}\left(x\right)=\begin{cases}
x\sin\left(\frac{1}{x}\right), & x\neq0\\
0, & x=0
\end{cases}$ & \color[HTML]{FFD400}{V} & & \\
        $f_{11}\left(x\right)=\boldsymbol{1}_{\mathbb{Z}}$ & & & \color[HTML]{0039BD}{V} \\
        $f_{12}\left(x\right)=\boldsymbol{1}_{\mathbb{Q}}$ (Dirichlet) & & & \\
        $f_{13}\left(x\right)=\sum_{n=1}^{\infty}\left(\frac{1}{2}\right)^{n}\cos\left(3^{n}x\right)$ (Weierstrass) & \color[HTML]{FFD400}{V} & & \color[HTML]{0039BD}{V} \\
        $f_{14}\left(x\right)=\begin{cases}
x+x\left|\sin\left(\frac{1}{x}\right)\right|, & x\neq0\\
0, & x=0
\end{cases}$ & \color[HTML]{FFD400}{V} & & \color[HTML]{0039BD}{V} \\
        $f_{15}\left(x\right)=\begin{cases}
\frac{1}{x}, & x\neq0\\
0, & x=0
\end{cases}$ & & & \color[HTML]{0039BD}{V} \\
        $f_{16}\left(x\right)=t\left(x-\sqrt{2}\right)$ (Thomae) & \color[HTML]{FFD400}{V} & & \\
        $f_{17}\left(x\right)=R\left(x\right)$ (Riemann) & \color[HTML]{FFD400}{V} & & \color[HTML]{0039BD}{V} \\
        \bottomrule
        \end{tabular}
\caption{Relation between selected functions' analytical properties at the point $x=0$, corresponding to figure \ref{calculus_structure_interval}, and some of these functions, along with their one-sided detachments, are plotted in figure \ref{functions_plots_with_detachments}.}
\label{pointwise_properties_examples}
\end{table}

\begin{table}[h!]
\centering
        \begin{tabular}{cccccc}
        \toprule
        \textbf{Function} & \color[HTML]{FFD400}{\textbf{Cont.}} & \color[HTML]{FFA006}{\textbf{Diff.}} & \color[HTML]{0039BD}{\textbf{Trend.}} & \color[HTML]{7800cf}{\textbf{Riemman}} & \color[HTML]{6ec1e4}{\textbf{Lebesgue}} \\ 
        \midrule
        $f_1\left(x\right)=\boldsymbol{1}_{\mathbb{Q}}$ & & & & & \color[HTML]{6ec1e4}{V}\\
        $f_2\left(x\right)=\begin{cases}
\sin\left(\frac{1}{x}\right), & x\neq0\\
0.5, & x=0
\end{cases}$ 
        & & & & \color[HTML]{7800cf}{V} & \color[HTML]{6ec1e4}{V} \\
        $f_3\left(x\right)=\begin{cases}
\sin\left(\frac{1}{x}\right), & x\neq0\\
0, & x=0
\end{cases}$ & \color[HTML]{FFD400}{V} & & & \color[HTML]{7800cf}{V} & \color[HTML]{6ec1e4}{V} \\
        $f_4\left(x\right)=\begin{cases}
x^2\sin\left(\frac{1}{x}\right), & x\neq0\\
0, & x=0
\end{cases}$ & \color[HTML]{FFD400}{V} & \color[HTML]{FFA006}{V} & & \color[HTML]{7800cf}{V} & \color[HTML]{6ec1e4}{V} \\
        $f_5\left(x\right)=x$ & \color[HTML]{FFD400}{V} & \color[HTML]{FFA006}{V} & \color[HTML]{0039BD}{V} & \color[HTML]{7800cf}{V} & \color[HTML]{6ec1e4}{V} \\
        $f_6\left(x\right)=\sqrt{\left|x\right|}$ & \color[HTML]{FFD400}{V} & & \color[HTML]{0039BD}{V} & \color[HTML]{7800cf}{V} & \color[HTML]{6ec1e4}{V} \\
        $f_{7}\left(x\right)=\text{sgn}\left(x\right)$ & & & \color[HTML]{0039BD}{V} & \color[HTML]{7800cf}{V} & \color[HTML]{6ec1e4}{V} \\
        $f_{8}\left(x\right)=\begin{cases}
\frac{1}{\sqrt{\left|x\right|}}, & x\neq0\\
0, & x=0
\end{cases}$ & & & \color[HTML]{0039BD}{V} & & \color[HTML]{6ec1e4}{V} \\
        $f_{9}\left(x\right)=\begin{cases}
\frac{1}{x}, & x\neq0\\
0, & x=0
\end{cases}$ & & & \color[HTML]{0039BD}{V} & &\\
        \bottomrule
        \end{tabular}
\centering
        \caption{Relation between selected functions' analytical properties at every point of the closed intervals $\left[-1,+1\right]$, corresponding to Figure \ref{calculus_structure_interval}. Cont., Diff., Trend., Riemann, and Lebesgue stand for continuity, differentiability, the existence of a local trend, Riemann integrability, and Lebesgue integrability, across the interval.}
\label{interval_properties_examples}
\end{table}

\index{Sign Continuity}
\subsection{Sign Continuity}

In the discussion below we do not restrict ourselves merely to continuous functions. It turns out, that the ability to extend results to discontinuity requires to handle various cases differently. The following definitions will be useful upon classifying these cases.

For example, we may be naturally inclined to introduce product and quotient rules for the detachment operator; the following example illustrates that this may not be as straight-forward as we would imagine.

\begin{exm}Let $f,g$ be functions defined at a point $x$ and in its neighborhood. If $f^{;}=g^{;}=+1$, and $f=g=0$ at a point $x$, then the detachment of the product is $\left(fg\right)^{;}\left(x\right)=+1$ as well. Following this simple scenario, one might intuitively think that the product rule for the detachment is simply $\left(fg\right)^{;}=f^{;}g^{;}$. However, upon shifting $g$ vertically downwards, that is, requiring that $g(x)<0$, we obtain a setting where $\left(fg\right)^{;}=-1$, although the detachments of $f,g$ remain as before. This means that the detachment of the product $\left(fg\right)^{;}\left(x\right)$ necessarily also depends on the sign of $f,g$ at the point of interest. Indeed, assuming $g$ is continuous, the product's detachment in this new setting is $-1$.

However, let us recall that in Semi-discrete Calculus we are not restricted to differentiable or even continuous functions. What if $g$ is discontinuous? As long as $g$ maintains its sign in a neighborhood of the point, it is possible to show that the detachment of the product remains $-1$. But if $g$ changes its sign, meaning there is a neighborhood of $x$ where $g$ is zeroed or one where $g$ is positive, then $\left(fg\right)^{;}\left(x\right)$ can be either $0$ or $+1$, respectively. This implies that we should somehow bound the functions' signs. It would be convenient to pay attention to the following intuitive trait of the functions $f,g$.
\end{exm}

\begin{definition}[Sign Continuity]
Given a function $f$, we say that it is sign-continuous (s.c.) at a point $x$ if $\text{sgn}\left(f\right)$ is continuous.
\end{definition}

\begin{figure}
\begin{tabular}{cc}
  \includegraphics[width=45mm]{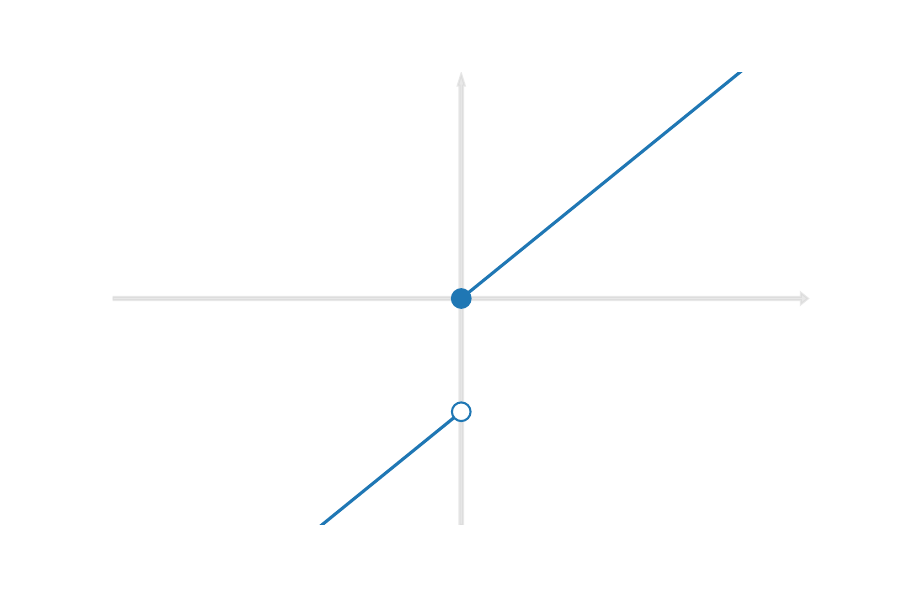} &   \includegraphics[width=45mm]{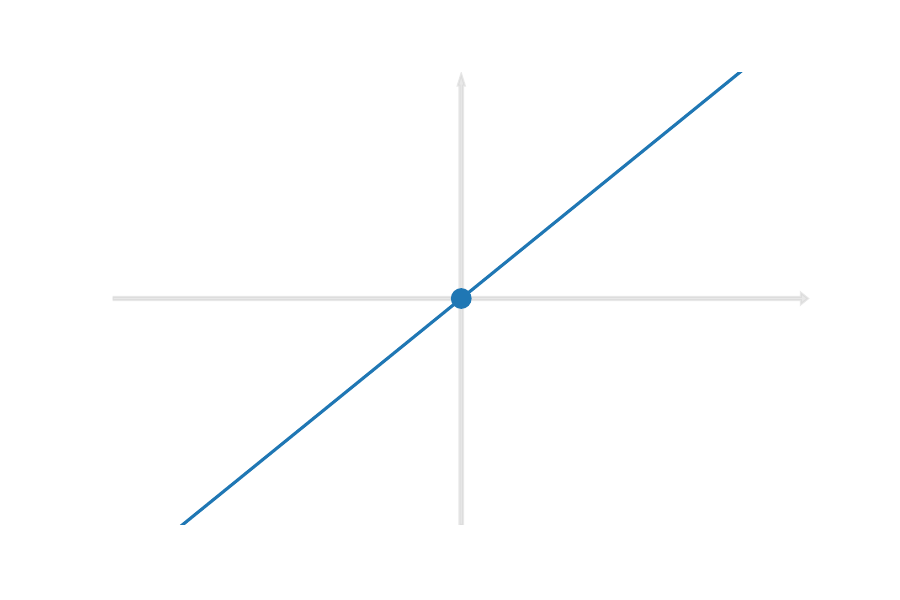}  \\
Discontinuous and sign-discontinuous & Continuous and sign-discontinuous \\[4pt]
 \includegraphics[width=45mm]{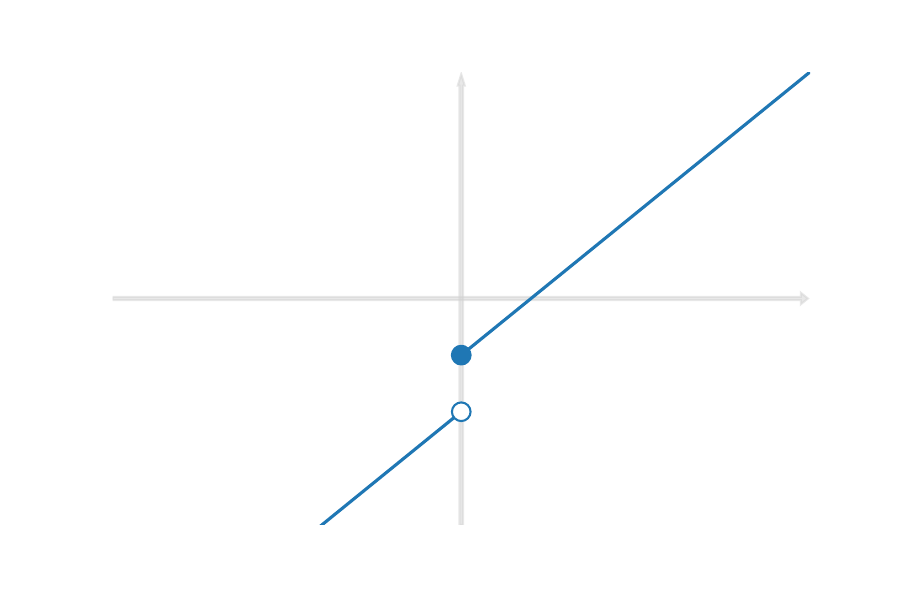} &   \includegraphics[width=45mm]{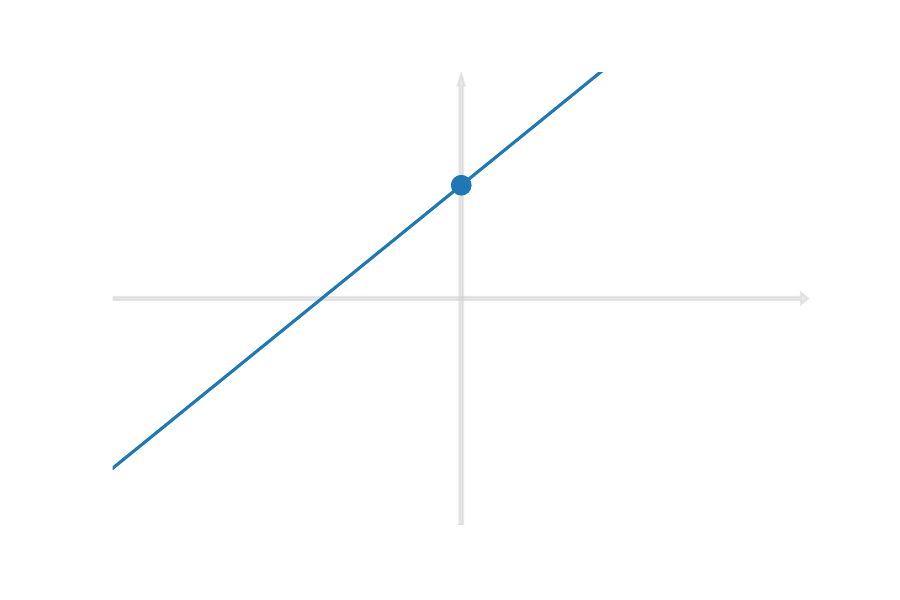} \\
Discontinuous and sign-continuous & Continuous and sign-continuous \\

\end{tabular}
\caption{Illustration of the relation between continuity and sign-continuity. Each function illustrates different constellations of these properties at $x=0$.}
\label{continuity_vs_sign_continuity}
\end{figure}

\begin{definition}[Sign-discontinuity] Given a function $f$, we say that it is sign-discontinuous (s.d.) at a point $x$  if all the partial limits of $\text{sgn}\left(f\right)$ are different than its sign at $x$.
\end{definition}
Observe that the function's sign-continuity at a point may be determined given its sign and detachment. For example, if $f^{;}\left(x\right)=0$ or $\left(ff^{;}\right)\left(x\right)>0$, then $f$ is sign-continuous. This observation suggests the following definition.

\begin{definition}[Inherent Sign-continuity (discontinuity)]
We say that $f$ is inherently sign-continuous (sign-discontinuous), abbreviated by i.s.c. (i.s.d.)  at a point $x$ if the knowledge of its sign and detachment at the point alone determine its sign-continuity (sign-discontinuity).
\end{definition}

For convenience, we limit the following discussion to sign-continuous or sign-discontinuous functions.

\begin{definition}[Sign-consistency]
We call a function that is either sign-continuous or sign-discontinuous "sign-consistent".
\end{definition}

\section{Semi-discrete Analogues to Calculus Theorems}

Equipped with a concise definition of the instantaneous trend of change, we may formulate analogs to Calculus theorems with the following trade-off.  Those simple corollaries inform us of the function's trend rather than its rate. In return, they hold for a broad set of detachable and non-differentiable functions.

\begin{figure}
\begin{tabular}{ccc}
  \includegraphics[width=45mm]{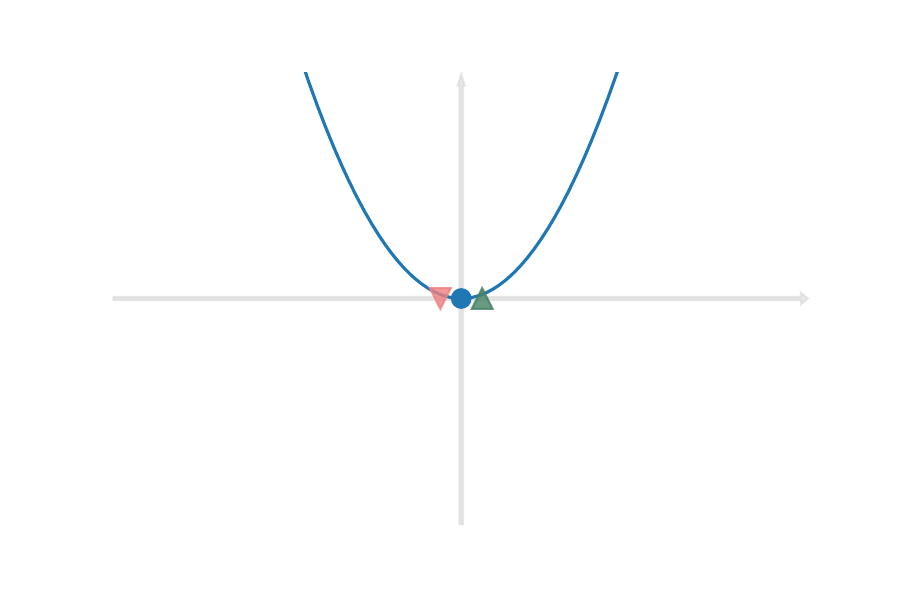} &   \includegraphics[width=45mm]{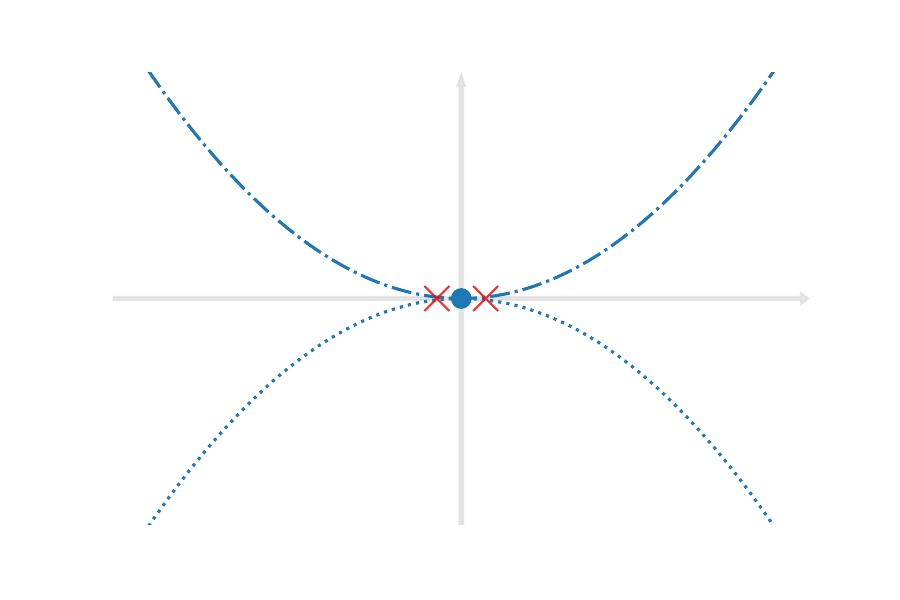} & \includegraphics[width=45mm]{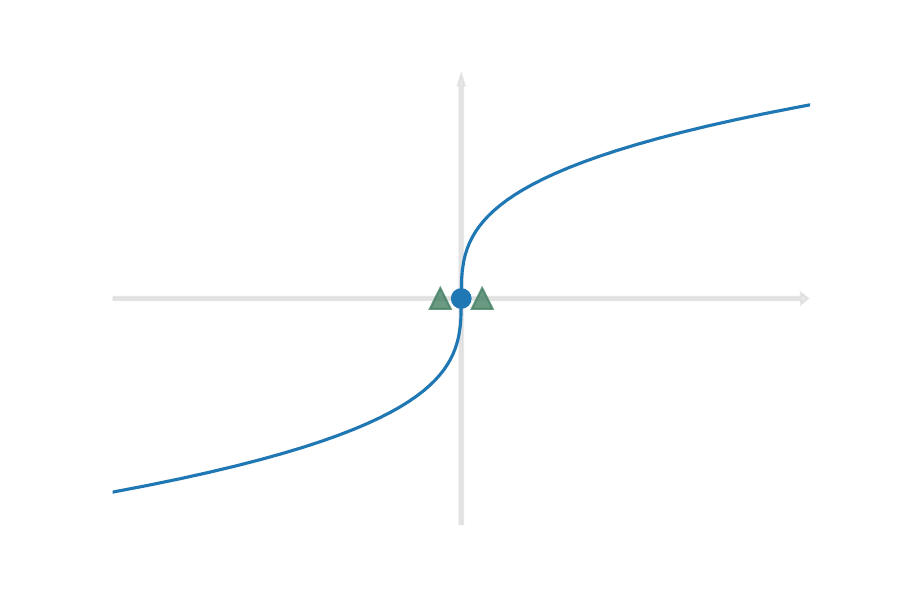} \\
$f_1\left(x\right)=x^2$ & $f_2\left(x\right)=\left(\frac{1}{2}-\boldsymbol{1}_{\mathbb{Q}}\right)x^{2}$ & $f_3\left(x\right)=x^{\frac{1}{3}}$ \\[4pt]
 \includegraphics[width=45mm]{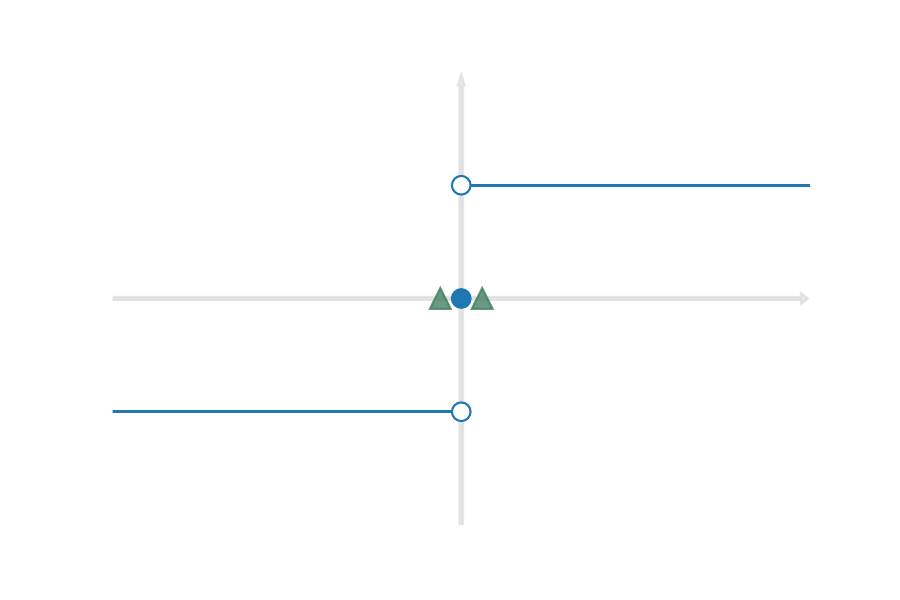} &   \includegraphics[width=45mm]{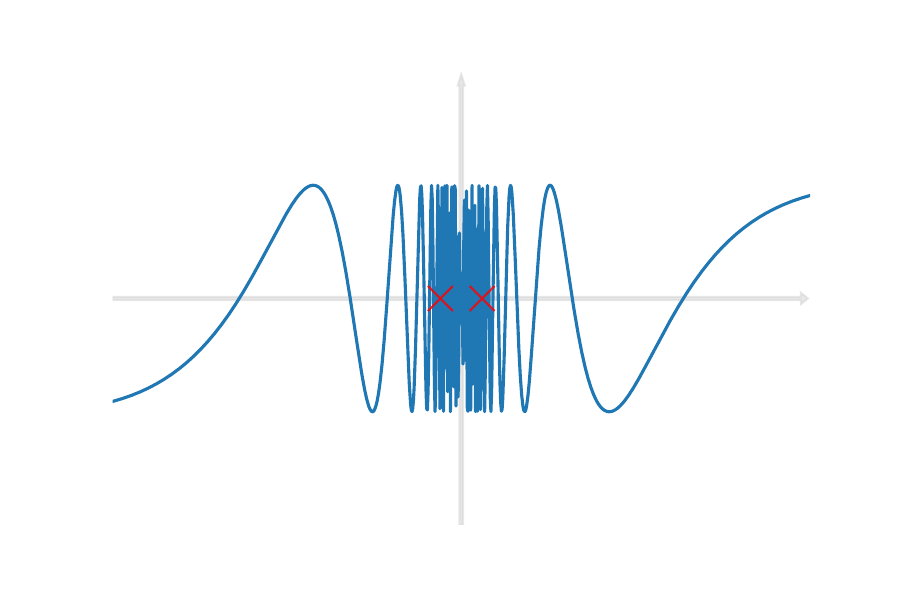} & \includegraphics[width=45mm]{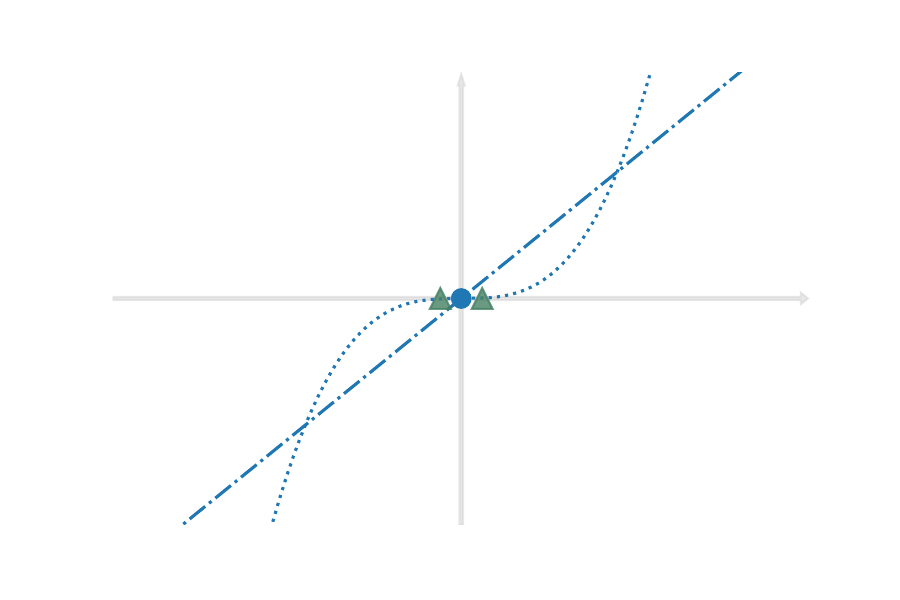}\\
$f_4\left(x\right)=\text{sgn}\left(x\right)$ & $f_5\left(x\right)=\begin{cases}
\sin\left(\frac{1}{x}\right), & x\neq0\\
0, & x=0
\end{cases}$ & $f_6\left(x\right)= x^{1+2\cdot\boldsymbol{1}_{\mathbb{Q}}}$ \\[6pt]
 \includegraphics[width=45mm]{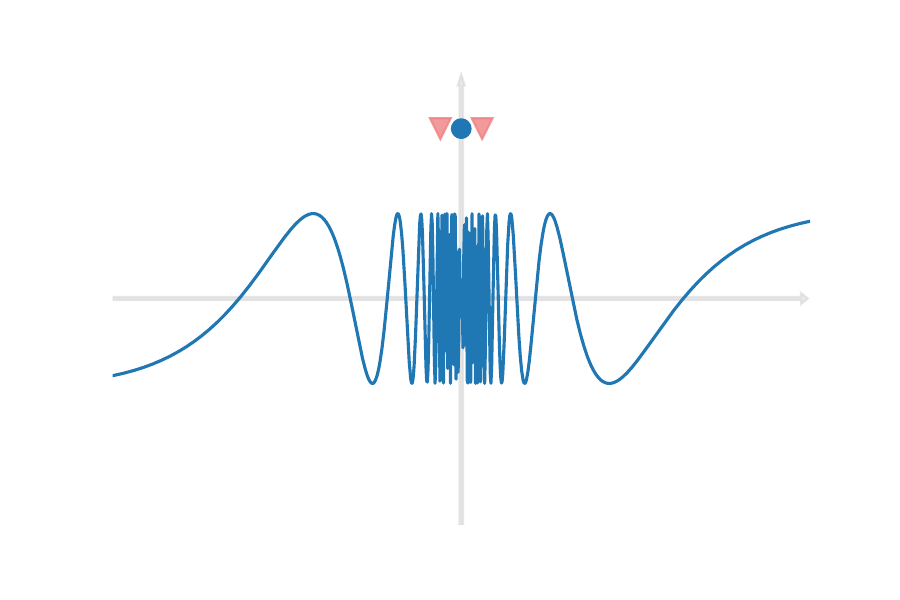} &   \includegraphics[width=45mm]{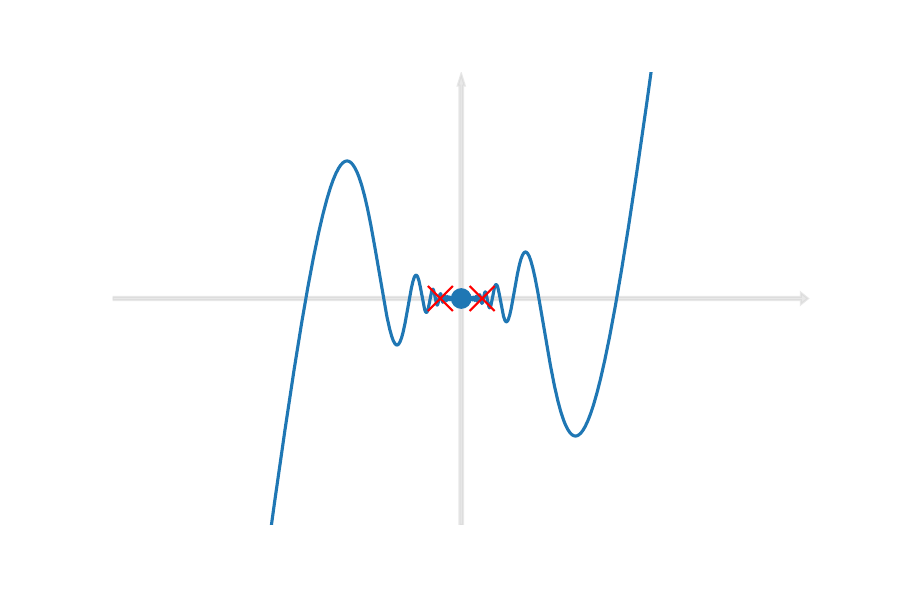} & \includegraphics[width=45mm]{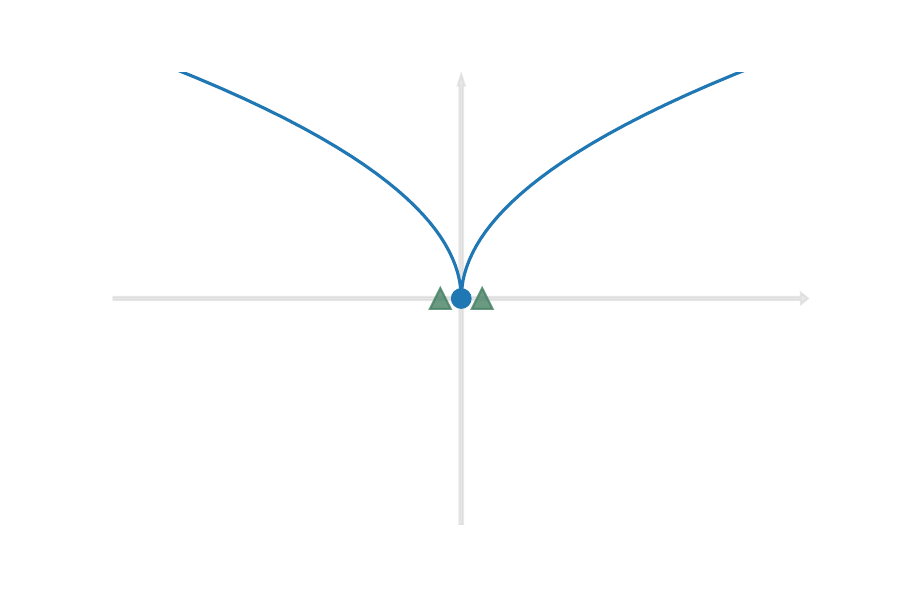}\\
$f_{7}\left(x\right)=\begin{cases}
\sin\left(\frac{1}{x}\right), & x\neq0\\
2, & x=0
\end{cases}$ & $f_{8}\left(x\right)=\begin{cases}
x^{2}\sin\left(\frac{1}{x}\right), & x\neq0\\
0, & x=0
\end{cases}$ & $f_{9}\left(x\right)=\sqrt{\left|x\right|}$ \\[6pt]
 \includegraphics[width=45mm]{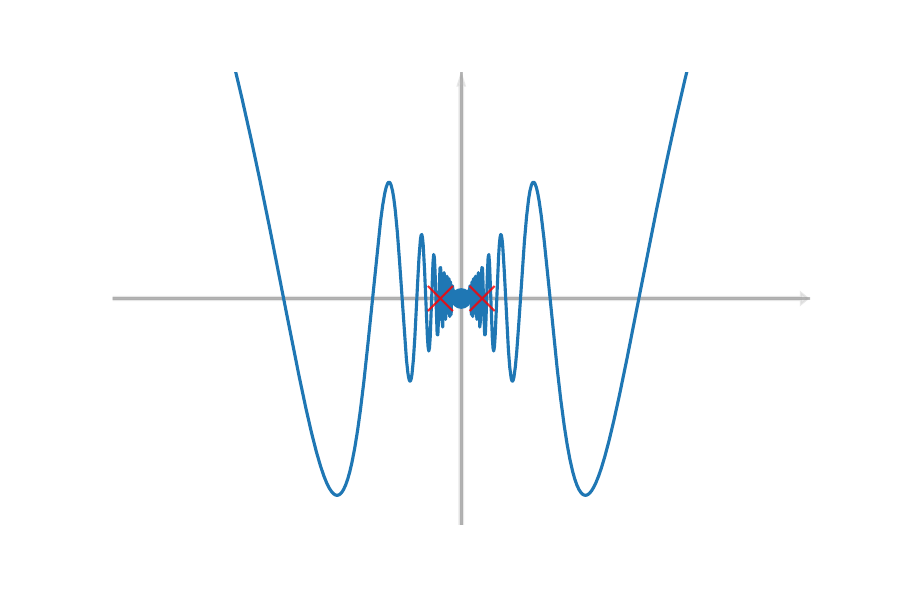} &   \includegraphics[width=45mm]{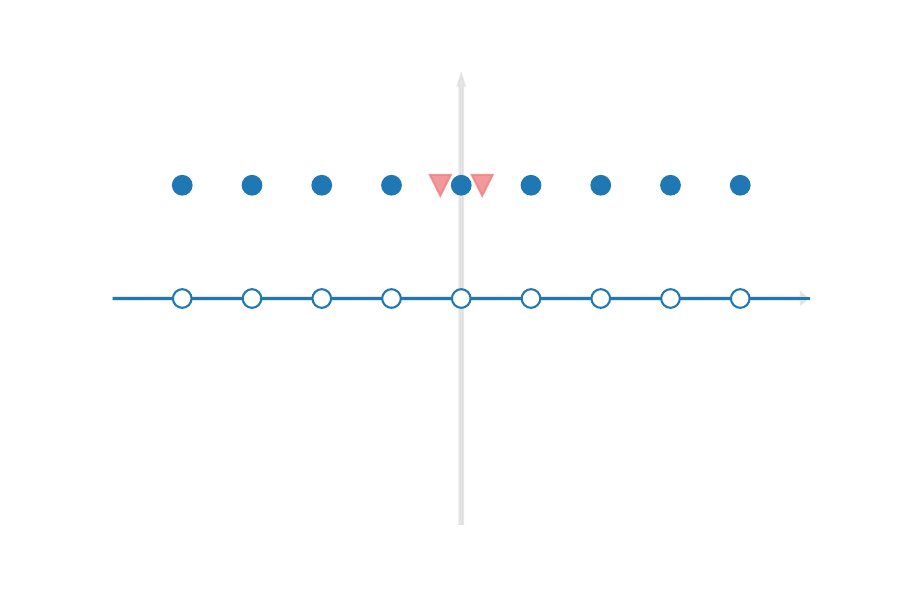} & \includegraphics[width=45mm]{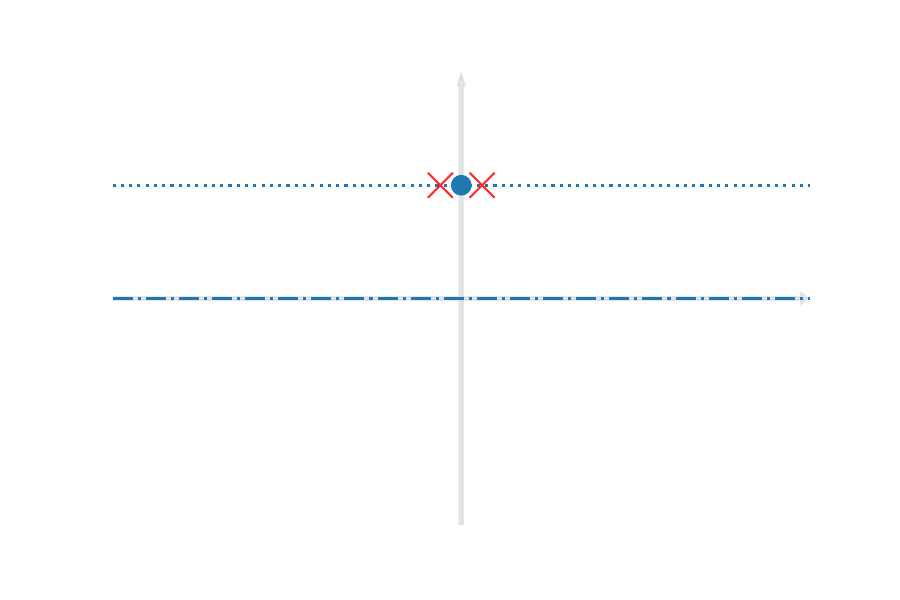}\\
$f_{10}\left(x\right)=\begin{cases}
x\sin\left(\frac{1}{x}\right), & x\neq0\\
0, & x=0
\end{cases}$ & $f_{11}\left(x\right)=\boldsymbol{1}_{\mathbb{Z}}$ & $f_{12}\left(x\right)=\boldsymbol{1}_{\mathbb{Q}}$ \\[6pt]
 \includegraphics[width=45mm]{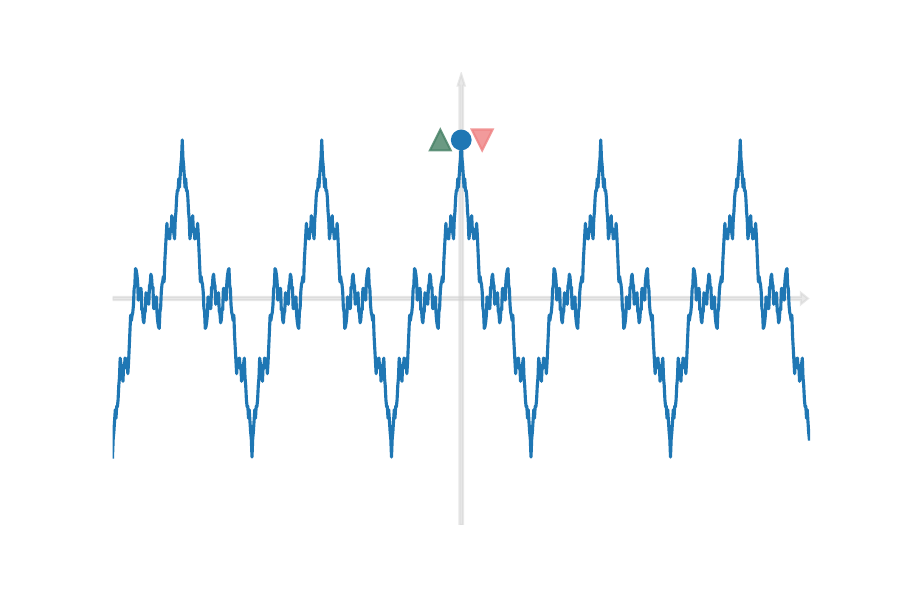} &   \includegraphics[width=45mm]{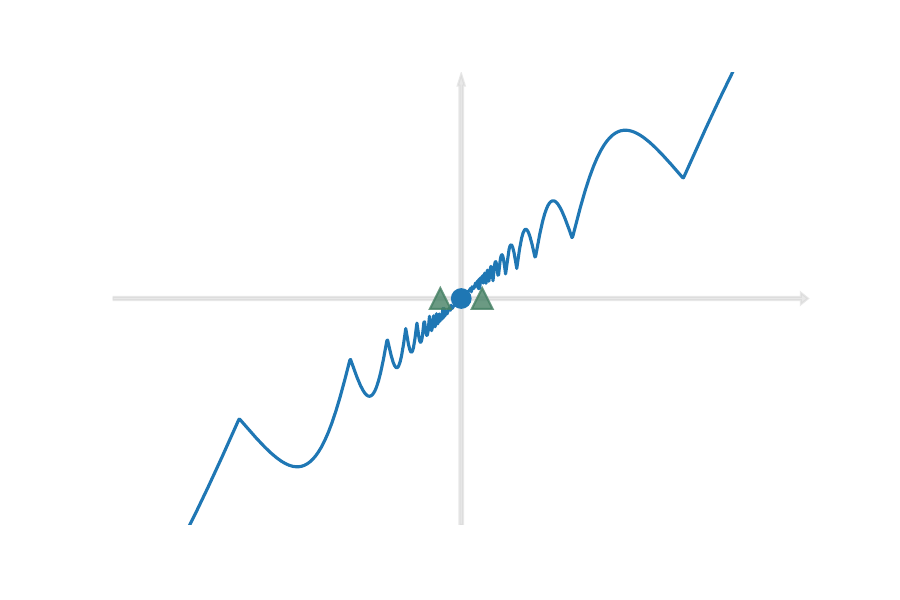} & \includegraphics[width=45mm]{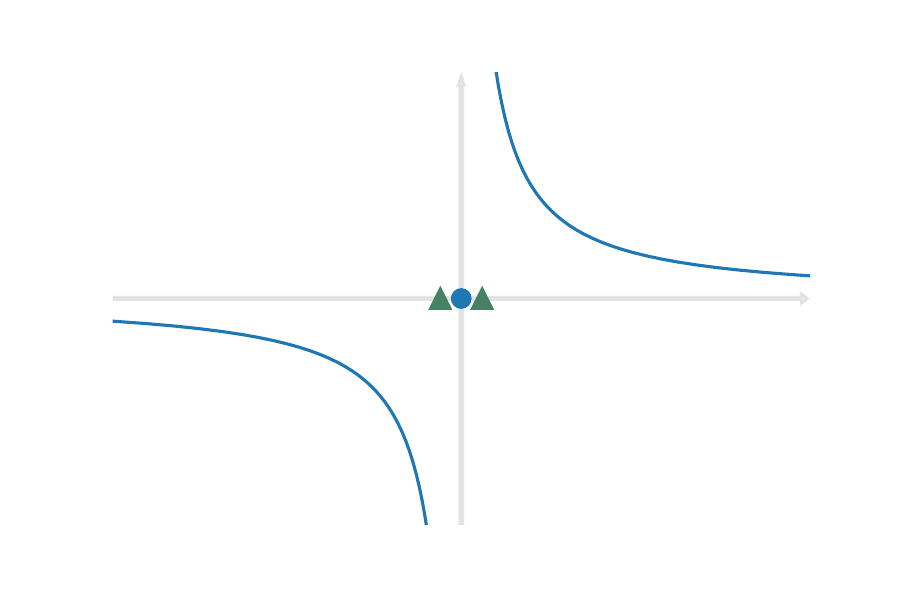}\\
$f_{13}\left(x\right)=\sum_{n=1}^{\infty}\left(\frac{1}{2}\right)^{n}\cos\left(3^{n}x\right)$& $f_{14}\left(x\right)=\begin{cases}
x+x\left|\sin\left(\frac{1}{x}\right)\right|, & x\neq0\\
0, & x=0
\end{cases}$ & $f_{15}\left(x\right)=\begin{cases}
\frac{1}{x}, & x\neq0\\
0, & x=0
\end{cases}$ \\[6pt]

\end{tabular}
\caption{The one-sided detachments of selected functions at $x=0$, denoted with upper or lower carets next to $f\left(0\right)$ if the detachment exists. Non detachable functions there are annotated with "X". Other properties of these functions are detailed in table \ref{pointwise_properties_examples}.}
\label{functions_plots_with_detachments}
\end{figure}

\subsection{Simple Algebraic Properties}
\index{Constant Multiple Rule}
\begin{clm}\label{constant_multiple_rule}\textbf{A semi-discrete constant multiple rule}. If $f$ is detachable at $x$, and $c\in \mathbb{R}$ is a constant, then $cf$ is also detachable and the following holds there:
$$\left(cf\right)^{;}=\text{sgn}\left(cf^{;}\right).$$
\end{clm}

\begin{proof}
\begin{align}
&\begin{aligned}\left(cf\right)_{\pm}^{;} &= \pm \underset{t\rightarrow x^{\pm}}{\lim}\text{sgn}\left[\left(cf\right)\left(x+h\right)-\left(cf\right)\left(x\right)\right] \\
&= \pm \underset{t\rightarrow x^{\pm}}{\lim}\text{sgn}\left(c\right)\text{sgn}\left[f\left(x+h\right)-f\left(x\right)\right]\\
&=\text{sgn}\left(c\right)f_{\pm}^{;}=\text{sgn}\left[cf_{\pm}^{;}\right].
\end{aligned}
\end{align}
\end{proof}

\index{Sum Rule}
\begin{clm}\label{sum_rule}\textbf{Semi-discrete sum and difference Rules}. If $f$ and $g$ are detachable at $x$ and $\left(f^{;} g^{;} \right) \left(x\right) \in \left\{0, \pm 1 \right \}$ (the plus or minus signs are for the sum and difference rules, respectively), then the following holds at $x$:

$$\left(f \pm g\right)^{;}=\text{sgn}\left( f^{;} \pm g^{;} \right).$$
\label{sum_rule}
\end{clm}

\begin{proof}Without loss of generality, let us focus on right-detachments. We will show that if $f^{;}_{+}\left(x\right)=g^{;}_{+}\left(x\right)$ then $\left(f+g\right)^{;}_{+}\left(x\right)=+1$, and the rest of the cases are proved similarly.

There is a right-neighborhood bounded by $\delta_{f}$ where:

$$0<\bar{x}-x<\delta_{f}\Longrightarrow \text{sgn}\left[f\left(\bar{x}\right)-f\left(x\right)\right]=+1\Longrightarrow f\left(\bar{x}\right) > f\left(x\right).$$

Similarly there exists a right-neighborhood bounded by $\delta_{g}$ where:

$$0<\bar{x}-x<\delta_{g}\Longrightarrow g\left(\bar{x}\right) > g\left(x\right).$$

Therefore there is a right-neighborhood bounded by $\delta_{f+g}\equiv\min\left\{ \delta_{f},\delta_{g}\right\}$ where:

$$0<\bar{x}-x<\delta_{f+g}:\,\, \text{sgn}\left[\left(f+g\right)\left(\bar{x}\right)-\left(f+g\right)\left(x\right)\right]=+1,$$

hence $$\underset{\bar{x}\rightarrow x^{+}}{\lim}\text{sgn}\left[\left(f+g\right)\left(\bar{x}\right)-\left(f+g\right)\left(x\right)\right]=+1.$$\end{proof}

\begin{clm}If $f$, $g$ and $f+g$ are detachable at $x$, and the one-sided detachments of $f$ and $g$ aren’t both zeroed, then the following holds at $x$:

$$f^{;}g^{;}=\left(f+g\right)^{;}\left(f^{;}+g^{;}\right)-1.$$
\end{clm}

\begin{proof}It is possible to show by separating to cases that for $A,B$ not both zero:$$\text{sgn}\left(A\right)\text{sgn}\left(B\right)=\text{sgn}\left(A+B\right)\left[\text{sgn}\left(A\right)+\text{sgn}\left(B\right)\right]-1.$$The result is obtained by taking $A=f\left(x+h\right)-f\left(x\right)$ and $B=g\left(x+h\right)-g\left(x\right)$, followed by applying the one-sided limit process to both sides.
\end{proof}

This sum rule holds only for functions whose detachments are not additive inverses. We will handle those cases, assuming differentiability, later on with Taylor series.

\subsection{Product and Quotient Rules}
In the process of formulating the product rule, we repeatedly conduct analyses similar to the above, with different combinations of $f^{;},g^{;}$, their signs, and assumptions on their sign-continuity or lack of.

The following link includes a simulation code for the creation of the data required to extract those rules. The final results obtained from the simulation are also available here.

In each of these cases, we conduct an $\epsilon-\delta$ analysis and prove that the product's detachment equals a value, in case it's indeed determined.

Recall that the product rule for derivatives dictates the following combination of the original functions' derivatives: $$\left(fg\right)'=f'g+fg'.$$ Evidently, the results we gather regarding the detachments product follow a similar rule in part of the cases and follow another intuitive formula in others. Recall that the following product and quotient rules hold for detachable, not necessarily continuous functions.

\begin{theorem}[A semi-discrete product rule]
Let $f$ and $g$ be detachable and sign-consistent at $x$. If one of the following holds there:
\begin{enumerate}
\item $ff^{;}gg^{;}\geq0$, where $f$ or $g$ is s.c., or
\item $f=g=0$
$ff^{;}gg^{;}<0$, where $f$ or $g$ is s.d.
\label{product_rule}
\end{enumerate}

Then $fg$ is detachable there, and: $$\left(fg\right)^{;}=\begin{cases}
\text{sgn}\left[f^{;}\text{sgn}\left(g\right)+g^{;}\text{sgn}\left(f\right)\right], & ff^{;}gg^{;}\geq0\text{, and }f\text{ or }g\text{ is s.c.}\\
f^{;}g^{;}, & \text{else}\text{.}
\end{cases}$$
\index{Product Rule}
\label{product_rule}
\end{theorem}
\begin{proof}For brevity, we will refer to the terms $\text{sgn}\left[f^{;}\text{sgn}\left(g\right)+g^{;}\text{sgn}\left(f\right)\right]$ and $f^{;}g^{;}$ from the theorem statement as the first and second formulas, respectively.

Let us distinguish between the following cases, according to the pointwise signs of $f,g$ and their detachments (the vector $\left(f^{;},\text{sgn}\left(f\right),g^{;},\text{sgn}\left(g\right)\right)\in\left\{ \pm1,0\right\} ^{4}$), and their inherent sign-continuity or lack of.

\begin{enumerate}

\item Assume $ff^{;}gg^{;}>0$. There are 8 such combinations of $\left(f^{;},\text{sgn}\left(f\right),g^{;},\text{sgn}\left(g\right)\right)$:

\begin{itemize}

\item
Assume that either $f$ or $g$ is inherently sign-continuous. There are 4 such cases, in which both $f,g$ adhere to the inherent sign-continuity property. Without loss of generality assume that $f^{;}=g^{;}=\text{sgn}\left(f\right)=\text{sgn}\left(g\right)=+1$. Then: $$\begin{cases} \exists\delta_{f}: & \forall\bar{x}\in B_{\delta_{f}}\left(x\right):f\left(\bar{x}\right)>f\left(x\right)>0\\ \exists\delta_{g}: & \forall\bar{x}\in B_{\delta_{g}}\left(x\right):g\left(\bar{x}\right)>g\left(x\right)>0 \end{cases}$$ hence for $\delta_{fg}\equiv\min\left\{ \delta_{f},\delta_{g}\right\}$ it holds that: $$\forall\bar{x}\in B_{\delta_{fg}}\left(x\right):\left(fg\right)\left(\bar{x}\right)=f\left(\bar{x}\right)g\left(\bar{x}\right)>f\left(x\right)g\left(x\right)=\left(fg\right)\left(x\right),$$ and  $\left(fg\right)^{;}\left(x\right)=+1=\text{sgn}\left[\left(+1\right)\cdot\left(+1\right)+\left(+1\right)\cdot\left(+1\right)\right]$, in accordance with the second formula.

\item Assume that neither $f$ nor $g$ is inherently sign-continuous. There are 4 such cases. Without loss of generality assume that $f^{;}=g^{;}=+1$ and $f,g<0$. Then: $$\begin{cases} \exists\delta_{f}^{\left(1\right)}: & \forall\bar{x}\in B_{\delta_{f}^{\left(1\right)}}\left(x\right):f\left(\bar{x}\right)>f\left(x\right)\\ \exists\delta_{g}: & \forall\bar{x}\in B_{\delta_{g}}\left(x\right):g\left(\bar{x}\right)>g\left(x\right)\\ & f\left(x\right)<0\\ & g\left(x\right)<0 \end{cases}$$ The continuity of $\text{sgn}\left(f\right)$ cannot be inferred directly from the definition of the detachment and its value's sign. Therefore, we will assume that $f$ is sign-continuous explicitly: $\exists\delta_{f}^{\left(2\right)}:\forall\bar{x}\in B_{\delta_{f}^{\left(2\right)}}\left(x\right):f\left(\bar{x}\right)<0,$ hence for $\delta_{fg}\equiv\min\left\{ \delta_{f^{\left(1\right)}},\delta_{f^{\left(2\right)}},\delta_{g}\right\}$ it holds that: $$\forall\bar{x}\in B_{\delta_{fg}}\left(x\right): \left(fg\right)\left(\bar{x}\right)=f\left(\bar{x}\right)g\left(\bar{x}\right) < f\left(\bar{x}\right)g\left(x\right) < f\left(x\right)g\left(x\right)=\left(fg\right)\left(x\right),$$ and $\left(fg\right)^{;}\left(x\right)=-1=\text{sgn}\left[\left(+1\right)\cdot\left(-1\right)+\left(+1\right)\cdot\left(-1\right)\right]$, in accordance with the second formula.
\end{itemize}

\item Assume $ff^{;}gg^{;}=0$. There are 65 such combinations of $\left(f^{;},\text{sgn}\left(f\right),g^{;},\text{sgn}\left(g\right)\right)$:

\begin{itemize}

\item Assume that $f$ or $g$ is sign-continuous. There are 61 such combinations:
Assume that one of $f,g$ is inherently sign-continuous and the other is inherently sign-discontinuous. There are 20 such combinations of $\left(f^{;},\text{sgn}\left(f\right),g^{;},\text{sgn}\left(g\right)\right)$. Without loss of generality, assume that $f^{;}=g^{;}=-1,f<0$ and $g=0$, where $f$ is inherently sign-continuous and $g$ is inherently sign-discontinuous. Then: $$\begin{cases} \exists\delta_{f}: & \forall\bar{x}\in B_{\delta_{f}}\left(x\right):f\left(\bar{x}\right) < f\left(x\right) < 0\\ \exists\delta_{g}: & \forall\bar{x}\in B_{\delta_{g}}\left(x\right):g\left(\bar{x}\right) < g\left(x\right)=0 \end{cases}$$ hence for $\delta_{fg}\equiv\min\left\{ \delta_{f},\delta_{g}\right\}$ it holds that: $$\forall\bar{x}\in B_{\delta_{fg}}\left(x\right):\left(fg\right)\left(\bar{x}\right)=f\left(\bar{x}\right)g\left(\bar{x}\right) > 0=\left(fg\right)\left(x\right),$$ and $\left(fg\right)^{;}\left(x\right)=+1=\text{sgn}\left[\left(-1\right)\cdot0+\left(-1\right)\cdot\left(-1\right)\right]$, in accordance with the second formula.

\item Assume that one of $f,g$ is inherently sign-continuous and the other is neither inherently sign-continuous nor inherently sign-discontinuous. There are 12 such combinations of $\left(f^{;},\text{sgn}\left(f\right),g^{;},\text{sgn}\left(g\right)\right)$. Without loss of generality, assume that $f^{;}=0,g^{;}=-1,f < 0$ and $g > 0$, where $f$ is inherently sign-continuous. Then: $$\begin{cases} \exists\delta_{f}: & \forall\bar{x}\in B_{\delta_{f}}\left(x\right):f\left(\bar{x}\right)=f\left(x\right) < 0\\ \exists\delta_{g}: & \forall\bar{x}\in B_{\delta_{g}}\left(x\right):g\left(\bar{x}\right)< g\left(x\right)\\ & g\left(x\right) > 0 \end{cases}$$ hence for $\delta_{fg}\equiv\min\left\{ \delta_{f},\delta_{g}\right\}$ it holds that: $$\forall\bar{x}\in B_{\delta_{fg}}\left(x\right):\left(fg\right)\left(\bar{x}\right)=f\left(\bar{x}\right)g\left(\bar{x}\right) > f\left(x\right)g\left(x\right)=\left(fg\right)\left(x\right),$$ and $\left(fg\right)^{;}\left(x\right)=+1=\text{sgn}\left[\left(0\right)\cdot\left(+1\right)+\left(-1\right)\cdot\left(-1\right)\right],$ in accordance with the second formula.
\item Assume that one of $f,g$ is inherently sign-discontinuous and the other is neither inherently sign-continuous nor inherently sign-discontinuous. There are 8 such combinations of $\left(f^{;},\text{sgn}\left(f\right),g^{;},\text{sgn}\left(g\right)\right)$. Without loss of generality, assume that $f^{;}=g^{;}=-1,f>0$ and $g=0$, where $g$ is inherently sign-discontinuous and $f$ is neither inherently sign-continuous nor sign-discontinuous. Then: $$\begin{cases} \exists\delta_{f^{\left(1\right)}}: & \forall\bar{x}\in B_{\delta_{f}^{\left(1\right)}}\left(x\right):f\left(\bar{x}\right) < f\left(x\right) < 0\\ \exists\delta_{g}: & \forall\bar{x}\in B_{\delta_{g}}\left(x\right):g\left(\bar{x}\right) < g\left(x\right)=0 \end{cases}$$ Let us assume the continuity of $f$ at $x$ explicitly: $\exists\delta_{f^{\left(2\right)}}:\forall\bar{x}\in B_{\delta_{f}^{\left(2\right)}}\left(x\right):f\left(\bar{x}\right) < 0,$ hence for $\delta_{fg}\equiv\min\left\{ \delta_{f}^{\left(1\right)},\delta_{f}^{\left(2\right)},\delta_{g}\right\}$ it holds that: $$\forall\bar{x}\in B_{\delta_{fg}}\left(x\right):\left(fg\right)\left(\bar{x}\right)=f\left(\bar{x}\right)g\left(\bar{x}\right)< 0=\left(fg\right)\left(x\right),$$ and $\left(fg\right)^{;}\left(x\right)=-1=\text{sgn}\left[\left(-1\right)\cdot0+\left(-1\right)\cdot\left(+1\right)\right]$, in accordance with the second formula.

\item Assume that both $f,g$ are inherently sign-continuous. There are 21 such combinations of $\left(f^{;},\text{sgn}\left(f\right),g^{;},\text{sgn}\left(g\right)\right)$. Without loss of generality, assume that $f^{;}=g^{;}=0,f< 0$ and $g<0$, where $f,g$ are both inherently sign-continuous. Then: $$\begin{cases} \exists\delta_{f}: & \forall\bar{x}\in B_{\delta_{f}}\left(x\right):f\left(\bar{x}\right)=f\left(x\right)< 0\\ \exists\delta_{g}: & \forall\bar{x}\in B_{\delta_{g}}\left(x\right):g\left(\bar{x}\right)< g\left(x\right)<0 \end{cases}$$ hence for $\delta_{fg}\equiv\min\left\{ \delta_{f},\delta_{g}\right\}$ it holds that: $$\forall\bar{x}\in B_{\delta_{fg}}\left(x\right):\left(fg\right)\left(\bar{x}\right)=f\left(\bar{x}\right)g\left(\bar{x}\right) > f\left(x\right)g\left(x\right)=\left(fg\right)\left(x\right),$$ and $\left(fg\right)^{;}\left(x\right)=0=\text{sgn}\left[\left(0\right)\cdot\left(-1\right)+\left(0\right)\cdot\left(-1\right)\right]$, in accordance with the second formula.

\item Assume that $f=g=0$. There are 4 such combinations of $\left(f^{;},\text{sgn}\left(f\right),g^{;},\text{sgn}\left(g\right)\right)$. In these cases, both $f,g$ are inherently sign-discontinuous. Without loss of generality, assume that $f^{;}=g^{;}=1,f=g=0$, where $f,g$ are both inherently sign-discontinuous. Then: $$\begin{cases} \exists\delta_{f}: & \forall\bar{x}\in B_{\delta_{f}}\left(x\right):f\left(\bar{x}\right)< f\left(x\right)=0\\ \exists\delta_{g}: & \forall\bar{x}\in B_{\delta_{g}}\left(x\right):g\left(\bar{x}\right)< g\left(x\right)=0 \end{cases}$$ hence for $\delta_{fg}\equiv\min\left\{ \delta_{f},\delta_{g}\right\}$ it holds that: $$\forall\bar{x}\in B_{\delta_{fg}}\left(x\right):\left(fg\right)\left(\bar{x}\right)=f\left(\bar{x}\right)g\left(\bar{x}\right)> 0=f\left(x\right)g\left(x\right)=\left(fg\right)\left(x\right),$$ and $\left(fg\right)^{;}\left(x\right)=+1=\left(+1\right)\cdot\left(+1\right)$, in accordance with the first formula.
\end{itemize}

\item Assume $ff^{;}gg^{;}<0$. There are 8 such combinations of $\left(f^{;},\text{sgn}\left(f\right),g^{;},\text{sgn}\left(g\right)\right)$. For each combination it holds that $ff^{;}>0$ or $gg^{;}>0$, hence either $f$ or $g$ is inherently sign-continuous. Without loss of generality assume that $f^{;}=+1,f>0,g^{;}=-1$ and $g>0$. Then: $$\begin{cases} \exists\delta_{f}: & \forall\bar{x}\in B_{\delta_{f}^{\left(1\right)}}\left(x\right):f\left(\bar{x}\right) > f\left(x\right) >0\\ \exists\delta_{g}^{\left(1\right)}: & \forall\bar{x}\in B_{\delta_{g}}\left(x\right):g\left(\bar{x}\right)< g\left(x\right)\\ & g\left(x\right)>0 \end{cases}$$ The continuity of $\text{sgn}\left(f\right)$ can be inferred directly from the definition of the detachment and its value's sign. However, $g$ is neither inherently sign-continuous nor inherently sign-discontinuous. Thus we will assume that $g$ is sign-discontinuous explicitly: $$\exists\delta_{g}^{\left(2\right)}:\forall\bar{x}\in B_{\delta_{g}^{\left(2\right)}}\left(x\right):g\left(\bar{x}\right)\leq0,$$ hence for $\delta_{fg}\equiv\min\left\{ \delta_{f},\delta_{g^{\left(1\right)}},\delta_{g^{\left(2\right)}}\right\}$ it holds that: $$\forall\bar{x}\in B_{\delta_{fg}}\left(x\right):\left(fg\right)\left(\bar{x}\right)=f\left(\bar{x}\right)g\left(\bar{x}\right)\leq0 < f\left(x\right)g\left(x\right)=\left(fg\right)\left(x\right),$$ and $\left(fg\right)^{;}\left(x\right)=-1=\left(+1\right)\cdot\left(-1\right)$, in accordance with the first formula.
\end{enumerate}
\end{proof}

Similarly to the product rule for detachment, the following quotient rule also incorporates a term that looks familiar from the nominator in its derivative counterpart, $f'g-gf'$.

\begin{theorem}[A semi-discrete quotient rule]Let $f$ and $g$ be detachable and sign-consistent at $x$, where $g\neq0$ across their definition domain. Assume that $f,g$ are either s.c. or s.d. If one of the following holds at:

\begin{enumerate}
\item $ff^{;}gg^{;}\leq0$, where $f$ or $g$ is s.c., or $f=0$
\item $ff^{;}gg^{;}>0$, where only one of $f,g$ is s.c.
\end{enumerate}

Then $\frac{f}{g}$ is detachable, and: $$\left(\frac{f}{g}\right)^{;}=\begin{cases}
\text{sgn}\left[f^{;}\text{sgn}\left(g\right)-g^{;}\text{sgn}\left(f\right)\right], & g\text{ is i.s.c., or }f\text{ and }g\text{ are s.c.},\text{or }f=0\text{ and }f\text{ or }g\text{ is s.c.}\\
\text{sgn}\left[f^{;}\text{sgn}\left(g\right)+g^{;}\text{sgn}\left(f\right)\right], & \text{else if }ff^{;}gg^{;}\geq0\text{, and }f\text{ or }g\text{ is s.c.}\\
f^{;}g^{;}, & \text{else.}
\end{cases}$$
\label{quotient_rule}
\index{Quotient Rule}
\end{theorem}

\begin{proof}For brevity, we will refer to the terms $\text{sgn}\left[f^{;}\text{sgn}\left(g\right)-g^{;}\text{sgn}\left(f\right)\right]$, $\text{sgn}\left[f^{;}\text{sgn}\left(g\right)+g^{;}\text{sgn}\left(f\right)\right]$ and $f^{;}g^{;}$ from the theorem statement as the first, second and third formulas and conditions, respectively.

We prove the claim by separating to cases. We will suggest a slightly shorter analysis with respect to the proof of the product rule, as the ideas are similar. We will survey a handful of representative cases. The rest of them are shown to hold analogously.

\begin{enumerate}

\item Assume that $g$ is i.s.c. Without lose of generality, assume $f^{;}=g^{;}=-1,f>0$ and $g<0$. There are two cases: either $f$ is s.d., or it is s.c.
\begin{itemize}

\item Assuming $f$ is s.d., then:$$\begin{cases}\exists\delta_{f}^{\left(1\right)}: & \forall\bar{x}\in B_{\delta_{f}^{\left(1\right)}}\left(x\right):f\left(\bar{x}\right)< f\left(x\right)\\\exists\delta_{f}^{\left(2\right)}: & \forall\bar{x}\in B_{\delta_{f}^{\left(2\right)}}\left(x\right):f\left(\bar{x}\right)\leq0\\\exists\delta_{g}: & \forall\bar{x}\in B_{\delta_{g}}\left(x\right):g\left(\bar{x}\right)< g\left(x\right)\\ & f\left(x\right)>0\\& g\left(x\right)<0,\end{cases}$$
hence for $\delta_{f/g}\equiv\min\left\{ \delta_{f}^{\left(1\right)},\delta_{f}^{\left(2\right)},\delta_{g}\right\}$ it holds that:
$$\forall\bar{x}\in B_{\delta_{f/g}}\left(x\right):\frac{f}{g}\left(\bar{x}\right)\in\left\{ 0,+1\right\},$$
however $\frac{f}{g}\left(x\right)<0$, therefore $\text{sgn}\left[\frac{f}{g}\left(\bar{x}\right)-\frac{f}{g}\left(x\right)\right]=+1$ in a neighborhood of $x$. Thus $$\left(\frac{f}{g}\right)^{;}\left(x\right)=+1=\text{sgn}\left[\left(-1\right)\cdot\left(-1\right)-\left(-1\right)\cdot\left(+1\right)\right]$$, in accordance with the first formula.
\item Assuming $f$ is s.c. while maintaining the conditions in the previous example, we have: $$\exists\tilde{\delta}_{f}^{\left(2\right)}:\forall\bar{x}\in B_{\tilde{\delta}_{f}^{\left(2\right)}}\left(x\right):f\left(\bar{x}\right)>0,$$
hence for $\tilde{\delta}{}_{f/g}\equiv\min\left\{ \delta_{f}^{\left(1\right)},\tilde{\delta}_{f}^{\left(2\right)},\delta_{g}\right\}$ it holds that:$$\forall\bar{x}\in B_{\delta_{f/g}}\left(\bar{x}\right):\left|\frac{f}{g}\left(\bar{x}\right)\right|<\left|\frac{f}{g}\left(x\right)\right|,$$ and since $\frac{f}{g}\left(\bar{x}\right),\frac{f}{g}\left(x\right)$ are both negative then $\frac{f}{g}\left(\bar{x}\right)>\frac{f}{g}\left(x\right)$ in that neighborhood of $x$, and again $\left(\frac{f}{g}\right)^{;}\left(x\right)=+1$.
\end{itemize}

\item Assume that $f,g$ are both s.c. Without lose of generality, assume $f^{;}=g^{;}=-1,f<0$ and $g>0$. Then $f$ is i.s.c., and we will impose the assumption that $g$ is s.c. Then:
$$\begin{cases}\exists\delta_{f}: & \forall\bar{x}\in B_{\delta_{f}}\left(x\right):f\left(\bar{x}\right)< f\left(x\right)\\ \exists\delta_{g}^{\left(1\right)}: & \forall\bar{x}\in B_{\delta_{g}^{\left(1\right)}}\left(x\right):g\left(\bar{x}\right)\leq g\left(x\right)\\ \exists\delta_{g}^{\left(2\right)}: & \forall\bar{x}\in B_{\delta_{g}^{\left(2\right)}}\left(x\right):g\left(\bar{x}\right)>0\\& f\left(x\right)>0\\& g\left(x\right)<0, \end{cases}$$

hence for $\delta_{f/g}\equiv\min\left\{ \delta_{f},\delta_{g}^{\left(1\right)},\delta_{g}^{\left(2\right)}\right\}$ it holds that: $$\forall\bar{x}\in B_{\delta_{f/g}}\left(x\right):\left|\frac{f}{g}\left(\bar{x}\right)\right|>\left|\frac{f}{g}\left(x\right)\right|,$$
and since $\frac{f}{g}\left(x\right),\frac{f}{g}\left(\bar{x}\right)$ are both negative, then $\frac{f}{g}\left(\bar{x}\right)<\frac{f}{g}\left(x\right)$ in that neighborhood of $x$. Thus $\left(\frac{f}{g}\right)^{;}\left(x\right)=-1=\text{sgn}\left[\left(-1\right)\cdot\left(+1\right)-\left(-1\right)\cdot\left(-1\right)\right]$, in accordance with the first formula.

\item Assume that the second conditions hold, that is, the conditions of the first formula do not hold, while $ff^{;}gg^{;}\geq0$ and $f$ or $g$ is s.c. There are two slightly different families of cases, where $ff^{;}gg^{;}$ is either zeroed or positive.

\begin{enumerate}
\item Assume first that $ff^{;}gg^{;}>0$. Without lose of generality, let $f^{;}=g^{;}=-1,f>0$ and $g>0$. Assume that $f$ is s.c. Then we can assume that $g$ is not s.c., because otherwise the first condition would hold. Then:
$$\begin{cases}\exists\delta_{f}^{\left(1\right)}: & \forall\bar{x}\in B_{\delta_{f}^{\left(1\right)}}\left(x\right):f\left(\bar{x}\right)< f\left(x\right)\\ \exists\delta_{f}^{\left(2\right)}: & \forall\bar{x}\in B_{\delta_{f}^{\left(2\right)}}\left(x\right):f\left(\bar{x}\right)>0\\\exists\delta_{g}^{\left(1\right)}: & \forall\bar{x}\in B_{\delta_{g}^{\left(1\right)}}\left(x\right):g\left(\bar{x}\right)< g\left(x\right)\\\exists\delta_{g}^{\left(2\right)}: & \forall\bar{x}\in B_{\delta_{g}^{\left(2\right)}}\left(x\right):g\left(\bar{x}\right)<0\\ & f\left(x\right)>0\\& g\left(x\right)>0,\end{cases}$$

hence for $\delta_{f/g}\equiv\min\left\{ \delta_{f}^{\left(1\right)},\delta_{f}^{\left(2\right)},\delta_{g}^{\left(1\right)},\delta_{g}^{\left(2\right)}\right\}$ it holds that:
$$\forall\bar{x}\in B_{\delta_{f/g}}\left(x\right):\frac{f}{g}\left(\bar{x}\right)<0,$$
and since $\frac{f}{g}\left(x\right)>0,$ then $\frac{f}{g}\left(\bar{x}\right)<\frac{f}{g}\left(x\right)$ in that neighborhood of $x$. Thus $\left(\frac{f}{g}\right)^{;}\left(x\right)=-1=\text{sgn}\left[\left(-1\right)\cdot\left(+1\right)+\left(-1\right)\cdot\left(+1\right)\right]=-1$, in accordance with the second formula.
\item Assume that $ff^{;}gg^{;}=0$. Without lose of generality, let $f^{;}=g^{;}=-1,f=0$ and $g>0$. Since $f$ is i.s.d. we will assume that $g$ is s.c. Then: $$\begin{cases}\exists\delta_{f}: & \forall\bar{x}\in B_{\delta_{f}}\left(x\right):f\left(\bar{x}\right)< f\left(x\right)\\\exists\delta_{g}^{\left(1\right)}: & \forall\bar{x}\in B_{\delta_{g}^{\left(1\right)}}\left(x\right):g\left(\bar{x}\right)< g\left(x\right)\\ \exists\delta_{g}^{\left(2\right)}: & \forall\bar{x}\in B_{\delta_{g}^{\left(2\right)}}\left(x\right):g\left(\bar{x}\right)>0\\& f\left(x\right)=0\\& g\left(x\right)>0,\end{cases}$$
hence for $\delta_{f/g}\equiv\min\left\{ \delta_{f},\delta_{g}^{\left(1\right)},\delta_{g}^{\left(2\right)}\right\}$ it holds that:
$$\forall\bar{x}\in B_{\delta_{f/g}}\left(x\right):\frac{f}{g}\left(\bar{x}\right)>0,$$
and since $\frac{f}{g}\left(x\right)>0$, then $\frac{f}{g}\left(\bar{x}\right)<\frac{f}{g}\left(x\right)$ in that neighborhood of $x$. Thus $\left(\frac{f}{g}\right)^{;}\left(x\right)=-1=\text{sgn}\left[\left(-1\right)\cdot\left(+1\right)+\left(-1\right)\cdot0\right]$, in accordance with the second formula.
\end{enumerate}

\item Assume that the third condition holds. There are two slightly different cases, where $f$ is either zeroed or not.
\begin{itemize}
\item First assume that $f\left(x\right)=0$. Without lose of generality, let $f^{;}=g^{;}=-1$, and $g>0$. We can assume that $g$ is s.d., because otherwise the first formula would hold. Then:
$$\begin{cases}\exists\delta_{f}^{\left(1\right)}: & \forall\bar{x}\in B_{\delta_{f}^{\left(1\right)}}\left(x\right):f\left(\bar{x}\right)< f\left(x\right)\\ \exists\delta_{f}^{\left(2\right)}: & \forall\bar{x}\in B_{\delta_{f}^{\left(2\right)}}\left(x\right):f\left(\bar{x}\right)>0\\\exists\delta_{g}^{\left(1\right)}: & \forall\bar{x}\in B_{\delta_{g}^{\left(1\right)}}\left(x\right):g\left(\bar{x}\right)< g\left(x\right)\\\exists\delta_{g}^{\left(2\right)}: & \forall\bar{x}\in B_{\delta_{g}^{\left(2\right)}}\left(x\right):g\left(\bar{x}\right)<0\\ & f\left(x\right)=0\\ & g\left(x\right)>0,\end{cases}$$
hence for $\delta_{f/g}\equiv\min\left\{ \delta_{f}^{\left(1\right)},\delta_{f}^{\left(2\right)},\delta_{g}^{\left(1\right)},\delta_{g}^{\left(2\right)}\right\}$ it holds that:
$$\forall\bar{x}\in B_{\delta_{f/g}}\left(x\right):\frac{f}{g}\left(\bar{x}\right)>0,$$
and since $\frac{f}{g}\left(x\right)>0,$then $\frac{f}{g}\left(\bar{x}\right)>\frac{f}{g}\left(x\right)$ in that neighborhood of $x$. Thus $\left(\frac{f}{g}\right)^{;}\left(x\right)=+1=\left(-1\right)\cdot\left(-1\right)$, in accordance with the third formula.
\item Assume that $f\left(x\right)\neq0$. Without lose of generality, let $f^{;}=g^{;}=-1,f<0$ and $g>0$. We can assume that $g$ is s.d., because $f$ is i.s.c. and if $g$ was also s.c. then the first formula would hold. Then:
$$\begin{cases}\exists\delta_{f}: & \forall\bar{x}\in B_{\delta_{f}}\left(x\right):f\left(\bar{x}\right)< f\left(x\right)\\\exists\delta_{g}^{\left(1\right)}: & \forall\bar{x}\in B_{\delta_{g}^{\left(1\right)}}\left(x\right):g\left(\bar{x}\right)< g\left(x\right)\\\exists\delta_{g}^{\left(2\right)}: & \forall\bar{x}\in B_{\delta_{g}^{\left(2\right)}}\left(x\right):g\left(\bar{x}\right)<0\\& f\left(x\right)<0\\ & g\left(x\right)>0,\end{cases}$$
hence for $\delta_{f/g}\equiv\min\left\{ \delta_{f},\delta_{g}^{\left(1\right)},\delta_{g}^{\left(2\right)}\right\} $ it holds that:
$$\forall\bar{x}\in B_{\delta_{f/g}}\left(x\right):\frac{f}{g}\left(\bar{x}\right)>0,$$
and since $\frac{f}{g}\left(x\right)<0,$ then $\frac{f}{g}\left(\bar{x}\right)>\frac{f}{g}\left(x\right)$ in that neighborhood of $x$. Thus $\left(\frac{f}{g}\right)^{;}\left(x\right)=+1=\left(-1\right)\cdot\left(-1\right)$, in accordance with the third formula.
\end{itemize}
\end{enumerate}
\end{proof}

\begin{remark}In the quotient rule, if $f,g$ are both continuous then, because $g\neq0$, $g$ has to be s.c. If $f$ is also s.c. then the first formula holds according to its second sub-condition. If $f$ is not s.c., then from its continuity, $f=0$ and the first formula holds according to its third sub-condition. Thus, the first formula holds for all the pairs of continuous functions subject to the proposition statement. However, in the product rule, the first formula does not hold for some continuous functions. For example, consider $f=g=x$ at $x=0$. Since both $f$ and $g$ are s.d., their product's detachment follows the second formula instead.
Upon formulating the aforementioned rules, there are other ways to bound the signs of the functions $f,g$ rather than inquiring about their sign continuity. For example, we could offer a precise bound on the signs of $f,g$ in a given neighborhood of $x$. A product rule constructed based on these traits holds in more cases than do claims 4 and 5. However, for consistency with Differential Calculus, I preferred to introduce in this chapter the intuitive trait of sign-continuity. This property corresponds with the traditional requirement of differentiability and continuity, and, as such, the first-time reader may feel slightly more comfortable with it.
We define sign-discontinuity as stated above to be able to bound the function's sign in the neighborhood of the point. The mere lack of sign-continuity does not necessarily impose such a bound.
\end{remark}

\begin{figure}[h!]
\includegraphics[scale=0.9]{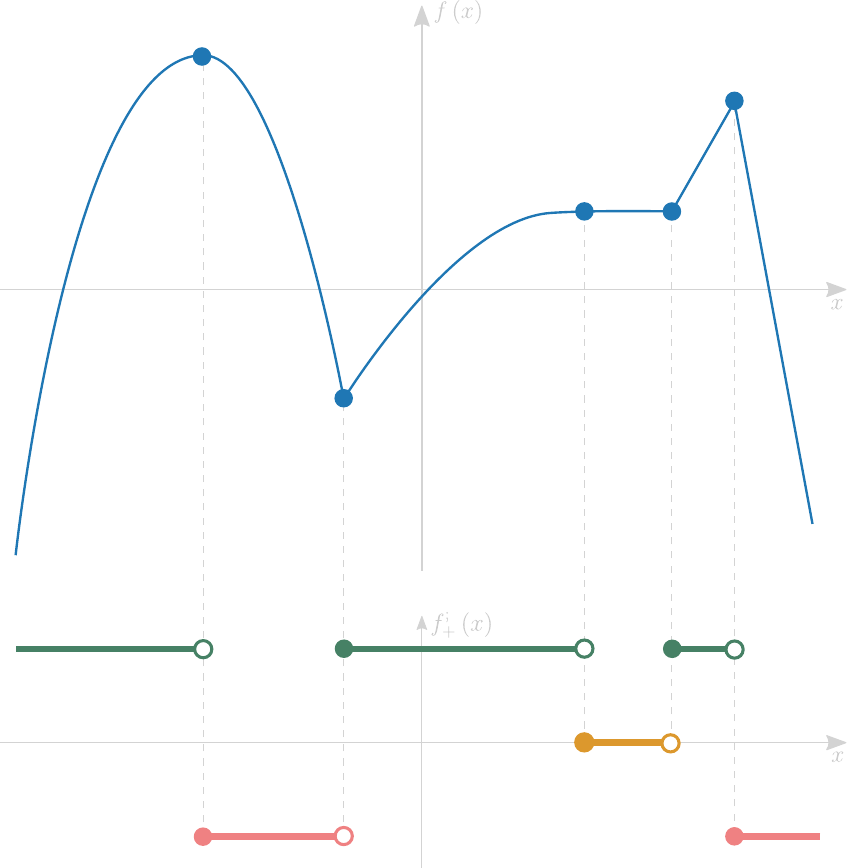}
\caption{An illustration of a function's right detachment across an interval. The function $f$'s right-detachment is illustrated below the function's graph.}
\end{figure}

\subsection{Mean Value Theorems}

In theorem \ref{fermat_semi_discrete}, we formulated an analogue to Fermat's stationary points theorem. A natural question we may ask is whether other classical results that are applicable to rates calculations, can be reformulated based on trends, with a similar trade off between the information level provided by the theorem and the (broader) set of functions to which it is applicable. Let us first recall the formulation of the following classical results:

\index{Rolle's Theorem}
\begin{theorem}[Rolle's theorem]Let $f:\left[a,b\right]\rightarrow\mathbb{R}$ be continuous in $\left[a,b\right]$ and differentiable in $\left(a,b\right)$, such that $f\left(a\right)=f\left(b\right).$ Then, there is a point $c\in\left(a,b\right)$ where:
$$\exists \underset{h\rightarrow0}{\lim}{{f\left(c+h\right)-f\left(c\right)}\over{h}}=0.$$
\label{rolle_theorem}
\end{theorem}

\index{Lagrange's Theorem}
\begin{theorem}[Lagrange's mean value theorem]Let $f$ be continuous in $\left[a,b\right]$ and differentiable in $\left(a,b\right)$. Then there is $c\in \left(a,b\right)$ with:
$$f'\left(c_{v}\right)=\frac{f\left(b\right)-f\left(a\right)}{b-a}.$$
\label{mvt_derivative_thm}
\end{theorem}

A similar version also exists for integrals:

\begin{theorem}[Mean value theorem for definite integrals]Let $f:\left[a,b\right]\rightarrow\mathbb{R}$ be a continuous
function. Then there is $c\in\left(a,b\right)$ such that:
\[
\int_{a}^{b}f\left(x\right)dx=f\left(c\right)\left(b-a\right).
\]
Since the mean value of $f$ on $\left[a,b\right]$ is defined as
\[
\frac{1}{b-a}\int_{a}^{b}f\left(x\right)dx,
\]
we can interpret the conclusion as $f$ achieves its mean value at
some $c\in\left(a,b\right)$.
\label{mvt_integral_thm}
\index{Mean Value Theorem for Integrals}
\end{theorem}

Inspired by these classical results, let us attempt to suggest their semi-discrete versions.
 
\begin{theorem}[A semi-discrete Rolle's theorem]Let $f:\left[a,b\right]\rightarrow\mathbb{R}$ be continuous in $\left[a,b\right]$ and detachable in $\left(a,b\right)$ such that $f\left(a\right)=f\left(b\right).$ Then, there is a point $c\in\left(a,b\right)$ where:
$$f_{+}^{;}\left(c\right)+f_{-}^{;}\left(c\right)=0$$
\label{rolle_theorem_detachment}
\end{theorem}
\begin{proof}$f$ is continuous in a closed interval, hence according to Weierstrass’s theorem, it receives a maximum $M$ and a minimum $m$. In case $m < M$, then, since it is given that $f\left(a\right)=f\left(b\right)$, one of the values $m$ or $M$ must be an image of one of the points in the open interval $\left(a,b\right)$. Let $c\in f^{-1}\left(\left\{ M,m\right\} \right)\backslash\left\{ a,b\right\}$. $f$ receives a local extremum at $c$. If it is strict, then according to theorem 1, $\exists\underset{h\rightarrow0}{\lim}\text{sgn}\left[f\left(c+h\right)-f\left(c\right)\right]\neq0$, hence:
$$f_{+}^{;}\left(c\right)+f_{-}^{;}\left(c\right)=\underset{{\scriptscriptstyle h\rightarrow0^{+}}}{\lim}\text{sgn}\left[f\left(c+h\right)-f\left(c\right)\right]-\underset{{\scriptscriptstyle h\rightarrow0^{-}}}{\lim}\text{sgn}\left[f\left(c+h\right)-f\left(c\right)\right]=0,$$and the claim holds. If the extremum is not strict, then from detachability $f_{+}^{;}\left(c\right)=0$ or $f_{-}^{;}\left(c\right)=0$. If both the one-sided detachments are zeroed then we are done. Otherwise, we assume without loss of generality that $f_{+}^{;}\left(c\right)=0$. Then, $f$ is constant in a right-neighborhood of $c$; hence, $\bar{c}$ for whom $f_{+}^{;}\left(\bar{c}\right)=f_{-}^{;}\left(\bar{c}\right)=0$, and the sum of the one-sided detachments is zeroed trivially. The latter condition also holds trivially in case $m=M$ (where the function is constant).
\end{proof}

\begin{theorem}[A semi-discrete Lagrange's mean value theorem]Let $f$ be continuous in $\left[a,b\right]$ and detachable in $\left(a,b\right)$. Assume $f\left(a\right)\neq f\left(b\right).$ Then for each $v\in\left(f\left(a\right),f\left(b\right)\right)$ there is $c_{v}\in f^{-1}\left(v\right)$ with:

$$f^{;}\left(c_{v}\right)=\text{sgn}\left[f\left(b\right)-f\left(a\right)\right].$$
\label{mvt_detachment_thm}
\end{theorem}

\begin{proof}Let $v\in\left(f\left(a\right),f\left(b\right)\right).$ Without loss of generality, let us assume that $f\left(a\right)<f\left(b\right)$ and show that there is a point $c_{v}\in f^{-1}\left(v\right)\bigcap\left(a,b\right)$ with $f^{;}_{+}\left(c_v\right)=+1.$ From the continuity of $f$ and according to the intermediate value theorem, $f^{-1}\left(v\right)\bigcap\left(a,b\right)\neq\emptyset. $ Assume, on the contrary, that $f_{+}^{;}\left(x\right)=-1$ for each $x\in f^{-1}\left(v\right)\bigcap\left(a,b\right).$ Let $x_{\sup}=\sup\left[f^{-1}\left(v\right)\bigcap\left(a,b\right)\right].$ The maximum is accepted since $f$ is continuous, thus $f\left(x_{\sup}\right)=v.$ According to our assumption, $f_{+}^{;}\left(x_{\sup}\right)=-1,$; hence there is a point $t_{1} > x_{\sup}$ such that $f\left(t_{1}\right) < f\left(x_{\sup}\right)=v.$ But $f$ is continuous in $\left[t_{1},b\right],$; hence, from the intermediate value theorem there is a point $s\in\left(t_{1},b\right)$ for which $f\left(s\right)=v,$ contradicting the selection of $x_{\sup}.$ Had we alternatively assumed that $f_{+}^{;}\left(x_{\sup}\right)=0,$, then there would be a point $t_{2} > x_{\sup}$ for which $f\left(t_{2}\right)=f\left(x_{\sup}\right)=v,$ which again contradicts the selection of $x_{\sup}.$; therefore, $f_{+}^{;}\left(x_{\sup}\right)=+1.$ The proof regarding one-sided left detachments symmetrically leverages the infimum rather than the supremum.
\end{proof}

\begin{figure}[htp]

\subfloat[Lagrange's Mean Value Theorem (theorem \ref{mvt_integral_thm}) cannot be applied to the illustrated function due to cusp points in its definition domain, where the function is not differentiable.]{
  \includegraphics[clip,width=0.8\columnwidth]{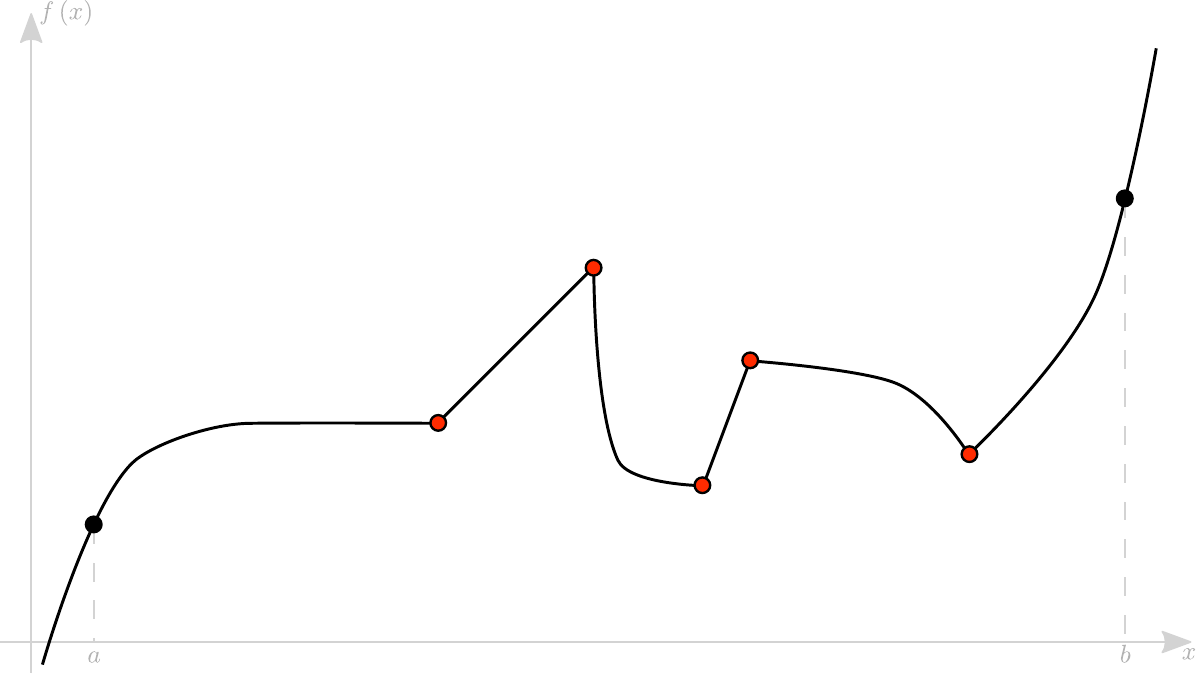}
}

\subfloat[The Mean Value Theorem for Integrals (theorem \ref{mvt_integral_thm}) can be applied in the domain, and the point $c$ guaranteed by the theorem's formulation is highlighted.]{%
  \includegraphics[clip,width=0.8\columnwidth]{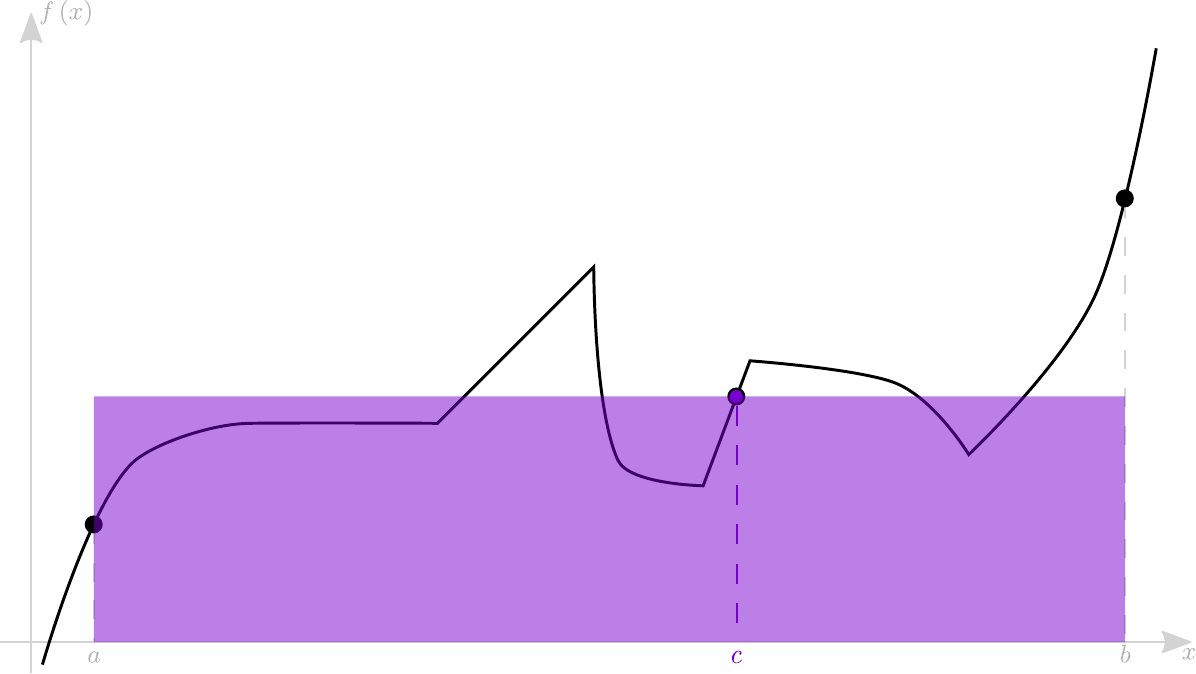}%
}

\subfloat[The Mean Value Theorem for Detachments (theorem \ref{mvt_integral_thm}) can be applied in the domain.]{%
  \includegraphics[clip,width=0.8\columnwidth]{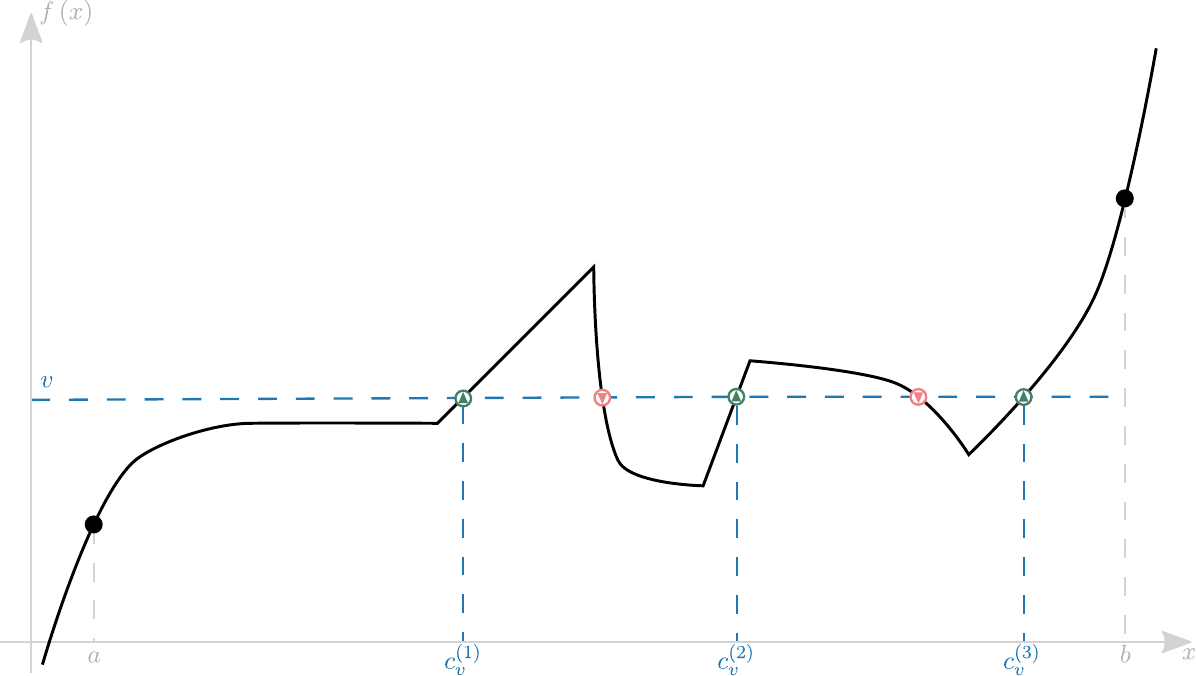}%
}

\caption{Mean Value Theorems for three concepts in Calculus: derivative, integral and detachment.}

\end{figure}

\subsection{Detachability and Monotonicity}
Detachability is tightly coupled with monotonicity when considered across an interval. The following lemma may be helpful in the below discussion.

\index{Monotonicity}
\begin{lem}\label{monotonicity}A function $f$ is strictly monotonic in an interval if and only if $f$ is continuously detachable and  $f^{;}\neq 0$. If the interval is closed with end points $a<b$, then $$f^{;}=f^{;}_{+}\left(a\right)=f^{;}_{-}\left(b\right).$$
\end{lem}
\begin{proof}First direction. Without loss of generality, assume that $f$ is strictly increasing in the interval. We will show that $f_{-}^{;}=f_{+}^{;}=+1.$ On the contrary, assume without loss of generality that there's $x$ in the interval for which $f_{+}^{;}\left(x\right)\neq+1.$ According to the definition of the one-sided detachment, it implies that there is a right neighborhood of $x$ such that $f\left(\bar{x}\right)\leq f\left(x\right).$ But $\bar{x}>x,$ contradicting the strict monotonicity.

Second direction. Without loss of generality, let us assume that $f^{;}\equiv+1$ in the interval; then, $f_{+}^{;}=+1$. It must also hold that $f_{-}^{;}=+1$ in the interval, as otherwise, there would exist a point with $f_{-}^{;}=0,$ and $f$ would be constant in the left neighborhood of that point; hence, there would be another point with $f_{+}^{;}=0.$
Let $x_{1},x_{2} \in \left( a,b \right)$ such that $x_{1}<x_{2}.$ We would like to show that $f\left(x_{1}\right)<f\left(x_{2}\right).$ From the definition of the one-sided detachment, there is a left neighborhood of $x_{2}$ such that $f\left(x\right)<f\left(x_{2}\right)$ for each $x$ in that neighborhood. Let $t\neq x_{2}$ be an element of that neighborhood. Let $s=\sup\left\{ x|x_{1}\leq x\leq t,f\left(x\right)\geq f\left(x_{2}\right)\right\}.$ On the contrary, let us assume that $f\left(x_{1}\right)\geq f\left(x_{2}\right).$ Then $s\geq x_{1},$ and the supremum exists. If $f\left(s\right)\geq f\left(x_{2}\right)$ (i.e., the supremum is accepted in the defined set), then since for any $x>s$ it holds that $f\left(x\right)<f\left(x_{2}\right)\leq f\left(s\right),$ then $f_{+}^{;}\left(s\right)=-1,$ contradicting $f_{+}^{;}\equiv+1$ in $\left(a,b\right).$ Hence the maximum is not accepted. This implies that $s\neq x_{1}.$ Therefore, according to the definition of the supremum, there is a sequence $x_{n}\rightarrow s$ with $\left\{ x_{n}\right\} _{n=1}^{\infty}\subset\left(x_{1},s\right)$ such that: $f\left(x_{n}\right)\geq f\left(x_{2}\right)>f\left(s\right),$ i.e., $f\left(x_{n}\right)>f\left(s\right),$ contradicting our assumption that $f^{;}\left(s\right)=+1$ (which implies that $f_{-}^{;}\left(s\right)\neq-1).$ Hence $f\left(x_{1}\right)<f\left(x_{2}\right).$

If these conditions hold and the interval is closed, then assume without loss of generality that the function strictly increases in the interval. Then, by the definition of the one-sided detachments,
$$f^{;}=f^{;}_{+}\left(a\right)=f^{;}_{-}\left(b\right)=+1.$$
\end{proof}

\subsection{Fundamental Theorem}
\index{The Fundamental Theorem of Calculus}
\label{fundamental_section}
Let us recall the classic Fundamental Theorem of Calculus.
\begin{theorem}[The fundamental theorem of Calculus]The following relations between the fundamental concepts in Calculus hold.
\begin{enumerate}
    \item 
    Let $f$ be a continuous real-valued function defined on a closed interval $[a,b]$. Let $F$ be the function defined, for all $x\in\left[a,b\right]$ by: 
$${\displaystyle F(x)=\int_{a}^{x}\!f(t)\,dt.}$$
Then $F$ is uniformly continuous on $[a,b]$ and differentiable on the open interval $(a,b),$ and
$${\displaystyle F'(x)=f(x),\,}\forall\,x\in\left(a,b\right).$$
    
    \item 
    Let $f$ be a real-valued function on a closed interval $\left[a,b\right]$ and $F$ an antiderivative of $f$ in $\left[a,b\right]$: $F'(x)=f(x)$.     If $f$ is Riemann integrable on $\left[a,b\right]$ then:
$$\int_{a}^{b}F'(x)\,dx=F(b)-F(a).$$
    \end{enumerate}
\label{ftc_1d}
\end{theorem}

While the detachment is clearly not invertible, it is directly related to the derivative and the integral. The following theorem states these connections and can be thought of as a semi-discrete extension to the fundamental theorem of Calculus.

\begin{theorem}[A semi-discrete extension to the fundamental theorem of Calculus]The following relations between the detachment and the fundamental concepts in Calculus hold.

\begin{enumerate}

\item Let $f$ be differentiable with a non-vanishing derivative at a point $x$. Then $f$ is detachable and the following holds at $x$:
$$f^{;}= \text{sgn}\left(f'\right).$$
\item Let $f$ be integrable in a closed interval $\left[a,b\right]$. Let $F\left(x\right)\equiv \int_a^x f\left(t\right)dt$. Let $x\in\left(a,b\right)$. Assume that $f$ is s.c. at $x.$ Then $F$ is detachable and the following holds at $x$:
$$F^{;}= \text{sgn}\left(f\right).$$

\item Let $f:\mathbb{R}\rightarrow\mathbb{R}$ be continuous in $\left[a,b\right]$ and detachable, then:

$$\forall v\in\left[f\left(a\right),f\left(b\right)\right]:\,\,\text{sgn}\left[\underset{c_{v}\in f^{-1}\left(v\right)}{\int}f_{\pm}^{;}\left(c_{v}\right)\right]=\text{sgn}\left[f\left(b\right)-f\left(a\right)\right].$$

\item Let $f:\left[a,b\right]\rightarrow\mathbb{R}$ be a continuous function where $f(a)\cdot f(b)\neq 0$. Let $F$ be an antiderivative of $f$ in $\left[a,b\right]$: $F'(x)=f(x)$. If $f$ is piecewise monotone on $\left[a,b\right]$, then:
$$\int_{a}^{b}F^{;}(x)\,dx=b\text{sgn}f\left(b\right)-a\text{sgn}f\left(a\right)-\underset{x_{i}\in f^{-1}\left(0\right)}{\sum}\left[f_{-}^{;}\left(x_{i}\right)+f_{+}^{;}\left(x_{i}\right)\right]x_{i}.$$
\end{enumerate}
\label{semi_discrete_fundamental}
\end{theorem}

\begin{proof}Let us show each of the statements separately.
\begin{enumerate}
\item Let us write the one-sided derivatives' sign as follows: 
\begin{align}
&\begin{aligned}
\text{sgn}\left[ f_{\pm}'\left(x\right) \right] &= \text{sgn}\left[\underset{h\rightarrow0^{\pm}}{\lim}\frac{f\left(x+h\right)-f\left(x\right)}{h}\right]
=\underset{h\rightarrow0^{\pm}}{\lim}\text{sgn}\left[\frac{f\left(x+h\right)-f\left(x\right)}{h}\right] \\
&=\pm \underset{h\rightarrow0^{\pm}}{\lim}\text{sgn}\left[f\left(x+h\right)-f\left(x\right)\right]
=f_{\pm}^{;}\left(x\right),
\end{aligned}
\end{align}
where the second transition follows from the fact that the derivative is not zeroed, and because the sign function is continuous at $\mathbb{R}\backslash\left\{ 0\right\}.$
\item Let us apply the following transitions:
\begin{align}
&\begin{aligned}\text{sgn}\left[f\left(x\right)\right] &=\underset{h\rightarrow0^{\pm}}{\lim}\text{sgn}\left[f\left(x+h\right)\right]=\pm \underset{h\rightarrow0^{\pm}}{\lim}\text{sgn}\left[\int_x^{x+h}f\left(t\right)dt\right] \\ &=\pm \underset{h\rightarrow0^{\pm}}{\lim}\text{sgn}\left[F\left(x+h\right)-F\left(x\right)\right]=F_{\pm}^{;}\left(x\right),
\end{aligned}
\end{align}
Where the first transition is due to the continuity of $\text{sgn}\left(f\right)$, and the second transition is explained as follows. Assuming $\underset{h\rightarrow0^{\pm}}{\lim}\text{sgn}\left[f\left(x+h\right)\right]=\Delta_{\pm}$, $f$ maintains the signs $\Delta_{\pm}$ in the one-sided $\delta$-neighborhoods of $x$. Measure theory tells us that the integral $\int_x^{x+\delta} f\left(t\right)dt$ also maintains the sign $\Delta_{\pm}$, and $$\underset{h\rightarrow0^{\pm}}{\lim}\text{sgn}\left[\int_x^{x+h}f\left(t\right)dt\right]=\pm\Delta_{\pm}.$$
\item If $f\left(a\right)<v<f\left(b\right)$, then consider the function $\widetilde{f}\left(x\right)=\begin{cases}
1, & f\left(x\right)>v\\
0, & f\left(x\right)\leq v
\end{cases}$. Sperner's lemma in one dimension says that if a discrete function takes only the values $0$ and $1$, begins at the value $0$ and ends at the value $1$, then it must switch values an odd number of times. From the continuity of $f$, however, $\widetilde{f}$ switches values in the intersections of $f$ with the line $f=v$. As long as $f\left(x\right)\leq v$, the sum of the one-sided detachments is either $0$ or $+1$. Once $\widetilde{f}$ switches value, the sum of the one-sided detachments is $+2$. Since the switching happens an odd number of times, the final sum is $+2$.

If $v=f\left(a\right)$ then the above argument yields that the one-sided detachments sum is $+1$.

If $v=f\left(b\right)$ then define $\utilde{f}\left(x\right)=\begin{cases}
1, & f\left(x\right)\geq v\\
0, & f\left(x\right)<v
\end{cases}$ and the above argument applied to $\utilde{f}$ yields that the one-sided detachments sum is $+1$.
\item We first show that given any piecewise continuously detachable function $g:\left[a,b\right]\rightarrow\mathbb{R}$, it holds that: $$\int_{a}^{b}g^{;}(x)dx=g_{-}^{;}\left(b\right)b-g_{+}^{;}\left(a\right)a+\underset{1<i<n}{\sum}\left[g_{-}^{;}\left(x_{i}\right)-g_{+}^{;}\left(x_{i}\right)\right]x_{i},$$ where $\left\{ x_{i}\right\} _{i=1}^{n}$ is the set of countable discontinuities of $g^{;}$. $g^{;}$ is integrable as a step function. According to lemma \ref{monotonicity}, the detachment is constant in each $\left(x_{i},x_{i+1}\right)$. Thus from known results on integration of step functions in Measure Theory and Calculus:$$\int_{a}^{b}g^{;}\left(x\right)dx=\underset{0\leq i\leq n}{\sum}\left(x_{i+1}-x_{i}\right)g_{i}{}^{;},$$
where $g_{i}{}^{;}$ is the detachment in the (open) $i^{th}$ interval and $x_{0}\equiv a,x_{n+1}\equiv b$. Rearranging the terms and applying the last part of lemma \ref{monotonicity} finalizes the proof. Because $F$ is piecewise monotone, then it is piecewise continuously detachable and we can assign it to $g$ in the later formula. The sign-discontinuities of $f$ are the discontinuities of $F^{;}$. Since $f$ is continuous, then its sign-discontinuities are its zeros. Thus, we also assign in the formula $x_{i}\in f^{-1}\left(0\right)$, the discontinities of $F^{;}$. Hence:
\begin{align}
&\begin{aligned}
&\int_{a}^{b}F^{;}\left(x\right)dx-b\text{sgn}f\left(b\right)+a\text{sgn}f\left(a\right)\\
&=F_{-}^{;}\left(b\right)b-F_{+}^{;}\left(a\right)a+\underset{x_{i}\in f^{-1}\left(0\right)}{\sum}\left[F_{-}^{;}\left(x_{i}\right)-F_{+}^{;}\left(x_{i}\right)\right]x_{i}
-b\text{sgn}f\left(b\right)+a\text{sgn}f\left(a\right) \\
&=-\underset{x_{i}\in f^{-1}\left(0\right)}{\sum}\left\{ \underset{h\rightarrow0^{-}}{\lim}\text{sgn}\left[F\left(x_{i}+h\right)-F\left(x_{i}\right)\right]+\underset{h\rightarrow0^{+}}{\lim}\text{sgn}\left[F\left(x_{i}+h\right)-F\left(x_{i}\right)\right]\right\} x_{i}\\
&= -\underset{x_{i}\in f^{-1}\left(0\right)}{\sum}\left\{ \underset{h\rightarrow0^{-}}{\lim}\text{sgn}\left[\int_{x_{i}}^{x_{i}+h}f\left(t\right)dt\right]+\underset{h\rightarrow0^{+}}{\lim}\text{sgn}\left[\int_{x_{i}}^{x_{i}+h}f\left(t\right)dt\right]\right\} x_{i}\\
&=-\underset{x_{i}\in f^{-1}\left(0\right)}{\sum}\left\{ \underset{t\rightarrow x_{i}^{+}}{\lim}\text{sgn}\left[f\left(t\right)\right]-\underset{t\rightarrow x_{i}^{-}}{\lim}\text{sgn}\left[f\left(t\right)\right]\right\} x_{i}\\
&=-\underset{x_{i}\in f^{-1}\left(0\right)}{\sum}\left\{ \underset{t\rightarrow x_{i}^{+}}{\lim}\text{sgn}\left[f\left(t\right)-f\left(x_{i}\right)\right]-\underset{t\rightarrow x_{i}^{-}}{\lim}\text{sgn}\left[f\left(t\right)-f\left(x_{i}\right)\right]\right\} x_{i}\\
&=-\underset{x_{i}\in f^{-1}\left(0\right)}{\sum}\left[f_{+}^{;}\left(x_{i}\right)+f_{-}^{;}\left(x_{i}\right)\right]x_{i},
\end{aligned}
\end{align}
where the second transition is due to the definition of the detachment and part 2 of this theorem; the third transition is due to the second part of the Fundamental Theorem of Calculus; the fourth transition is due to considerations similar to these applied in the proof of this theorem's second part; the fifth transition is due to $f$ being zeroed at the points $x_{i}$; and the sixth transition is due to the definition of the detachment. Note that, for convenience, in the proof above we assumed that the function's zeros is a countable set. However, the formula clearly holds also for functions with uncountable zeros.
\end{enumerate}
\end{proof}
\begin{remark}The first part of theorem \ref{semi_discrete_fundamental} shows that the detachment operator coincides with the derivative sign whenever a function has a non-zeroed derivative; hence this operator's mathematical power lies mostly in the cases where the derivative is zeroed or undefined.
\end{remark}
\begin{remark}The fourth part of theorem \ref{semi_discrete_fundamental} is the only result in this book that incorporates all these concepts together: differentiability, integrability, sign-continuity, monotonicity, and detachability.
\end{remark}

\subsection{Composite and Inverse Functions Rules}

Recall the chain rule in Differential Calculus:
\index{Chain Rule}
\begin{clm}\textbf{Chain rule. }If $g$ is a function that is differentiable at a point $c$ (i.e. the derivative $g'\left(c\right)$ exists) and $f$ is a function that is differentiable at $g\left(c\right)$, then the composite function $f\circ g$ is differentiable at $c$, and:
$$\left(f\circ g\right)'\left(c\right)=f'\left(g\left(c\right)\right)\cdot g'\left(c\right).$$
\label{chain_rule_derivative}
\end{clm}
Let us formulate its semi-discrete analogue that allows to calculate trends of composite, not necessarily differentiable, functions.
\begin{clm}\textbf{A semi-discrete chain rule.} If $g$ is continuous and detachable at $x$, and $f$ is detachable at $g\left(x\right)$, then $f\circ g $ is detachable at $x$ and the following holds there:
$$\left( f \circ g \right) ^{;}=\left(f^{;}\circ g\right)g^{;}.$$
\label{chain_rule_detachment}
\end{clm}

\begin{proof}By separating to cases. Without loss of generality, let us limit our discussion to right-detachments, and assume $g_{+}^{;}\left(x\right)=f_{+}^{;}\left(g\left(x\right)\right)=+1.$

According to the definition of the detachment of $f$ at $g\left(x\right)$:
$$\exists\delta_{f}:0<\bar{t}-g\left(x\right)<\delta_{f}\Longrightarrow f\left(g\left(x\right)\right)<f\left(\bar{t}\right).$$

Further, according to the definition of the detachment and the continuity of $g$:

$$\forall\epsilon,\,\exists\delta_{g}:0<\bar{x}-x<\delta_{g}\Longrightarrow0<g\left(\bar{x}\right)-g\left(x\right)<\epsilon.$$

Therefore, for $\epsilon\equiv\delta_{f}$ it holds that:
$$\exists\delta_{g}:0<\bar{x}-x<\delta_{g}\Longrightarrow0<g\left(\bar{x}\right)-g\left(x\right)<\delta_{f}\Longrightarrow f\left(g\left(x\right)\right)<f\left(g\left(\bar{x}\right)\right),$$

hence $\left(f\left(g\left(x\right)\right)\right)^{;}=+1.$ The rest of the cases are handled similarly.
\end{proof}

Recall the inverse function rule for derivatives:

\index{Inverse Function Rule}
\begin{clm}\textbf{Inverse function rule. }If $g$ is a function that is differentiable at a point $c$ (i.e. the derivative $g'\left(c\right)$ exists) and $f$ is a function that is differentiable at $g\left(c\right)$, then the composite function $f\circ g$ is differentiable at $c$, and:
$$\left(f\circ g\right)'\left(c\right)=f'\left(g\left(c\right)\right)\cdot g'\left(c\right).$$
\label{inverse_function_rule_derivative}
\end{clm}

Let us suggest a semi-discrete analogue for the detachment operator:
\begin{clm}\textbf{A semi-discrete inverse function rule.} A function $f:A\rightarrow \mathbb{R}$ is continuously detachable at $a\in A$ and $f^{;} \left(a\right)\neq0$, if and only if $f$ is invertible in a neighborhood of $a$. It then holds that $f^{-1}$ is continuously detachable at a neighborhood of $f\left(a\right)$ and:

$$\left(f^{-1}\right)^{;}\left(f\left(a\right)\right)=f^{;}\left(a\right)$$
\label{inverse_function_rule_detachment}
\end{clm}
\begin{proof}First direction. Since $f^{;}\neq0$ is continuous then either $f^{;}\equiv+1$ or $f^{;}\equiv-1$ in a neighborhood of $a.$ According to lemma \ref{monotonicity}, $f$ is thus strictly monotonic in that neighborhood. Without loss of generality assume that $f$ is strictly increasing. Then $x < y\Longleftrightarrow f\left(x\right) < f\left(y\right).$ Thus $f^{-1}\left(x\right) < f^{-1}\left(y\right)\Longleftrightarrow x < y.$ According to the second direction of lemma \ref{monotonicity}, $\left(f^{-1}\right)^{;}\equiv+1.$

Second direction. $f$ is invertible, hence strictly monotonic. According to lemma \ref{monotonicity}, $f^{;}\neq 0$ is continuous.

Under those conditions, both $f$ and $f^{\left(-1\right)}$ are monotonic in the neighborhood of the points at stake, and, according to lemma \ref{monotonicity}, they are continuously detachable.
\end{proof}

Another useful result is the following functional power rule in Differential Calculus:
\begin{clm}\textbf{Functional Power Rule. }Given a base function $f$ and an exponent function $g$, if:
\begin{enumerate}
\item The power function $f^{g}$ is well-defined at a point $x$ (particularly $f\left(x\right)>0$)
\item $f$ and $g$ are both differentiable
\end{enumerate}
Then the function $f^{g}$ is differentiable and:
$$\left(f^{g}\right)'=f^{g}\left(g'\ln f+f'\frac{g}{f}\right).$$
\label{functional_power_rule_derivative}
\end{clm}

Let us suggest its semi-discrete analogue as follows:
\begin{clm}\textbf{A semi-discrete functional power rule.} Given a base function $f$, an exponent function $g$, and a point $x$ in their definition domain, if the following conditions hold:

\begin{enumerate}
\item The power function $f^{g}$ is well-defined (particularly $f\left(x\right)>0$)
\item $f$ and $g$ are both detachable and sign-consistent
\item Either:
\begin{itemize}
\item $f^{;}g^{;}\left(f-1\right)g\geq0$, where $f$ or $g$ is s.c., or $f=1$ and $g=0$, or:
\item $f^{;}g^{;}\left(f-1\right)g<0$, where $f$ or $g$ is s.d.
\end{itemize}
\label{functional_power_rule_detachment}
\end{enumerate}

Then the function $f^{g}$ is detachable and:
$$\left(f^{g}\right)^{;}=\begin{cases}
\left[\left(f-1\right)g\right]^{;}, & gg^{;}\left(f-1\right)f^{;}\geq0,\text{ and }f-1\text{ or }g\text{ is s.c.}\\
f^{;}g^{;}, & \text{else.}
\end{cases}$$
\end{clm}

\begin{proof}The claim follows from the following transitions: \begin{align}
&\begin{aligned}
\left(f^{g}\right)^{;} & =\left(e^{g\ln f}\right)^{;}=\left(g\ln f\right)^{;}\\
& =\begin{cases}
\text{sgn}\left[g^{;}\text{sgn}\left(\ln f\right)+\left(\ln f\right)^{;}\text{sgn}\left(g\right)\right], & gg^{;}\ln f\left(\ln f\right)^{;}\geq0\text{, and }g\text{ or }\ln f\text{ is s.c.}\\
f^{;}g^{;}, & \text{else}
\end{cases}\\
& =\begin{cases}
\text{sgn}\left[g^{;}\text{sgn}\left(f-1\right)+f^{;}\text{sgn}\left(g\right)\right], & gg^{;}\left(f-1\right)f^{;}\geq0\text{, and }g\text{ or }f-1\text{ is s.c.}\\
f^{;}g^{;}, & \text{else}
\end{cases}\\
& =\begin{cases}
\text{sgn}\left[g^{;}\text{sgn}\left(f-1\right)+\left(f-1\right)^{;}\text{sgn}\left(g\right)\right], & gg^{;}\left(f-1\right)f^{;}\geq0\text{, and }g\text{ or }f-1\text{ is s.c.}\\
f^{;}g^{;}, & \text{else}
\end{cases}\\
& =\begin{cases}
\left[\left(f-1\right)g\right]^{;}, & gg^{;}\left(f-1\right)f^{;}\geq0,\text{ and }f-1\text{ or }g\text{ is s.c.}\\
f^{;}g^{;}, & \text{else.}
\end{cases}
\end{aligned}
\end{align}
where the second transition is due to the strict monotonicity of the exponent function, the third transition is due to the product rule (claim \ref{product_rule}), the fourth is since $\text{sgn}\left[\ln\left(f\right)\right]=\text{sgn}\left(f-1\right)$, and due to the strict monotonicity of the natural logarithm function, the fifth is due to claim \ref{constant_multiple_rule}, and the sixth is due to claim \ref{product_rule} again.
\end{proof}

A simulation of the functional power rule is available. Note that this rule forms another example of the detachment's numerical efficiency with respect to that of the derivative sign. The functional power rule for derivatives yields a formula that involves logarithms and division. Instead, the rule above, while appears to be more involved on the paper because of the conditions, may be more efficient computationally-wise, depending on the setting.

\section{Theoretical Applications}
\subsection{Limits and Trends Evaluation Tools}

Let us recall the Taylor series formulation.
\begin{theorem}[Taylor series]The Taylor series of a real or complex-valued function $f$ that is infinitely differentiable at a real or complex number $a$ is the power series

$$f\left(x\right)=f\left(a\right)+\frac{f'\left(a\right)}{1!}\left(x-a\right)+\frac{f''\left(a\right)}{2!}\left(x-a\right)^{2}+\frac{f'''\left(a\right)}{3!}\left(x-a\right)^{3}+\ldots=\underset{n}{\sum}\frac{f^{\left(n\right)}\left(a\right)}{n!}\left(x-a\right)^{n},$$

where $f^{\left(n\right)}\left(a\right)$ denotes the $n^{th}$ derivative of $f$ evaluated at the point $a$.
\label{taylor_series}
\end{theorem}
The following semi-discrete reformulation of Taylor's theorem expresses the trend given rates concisely. It may also be thought of as an interim step towards proving known Calculus claims about stationary points classification.

\index{Taylor Series}
\begin{crl}\label{taylor}\textbf{A semi-discrete rearrangement of Taylor series.} If $f\not\equiv0$ is differentiable infinitely many times at $x$, and detachable, then the detachment of $f$ is calculated as follows:

$$f_{\pm}^{;}\left(x\right)=\left(\pm1\right)^{k+1}\text{sgn}\left[f^{\left(k\right)}\left(x\right)\right],$$

where $f^{\left(i\right)}$ represents the $i^{th}$ derivative, and $k=\min\left\{ i\in\mathbb{N}|f^{\left(i\right)}\left(x\right)\neq0\right\}$, if there exists such $k$.
\label{semi_discrete_taylor}
\end{crl}
\begin{proof}The first equality is obtained from the Taylor series: $$f(x+h)=f(x)+hf'(x)+\frac{h^{2}}{2}f''(x)+\frac{h^{3}}{6}f^{(3)}(x)+\ldots$$

by simple algebraic manipulations followed by applying the limit process.

The second equality holds due to the following analysis:
\begin{align}
&\begin{aligned}
\underset{{\scriptscriptstyle h\rightarrow0^{\pm}}}{\lim}\text{sgn}\left[\underset{{\scriptscriptstyle i\in\mathbb{N}}}{\overset{}{\sum}}\frac{h^{i-1}}{i!}f_{\pm}^{\left(i\right)}\left(x\right)\right] &= \underset{{\scriptscriptstyle h\rightarrow0^{\pm}}}{\lim}\text{sgn}\left[\frac{h^{k-1}}{k!}f^{\left(k\right)}\left(x\right)+\mathcal{O}\left(h^{k}\right)\right] \\ 
&=\underset{{\scriptscriptstyle h\rightarrow0^{\pm}}}{\lim}\text{sgn}\left[\frac{h^{k-1}}{k!}f^{\left(k\right)}\left(x\right)\right]=\left(\pm1\right)^{k+1}\text{sgn}\left[f^{\left(k\right)}\left(x\right)\right],
\end{aligned}
\end{align}

where the second equality holds because for a sufficiently small $h$, we have that $$\text{sgn}\left[\frac{h^{k-1}}{k!}f^{\left(k\right)}\left(x\right)+\mathcal{O}\left(h^{k}\right)\right]=\text{sgn}\left[\frac{h^{k-1}}{k!}f^{\left(k\right)}\left(x\right)\right].$$
The final step is obtained by keeping in mind the limit’s side and the parity of $k$.
\end{proof}

\begin{remark}Note that claim \ref{sum_rule} and theorems \ref{product_rule} and \ref{quotient_rule} (the sum and difference, product and quotient rules, respectively) impose varied conditions. However, in the special case that the functions $f,g$ are both detachable and differentiable infinitely many times, the following corollaries from claim \ref{taylor} hold independently of these conditions:

\begin{enumerate}
\item If $f+g$ is detachable, then $$\left(f+g\right)_{\pm}^{;}=\left(\pm1\right)^{k+1}\text{sgn}\left[f^{\left(k\right)}+g^{\left(k\right)}\right],$$ where $k\equiv\min\left\{ i\in\mathbb{N}|f^{\left(i\right)}\neq-g^{\left(i\right)}\right\}$ . An analogous statement holds for the difference $f-g$.
\item If $fg$ is detachable, then $$\left(fg\right)_{\pm}^{;}=\left(\pm1\right)^{k+1}\text{sgn}\left[f^{\left(k\right)}g+fg^{\left(k\right)}\right],$$ where $k\equiv\min\left\{ i\in\mathbb{N}|f^{\left(i\right)}g\neq-fg^{\left(i\right)}\right\}.$  An analogous statement holds for the difference $\frac{f}{g}$.
\end{enumerate}

For example, consider the functions $f\left(x\right)=x^{2},g\left(x\right)=-x^{4}$ at $x=0$. Rule 3 does not yield an indication regarding $\left(f+g\right)^{;}$ since $f^{;}g^{;}=-1\notin\left\{ 0,1\right\}$. However, the aforementioned statement lets us know that $\left(f+g\right)_{\pm}^{;}\left(x\right)=\left(\pm1\right)^{2+1}\text{sgn}\left[2+0\right]=\pm1$.
\end{remark}

Recall L'Hôpital Rule that serves as a convenient limit evaluations tool.
\index{L'Hôpital Rule}
\begin{theorem}[L'Hôpital rule]Given functions $f$ and $g$ which are differentiable on an open interval $I$ except possibly at a point $c$ contained in $I$, if for all $x$ in $I$ with $x \neq c$:
\begin{enumerate}
    \item $\lim _{x\to c}f(x)=\lim _{x\to c}g(x)=0{\text{ or }}\pm \infty$
    \item $g'(x)\neq 0$
    \item $\exists \lim _{x\to c}{\frac {f'(x)}{g'(x)}}$
\end{enumerate}
Then:
$$\lim _{x\to c}{\frac {f(x)}{g(x)}}=\lim _{x\to c}{\frac {f'(x)}{g'(x)}}.$$
\label{lhopital_derivative}
\end{theorem}
Let us suggest its semi-discrete analogue as follows.
\begin{theorem}[A semi-discrete L'Hôpital rule]Let $f,g:\mathbb{R}\rightarrow\mathbb{R}$ be a pair of functions and a point $x$ in their definition domain. Assume that $\underset{t\rightarrow x^{\pm}}{\lim}f^{;}\left(t\right)$ and $\underset{t\rightarrow x^{\pm}}{\lim}g^{;}\left(t\right)$ exist. If $\underset{t\rightarrow x^{\pm}}{\lim}\left|f\left(t\right)\right|=\underset{t\rightarrow x^{\pm}}{\lim}\left|g\left(t\right)\right|\in\left\{ 0,\infty\right\},$ then:
$$\underset{t\rightarrow x^{\pm}}{\lim}\text{sgn}\left[f\left(t\right)g\left(t\right)\right]=\underset{t\rightarrow x^{\pm}}{\lim}f^{;}\left(t\right)g^{;}\left(t\right).$$
\label{lhopital}
\end{theorem}
\begin{proof}
We prove a slightly more generic claim: Let $\left\{ f_{i}:\mathbb{R}\rightarrow\mathbb{R}\right\} _{1\leq i\leq n}$ be a set of functions and a point $x$ in their definition domain. Assume that $\underset{t\rightarrow x^{\pm}}{\lim}f_{i}^{;}\left(t\right)$ exist for each $i.$ If $\underset{t\rightarrow x^{\pm}}{\lim}\left|f_{i}\left(t\right)\right|=L\in\left\{ 0,\infty\right\},$ then we will show that:
$$\underset{t\rightarrow x^{\pm}}{\lim}\text{sgn}\underset{i}{\prod}f\left(t\right)=\left(\pm C\right)^{n}\underset{t\rightarrow x^{\pm}}{\lim}\underset{i}{\prod}f^{;}\left(t\right),$$

where $C$ equals $+1$ or $-1$ if $\underset{t\rightarrow x^{\pm}}{\lim}\left|f_{i}\left(t\right)\right|$ is $0$ or $\infty,$ to which we refer below as part 1 and 2, respectively.

We apply induction on $n.$ Let $n=1,$ and for simplicity denote $f=f_{1}.$ Without loss of generality we focus on right limits and assume that $\underset{t\rightarrow0^{+}}{\lim}f^{;}\left(t\right)=+1.$ Then $f^{;}=+1$ for each point in a right $\delta$-neighborhood of $x.$ According to lemma \ref{monotonicity}, $f$ is strictly increasing in $\left(x,x+\delta\right).$ Therefore:
$$\inf\left\{ f\left(t\right)|t\in\left(x,x+\delta\right)\right\} =\underset{t\rightarrow x^{+}}{\lim}f\left(t\right).$$

Proof of Part 1. According to our assumption, $\inf\left\{ f\left(t\right)|t\in\left(x,x+\delta\right)\right\} = 0.$ Thus $f\left(t\right)\geq0$ for $t\in\left(x,x+\delta\right).$ Clearly, $f$ cannot be zeroed in $\left(x,x+\delta\right)$ because that would contradict the strict monotony. Thus $f>0$ there, and $\underset{t\rightarrow x^{+}}{\lim}\text{sgn}f\left(t\right)=\underset{t\rightarrow x^{+}}{\lim}f^{;}\left(t\right)=+1.$ If $\underset{t\rightarrow x^{+}}{\lim}f^{;}\left(t\right)=0,$ then $f$ is constant in a right-neighborhood of $x,$ and from the continuity $f\equiv 0$. Thus $\underset{t\rightarrow x^{+}}{\lim}\text{sgn}f\left(t\right)=\underset{t\rightarrow x^{+}}{\lim}f_{+}^{;}\left(t\right)=0.$ The signs switch for left-limits, hence the $\pm$ coefficient in the right handside.

Proof of Part 2. Since $f$ is strictly increasing in a right-neighborhood of $x,$, then $\underset{t\rightarrow x^{+}}{\lim}f\left(t\right)=-\infty,$ and $\underset{t\rightarrow x^{+}}{\lim}\text{sgn}f\left(t\right)=-\underset{t\rightarrow x^{+}}{\lim}f^{;}\left(t\right)=-1.$ The signs switch for left-limits, hence the $\mp$ coefficient in the right handside.

Assume that the theorem holds for $n,$ and we show its correctness for $n+1$:

\begin{align}
&\begin{aligned}\underset{t\rightarrow x^{\pm}}{\lim}\text{sgn}\underset{i}{\prod}f_{i}\left(t\right) &=\underset{t\rightarrow x^{\pm}}{\lim}\text{sgn}\underset{i}{\prod}f_{i}\left(t\right)\cdot\underset{t\rightarrow x^{\pm}}{\lim}\text{sgn}f_{n+1}\left(t\right) \\
&=\left(\pm C\right)^{n}\underset{t\rightarrow x^{\pm}}{\lim}\underset{i}{\prod}f_{i}^{;}\left(t\right)\cdot\underset{t\rightarrow x^{\pm}}{\lim}\text{sgn}f_{n+1}\left(t\right)\\
&=\left(\pm C\right)^{n+1}\underset{t\rightarrow x^{\pm}}{\lim}\underset{i}{\prod}f_{i}^{;}\left(t\right),
\end{aligned}
\end{align}
where the second transition follows from the induction hypothesis, and the third follows from the induction base.
\end{proof}

\begin{remark}Without assuming that the conditions stated in claim \ref{sum_rule} and \ref{product_rule}, detachable functions' sum and product are not even guaranteed to be detachable.
For example, consider the right-detachable pair of functions at $x=0$:
$$g_{1}\left(x\right)=\begin{cases}
1, & x>0\\
0, & x=0
\end{cases},\qquad,g_{2}\left(x\right)=\begin{cases}
1, & x\in\mathbb{Q^{+}}\\
-1, & x\in\mathbb{R^{+}\backslash\mathbb{Q}}\\
2, & x=0,
\end{cases}$$
whose sum and product are not detachable at zero. Counterexamples exist even if we assume differentiability on top of detachability.
Discarding the continuity assumption on $g$ in claim \ref{chain_rule_detachment} may result in a non-detachable $f \circ g$, for example at $x=0$ for the pair of functions $f\left(x\right)=x^{2}$ and: $$g\left(x\right)=\begin{cases}\left|\sin\left(\frac{1}{x}\right)\right|, & x\neq0,\\-\frac{1}{2}, & x=0.\end{cases}$$
\end{remark}

We finalize this subsection with a simple criterion for constancy in an interval containing a point.

\index{Monotonicity}
\begin{clm}\label{constancy_condition}$f:\mathbb{R}\rightarrow\mathbb{R}$ is constant in the neighborhood
of $x_{0}$ if and only if $f_{\pm}^{;}\left(x_{0}\right)=0$.
\end{clm}

\begin{proof}First direction. Assume that $f_{\pm}^{;}=0$. Then:
\[
\forall\epsilon>0,\exists\delta>0:\quad\left|x-x_{0}\right|<\delta\quad\Longrightarrow\quad\left|\text{sgn}\left[f\left(x\right)-f\left(x_{0}\right)\right]-0\right|<\epsilon,
\]
hence $$\left|\text{sgn}\left[f\left(x\right)-f\left(x_{0}\right)\right]\right|<\epsilon,$$
which particularly holds for $\epsilon=\frac{1}{2}$. The only option
is $\text{sgn}\left[f\left(x\right)-f\left(x_{0}\right)\right]=0$, which
implies $f\left(x\right)=f\left(x_{0}\right)$ for the $\delta$-neighborhood
of $x_{0}$, resulting with $f$ constant.

Second direction. Let $\delta$ be small enough such that $f$ is
constant in $B_{\delta}\left(x_{0}\right)=\left(x_{0}-\delta,x_{0}+\delta\right)$,
and let $\epsilon>0$. Let $x\in B_{\delta}\left(x_{0}\right)$. Then:
\[
\left|\text{sgn}\left[f\left(x\right)-f\left(x_{0}\right)\right]-0\right|=0<\epsilon.
\]
\end{proof}

\subsection{Detachment Test}

In this subsection we mirror several derivative tests in terms of the detachment and suggest that the detachment version may be occasionally beneficial, for example, in case a given function is detachable but not differentiable.

Recall the first derivative test:

\begin{clm}\textbf{First derivative test. }Suppose $f$ is a real-valued function
of a real variable defined on some interval containing the critical
point $a$. Furthermore, suppose that $f$ is continuous at $a$ and differentiable
on some open interval containing $a$, except, perhaps, at $a$ itself.
\begin{itemize}
\item If there is a positive number $r>0$ such that for every $x\in\left(a-r,a\right)$
we have $f'\left(x\right)\geq0$ ($f'\left(x\right)\geq0$), and for
every $x\in\left(a,a+r\right)$ we have $f'\left(x\right)\leq0$ ($f'\left(x\right)\leq0$),
then $f$ has a local maximum (minimum) at $a$.
\item If there is a positive number $r>0$ such that for
every x in $\left(a-r,a\right)\bigcup\left(a,a+r\right)$ we have
$f'\left(x\right)>0$ ($f'\left(x\right)<0$), then $f$ is strictly
increasing (decreasing) at $a$ and has neither a local maximum nor
a local minimum.
\item If none of the above conditions hold, then the test fails.
\end{itemize}
\label{first_derivative_test}
\end{clm}

A detachment-based version of the first derivative test does not require
continuity and provides both sufficient and necessary conditions.
\begin{clm}\textbf{First detachment test. }Suppose $f$ is a real-valued function
of a real variable defined on some interval containing the point $a$.
Furthermore, suppose that $f$ is detachable at $a$. Then:
\begin{itemize}
\item $f_{\pm}^{;}\left(x\right)\in\left\{ 0,\mp1\right\} $ ($f_{\pm}^{;}\left(x\right)\in\left\{ 0,\pm1\right\} $)
respectively, if and only if $f$ has a local maximum (minimum) at
$a$.
\item $f^{;}\left(x\right)\equiv+1$ ($f^{;}\left(x\right)=-1$) respectively
in a neighborhood of $a$, if and only if $f$ is strictly increasing
(decreasing) at $a$ and has neither a local maximum nor a local minimum.
\end{itemize}
\label{first_detachment_test}
\end{clm}
\begin{proof}Without loss of generality, we prove the claim for maxima and
strict increasing points.
\begin{itemize}
\item The condition $f_{\pm}^{;}\left(a\right)\in\left\{ 0,\mp1\right\} $,
is equivalent, by the definition on the one-sided detachment, to the
existence of left and right neighborhoods of $a$, $B_{r^{-}}\left(a\right)$
and $B_{r^{+}}\left(a\right)$ respectively, such that:
\begin{align*}
\forall x_{-}\in B_{r^{-}}\left(a\right):\quad & -sgn\left[f\left(x_{-}\right)-f\left(a\right)\right]\in\left\{ 0,+1\right\} \\
\forall x_{+}\in B_{r^{+}}\left(a\right):\quad & sgn\left[f\left(x_{+}\right)-f\left(a\right)\right]\in\left\{ 0,-1\right\} ,
\end{align*}
which is equivalent to the following condition:
\begin{align*}
\forall x\in B_{r^{\pm}}\left(a\right):\quad & f\left(x\right)\leq f\left(a\right),
\end{align*}
that coincides with the definition of a local maximum.
\item By the lemma \ref{monotonicity}, $f^{;}\left(x\right)\equiv+1$ in an interval if
and only if $f$ is strictly increasing. Therefore, this property
in an open neighborhood of $a$ is a sufficient and necessary condition
for $f$ to strictly increase at $a$.
\end{itemize}
\end{proof}

Similarly, we may mirror the second derivative test.
\begin{clm}\textbf{Second-derivative test. }If the function $f$ is twice-differentiable
at a critical point $x$ (i.e. a point where $f'\left(x\right)=0$),
then:
\begin{itemize}
\item If $f''\left(x\right)<0$, then $f$ has a local maximum at $x$.
\item If $f''\left(x\right)>0$, then $f$ has a local minimum at $x$.
\item If $f''\left(x\right)=0$, the test is inconclusive.
\end{itemize}
\label{second_derivative_test}
\end{clm}
\begin{clm}\textbf{Second-detachment test. }If the derivative of the function $f$
is detachable at a critical point $x$, then:
\begin{itemize}
\item If $f'^{;}\left(x\right)=-1$, then $f$ has a local maximum at $x$.
\item If $f'^{;}\left(x\right)=+1$, then $f$ has a local minimum at $x$.
\item If $f'^{;}\left(x\right)=0$, then $f$ has both a local minimum and
maximum at $x$ (it is constant in its neighborhood).
\end{itemize}
\label{second_detachment_test}
\end{clm}
\begin{proof}Let us prove the first and third statements:
\begin{itemize}
\item $f'^{;}\left(x\right)=-1$ implies that $\pm\underset{x\to x_{0}^{\pm}}{\lim}\text{sgn}\left[f'\left(x\right)-f'\left(x_{0}\right)\right]=-1$.
Since $x_{0}$ is a critical point, $f'\left(x_{0}\right)=0$ (by
definition). Therefore according to the limit definition, for all
$x_{-}$ in some left-neighborhood of $x_{0}$, it holds that $\text{sgn}\left[f'\left(x_{-}\right)\right]=+1$,
or $f'\left(x_{-}\right)>0$; and for all $x_{+}$ in some right-neighborhood
of $x_{0}$, it holds that $\text{sgn}\left[f'\left(x_{+}\right)\right]=-1$,
or $f'\left(x_{+}\right)<0$. Hence, according to the first derivative
test, $f$ has a local maximum at $x$.
\item $f'^{;}\left(x\right)=0$ implies that $\underset{x\to x_{0}^{\pm}}{\lim}\text{sgn}\left[f'\left(x\right)\right]=0$,
hence $f'$ is zeroed in a neighborhood of $x$, and $f$ is thus
constant.
\end{itemize}
\end{proof}

Similarly, one may apply the one-sided detachments as a concise
sufficient condition upon analyzing inflection points. As stated in
\cite{bronshtein2013handbook}, the existence of an inflection point $x_{0}$ is guaranteed if
the sign of the second derivative changes while traversing from the
left neighborhood of $x_{0}$ to the right, if there is also a tangent
at the point. As such, the question of whether the curve has an inflection
point can be answered by checking the sign of the second derivative
traversing the considered point. This condition may be further alleviated:
we do not require that the function is differentiable twice, but solely
that its derivative is detachable. We add to that the following condition:
\[
\begin{cases}
\left(\frac{df}{dx}\right)_{+}^{;}\left(x_{0}\right)+\left(\frac{df}{dx}\right){}_{-}^{;}\left(x_{0}\right)=0\\
\left(\frac{df}{dx}\right)_{+}^{;}\left(x_{0}\right)\cdot\left(\frac{df}{dx}\right)_{-}^{;}\left(x_{0}\right)\neq0.
\end{cases}
\]

\subsection{Calculating Instantaneous Trends in Practice}

Let us calculate the detachments directly according to the detachment definition, or according to corollary \ref{semi_discrete_taylor}. In the following examples, we scrutinize cases where the detachable functions are either continuous but not differentiable, discontinuous, or differentiable. As a side note, we also examine a case where the trend does not exist.

Let $g\left(x\right)=\sqrt{\left|x\right|}$ which is not differentiable at zero. Then we can calculate the trend directly from the detachment definition:

\begin{align}
&\begin{aligned}
g_{\pm}^{;}\left(x\right) & =\pm \underset{h\rightarrow0^{\pm}}{\lim}\text{sgn}\left(\sqrt{\left|x+h\right|}-\sqrt{\left|x\right|}\right)=\pm \underset{h\rightarrow0^{\pm}}{\lim}\text{sgn}\left[\left(x+h\right)^{2}-x^{2}\right]\\
& =\pm \underset{h\rightarrow0^{\pm}}{\lim}\text{sgn}\left[h\left(2x+h\right)\right]=\begin{cases}
\text{sgn}\left(x\right), & x\neq0\\
\pm 1, & x=0
\end{cases}
\end{aligned}
\end{align}

That is, the one-sided detachments are positive at zero, indicating a minimum. At points other than zero, we see that the detachment’s values agree with the derivative’s sign, as expected from part 1 of theorem claim \ref{semi_discrete_fundamental}. Weirstrass function, which is nowhere differentiable, can be shown to be detachable at infinitely many points with similar means.

Next, let the sign function $\ell\left(x\right)=\text{sgn}\left(x\right)$ (not to be confused with the definition of the detachment), which is discontinuous at $x=0$. Then, its trends can be concisely evaluated by the definition:
$$\ell_{\pm}^{;}\left(x\right)=\pm \underset{{\scriptscriptstyle h\rightarrow0^{\pm}}}{\lim}\text{sgn}\left[\text{sgn}\left(x+h\right)-\text{sgn}\left(x\right)\right]=\begin{cases}
0, & x\neq0\\
\pm \underset{{\scriptscriptstyle h\rightarrow0^{\pm}}}{\lim}\text{sgn}\left(h\right)=+1, & x=0
\end{cases}$$

Finally, let us calculate trends based on claim \ref{taylor} (in case the function is differentiable infinitely many times). Those are the explicit calculations that are otherwise obfuscated by Taylor series based theorems on critical points classification. For instance, consider the function $f\left(x\right) = -3x^{5}+5x^{3},$ whose critical points are at $0, \pm 1$:

\begin{align}
&\begin{aligned}
f_{\pm}^{;}\left(x\right) & =\left(\pm1\right)^{k+1}\text{sgn}\left[f^{\left(k\right)}\left(x\right)\right]\\
f_{\pm}^{;}\left(0\right) & =\left(\pm1\right)^{3+1}\text{sgn}\left(5\right)=+1\\
f_{\pm}^{;}\left(-1\right) & =\left(\pm1\right)^{2+1}\text{sgn}\left(15\right)=\pm1\\
f_{\pm}^{;}\left(1\right) & =\left(\pm1\right)^{2+1}\text{sgn}\left(-15\right)=\mp1,
\end{aligned}
\end{align}

where the transition on the first raw is due to claim \ref{semi_discrete_taylor}. We know that $0, +1$ and $-1$ are inflection, maximum and minimum points, respectively.

\begin{remark}For completeness, let us show that the trend does not exist for the function $$s\left(x\right)=\begin{cases}
\sin\left(\frac{1}{x}\right), & x\neq0\\
0, & x=0,
\end{cases}$$
at $x=0$. We present two different sequences whose limit is zero: $$a_{n}=\left\{ \frac{2}{\pi\left(1+4n\right)}\right\} ,b_{n}=\left\{ \frac{1}{\pi\left(1+2n\right)}\right\}.$$
As the sine function is constant for both sequences ($1$ and $0$ respectively), then so is the limit of the change’s sign, which equals $+1$ and $0$ for $a_{n}$ and $b_{n}$, respectively. Heine’s limit definition doesn’t hold, so $s$ isn’t detachable. Indeed, this function’s instantaneous trend is non existent at $x=0$.
\end{remark}

\section{Exercises}
Prove or refute the following claims, that are based on
the counter examples illustrated in \cite{klymchuk2010counterexamples}. Let us assume in all the below
examples that $f,g$ are functions whose domains and images are $\mathbb{R}$
unless specified otherwise. The solutions are listed in the following
subsection.

\begin{exercise}\label{ex_1}
If $\left|f\right|$ is detachable at $x$ then so is $f$.
\end{exercise}

Counter example: $f\left(x\right)=\left(-1\right)^{\boldsymbol{1}_{\mathbb{Q}}}$.
$\left|f\right|\equiv+1$ is detachable, but $f$ oscillates indefinitely
between $-1$ and $+1$ and is, therefore, particularly nowhere detachable.

\begin{figure}[h!]
\includegraphics[scale=0.7]{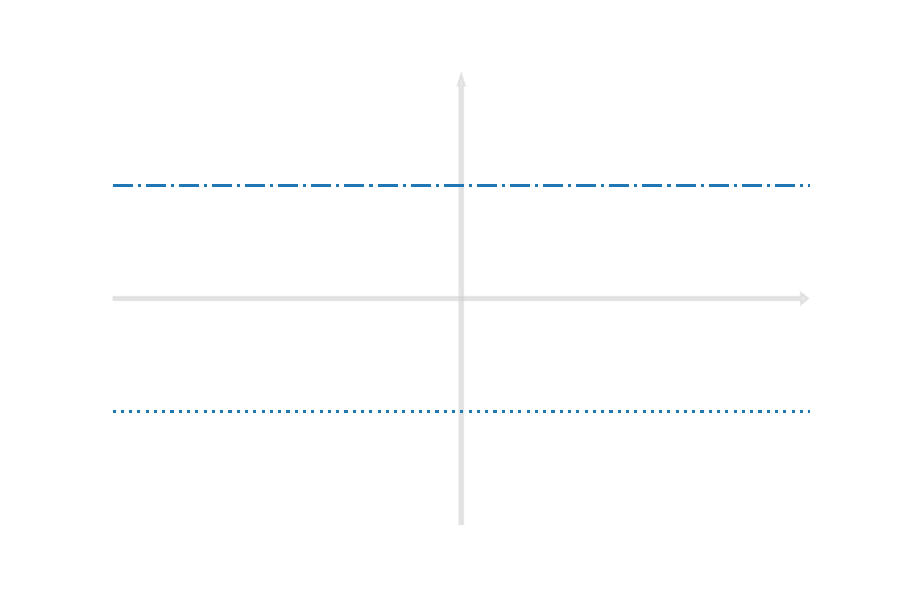}
\caption{An illustration of a function refuting the claim in exercise \ref{ex_1}}
\label{minus_one_in_irrationals}
\end{figure}

\begin{exercise}If $f,g$ are not detachable at $x$ then so is $f+g$.
\end{exercise}

Counter example: Any non-detachable function $f$ and its additive
inverse $-f$ refute the claim.

\begin{exercise}If $f,g$ are not detachable at $x$ so is $f\cdot g$.
\end{exercise}

Counter example: $f\left(x\right)=\left(-1\right)^{\boldsymbol{1}_{\mathbb{Q}}}$
and $g\left(x\right)=\left(-1\right)^{1+\boldsymbol{1}_{\mathbb{Q}}}$
are not detachable, however $\left(f\cdot g\right)\left(x\right)\equiv-1$
is detachable.

\begin{exercise}\label{ex_4}If $f$ is detachable in the neighborhood of $a\in\mathbb{R}$,
including at $a$, and:
\[
f^{;}\left(x\right)=\begin{cases}
+1, & x<a\\
-1, & x>a,
\end{cases}
\]
then $f$ incurs a local maximum at $x=a$.
\end{exercise}

Counter example: $a=0$ with the following function:
\[
f\left(x\right)=\begin{cases}
\frac{1}{x^{2}}, & x\neq0\\
0, & x=0.
\end{cases}
\]

\begin{figure}[h!]
\includegraphics[scale=0.7]{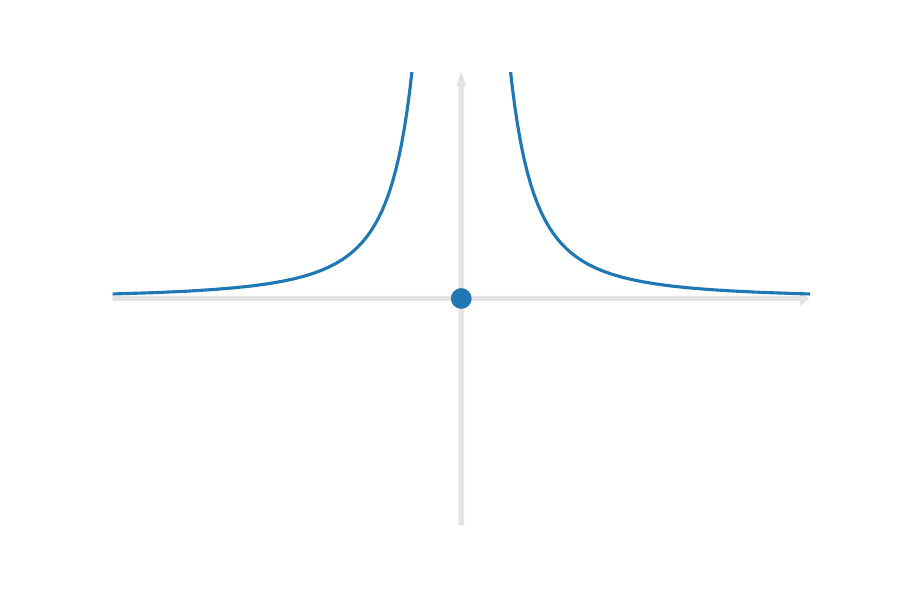}
\label{one_over_x_square}
\caption{An illustration of a function refuting the claim in exercise \ref{ex_4}}
\end{figure}

\begin{exercise}$f$ cannot be detachable only at a single point $a$.
\end{exercise}

Counter example: The following function is detachable only at $a=0$:
\[
f\left(x\right)=\begin{cases}
\left(-1\right)^{\boldsymbol{1}_{\mathbb{Q}}}, & x\neq0\\
2, & x=0.
\end{cases}
\]

\begin{exercise}\label{ex_6}If $f,g$ are detachable and $f>g$ in the interval $\left(a,b\right)$
the $f^{;}\geq g^{;}$.
\end{exercise}

Counter example:
\begin{align*}
f\left(x\right) & =-x\\
g\left(x\right) & =x,
\end{align*}
in the interval $\left(-\infty,0\right)$.
\begin{figure}[h!]
\begin{tabular}{cc}
  \includegraphics[width=60mm]{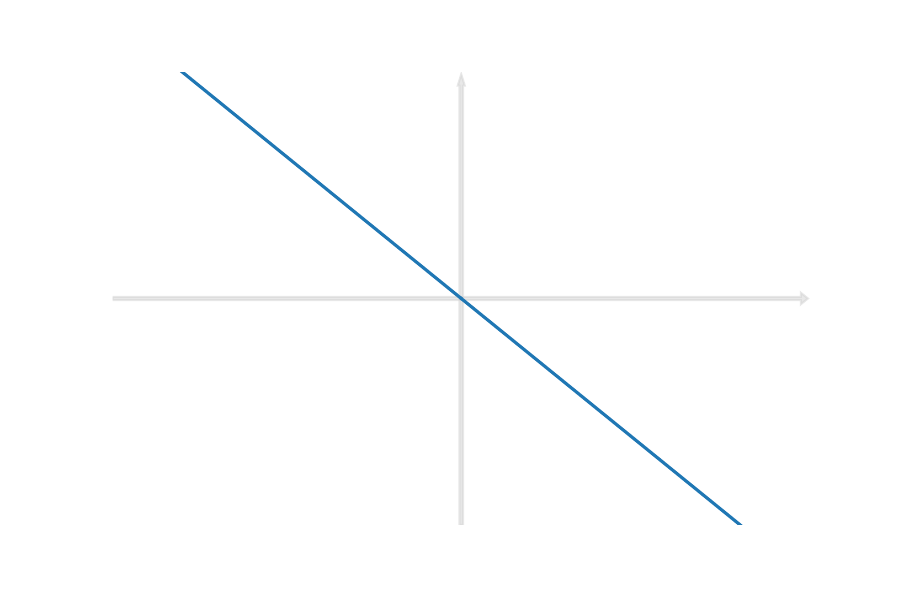} &   \includegraphics[width=60mm]{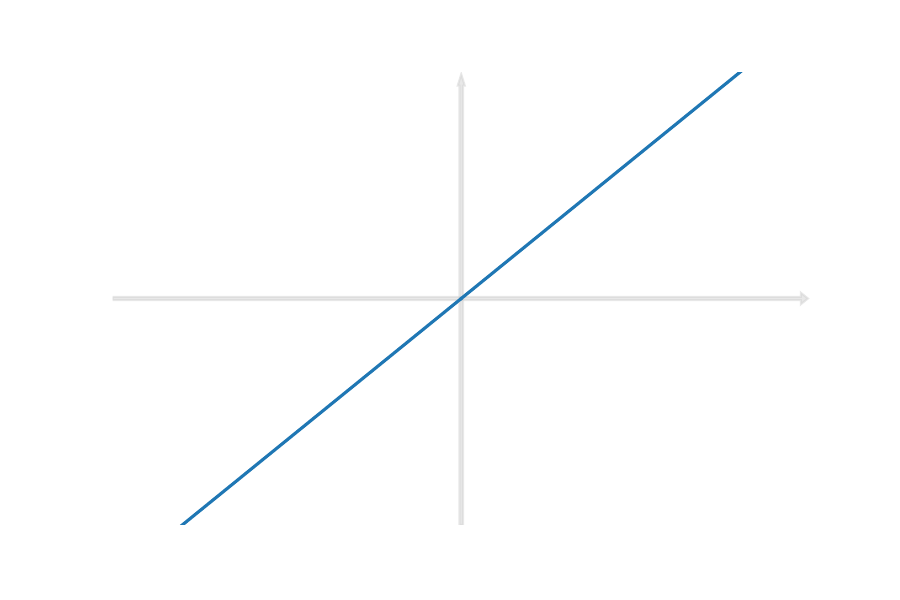} \\
\end{tabular}
\caption{An illustration of a pair of functions refuting the claim in exercise \ref{ex_6}}
\end{figure}

\begin{exercise}\label{ex_7}If $f$ is non-linear, differentiable, and $f^{;}$ is constant
in $\left(a,\infty\right)$, then its derivative's detachment is also
constant.
\end{exercise}

Counter example: $f\left(x\right)=1.1x+\sin\left(x\right)$, for
which $f^{;}\left(x\right)\equiv+1$ but its derivative $\frac{df}{dx}\left(x\right)=1.1+\cos\left(x\right)$,
and $\left(\frac{df}{dx}\right)^{;}\in\left\{ \pm1\right\} $.

\begin{figure}[h!]
\includegraphics[scale=0.7]{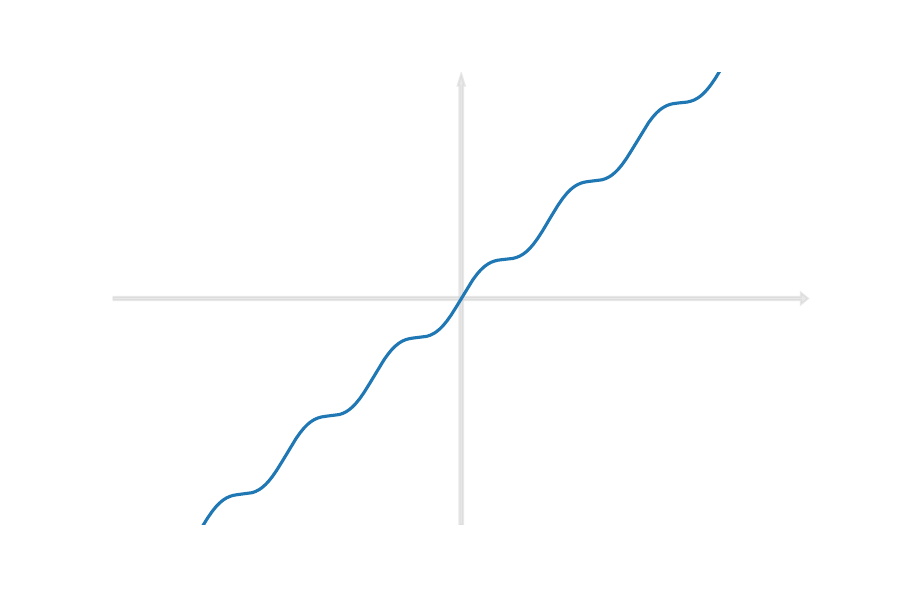}
\label{counter_example7}
\caption{An illustration of a function refuting the claim in exercise \ref{ex_7}}
\end{figure}

\begin{exercise}If $f$ is continuous or differentiable at a point, then it is
also detachable.
\end{exercise}

Counter example: $f\left(x\right)=\begin{cases}
x\sin\left(\frac{1}{x}\right), & x\neq0\\
0, & x=0
\end{cases}$ and $g\left(x\right)=\begin{cases}
x^{2}\sin\left(\frac{1}{x}\right), & x\neq0\\
0, & x=0
\end{cases}$ are continuous and differentiable at $x=0$ respectively, and neither is detachable.

\begin{exercise}If $f$ is decreasing in $\left(a,b\right)$ then $f^{;}=-1$.
\end{exercise}

This is true, according to lemma \ref{monotonicity}.
Note that a similar claim for derivatives is not true: $g\left(x\right)=-x^{3}$
is decreasing everywhere but $f'\left(0\right)=0$.

\begin{exercise}\label{ex_10}If $f$ satisfies $f_{+}^{;}=+1$ everywhere in an interval $\left(a,b\right)$,
then it is increasing.
\end{exercise}

Counter example: let
\[
f\left(x\right)=\begin{cases}
-\frac{1}{x}, & x\neq0\\
x, & x\geq0,
\end{cases}
\]

then $f\left(-0.1\right)<f\left(1\right)$, contradicting the definition
of a function increasing in an interval. Therefore, $f$ is not increasing
in $\mathbb{R}$, although $f_{+}^{;}\equiv+1$.

\begin{figure}[h!]
\includegraphics[scale=0.7]{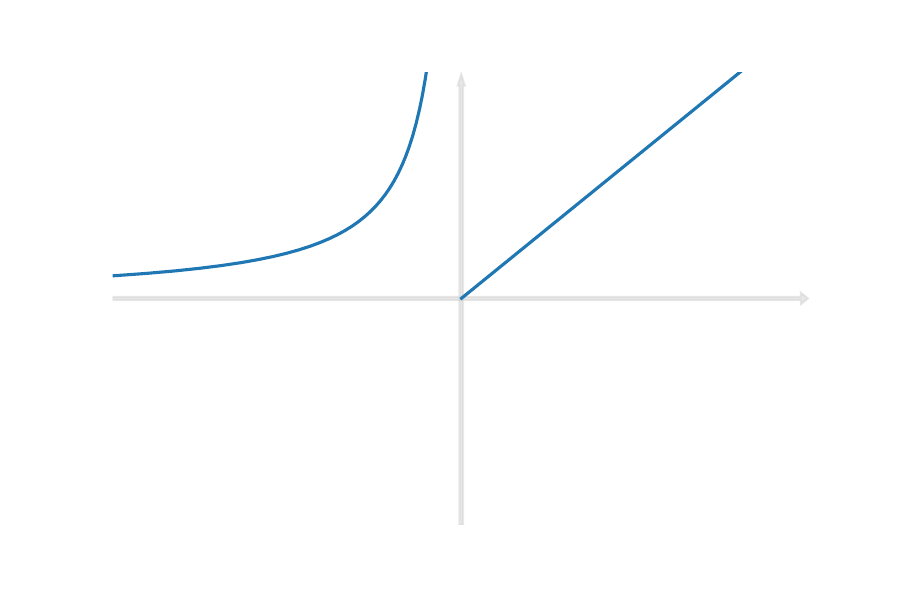}
\label{counter_example7}
\caption{An illustration of a function refuting the claim in exercise \ref{ex_10}}
\end{figure}

\begin{exercise}\label{ex_11}If $f_{\pm}^{;}\left(x\right)=\mp1$, then $f$ is increasing and
decreasing in a left- and right-neighborhood of $x$, respectively.
\end{exercise}

Counter example: Let
\[
f\left(x\right)=\begin{cases}
x^{2}, & x\neq0\\
+1, & x=0,
\end{cases}
\]
then $f_{\pm}^{;}\left(x\right)=\mp1$ at $x=0$, which is a strict
local maximum, but $f$ is decreasing and increasing in the left-
and right-neighbourhoods of $x$, respectively.

\begin{figure}[h!]
\includegraphics[scale=0.7]{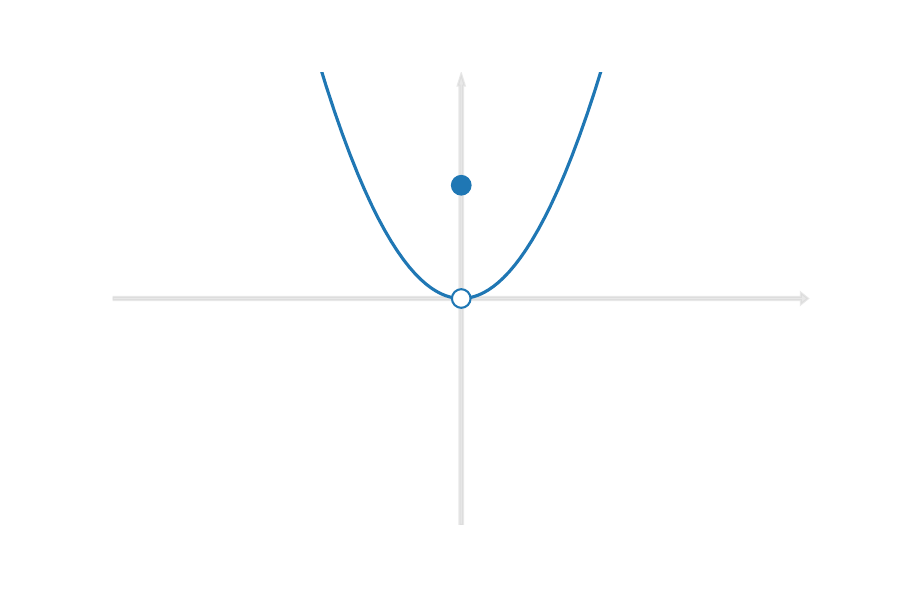}
\label{counter_example11}
\caption{An illustration of a function refuting the claim in exercise \ref{ex_11}}
\end{figure}

\begin{exercise}If $f$ is detachable for each $x\in\mathbb{R}$ and $f\left(0\right)=f_{\pm}^{;}\left(0\right)=0$,
then $f\left(x\right)=0$ for each $x\in\mathbb{R}$. 
\end{exercise}

Counter example: Any detachable function that is zeroed only in
a neighbourhood of $x=0$.

\begin{exercise}If $f,g$ are detachable in $\left(a,b\right)$ and intersect there,
then the function $\max\left\{ f\left(x\right),g\left(x\right)\right\} $
is not detachable at points where $f\left(x\right)=g\left(x\right)$.
\end{exercise}

Counter example: $f\left(x\right)=x^{3},\quad g\left(x\right)=x^{4}$
are both detachable at $x=0$ and so is $\max\left\{ f,g\right\} $
.

\begin{exercise}If the derivative of $f$ is detachable at a maximum (or minimum)
point of $f$, then its derivative's detachment is negative (or positive).
\end{exercise}

Counter example: let $f\left(x\right)=x^{n}$, where $n\in\mathbb{N}$.
Then:
\[
\left[\frac{d\left(x^{n}\right)}{dx}\right]_{\pm}^{;}\biggm\lvert_{x=0}=\begin{cases}
\pm1, & n\text{ is even}\\
+1, & n\text{ is odd}.
\end{cases}
\]

\begin{exercise}If $f$ is detachable and $g$ is not detachable at $x=a$, then
$fg$ is not detachable.
\end{exercise}

Counter example: $f\left(x\right)=x$ is detachable at $a=0$,
$g\left(x\right)=x^{2\cdot\boldsymbol{1}_{\mathbb{Q}}}$ is not detachable. However, the product:
\[
\left(fg\right)\left(x\right)=x^{1+2\cdot\boldsymbol{1}_{\mathbb{Q}}}
\]
is detachable at $a=0$.

\begin{exercise}If $g$ is detachable at $x=a$ and $f$ is not detachable at
$g\left(a\right)$, then $F\left(x\right)=f\left(g\left(x\right)\right)$
is not detachable at $x=a$.
\end{exercise}

Counter example: Let $g\left(x\right)\equiv0$ and $f\left(x\right)=\frac{1}{2}-\boldsymbol{1}_{\mathbb{Q}}$.
Then $g$ is detachable at $a=0$, $f$ is not detachable anywhere
and particularly at $g\left(0\right)=0$, and $f\left(g\right)\equiv\frac{1}{2}$
is detachable at $a=0$.

\begin{exercise}If $g$ is not detachable at $x=a$ and $f$ is detachable at
$g\left(a\right),$then $F\left(x\right)=f\left(g\left(x\right)\right)$
is not detachable at $x=a$.
\end{exercise}

Counter example: Let $f\left(x\right)=x^{2}$ and $g\left(x\right)=\left(-1\right)^{\boldsymbol{1}_{\mathbb{Q}}}$.
Then $g$ is not detachable at $a=0$, $f$ is detachable at $g\left(0\right)=-1$,
and $f\left(g\right)=+1$ is detachable everywhere.

\begin{exercise}If $g$ is not detachable at $x=a$ and $f$ is not detachable
at $g\left(a\right),$then $F\left(x\right)=f\left(g\left(x\right)\right)$
is not detachable at $x=a$.
\end{exercise}

Counter example: Let $f\left(x\right)=\boldsymbol{1}_{\mathbb{Q}}$
and $g\left(x\right)=\boldsymbol{1}_{\mathbb{Q}}$, then neither are
detachable anywhere, but $f\left(g\right)\equiv+1$ is detachable
everywhere.

\begin{exercise}If $f$ is detachable in $\left(a,b\right)$ and $f\left(a\right)=f\left(b\right)$,
then there is a point $c\in\left(a,b\right)$ such that $f_{+}^{;}\left(c\right)f_{-}^{;}\left(c\right)=-1$.
\end{exercise}

Counter example: Let
\[
f\left(x\right)=\begin{cases}
x, & x\notin\left\{ \pm1\right\} \\
0, & x\in\left\{ \pm1\right\} .
\end{cases}
\]
Then $f_{\pm}^{;}\equiv+1$ in $\left(-1,1\right)$.

\begin{exercise}If $f,g$ are detachable at a point $a$, then:
\[
\underset{t\rightarrow x^{\pm}}{\lim}\text{sgn}\left(f\cdot g\right)\left(t\right)=\underset{t\rightarrow x^{\pm}}{\lim}\left(f_{\pm}^{;}\cdot g_{\pm}^{;}\right)\left(t\right).
\]
\end{exercise}

Counter example: $f\left(x\right)\equiv+1,\quad g\left(x\right)\equiv-1$.
Then $\underset{t\rightarrow x^{\pm}}{\lim}\text{sgn}\left(f\cdot g\right)\left(t\right)=-1$
but $\underset{t\rightarrow x^{\pm}}{\lim}\left(f_{\pm}^{;}\cdot g_{\pm}^{;}\right)\left(t\right)=0$.
However, note that if $\underset{t\rightarrow x^{\pm}}{\lim}\left|f\left(t\right)\right|,\underset{t\rightarrow x^{\pm}}{\lim}\left|g\left(t\right)\right|$
are both in $\left\{ 0,\pm\infty\right\} $, then the statement holds
based on theorem \ref{lhopital}.

\begin{exercise}\label{ex_21}If $f$ is detachable and $\underset{x\rightarrow\infty}{\lim}f\left(x\right)$
exists, then $\underset{x\rightarrow\infty}{\lim}f^{;}\left(x\right)$
also exists.
\end{exercise}

Counter example: Let $f:\left(0,\infty\right)\rightarrow\mathbb{R}$
be defined as: $f\left(x\right)=\frac{\sin\left(x^{2}\right)}{x}$.
Then:
\[
\underset{x\rightarrow\infty}{\lim}f'\left(x\right)=\underset{x\rightarrow\infty}{\lim}\frac{2x^{2}\cos\left(x^{2}\right)-\sin\left(2x\right)}{x^{2}}
\]
does not exist. The sign of $f'\left(x\right)$ is unstable
in the limit, switching between $0,\pm1$. Since whenever a function's
derivative exists and is not zeroed, then its sign equals the detachment and
we deduce that $\underset{x\rightarrow\infty}{\lim}f^{;}\left(x\right)$
does not exist as well.

\begin{figure}[h!]
\includegraphics[scale=0.7]{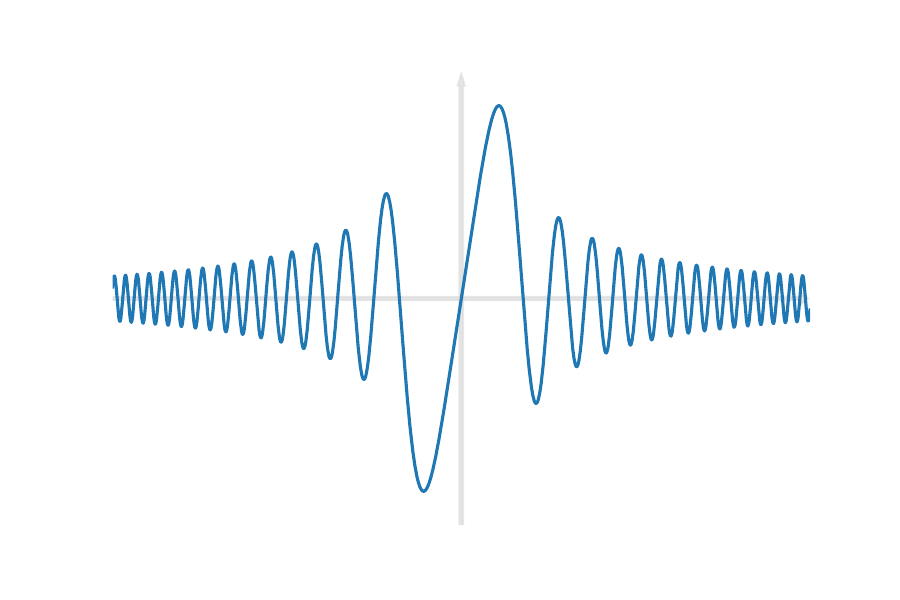}
\label{counter_example21}
\caption{An illustration of a function refuting the claim in exercise \ref{ex_21}}
\end{figure}

\begin{exercise}If $f$ is detachable and bounded in $\left(0,\infty\right)$,
and $\underset{x\rightarrow\infty}{\lim}f_{+}^{;}\left(x\right)$
exists, then $\underset{x\rightarrow\infty}{\lim}f\left(x\right)$
also exists.
\end{exercise}

Counter example: The function $f\left(x\right)=x-\lfloor k\rfloor$,
for $x\in\left[k,k+1\right)$, where $k\in\mathbb{N}$ contradicts
the claim, since it is everywhere increasing from right ($f_{+}^{;}\equiv+1$),
and particularly $\underset{x\rightarrow\infty}{\lim}f_{+}^{;}\left(x\right)=+1$.
However, $\underset{x\rightarrow\infty}{\lim}f\left(x\right)$ does
not exist. Nevertheless, if $\underset{x\rightarrow\infty}{\lim}f^{;}\left(x\right)$
exists (in the sense that $f_{\pm}^{;}$ maintains a stable sign at
infinity), then $f$ is continuously detachable, implying (according
to lemma \ref{monotonicity}, that $f$ is either constant
or strictly monotone. Since it is also assumed to be bounded, then
it converges and the claim is correct.

\begin{exercise}If $f$ is detachable at $x=a$ and in its neighborhood, then
its detachment is continuous.
\end{exercise}

Counter example: let $a=0$ and $f$ be defined as follows:
\[
f\left(x\right)=\begin{cases}
-x+1, & x>0\\
0, & x\leq0,
\end{cases}
\]
then $f_{+}^{;}\left(0\right)=+1$ but $f_{\pm}^{;}=-1$ for $x>0$
and $f_{\pm}^{;}=0$ for $x\leq0$. Therefore, the limit of $f^{;}$
does not exists at $x=a$, and $f^{;}$ is discontinuous.

\begin{exercise}If $f$ is not detachable in infinitely many points in any neighborhood
of the point $x=a$, then $f$ is also not detachable at $x=a$.
\end{exercise}

Counter example: consider Weierstrass function, \[ f\left(x\right)=\sum_{n=1}^{\infty}\left(\frac{1}{2}\right)^{n}\cos\left(3^{n}x\right).\]
Any local extremum point (such as $a=0$) satisfies the claim's conditions:
the function is detachable and there are infinitely many
points in each neighbourhood where it is not detachable.

\chapterimage{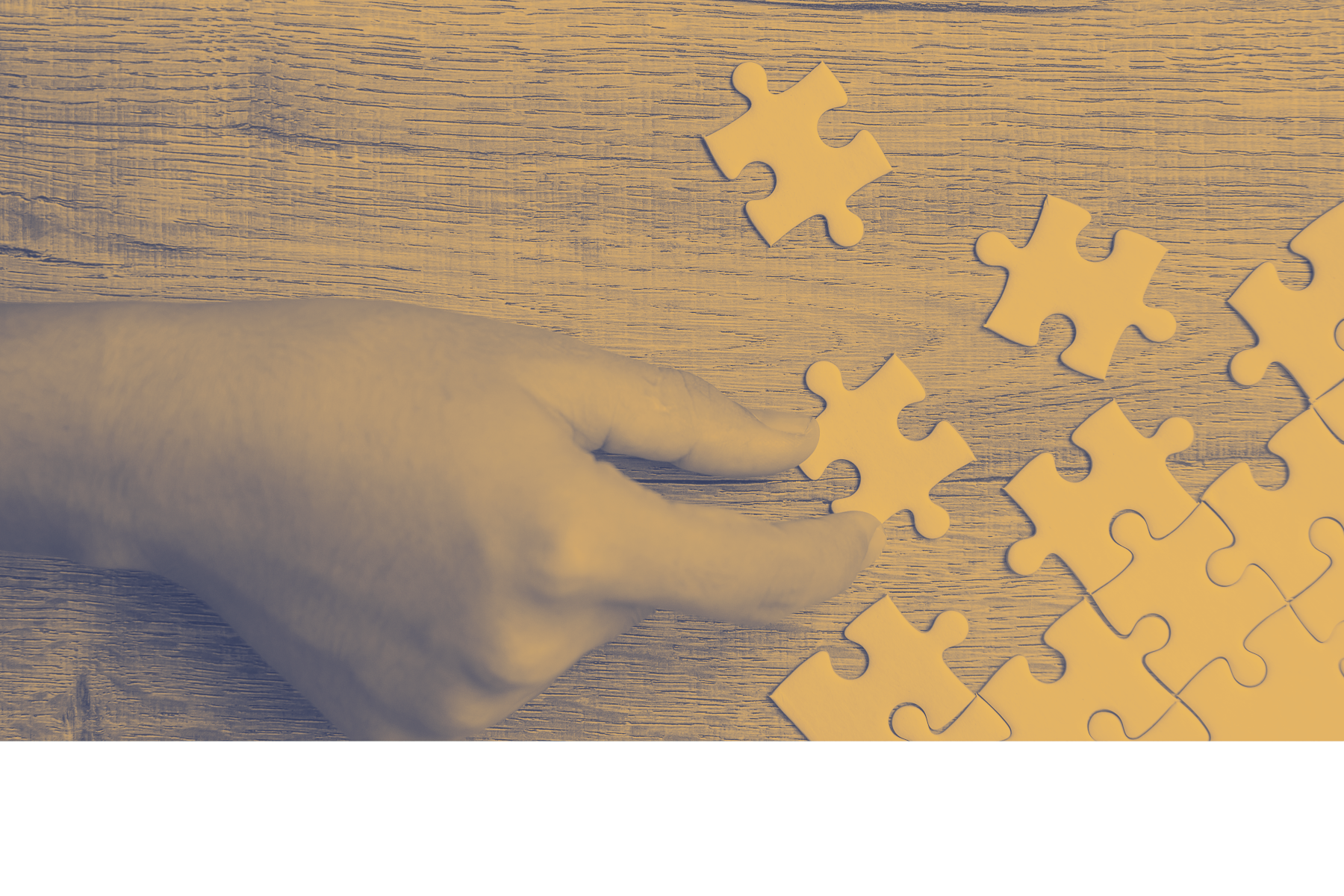}
\chapter{Multivariable Semi-discrete Calculus}

\section{Previous Work and Motivation}
The detachment indirectly serves to generalize a familiar integration algorithm (originated in Computer Vision), to generic continuous domains. The Integral Image algorithm calculates sums of rectangles in images efficiently. It was introduced in 2001, in Viola and Jobes' prominent "Integral Image" paper. The algorithm states that to calculate the integral of a function over a rectangle in the plane, it is possible to pre-calculate the antiderivative; then, in real-time, summarize its values in the rectangle’s corners, signed alternately. This approach can be thought of as an extension of the Fundamental Theorem of Calculus to the plane. 

The theorem proposed by (\cite{wang2007shape}) in 2007 further generalized the algorithm. While in our preliminary discussion we introduced works in which the instantaneous trend of change has been applied explicitly (either numerically or analytically), in the following theorem the detachment is leveraged only implicitly.

\index{Wang et al.'s Theorem}
\begin{theorem}[Wang et al.]Let $D\subset\mathbb{R}^{2}$ be a generalized rectangular domain (polyomino), and let $f:\mathbb{R}^{2}\rightarrow\mathbb{R}$ admit an antiderivative $F.$ Then:

$$\underset{{\scriptscriptstyle D}}{\iint}f\boldsymbol{dx}=\underset{{\scriptscriptstyle x\in\nabla D}}{\sum}\alpha_{D}\left(x\right)F\left(x\right),$$

where $\nabla D$ is the set of corners of $D$, and $\alpha_{D}$ accepts values in $\left\{ 0,\pm1\right\} $ and is determined according to the type of corner that $x$ belongs to.
\label{wang}
\end{theorem}

\begin{figure}
\includegraphics[scale=1.1]{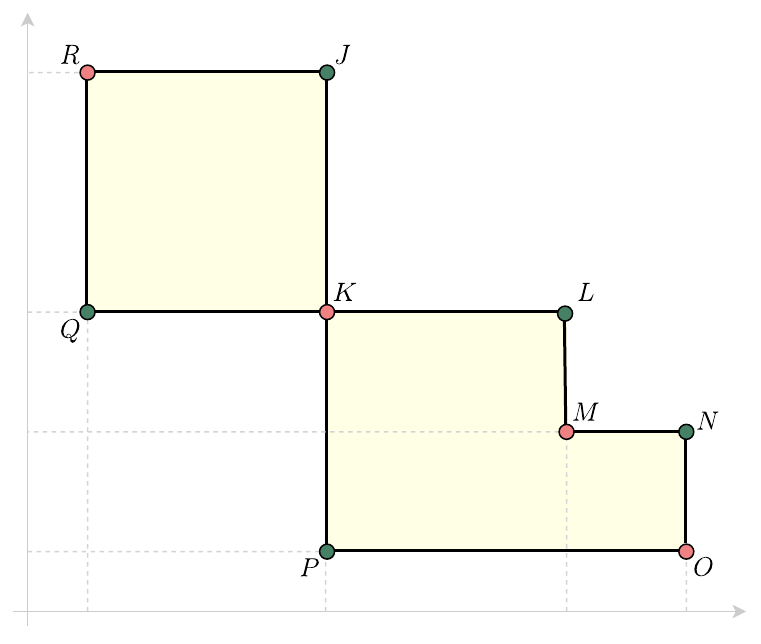}
\caption{An illustration of theorem \ref{wang}. The theorem states that given an integrable function $f$, its antiderivative $F\left(x,y\right)\equiv\iint f dx$, and a rectangular domain (highlighted in the above figure), then: $\iint_{D}f(x,y)dxdy=F(J)-2F(K)+F(L)-F(M)+F(N)-F(O)+F(P)+F(Q)-F(R).$ The coefficients of the antiderivative at the corner points are the parameter $\alpha_{D}$ from the formulation of the theorem, and they are uniquely determined according to the corner type.}
\end{figure}

The derivative sign does not reflect trends at cusp points such as corners; therefore, it does not classify corner types concisely. In contrast, it turns out that the detachment operator classifies corners (and thus defines the parameter $\alpha_{D}$ coherently) independently of the curve’s parameterization; this is a multidimensional generalization of our above discussion regarding the relation between both operators. The theorem utilizes the inclusion-exclusion principle as the series of domains finally add up to the domain bounded by the vertices. While this theorem is defined over continuous domains, it is limited to generalized rectangular ones. Let us attempt to alleviate this limitation.

In the proceeding subsections, we leverage the detachment to define a novel integration method that extends theorem \ref{wang} to non-rectangular domains by combining it with the mean value theorem, as follows. Given a simple closed curve $C$, we will define what it means that it is detachable. Then, let $D$ be the region bounded by the curve. Let $R\subseteq D$ be a rectangular domain circumscribed by $C$. Let ${C_i}$ be the sub-curves of $C$ between each pair of its consecutive touching points with $R.$ Let $D_{i}\subset D\backslash R$ be the sub-domain bounded between $C_i$ and $R$. Let $\partial D_i$ be the boundary of $D_i$, and let $\nabla D_{i}$ be the intersection between $D_i$ and the vertices of $R$. According to the mean value theorem in two dimensions, $\forall i:\exists x_{i}\in D_{i}$ such that $\underset{{\scriptscriptstyle D_{i}}}{\iint}f\boldsymbol{dx}=\beta_{i}f\left(x_{i}\right)\equiv\boldsymbol{\beta}\cdot\boldsymbol{f\left(x_{\beta}\right)},$ where $\beta_{i}$ is the area of $D_{i}.$

The semi-discrete integration method we will introduce accumulates a linear combination of the function and its antiderivative along the sub-domains $D_{i}$, and yields the following simple result:

\begin{corollary}\label{semi_discrete_green}Let $D\subset\mathbb{R}^{2}$ be a closed, bounded, and connected domain whose edge is detachable. Let $f:\mathbb{R}^{2}\rightarrow\mathbb{R}$ be a continuous function that admits an antiderivative $F$. Then:
$$\underset{{\scriptscriptstyle D}}{\iint}f\boldsymbol{dx}=\boldsymbol{\alpha}\cdot\boldsymbol{F\left(x_{\alpha}\right)}+\boldsymbol{\beta}\cdot\boldsymbol{f\left(x_{\beta}\right)},$$
where $\boldsymbol{\alpha}\cdot\boldsymbol{F\left(x_{\alpha}\right)}+\boldsymbol{\beta}\cdot\boldsymbol{f\left(x_{\beta}\right)}\equiv\sum\left[\boldsymbol{\alpha_{i}}\cdot F\left(\boldsymbol{x_{\alpha,i}}\right)+\beta_{i}f\left(x_{i}\right)\right]$ and $\boldsymbol{x_{\alpha,i}}=\left(F\left(x\right)\,|\,x\in\nabla D_{i}\right)^T$ is the vector of antiderivative values at the vertices of the subdomain $D_{i}$, $\boldsymbol{\alpha_{i}}=\left(\alpha\left(\partial D_{i}^{;},x\right)\,|\,x\in\nabla D_{i}\right)^T$ is a vector containing the results of applying a function $\alpha$ to the detachments of the curve $\partial D_{i}^{;}$ at its vertices $\nabla D_{i}$, and $\beta_i$ is the area of $D_i$, which we incorporate as part of the mean value theorem for integrals along with it matching point in $D_i$ denoted by $x_i$. The function $\alpha$ is constructed within the integration method based on the curve's pointwise detachments, as we illustrate in the following subsections.
\end{corollary}

To establish result \ref{semi_discrete_green} rigorously, let us define the parameter $\alpha_i$ with the notion of the curve's detachment, and introduce an integration method along curves that yields the above result as a relation between the double integral inside the domain and the novel line integral along its boundary, similarly to Green's theorem.

\begin{figure}
\includegraphics[scale=0.8]{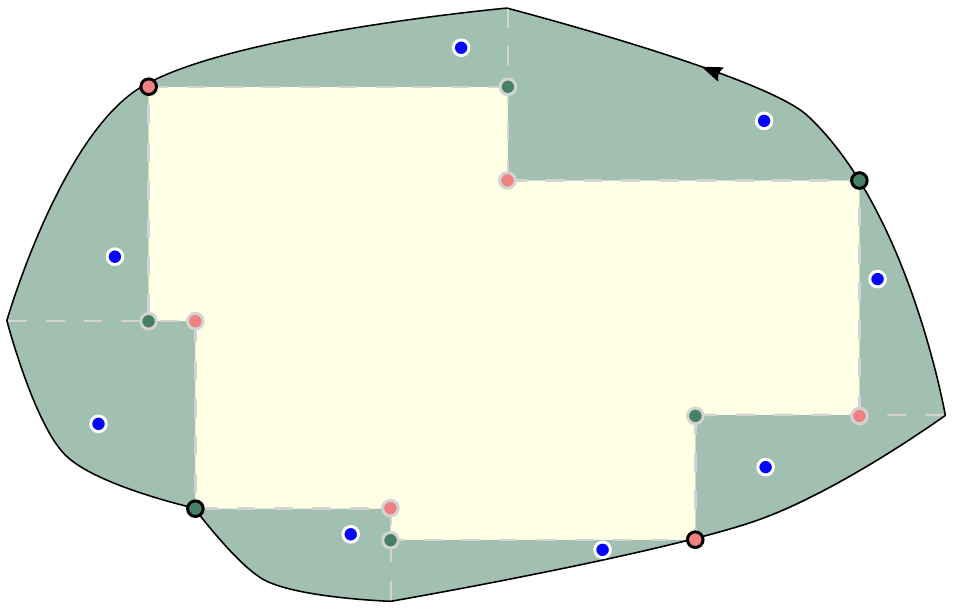}
\caption{An illustration of corollary \ref{semi_discrete_green}. The double integral of a function $f$ defined over the domain equals a sum of two linear combinations: that of the antiderivative $F$ along the inner rectangular domain's corners (the green and red points, whose coefficients $\boldsymbol{\alpha}$ are determined based on the domain's detachments), and that of the function $f$ at the interior of the subdomains outside the rectangular domains (the blue points, whose coefficients $\boldsymbol{\beta}$ are determined based on each green subdomain's area, due to the mean value theorem for integrals in $\mathbb{R}^2$).}
\label{ftc_on_grd}
\end{figure}

\index{Monotonicity}
\section{Classifying a Curve's Monotony}
For the simplicity of our discussion we focus on $\mathbb{R}^2$, and assume that all the curves are continuous, simple, finite, oriented and non oscillating (in a sense that will be clear later). Moreover, we assume that the closed curves are positively oriented, and the domains are simply connected. Having said that, the discussion can be naturally extended to higher dimensions and to curves with more general attributes. Hence, we would like to introduce a tool for classification of corners along a curve, which is a special case of classifying the monotony of the curve. For the sake of coherency and a proper classification, we will require from the classification tool to be dependent of the curve’s spatial representation and orientation, and independent of its parametrization.Let us analyze different parametrizations of the same curve and watch how the derivative yields different results for each of them.

\begin{exm}Let us examine a curve $C$ that consists of two line segments, $$C: \left(1,0\right)\longrightarrow\left(0,0\right)\longrightarrow\left(0,1\right).$$Let us evaluate its derivative at the corner point $\left(0,0\right)$ for different parametrizations, determined by the value of a parameter $k >0$:

\begin{align}
&\begin{aligned}
 & C:\gamma_{k}=\left(x_{k}\left(t\right),y_{k}\left(t\right)\right),\quad0\leq t\leq2\\
 & \gamma_{k}\left(t\right)=\begin{cases}
\left(\left(1-t\right)^{k},0\right), & 0\leq t\leq1\\
\left(0,\left(1-t\right)^{k}\right), & 1\leq t\leq2
\end{cases}
\end{aligned}
\end{align}
\begin{itemize}
\item For $k=1$, the curve's one-sided derivatives are 
\[
\partial_{+}x_{1}\left(1\right)=0,\quad\partial_{-}x_{1}\left(1\right)=-1,\quad\partial_{+}y_{1}\left(1\right)=+1\quad\text{and}\quad\partial_{-}y_{1}\left(1\right)=0.
\]
\item For $k>1$, the curve's one-sided derivatives are all
zeroed at $\left(0,0\right).$
\item For $k<1$, the curve's one-sided derivatives do not
exist at $\left(0,0\right)$.
\end{itemize}
Therefore, the vector $\left(\partial_{+}x,\partial_{-}x,\partial_{+}y,\partial_{-}y\right)\biggm\lvert_{t=t_{0}}$
is not a valid tool for the classification of corners (since this vector
is dependent of the curve's parametrization). It should
not come to us by surprise, as the derivative is a tool measuring
velocity, and the curve's parametrizations are descriptions of different
speeds of movement along the curve.
\end{exm}

\index{Detachments Vector}
\section{Detachments Vector of a Curve}

In this subsection, we propose that the vector of the curve's
detachments is a coherent tool for classifying the monotony of a broad
set of curves, in the sense that it is independent of the curve's
parametrization.

\begin{definition}[Detachable curve]
We say that a curve $C$ is detachable
at a point $z\in C$ if there is a continuous parametrization
$\gamma\left(t\right)=\left(x\left(t\right),y\left(t\right)\right),0\leq t\leq1$
of $C$ for which $z=\left(x\left(t_{0}\right),y\left(t_{0}\right)\right)$
such that $x,y$ are detachable at $t_{0}$.
\end{definition}

We show that in case a
curve is detachable at a point, then $x_{\pm}^{;},y_{\pm}^{;}$
are independent of its parametrization. To that end, let us define the
following spatial property of the curve.

\begin{definition}[Neighborhood system of a point]
Given a point $z=\left(x_{0},y_{0}\right)$
in the plane, we define its neighbourhood system as the following
eight domains:
\[
O_{s_{1},s_{2}}\left(z\right)\equiv\left\{ \left(x,y\right)\biggm\lvert \text{sgn}\left(x-x_{0}\right)=s_{1},\quad \text{sgn}\left(y-y_{0}\right)=s_{2}\right\} _{s_{1},s_{2}\in\left\{ 0,\pm1\right\} },
\]
where we omit the degenerated domain $O_{0,0}=\left\{ z\right\} $
from the neighborhood system.
\end{definition}

Let $z$ be a point on a detachable curve $C$. Assuming $C$ is oriented, we can refer to the set of points on $C$ that are preceding or following $z$ with respect to the orientation. Let us
denote those points by $C\bigm\lvert_{z^{-}}$ and $C\bigm\lvert_{z^{+}}$
respectively.

We claim that $C$ is detachable at $z$ if and only if there is
a small enough neighborhood of $z$ such that $C\bigm\lvert_{z^{-}}$
and $C\bigm\lvert_{z^{+}}$ are each contained in the same domain
out of the neighborhood system of $z$. Let us formulate it
rigorously. Let $\gamma\left(t\right)=\left(x\left(t\right),y\left(t\right)\right)$
be any continuous parametrization of a detachable curve $C$ such
that $\gamma\left(t_{0}\right)=z$. Then:
\[
\label{curve_detachment_condition}\begin{array}{cc}
 & x_{+}^{;}\left(t_{0}\right)=\delta_{1}\land x_{-}^{;}\left(t_{0}\right)=\delta_{2}\land y_{+}^{;}\left(t_{0}\right)=\delta_{3}\land y_{-}^{;}\left(t_{0}\right)=\delta_{4}\\
\Longleftrightarrow & \begin{cases}
\text{sgn}\left[x\left(t_{0}+h\right)-x\left(t_{0}\right)\right]=\delta_{1}, & h\rightarrow0^{+}\\
\text{sgn}\left[x\left(t_{0}\right)-x\left(t_{0}+h\right)\right]=\delta_{2}, & h\rightarrow0^{-}\\
\text{sgn}\left[y\left(t_{0}+h\right)-y\left(t_{0}\right)\right]=\delta_{3}, & h\rightarrow0^{+}\\
\text{sgn}\left[y\left(t_{0}\right)-y\left(t_{0}+h\right)\right]=\delta_{4}, & h\rightarrow0^{-}
\end{cases}\\
\Longleftrightarrow & \begin{cases}
\exists r^{+}>0:\quad B_{r^{+}}\left(z\right)\bigcap C\bigm\lvert_{z^{+}}\subseteq O_{\delta_{1},\delta_{3}}\left(z\right)\\
\exists r^{-}>0:\quad B_{r^{-}}\left(z\right)\bigcap C\bigm\lvert_{z^{-}}\subseteq O_{\delta_{2},\delta_{4}}\left(z\right).
\end{cases}
\end{array}
\]
Since condition \ref{curve_detachment_condition} is independent of the curve's parametrization,
we conclude that so is the vector $\left\{ \delta_{i}\right\} _{1\leq i\leq4}$.
Hence the following definition.

\begin{definition}[Detachments vector of a detachable curve]
Suppose that a curve $C$ is detachable at $z\in C$. We define the
curve's detachments vector at $z$ as:
\[
\delta C\left(z\right)\equiv\left(x_{+}^{;},x_{-}^{;},y_{+}^{;},y_{-}^{;}\right)\biggm\lvert_{t_{0}},
\]
where $\left(x\left(t\right),y\left(t\right)\right)$ is any continuous
parametrization of $C$ for which $z=\left(x\left(t_{0}\right),y\left(t_{0}\right)\right)$.
\end{definition}

In the following discussion we omit the notion of the curve's parametrization whenever possible, since a curve's detachments vector is independent of its parametrization.

\begin{remark}\label{curve_detachment_vector_possible_values}Let $C$ be a curve, and let $z$ be a point on $C$. Then:
\begin{enumerate}
\item If $C$ admits a constant detachments vector in the neighbordhood
of $z$, then $\delta C\left(z\right)$ is one of the following values:
\begin{align}
&\begin{aligned}
 & \left\{ \left(-1,+1,0,0\right),\left(0,0,+1,-1\right),\left(+1,-1,+1,-1\right),\left(+1,-1,-1,+1\right)\right\} \\
\bigcup & \left\{ \left(-1,+1,-1,+1\right),\left(-1,+1,+1,-1\right),\left(+1,-1,0,0\right),\left(0,0,-1,+1\right)\right\} 
\end{aligned}
\end{align}
\item If $z$ is a right corner then $\delta C\left(z\right)$ accepts one
of the following values:
\begin{align}
&\begin{aligned}
 & \left\{ \left(+1,0,0,-1\right),\left(-1,0,0,-1\right),\left(0,-1,+1,0\right),\left(0,+1,+1,0\right)\right\} \\
\bigcup & \left\{ \left(0,+1,-1,0\right),\left(0,-1,-1,0\right),\left(-1,0,0,+1\right),\left(+1,0,0,+1\right)\right\} 
\end{aligned}
\end{align}
\end{enumerate}
\end{remark}

\index{Detachment}
\section{A Curve's Detachment}In this subsection we introduce an attribute of a curve at a point, namely its detachment, which is determined according to the detachment vector, or equivalently, according to the domains (out of the point's neighborhood system) in which the curve resides close to the point.
A possible geometric interpretation of the curve's detachment at a point $z\in C$ is the sign of the sum of
\[
\alpha_{O_{s_{1},s_{2}}^{\pm}\left(z\right)},
\]
where $O_{s_{1},s_{2}}^{\pm}\left(z\right)$ are the domains (out
of the point's neighborhood system) that reside left to the curve in $C\bigm\lvert_{z^{\pm}}$ respectively, and $\alpha_{D}$
is determined as in theorem \ref{wang}. This definition extends the parameter
$\alpha_{D}$, in the sense that it coalesces with it at the right corners.
The geometric interpretation of a curve's detachment will become clearer once we apply it to extend that theorem to more general domains in subsection \ref{semi_discrete_line_integral_section}, to obtain a rigorous version of corollary \ref{semi_discrete_green} above.

We extract definition \ref{curve_detachment} by reverse engineering the possible values
of the curve's detachment given this geometric interpretation.

\begin{definition}[Detachment of a curve]
Let $C$ be a detachable curve. We
define the detachment of the curve $C$ at the point $z\in C$ as
a function of its detachments vector $\delta C\left(z\right)=\left(x^;_+,x^;_-,y^;_+,y^;_-\right)$
as follows:
$$C^{;}\left(z\right)\equiv y_{-}^{;}\text{sgn}\left(y_{-}^{;}-x_{-}^{;}\right)-y_{+}^{;}\text{sgn}\left(y_{+}^{;}+x_{+}^{;}\right).$$

\label{curve_detachment}
\end{definition}
See the exact values in table \ref{detachments_table}. We will clarify this closed algebraic formula for the detachment via its geometric meaning in the below discussion.

\begin{table}
  \color[HTML]{ef8182}
            \footnotesize
                \begin{tabular}{cccc}
        \toprule
        $\boldsymbol{x_{+}^{;}}$ & $\boldsymbol{x_{-}^{;}}$ & $\boldsymbol{y_{+}^{;}}$ & $\boldsymbol{y_{-}^{;}}$\\
        \midrule
        $+1$ & $-1$ & $+1$ & $-1$\\
        $-1$ & $+1$ & $-1$ & $+1$\\
        $0$ & $-1$ & $-1$ & $0$\\
        $0$ & $+1$ & $+1$ & $0$\\
        $0$ & $-1$ & $+1$ & $0$\\
        $0$ & $+1$ & $-1$ & $0$\\
        $0$ & $+1$ & $+1$ & $+1$\\
    $-1$ & $-1$ & $-1$ & $0$\\
        $+1$ & $+1$ & $+1$ & $0$\\
        $0$ & $-1$ & $-1$ & $-1$\\
        $-1$ & $+1$ & $-1$ & $0$\\
        $0$ & $+1$ & $-1$ & $+1$\\
        $0$ & $-1$ & $+1$ & $-1$\\
        $+1$ & $-1$ & $+1$ & $0$\\
        $+1$ & $+1$ & $+1$ & $+1$\\
        $-1$ & $-1$ & $-1$ & $-1$\\
        \bottomrule
        \end{tabular}
                \hfill                      \color[HTML]{dc982d}
                \begin{tabular}{cccc}
        \toprule
        $\boldsymbol{x_{+}^{;}}$ & $\boldsymbol{x_{-}^{;}}$ & $\boldsymbol{y_{+}^{;}}$ & $\boldsymbol{y_{-}^{;}}$\\
        \midrule
        $-1$ & $-1$ & $-1$ & $+1$\\
        $+1$ & $+1$ & $-1$ & $+1$\\
        $+1$ & $-1$ & $-1$ & $-1$\\
        $+1$ & $-1$ & $+1$ & $+1$\\
        $+1$ & $+1$ & $+1$ & $-1$\\
    $-1$ & $-1$ & $+1$ & $-1$\\
        $-1$ & $+1$ & $-1$ & $-1$\\
        $-1$ & $+1$ & $+1$ & $+1$\\
        $0$ & $-1$ & $+1$ & $+1$\\
        $-1$ & $0$ & $-1$ & $-1$\\
        $+1$ & $+1$ & $0$ & $+1$\\
        $-1$ & $-1$ & $+1$ & $0$\\
    $+1$ & $+1$ & $-1$ & $0$\\
        $-1$ & $-1$ & $0$ & $-1$\\
        $+1$ & $0$ & $+1$ & $+1$\\
        $0$ & $+1$ & $-1$ & $-1$\\
        $-1$ & $+1$ & $+1$ & $0$\\
        $-1$ & $+1$ & $0$ & $+1$\\
        $0$ & $-1$ & $-1$ & $+1$\\
    $-1$ & $0$ & $-1$ & $+1$\\
        $+1$ & $0$ & $+1$ & $-1$\\
        $0$ & $+1$ & $+1$ & $-1$\\
        $+1$ & $-1$ & $0$ & $-1$\\
        $+1$ & $-1$ & $-1$ & $0$\\
        $+1$ & $+1$ & $0$ & $0$\\
        $-1$ & $-1$ & $0$ & $0$\\
        $0$ & $0$ & $-1$ & $-1$\\
        $0$ & $0$ & $+1$ & $+1$\\
        \bottomrule
        \end{tabular}
                \hfill
\color[HTML]{468165}                
                \begin{tabular}{cccc}
        \toprule
        $\boldsymbol{x_{+}^{;}}$ & $\boldsymbol{x_{-}^{;}}$ & $\boldsymbol{y_{+}^{;}}$ & $\boldsymbol{y_{-}^{;}}$\\
        \midrule
        $+1$ & $-1$ & $-1$ & $+1$\\
        $-1$ & $+1$ & $+1$ & $-1$\\
        $+1$ & $0$ & $0$ & $-1$\\
        $-1$ & $0$ & $0$ & $+1$\\
        $+1$ & $0$ & $0$ & $+1$\\
    $-1$ & $0$ & $0$ & $-1$\\
        $+1$ & $0$ & $0$ & $-1$\\
        $+1$ & $+1$ & $0$ & $-1$\\
        $-1$ & $-1$ & $0$ & $+1$\\
        $-1$ & $0$ & $+1$ & $+1$\\
        $-1$ & $+1$ & $0$ & $-1$\\
        $+1$ & $0$ & $-1$ & $+1$\\
        $-1$ & $0$ & $+1$ & $-1$\\
    $+1$ & $-1$ & $0$ & $+1$\\
        $+1$ & $+1$ & $-1$ & $-1$\\
        $-1$ & $-1$ & $+1$ & $+1$\\ 
        \bottomrule
        \end{tabular}
\bigskip
\caption{Tables defining the term $C^;$ from definition \ref{curve_detachment}. The left, middle and right tables correspond to detachment vectors yielding $C^;$ of $-1,0$ and $+1$, respectively.}
\label{detachments_table}
\end{table}
\index{Semi-discrete Line Integral}
\section{Defining the Semi-discrete Line Integral}\label{semi_discrete_line_integral_section}In this subsection we introduce
an integration method whose aim is to naturally extend theorem \ref{wang}
to non-rectangular domains, as in corollary \ref{semi_discrete_green}. In \ref{semi_discrete_line_integral_monotonic} we define the
integration method for monotonic curves, in subsection \ref{semi_discrete_line_integral_algbebra} we formulate
some of its algebraic properties, and in subsection \ref{semi_discrete_line_integral_general} we extend the integration method to general curves and deduce corollary \ref{semi_discrete_green}.

\subsection{Defining the Semi-discrete Line Integral for Monotonic Curves} \label{semi_discrete_line_integral_monotonic}
\begin{definition}[Monotonic curve]
We say that a curve is monotonic if its detachments vector is constant, except, perhaps, at its endpoints.  The detachment of a monotonic curve is defined as its (constant) detachment at internal points.
\end{definition}

\begin{corollary}\label{corollary_about_detachment_vector}Let $C$ be a monotonic curve. Then $C$ is entirely contained in a (possibly degenerated) rectangle whose opposite vertices are the endpoints of $C$.
\end{corollary}
\begin{proof}Let $\gamma\left(t\right)=\left(x\left(t\right),y\left(t\right)\right),\quad 0\leq t\leq 1$ be any continuous parametrization of $C.$
According to remark \ref{curve_detachment_vector_possible_values} combined with lemma \ref{monotonicity}, both the functions $x$ and $y$ are strictly monotonous. Without loss of generality, let us analyze two possible values of the curve’s constant detachments vector $\delta\left(C\right).$
In case $\delta\left( C \right) = \left( +1,-1,+1,-1 \right)$, then for each $0<t<1$, it holds that $x\left(0\right)< x\left(t\right)<x\left(1\right)$ and $y\left(0\right)<y\left(t\right)<y\left(1\right),$
hence the curve’s points are fully contained in the square $\left[x\left(0\right),y\left(0\right)\right]\times\left[x\left(1\right),y\left(1\right)\right].$
In case $\delta\left(C\right) = \left(-1,+1,0,0\right)$, then for each $0< t <1$, it holds that $x\left(1\right)< x\left(t\right)< x\left(0\right)$ and $y\left(0\right) =y\left(t\right) =y\left(1\right),$ and the statement holds.
\end{proof}

\begin{definition}[Straight path of a pair of points]
Given a pair of points,
$$\left\{x=\left(a,b\right), y=\left(c,d\right)\right\}\subset \mathbb{R}^2,$$
we define the following paths:
\begin{align}
\begin{aligned}
 & \gamma^{+}\left(x,y\right): & \left(a,b\right)\longrightarrow\left(a,d\right)\longrightarrow\left(c,d\right)\\
 & \gamma^{-}\left(x,y\right): & \left(a,b\right)\longrightarrow\left(c,b\right)\longrightarrow\left(c,d\right)
\end{aligned}
\end{align}

as the positive and negative straight paths of $\left\{x,y\right\}$, respectively. We denote the mentioned points along the paths as follows:
$$\gamma_1\left(x,y\right) = \left(a,b\right),\quad \gamma^+_2\left(x,y\right) = \left(a,d\right),\quad \gamma^-_2\left(x,y\right) = \left(c,b\right),\quad \gamma_3\left(x,y\right) = \left(c,d\right).$$
If we omit the sign subscript, then a positive sign is assumed, for example $\gamma\left(x,y\right)\equiv\gamma^+\left(x,y\right)$ and $\gamma_2\left(x,y\right)\equiv\gamma^+_2\left(x,y\right).$
\label{straight_path_points_pair}
\end{definition}

\index{Paths of a Curve}
\begin{definition}[Paths of a curve]
Let $C$ be a given monotonic curve with endpoints $\left\{z_0,z_1\right\}.$ We define the curve’s positive and negative paths,denoted by $C^\pm$, respectively, as the straight paths between the endpoints:
$$C^\pm\equiv\gamma^\pm\left(z_0,z_1\right).$$
We denote the points along the curve’s paths by $C_i^\pm\left(z_0,z_1\right)=\gamma_i^\pm\left(z_0,z_1\right)$  (see definition \ref{straight_path_points_pair} above).  If we omit the sign subscript, then a positive sign is assumed. Thus, $C_i\equiv C^+_i.$
\end{definition}

\index{Local Domains}
\begin{definition}[Local domains]
Given a monotonic curve $C$, we define the positive and negative local domains of $C$, namely $D^\pm\left(C\right)$, as the closed domains whose boundaries satisfy:
$$\partial D^\pm\left(C\right)\equiv C^\pm,$$
where $C^\pm$ are the positive and negative paths of $C$, respectively. If we omit the sign subscript then a positive sign is assumed. Thus, $D\left(C\right)\equiv D^+\left(C\right).$
\end{definition}

\begin{figure}
\begin{tabular}{cc}
  \includegraphics[width=70mm]{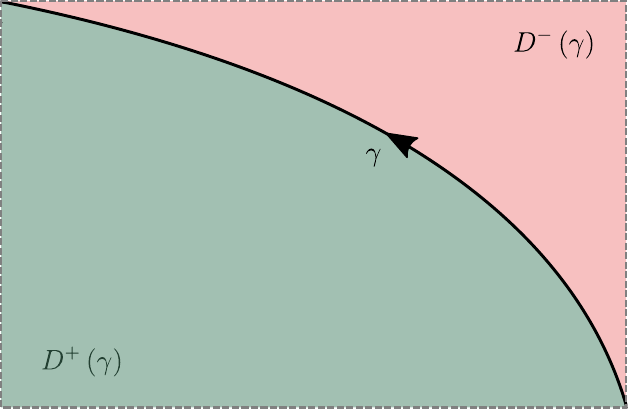} &   \includegraphics[width=70mm]{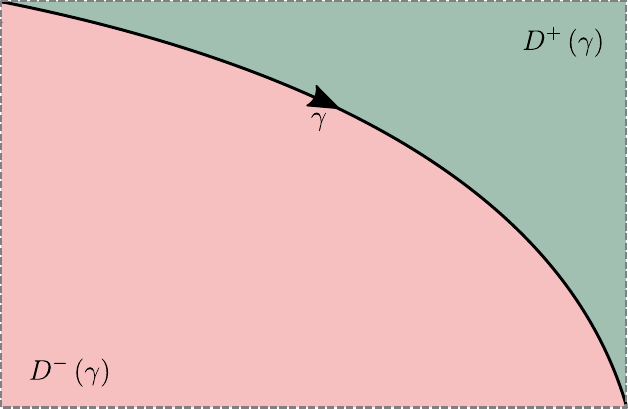}  \\
Positively oriented curve & Negatively oriented curve \\
\end{tabular}
\caption{Local domains of a monotonic curve. The positive domain is the left hand-side of the curve that is bounded by its positive straight path. Note that as the curve’s orientation flips, so do the signs of the local domains.}
\label{positive_domain}
\end{figure}

A monotonic curve’s local domains may be degenerated in case some entries of the curve’s detachments vector are zeroed.

\index{Semi-discrete Line Integral}
\begin{definition}[Semi-discrete line integral over a monotonic curve]
Let $C\subset\mathbb{R}^2$ be a curve, and let $\ell\subset C$ be a monotonic subcurve of $C$, whose constant detachment is $\ell^{;}$. Let us consider a function $f:\mathbb{R}^2\rightarrow\mathbb{R}$ that admits an antiderivative $F$. We define the semi-discrete line integral of $F$ along the curve $\ell$ in the context of the curve $C$ as follows:
$$\underset{\ell\subset C}{\sqint}F\equiv\underset{D\left(\ell\right)}{\iint}f\boldsymbol{dx}-\ell^{;} F\left(\ell_{2}\right)+\frac{1}{2}\left[\ell_{1}^{;}F\left(\ell_{1}\right)+\ell_{3}^{;}F\left(\ell_{3}\right)\right],$$
where $\ell_i^{;}=C^{;}\left(\ell_i\right)$.
If the context $C$ is clear, then we denote the term $\underset{\ell\subset C}\sqint F$ as $\underset{\ell}\sqint F$.
\label{semi_discrete_line_integral_monotonic}
\end{definition}

\begin{figure}
\includegraphics[scale=0.75]{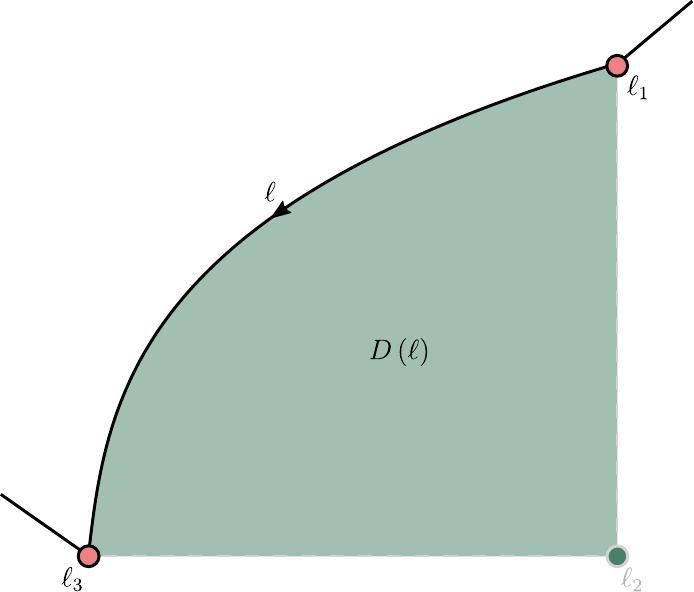}
\caption{An illustration of the definition of the semi-discrete line integral
over a monotonic curve. In this example, the curve $C$ has a highlighted
subcurve, denoted by $\ell$. The detachments vector of $\ell$ is constant
($\delta\ell=\left(-1,+1,-1,+1\right)$), and its detachment is $\ell^{;}=-1$.
The detachment at the subcurve's endpoints
is $\ell_{1}^{;}=-1$ and $\ell_{3}^{;}=-1$ for the point $\ell_{1}$
and $\ell_{3}$, respectively. Hence, according to the definition: $\underset{\ell}{\sqintop}F=\underset{D\left(\ell\right)}{\iint}fdx+F\left(\ell_{2}\right)-\frac{1}{2}\left[F\left(\ell_{1}\right)+F\left(\ell_{3}\right)\right].$}
\end{figure}
\subsection{Algebraic Properties of the Semi-discrete Line Integral}\label{semi_discrete_line_integral_algbebra}

We show that the semi-discrete line integral is a linear operator in the following sense.

\begin{clm}\textbf{Constant multiplication rule.} Let $c\in\mathbb{R}$ and $F$ be an antiderivative defined over the detachable curve $\ell\subset C$. Then:
$$\underset{\ell}{\sqint}\left(c F\right) = c\underset{\ell}{\sqint} F.$$
\end{clm}
\begin{proof}From the linearity of the double integral, if the antiderivative of $f$ is $F$, then the antiderivative of $c f$ is $c F$. Then:

\begin{align}
&\begin{aligned}
\underset{\ell}{\sqint}\left(cF\right) & \equiv\underset{D\left(\ell\right)}{\iint}\left(cf\right)\boldsymbol{dx}-\ell^{;}\left(cF\right)\left(\ell_{2}\right)+\frac{1}{2}\left[\ell_{1}^{;}\left(cF\right)\left(\ell_{1}\right)+\ell_{3}^{;}\left(cF\right)\left(\ell_{3}\right)\right]\\
 & =c\underset{D\left(\ell\right)}{\iint}f\boldsymbol{dx}-c\ell^{;}F\left(\ell_{2}\right)+\frac{1}{2}\left[c\ell_{1}^{;}F\left(\ell_{1}\right)+c\ell_{3}^{;}F\left(\ell_{3}\right)\right]\\
 & =c\left\{\underset{D\left(\ell\right)}{\iint}f\boldsymbol{dx}-\ell^{;}F\left(\ell_{2}\right)+\frac{1}{2}\left[\ell_{1}^{;}F\left(\ell_{1}\right)+\ell_{3}^{;}F\left(\ell_{3}\right)\right]\right\} \equiv c\underset{\ell}{\sqint}F.
\end{aligned}
\end{align}

\end{proof}

\index{Semi-discrete Line Integral}
\begin{lem}\label{semi_discrete_line_integral_additivity}Let $C$ be a curve, and let $\alpha,\beta$ be monotonic subcurves of $C$ such that $\alpha\bigcup\beta$ is monotonic and continuous. Let $f:\mathbb{R}^2\rightarrow\mathbb{R}$ be a function that admits an antiderivative $F$. Then:
$$\underset{\alpha\bigcup\beta}{\sqint}F=\underset{\alpha}{\sqint}F+\underset{\beta}{\sqint}F$$
\end{lem}

\begin{figure}
\begin{tabular}{cc}
  \includegraphics[width=63mm]{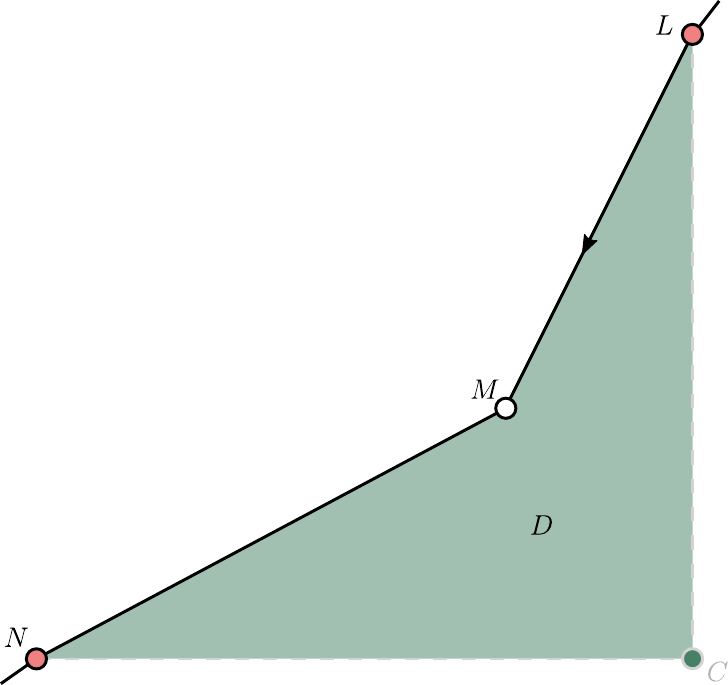} &   \includegraphics[width=63mm]{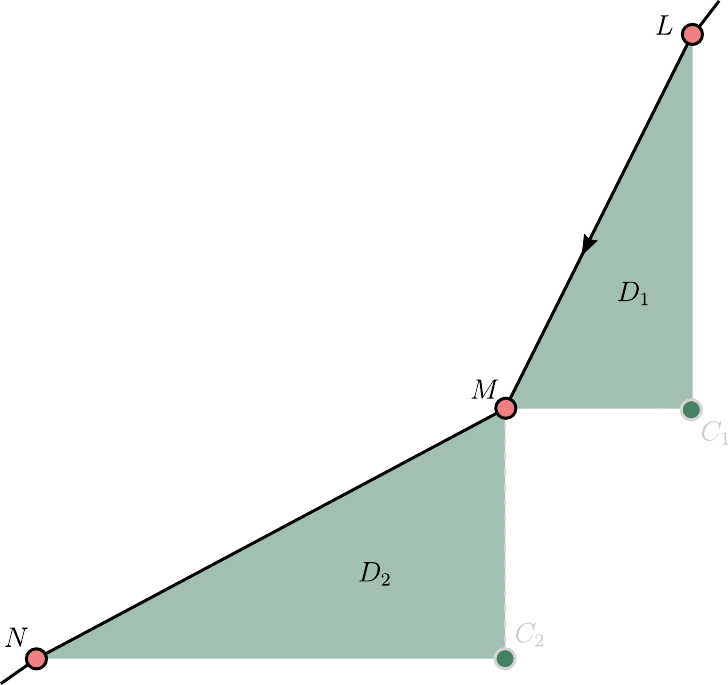}  \\
Semi-discrete line integral on $LMN$ & Semi-discrete line integrals on $LM$, $MN$ \\
\end{tabular}
\caption{An illustration of the proof of lemma \ref{semi_discrete_line_integral_additivity}. On the right: the semi-discrete
line integral over the curve $LMN$ is $\underset{LMN}{\sqint}F\equiv\underset{D}{\iint}f\left(x,y\right)dxdy+F\left(C\right)-\frac{1}{2}\left[F\left(L\right)+F\left(N\right)\right].$
On the left: the semi-discrete line integral over $LM$ is $\underset{LM}{\sqint}F\equiv\underset{D_{1}}{\iint}f\left(x,y\right)dxdy+F\left(C_{1}\right)-\frac{1}{2}\left[F\left(L\right)+F\left(M\right)\right]$,
and that over $MN$ is $\underset{MN}{\sqint}F\equiv\underset{D_{2}}{\iint}f\left(x,y\right)dxdy+F\left(C_{2}\right)-\frac{1}{2}\left[F\left(M\right)+F\left(N\right)\right]$.
Hence by applying theorem \textbackslash wang to the rectangular
domain $MC_{1}CC_{2}$, we obtain $\underset{LM}{\sqintop}F+\underset{MN}{\sqintop}F=\underset{LMN}{\sqintop}F.$}
\label{positive_domain}
\end{figure}

\begin{proof}For the sake of convenience, let us denote $\gamma=\alpha\bigcup\beta.$ Let us assume that $\alpha\bigcap\beta=\left\{\alpha_3\right\}=\left\{\beta_1\right\}$ and that $\gamma_1=\alpha_1,\quad\gamma_3=\beta_3.$ Since $\delta\left(\alpha\right)=\delta\left(\beta\right)=\delta\left(\gamma\right),$ then the curves also share a constant detachment $\gamma^;$. According to definition \ref{semi_discrete_line_integral_monotonic}, we have:

\begin{align}
&\begin{aligned}
\underset{\alpha}{\fint}F & \equiv\underset{D\left(\alpha\right)}{\iint}f\boldsymbol{dx}-\gamma^{;}F\left(\alpha_{2}\right)+\frac{1}{2}\left[\alpha_{1}^{;}F\left(\alpha_{1}\right)+\alpha_{3}^{;}F\left(\alpha_{3}\right)\right]\\
\underset{\beta}{\fint}F & \equiv\underset{D\left(\beta\right)}{\iint}f\boldsymbol{dx}-\gamma^{;}F\left(\beta_{2}\right)+\frac{1}{2}\left[\alpha_{3}^{;}F\left(\alpha_{3}\right)+\beta_{3}^{;}F\left(\beta_{3}\right)\right]\\
\underset{\gamma}{\fint}F & \equiv\underset{D\left(\gamma\right)}{\iint}f\boldsymbol{dx}-\gamma^{;}F\left(\gamma_{2}\right)+\frac{1}{2}\left[\alpha_{1}^{;}F\left(\alpha_{1}\right)+\beta_{3}^{;}F\left(\beta_{3}\right)\right].
\end{aligned}
\end{align}

Applying theorem \ref{wang} to the rectangle $\alpha_2\alpha_3\beta_2\gamma_2$ results with:
$$\underset{D\left(\gamma\right)}{\iint}f\boldsymbol{dx}  \equiv\underset{D\left(\alpha\right)}{\iint}f\boldsymbol{dx}+\underset{D\left(\beta\right)}{\iint}f\boldsymbol{dx}+\gamma^{;}\left\{ \left[F\left(\alpha_{2}\right)+F\left(\beta_{2}\right)\right]-\left[F\left(\alpha_{3}\right)+F\left(\gamma_{2}\right)\right]\right\} ,$$
and by inspecting all the possible values for $\alpha_1^;,\alpha_3^;,\beta_3^;$ and $\gamma^;$, we conclude the statement's correctness.
\end{proof}

\index{Semi-discrete Line Integral}
\begin{corollary}Let $C$ be a detachable curve and let $\ell$ be a monotonic subcurve of $C$ whose detachment is $\ell^;$. Denote by $-\ell$ and $-C$ the curves $\ell$ and $C$ with flipped orientations, respectively.
Suppose that the detachments along the curves $\pm\ell$ are nowhere zeroed, not even at their endpoints $\ell_1,\ell_3$. Let $f:\mathbb{R}^2\rightarrow\mathbb{R}$ be a function that admits an antiderivative $F$. Then it holds that:
$$\underset{-\ell\subset-C}{\sqint}F=-\underset{\ell\subset C}{\sqint}F.$$
\end{corollary}
\begin{proof}Applying theorem \ref{wang} to the rectangle whose boundary is $\ell_1\ell_2^+\ell_3\ell_2^-$ results with:
$$\label{proof_negative_semi_discrete_line_integral}\underset{D^{+}\left(\ell\right)}{\iint}f\boldsymbol{dx}+\underset{D^{-}\left(\ell\right)}{\iint}f\boldsymbol{dx}=\ell^{;}\left\{ F\left(\ell_{1}\right)+F\left(\ell_{3}\right)-\left[F\left(\ell_{2}^{+}\right)+F\left(\ell_{2}^{-}\right)\right]\right\}.$$

The corollary follows by combining equation \ref{proof_negative_semi_discrete_line_integral} with the definitions of $\underset{\pm\ell\subset \pm C}{\sqint}F,$ while considering all the cases of $\ell^;$ and $\delta\left(\ell\right)\bigm\lvert_{\ell_1,\ell_3}$, rearranging the terms and applying the definition of the semi-discrete line integral for $\ell$.
\end{proof}

\index{Semi-discrete Line Integral}
\begin{corollary}Let $\ell$ be a monotonic curve such that $\ell^;\equiv0$, also at the curve’s endpoints $\ell_1,\ell_3$. Then for any function $F$ it holds that $\underset{\ell}{\sqint}F=0.$
\end{corollary}
\begin{proof}According to remark \ref{curve_detachment_vector_possible_values}, and since the curve’s detachment is zeroed, the curve’s detachments vector satisfies:
$$\delta\left(\ell\right)\in\left\{\left(+1,-1,0,0\right),\left(-1,+1,0,0\right),\left(0,0,+1,-1\right),\left(0,0,-1,+1\right)\right\}.$$
Thus, the (positive) local domain of $\ell$ is degenerated; hence, the integral over it is zeroed. Furthermore, all the terms that involve the detachment at the definition of the semi-discrete line integral are zeroed as well. Hence, the semi-discrete line integral of any function along $\ell$ is zeroed.
\end{proof}

\subsection{Defining the Semi-discrete Line Integral for General Curves}
\begin{figure}
\includegraphics[scale=0.75]{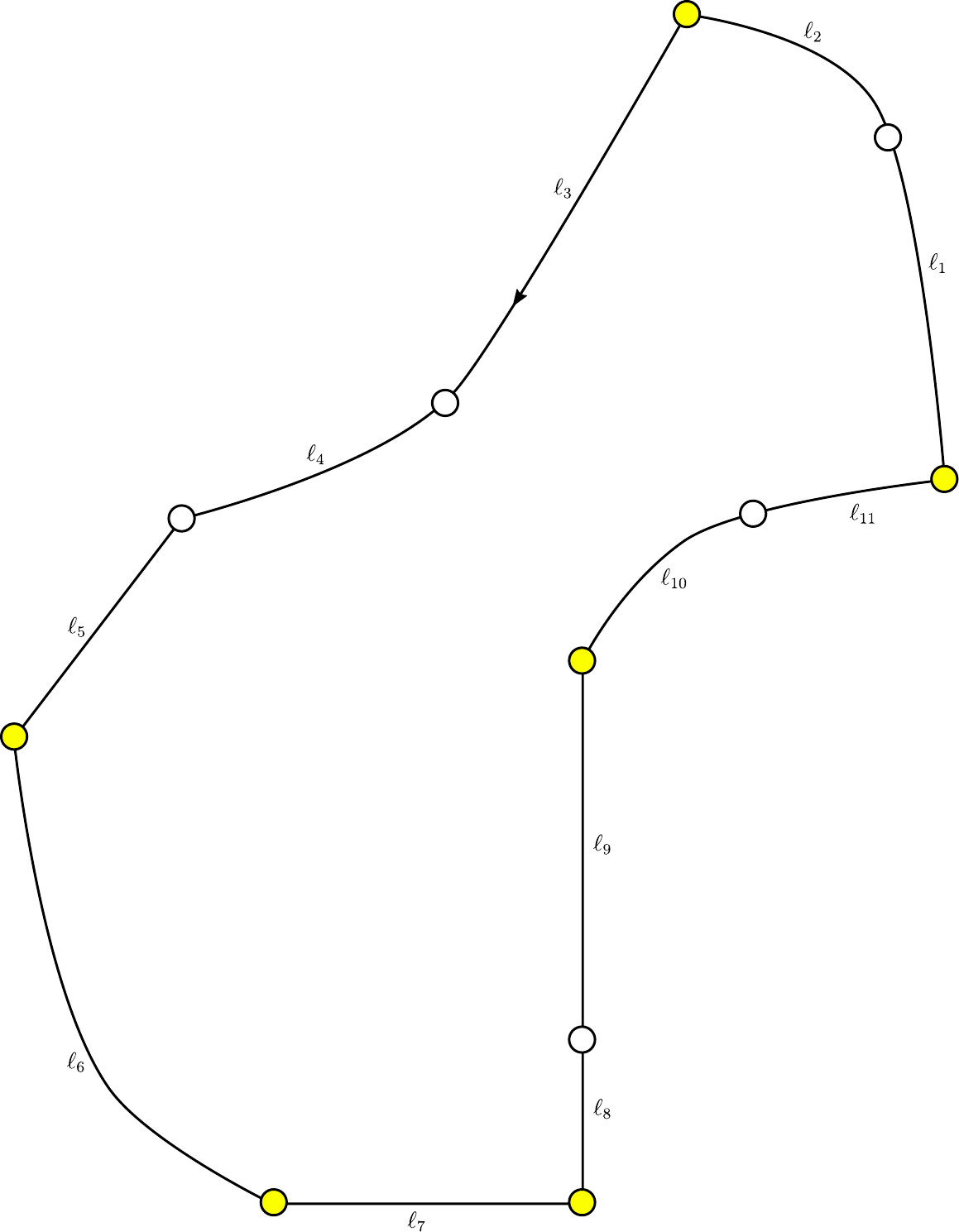}
\caption{An illustration of a monotonic division of a detachable curve. The points defining the curve's mininal division are highlighted. The empty points are not part of the mininal division, in the sense that we may coalesce the following curves and remain with a monotonic division (since their detachments vectors are equal): $\ell_1$ with $\ell_2$; $\ell_3$ with $\ell_4$ and $ell_5$;  $ ell_8$ with $\ell_9$; and $\ell_{10}$ with $\ell_{11}$. We illustrate the semi-discrete line integral along one of the induced monotonic divisions in figure \ref{monotonic_division_with_domains_figure}.}\label{monotonic_division_figure}
\end{figure}

\index{Semi-discrete Line Integral}
\label{semi_discrete_line_integral_general}
Let us extend the definition of the semi-discrete line integral over monotonic curves to unions of such curves.
\index{Monotonic Division}
\index{Monotonicity}
\begin{definition}[Monotonic division of a curve]Let $C$ be a detachable curve in $\mathbb{R}^2$. A monotonic division of $C$ is an ordered set of subcurves $\left\{\ell_i\right\}_{1\leq i\leq n}$, such that each $\ell_i$ is a monotonic subcurve of $C$ whose detachment $\ell_i^;$ is constant, and:
\[C=\bigcup_{i=1}^{n}\ell_{i}.\]
\label{monotonic_division}
\end{definition}
Definition \ref{monotonic_division} is illustrated in Figure \ref{monotonic_division_figure}.
\begin{definition}[Discrete line integral over a detachable curve]
Let $C$ be a detachable curve, and let $\left\{\ell_i\right\}_{1\leq i\leq n}$ be a monotonic division of a subcurve $\ell\subset C$.  Let us consider a function $f:\mathbb{R}^2\rightarrow\mathbb{R}$ that admits an antiderivative $F$. Then the semi-discrete line integral of $F$ over $\ell$ in the context of the curve $C$ is defined as follows:
$$\underset{\ell\subset C}{\sqint}F\equiv\underset{i}{\sum}\underset{\ell_{i}\subset C}{\sqint}F,$$
where each $\underset{\ell_{i}\subset C}{\sqint}F$ is calculated according to the definition of the semi-discrete line integral over monotonic curves, as in definition \ref{semi_discrete_line_integral_monotonic}.
\end{definition}

Note that the notion of semi-discrete line integral over a detachable curve is well-defined because the right hand-side is independent of the curve’s division, due to the additivity of the semi-discrete line integral over monotonic curves (lemma \ref{semi_discrete_line_integral_additivity}).

\begin{figure}
\includegraphics[scale=0.75]{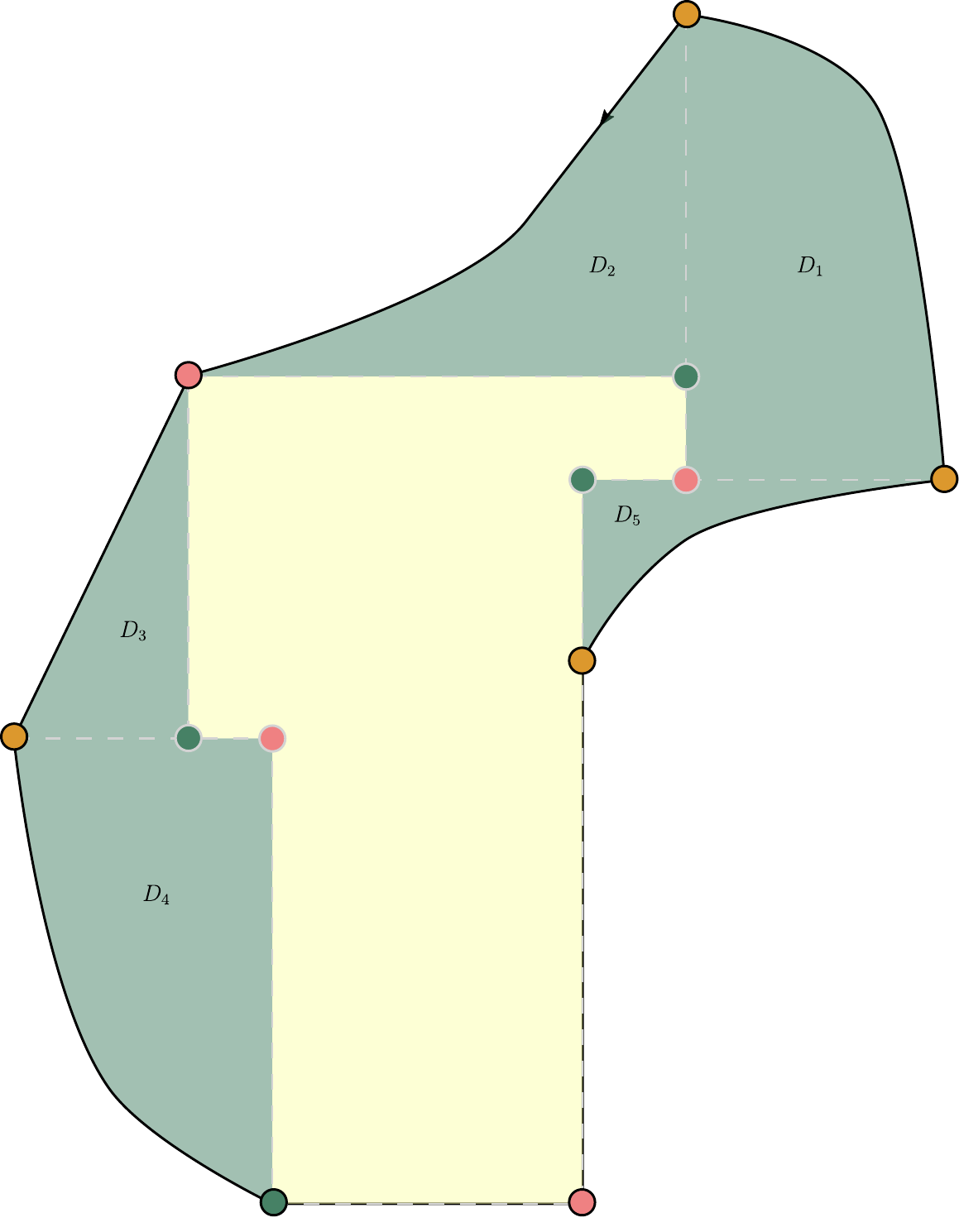}
\caption{An illustration of the semi-discrete line integral applied to each segment of a monotonic division of the detachable curve from figure \ref{monotonic_division_figure}. The $D_1=D^+\left(\ell_1\bigcup\ell_2\right)$, $D_2=D^+\left(\ell_3\bigcup\ell_4\right)$, $D_3=D^+\left(\ell_5\right)$, $D_4=D^+\left(\ell_6\right)$ and $D_5=D^+\left(\ell_{10}\bigcup\ell_{11}\right)$ are the positive domains of the curve's respective subsegments.}
\label{monotonic_division_with_domains_figure}
\end{figure}

\section{Establishing a Semi-discrete Version of Green's Theorem} 
In this section we apply the definition of the semi-discrete line integral to extend theorem \ref{wang} to general domains rather than merely rectangular ones, as in  corollary \ref{semi_discrete_green}. We already saw the final result in corollary \ref{semi_discrete_green}, and we will aim to arrive at that formula via the definition of the integral, to obtain a semi-discrete version of Green's theorem.
We begin by formulating the following lemma.

\begin{lem}\label{ftc_lemma}Let $\left\{ \ell_{i}\right\} _{1\leq i\leq n}$ be a monotonic
division of a closed and detachable curve $\gamma$. Let $f:\mathbb{R}^{2}\rightarrow\mathbb{R}$
be a function that admits an antiderivative $F$. Let $M,N,O$ be
the endpoints of the curves $\ell_{1},\ell_{2}$ respectively (where
$M=\ell_{1}\bigcap\ell_{2}$). Let:
\[
\alpha\equiv\ell_{1}\bigcup\ell_{2}\bigcup\overrightarrow{MO},
\]
and:
\[
\beta\equiv \bigcup_{i=1}^{n} \ell_{i}\bigcup\overrightarrow{OM}.
\]

Then:
\[
\underset{\gamma}{\sqint}F=\underset{\alpha}{\sqint}F+\underset{\beta}{\sqint}F.
\]
\end{lem}

\begin{proof}. We know that $\underset{\gamma}{\sqint}F-\underset{\beta}{\sqint}F-\underset{\alpha}{\sqint}F=0.$
Let us evaluate these terms:
\begin{align}\label{ftc_lemma_equations}
\begin{aligned}
\underset{\alpha}{\sqint}F & =\underset{\ell_{1}\subset\alpha}{\sqint}F+\underset{\ell_{2}\subset\alpha}{\sqint}F+\underset{\overrightarrow{MO}\subset\alpha}{\sqint}F,\\
\underset{\beta}{\sqint}F & =\underset{\overrightarrow{OM}\subset\alpha}{\sqint}F+\underset{\ell_{3}\subset\beta}{\sqint}F+\underset{\bigcup_{i=4}^{n-1}\ell_{i}\subset\beta}{\sqint}F+\underset{\ell_{n}\subset\beta}{\sqint}F,\\
\underset{\gamma}{\sqint}F & =\underset{\ell_{1}\subset\gamma}{\sqint}F+\underset{\ell_{2}\subset\gamma}{\sqint}F+\underset{\ell_{3}\subset\gamma}{\sqint}F+\underset{\bigcup_{i=4}^{n-1}\ell_{i}\subset\gamma}{\sqint}F+\underset{\ell_{n}\subset\gamma}{\sqint}F.
\end{aligned}
\end{align}
For readability and without loss of generality, let us assume that
$\gamma$ is structured as depicted in Figure \ref{ftc_lemma}.

By definition, the semi-discrete line integral of $F$ over $\ell_{1}$
in the context of the curve $\gamma$ equals:
\[
\underset{\ell_{1}\subset\gamma}{\sqint}F=\underset{D\left(\ell_{1}\right)}{\iint}f\boldsymbol{dx}-\ell_{1}^{;}F\left(O'\right)+\frac{1}{2}\left[\gamma^{;}\left(N\right)F\left(N\right)+\gamma^{;}\left(O\right)F\left(O\right)\right],
\]
where $\ell_{1}^{;}$ is the detachment of the monotonic curve $\ell_{1}$,
and $\gamma^{;}\left(N\right),\gamma^{;}\left(O\right)$ are the detachments
of the points $N$ and $O$ in the context of the curve $\gamma$,
respectively.

Similarly, the discrete line integral of $F$ over $\ell_{1}$ in
the context of the curve $\alpha$ is:
\[
\underset{\ell_{1}\subset\alpha}{\sqint}F=\underset{D\left(\ell_{1}\right)}{\iint}f\boldsymbol{dx}-\ell_{1}^{;}F\left(O'\right)+\frac{1}{2}\left[\alpha^{;}\left(N\right)F\left(N\right)+\alpha^{;}\left(O\right)F\left(O\right)\right],
\]
where $\alpha^{;}\left(N\right),\alpha^{;}\left(O\right)$ are the
detachments at the points $N$ and $O$ in the context of the curve
$\alpha$, respectively.

Hence:
\[
\underset{\ell_{1}\subset\gamma}{\sqint}F-\underset{\ell_{1}\subset\alpha}{\sqint}F=\frac{1}{2}\left[\gamma^{;}\left(O\right)-\alpha^{;}\left(O\right)\right]F\left(O\right)=\frac{1}{2}\left(+1-0\right)F\left(O\right)=\frac{1}{2}F\left(O\right).
\]

Similarly:
\begin{align}
\begin{aligned}
\underset{\ell_{1}\subset\gamma}{\sqint}F-\underset{\ell_{2}\subset\alpha}{\sqint}F & =\frac{1}{2}\left[\gamma^{;}\left(M\right)-\alpha^{;}\left(M\right)\right]F\left(M\right)=\frac{1}{2}\left[-1-\left(-1\right)\right]F\left(M\right)=0,\\
\underset{\bigcup_{i=4}^{n-1}\ell_{i}\subset\gamma}{\sqint}F-\underset{\bigcup_{i=4}^{n-1}\ell_{i}\subset\beta}{\sqint}F & =0,\\
\underset{\ell_{3}\subset\gamma}{\sqint}F-\underset{\ell_{3}\subset\beta}{\sqint}F & =\frac{1}{2}\left[\gamma^{;}\left(M\right)-\beta^{;}\left(M\right)\right]F\left(M\right)=\frac{1}{2}\left[-1-\left(-1\right)\right]F\left(M\right)=0\\
\underset{\ell_{n}\subset\gamma}{\sqint}F-\underset{\ell_{n}\subset\alpha}{\sqint}F & =\frac{1}{2}\left[\gamma^{;}\left(O\right)-\beta^{;}\left(O\right)\right]F\left(O\right)=\frac{1}{2}\left[+1-0\right]F\left(O\right)=0.
\end{aligned}
\end{align}

Thus, upon placing these values in equations \ref{ftc_lemma_equations}, we have:
\[
\underset{\gamma}{\sqint}F-\underset{\beta}{\sqint}F-\underset{\alpha}{\sqint}F=F\left(O\right)-\left(\underset{\overrightarrow{OM}\subset\beta}{\sqint}F+\underset{\overrightarrow{MO}\subset\alpha}{\sqint}F\right).\label{ftc_proof4}
\]
Once again, according to the definition of the semi-discrete line integral:
\begin{align}
\begin{aligned}
\underset{\overrightarrow{OM}\subset\beta}{\sqint}F & =\underset{OMM''}{\iint}f\boldsymbol{dx}-\left(\overrightarrow{OM}\right)^{;}F\left(M''\right)+\frac{1}{2}\left[\beta^{;}\left(O\right)F\left(O\right)+\beta^{;}\left(M\right)F\left(M\right)\right],\\
\underset{\overrightarrow{MO}\subset\alpha}{\sqint}F & =\underset{OMM''}{\iint}f\boldsymbol{dx}-\left(\overrightarrow{MO}\right)^{;}F\left(M'\right)+\frac{1}{2}\left[\alpha^{;}\left(O\right)F\left(O\right)+\alpha^{;}\left(M\right)F\left(M\right)\right].
\end{aligned}
\end{align}
Hence:
\[
\underset{\overrightarrow{OM}\subset\beta}{\sqint}F+\underset{\overrightarrow{MO}\subset\alpha}{\sqint}F=\underset{OM'MM''}{\iint}f\boldsymbol{dx}+F\left(M'\right)+F\left(M''\right)-F\left(M\right).
\]
Placing this result in equation \ref{ftc_proof4} yields:
\[
\underset{\gamma}{\sqint}F-\underset{\beta}{\sqint}F-\underset{\alpha}{\sqint}F=F\left(O\right)-F\left(M'\right)-F\left(M''\right)+F\left(M\right)-\underset{OM'MM''}{\iint}f\boldsymbol{dx}=0,
\]
where the last transition is due to theorem \ref{wang}.
This shows that, under our assumption, $\underset{\gamma}{\sqint}F=\underset{\alpha}{\sqint}F+\underset{\beta}{\sqint}F$,
and we are done.

\begin{figure}[H]
\includegraphics[scale=0.85]{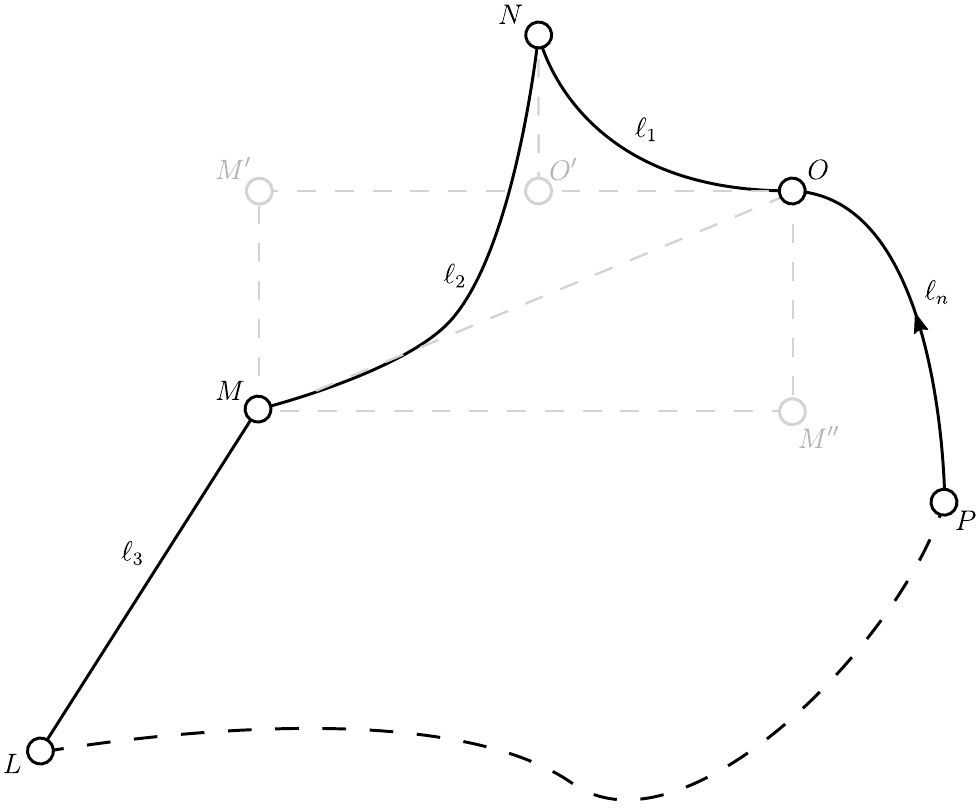}
\caption{An illustration of the proof of lemma \ref{ftc_lemma}}
\label{ftc_lemma_figure}
\end{figure}

We made the following assumptions throughout the proof:
\[
\gamma^{;}\left(O\right)=+1,\quad\gamma^{;}\left(M\right)=-1,\quad\delta\ell_{1}=\left(-1,+1,+1,-1\right),\quad\delta\ell_{2}=\left(-1,+1,-1,+1\right).
\]
Other cases left to be handled are different versions of the
curve $\gamma$, that vary for different values of $\delta\ell_{1},\delta\ell_{2}$
(each accepts only a few values according to corollary \ref{corollary_about_detachment_vector}) and $\gamma^{;}\left(M\right),\gamma^{;}\left(O\right)$.
There are $\frac{8^{2}\cdot3^{2}}{4}=144$ different cases, where
the division is due to symmetry-induced redundancy. The proof is complete
by a computerized inspection of the other cases which is similar to
the analysis we introduced.
\end{proof}

Now, we may formulate the extension of theorem \ref{wang}
in $\mathbb{R}^{2}$ that relies on the definition of the semi-discrete
line integral. Let us get inspired by the formulation of Green's theorem:
\index{Green's Theorem}
\begin{theorem}[Green's theorem]Let $C$ be a positively oriented, piecewise smooth, simple closed
curve in a plane, and let $D$ be the region bounded by $C$. If $L$
and $M$ are functions of $\left(x,y\right)$ defined on an open region
containing $D$ and having continuous partial derivatives, then

\[
\underset{C}{\ointctrclockwise}{\displaystyle \left(L\,dx+M\,dy\right)}{\displaystyle =\underset{D}{\iint}\left(\frac{\partial M}{\partial x}-\frac{\partial L}{\partial y}\right)dx\,dy},
\]
 where the path of integration along $C$ is anticlockwise.
\label{green_theorem}
\end{theorem}

Let us suggest the following semi-discrete line integral-based analogue to Green's theorem.
\begin{theorem}[A semi-discrete Green's theorem]Let $D\subset\mathbb{R}^{2}$ be a domain whose edge is detachable.
Let $f:\mathbb{R}\rightarrow\mathbb{R}$ be a function that admits
an antiderivative $F$. Then:
\[
\underset{D}{\iint}f\boldsymbol{dx}=\underset{\partial D}{\sqint}F.
\]
\label{ftc_theorem}
\end{theorem}

\begin{proof}Let $\left\{ \ell_{i}\right\} _{1\leq i\leq n}$ be the minimal
monotonic division of $\partial D$. Let us suggest a proof by induction
on $n$. The case where $n=1$ is degenerated.

\begin{figure}
\includegraphics[scale=0.9]{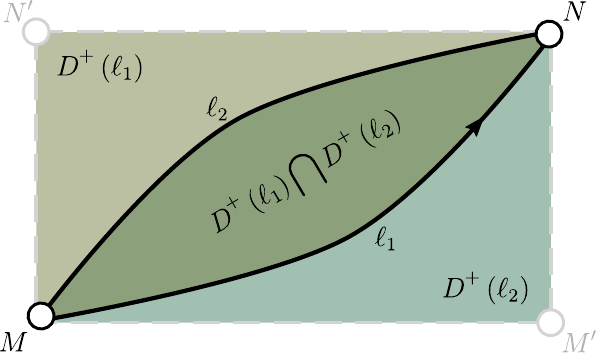}
\caption{An illustration of the proof of theorem \ref{ftc_theorem} for $n=2$, where the domain’s edge is $\ell_{1}\bigcup\ell_{2}$.}
\label{ftc_proof1_fig}
\end{figure}

For $n=2$, without loss of generality let us assume the case illustrated
in figure \ref{ftc_proof1_fig}. We have that:
\begin{align*}
\underset{\ell_{1}}{\sqint}F & =\underset{D\left(\ell_{1}\right)}{\iint}f\boldsymbol{dx}-\ell_{1}^{;}F\left(N'\right)+\frac{1}{2}\left[\gamma^{;}\left(M\right)F\left(M\right)+\gamma^{;}\left(N\right)F\left(N\right)\right],\\
\underset{\ell_{2}}{\sqint}F & =\underset{D\left(\ell_{2}\right)}{\iint}f\boldsymbol{dx}-\ell_{2}^{;}F\left(M'\right)+\frac{1}{2}\left[\gamma^{;}\left(M\right)F\left(M\right)+\gamma^{;}\left(N\right)F\left(N\right)\right].
\end{align*}

Hence, in the illustrated case, we have that:
\[
\underset{D}{\iint}f\boldsymbol{dx}=\underset{\ell_{1}}{\sqint}F+\underset{\ell_{2}}{\sqint}F=\underset{NN'MM'}{\iint}f\boldsymbol{dx}+\underset{D}{\iint}f\boldsymbol{dx}+F\left(M'\right)+F\left(N'\right)-F\left(M\right)-F\left(N\right).
\]
However, according to theorem \ref{wang}, $\underset{NN'MM'}{\iint}f\boldsymbol{dx}=F\left(M\right)+F\left(N\right)-F\left(N\right)-F\left(M'\right)-F\left(N'\right)$,
hence:
\[
\underset{D}{\iint}f\boldsymbol{dx}=\underset{\partial D}{\sqint}F.
\]

For $n=3$, without loss of generality, we consider three cases. The
rest are handles similarly.

\begin{figure}
\includegraphics[scale=0.8]{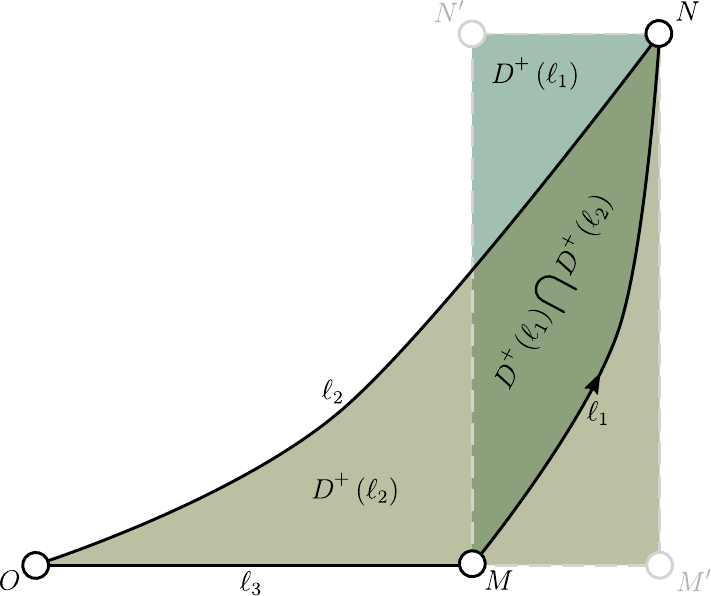}
\caption{An illustration of the proof of theorem \ref{ftc_theorem} for $n=3$ (the first case), where the domain’s edge is $\bigcup_{i=1}^{3}\ell_{i}$.}
\label{ftc_proof2_fig}
\end{figure}

In case 1, as depicted in figure \ref{ftc_proof2_fig}, we have that:
\begin{align}
\begin{aligned}
\underset{\ell_{1}}{\sqint}F & =\underset{D\left(\ell_{1}\right)}{\iint}f\boldsymbol{dx}-\ell_{1}^{;}F\left(N'\right)+\frac{1}{2}\left[\gamma^{;}\left(M\right)F\left(M\right)+\gamma^{;}\left(N\right)F\left(N\right)\right],\\
\underset{\ell_{2}}{\sqint}F & =\underset{D\left(\ell_{2}\right)}{\iint}f\boldsymbol{dx}-\ell_{1}^{;}F\left(M'\right)+\frac{1}{2}\left[\gamma^{;}\left(N\right)F\left(N\right)+\gamma^{;}\left(O\right)F\left(O\right)\right],\\
\underset{\ell_{3}}{\sqint}F & =\frac{1}{2}\left[\gamma^{;}\left(O\right)F\left(O\right)+\gamma^{;}\left(M\right)F\left(M\right)\right].
\end{aligned}
\end{align}
Hence, in the illustrated case, it holds that:
\begin{align}
\begin{aligned}
\underset{\partial D}{\sqint}F & =\underset{\ell_{1}}{\sqint}F+\underset{\ell_{2}}{\sqint}F+\underset{\ell_{3}}{\sqint}F\\
 & =\underset{D\left(\ell_{1}\right)\bigcup D\left(\ell_{2}\right)}{\iint}f\boldsymbol{dx}-\left[F\left(N\right)-F\left(N'\right)+F\left(M\right)-F\left(M'\right)\right]\\
 & =\underset{D}{\iint}f\boldsymbol{dx},
\end{aligned}
\end{align}
where the last transition is, once again, due to theorem \ref{wang}.

\begin{figure}
\includegraphics[scale=0.8]{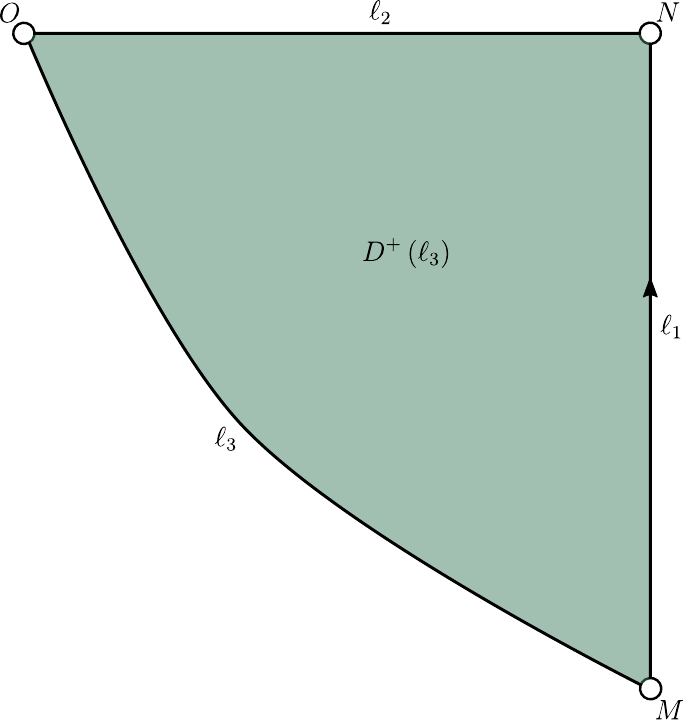}
\caption{An illustration of the proof of theorem \ref{ftc_theorem} for $n=3$ (the second case), where the domain’s edge is $\bigcup_{i=1}^{3}\ell_{i}$.}
\label{ftc_proof3_fig}
\end{figure}

In case 2, as depicted in Figure \ref{ftc_proof3_fig}, we have that:
\begin{align}
\begin{aligned}
\underset{\ell_{1}}{\sqint}F & =\frac{1}{2}\left[\gamma^{;}\left(M\right)F\left(M\right)+\gamma^{;}\left(N\right)F\left(N\right)\right],\\
\underset{\ell_{2}}{\sqint}F & =\frac{1}{2}\left[\gamma^{;}\left(N\right)F\left(N\right)+\gamma^{;}\left(O\right)F\left(O\right)\right],\\
\underset{\ell_{3}}{\sqint}F & =\underset{D\left(\ell_{3}\right)}{\iint}f\boldsymbol{dx}-\ell_{3}^{;}F\left(N\right)+\frac{1}{2}\left[\gamma^{;}\left(O\right)F\left(O\right)+\gamma^{;}\left(M\right)F\left(M\right)\right].
\end{aligned}
\end{align}
Hence, in the illustrated case, it holds that:
\[
\underset{\partial D}{\sqint}F=\underset{\ell_{1}}{\sqint}F+\underset{\ell_{2}}{\sqint}F+\underset{\ell_{3}}{\sqint}F=\underset{D\left(\ell_{3}\right)}{\iint}f\boldsymbol{dx}+f\left(N\right)-F\left(N\right)=\underset{D}{\iint}f\boldsymbol{dx}.
\]

\begin{figure}
\includegraphics[scale=1.1]{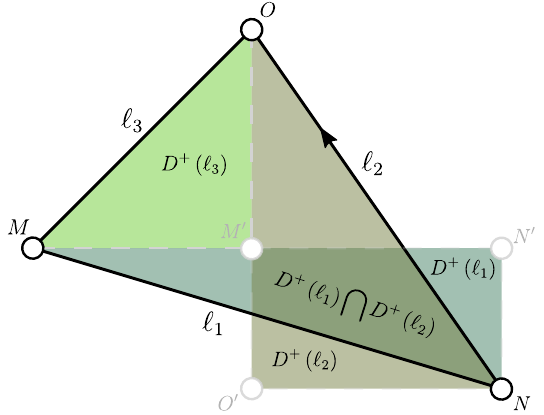}
\caption{An illustration of the proof of theorem \ref{ftc_theorem} for $n=3$ (the third case), where the domain’s edge is $\bigcup_{i=1}^{3}\ell_{i}$.}
\label{ftc_proof4_fig}
\end{figure}

In case 3, as depicted in figure \ref{ftc_proof4_fig}, we have:
\begin{align}
\begin{aligned}
\underset{\ell_{1}}{\sqint}F & =\underset{D\left(\ell_{1}\right)}{\iint}f\boldsymbol{dx}-\ell_{1}^{;}F\left(N'\right)+\frac{1}{2}\left[\gamma^{;}\left(M\right)F\left(M\right)+\gamma^{;}\left(N\right)F\left(N\right)\right],\\
\underset{\ell_{2}}{\sqint}F & =\underset{D\left(\ell_{2}\right)}{\iint}f\boldsymbol{dx}-\ell_{2}^{;}F\left(O'\right)+\frac{1}{2}\left[\gamma^{;}\left(N\right)F\left(N\right)+\gamma^{;}\left(O\right)F\left(O\right)\right],\\
\underset{\ell_{3}}{\sqint}F & =\underset{D\left(\ell_{3}\right)}{\iint}f\boldsymbol{dx}-\ell_{3}^{;}F\left(M'\right)+\frac{1}{2}\left[\gamma^{;}\left(O\right)F\left(O\right)+\gamma^{;}\left(M\right)F\left(M\right)\right].
\end{aligned}
\end{align}
Hence, in the illustrated case it holds that:
\[
\underset{\partial D}{\sqint}F=\underset{\ell_{1}}{\sqint}F+\underset{\ell_{2}}{\sqint}F+\underset{\ell_{3}}{\sqint}F=\underset{D\left(\ell_{1}\right)\bigcup D\left(\ell_{2}\right)}{\iint}f\boldsymbol{dx}-\underset{D\left(\ell_{3}\right)}{\iint}f\boldsymbol{dx}=\underset{D}{\iint}f\boldsymbol{dx}.
\]

Let us now apply the induction's step. Suppose that the theorem holds
for any domain whose boundary consists of less than $n$ monotonic
subcurves. Let $D$ be a domain whose boundary, $\partial D$, is
written as a minimal monotonic division of $n+1$ monotonic subcurves.
Let us divide $D$ into two subdomains, $D_{\alpha}$ and $D_{\beta}$
, by connecting (via a straight line) between the disjoint endpoints
of two adjacent monotonic subcurves of the division (such as the line
$\overrightarrow{OM}$ in figure \ref{ftc_lemma_figure}). According to lemma \ref{ftc_lemma}, it holds
that:
\[
\underset{\partial D}{\sqint}F=\underset{\partial D_{\alpha}}{\sqint}F+\underset{\partial D_{\beta}}{\sqint}F,
\]

however, according to the induction hypothesis, and since $\partial D_{\alpha}$
and $\partial D_{\beta}$ both consist of at most $n$ monotonic subcurves,
it holds that:
\[
\underset{\partial D_{\alpha}}{\sqint}F=\underset{D_{\alpha}}{\iint}f\boldsymbol{dx},\quad\underset{\partial D_{\beta}}{\sqint}F=\underset{D_{\beta}}{\iint}f\boldsymbol{dx}.
\]
Furthermore, according to the definition, $D_{\alpha}\bigcup D_{\beta}=D$,
hence:
\[
\underset{\partial D}{\sqint}F=\underset{\partial D_{\alpha}}{\sqint}F+\underset{\partial D_{\beta}}{\sqint}F=\underset{D\left(\ell_{1}\right)\bigcup D\left(\ell_{2}\right)}{\iint}f\boldsymbol{dx}=\underset{D}{\iint}f\boldsymbol{dx}.
\]
\end{proof}

\begin{remark}Note that the formulation of theorem \ref{ftc_theorem} is a semi-discrete
analog of Green's theorem; it relates the double integral of a function
in a domain with the semi-discrete line integral of the antiderivative
along the domain's edge. This is a concise version of corollary \ref{semi_discrete_green}.
One may opt to apply either version, according to the use case and
requirements. The formulation of the former is concise due to the
encapsulation embodied in the semi-discrete line integral. The later,
however, is slightly more verbose; however, it may be more convenient
in cases where one prefers to refrain from applying this new integration
method.
\end{remark}

\begin{exm}Let us apply the semi-discrete line integral to a domain
formed by a finite unification of rectangles, and see in which sense
theorem \ref{ftc_theorem} consolidates with theorem \ref{wang} for such
domains. Let $f:\mathbb{R}^{2}\rightarrow\mathbb{R}$ be a function
that admits an antiderivative $F$.

For a rectangle $ABCD$ whose edges are parallel to the axes, it holds
that:
\begin{align}
\begin{aligned}
\underset{BADC}{\sqint}F & =\underset{\overrightarrow{BA}}{\sqint}F+\underset{\overrightarrow{AD}}{\sqint}F+\underset{\overrightarrow{DC}}{\sqint}F+\underset{\overrightarrow{CB}}{\sqint}F\\
 & =\frac{1}{2}\left[+F\left(B\right)-F\left(A\right)\right]+\frac{1}{2}\left[+F\left(D\right)-F\left(A\right)\right]\\
 & \quad+\frac{1}{2}\left[+F\left(D\right)-F\left(C\right)\right]+\frac{1}{2}\left[+F\left(B\right)-F\left(C\right)\right]\\
 & =F\left(B\right)+F\left(D\right)-\left[F\left(A\right)+F\left(C\right)\right].
\end{aligned}
\end{align}

More generally, applying the semi-discrete line integral to the edge
of a rectangular domain results in a linear combination of the antiderivative
at the corners, where each coefficient (that equals the value of $\alpha_{D}$
) is determined according to the detachments (each half is provided
by the discrete line integral over curves from either side of the
corner). This case is depicted in figure \ref{ftc_on_grd}.
\end{exm}

\begin{figure}
\includegraphics[scale=0.8]{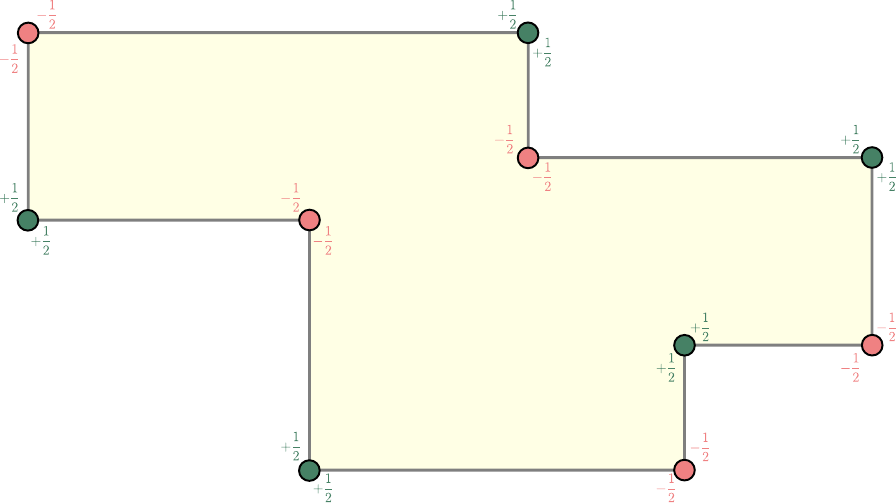}
\caption{An illustration of theorem \ref{ftc_theorem} for a finite unification of rectangles}
\label{ftc_on_grd}
\end{figure}

\begin{exm}Finally, let us verify the theorem's correctness for non-convex domains. Let us apply the semi-discrete line integral to a detachable
curve $\gamma$ (where $\gamma=\partial D$), as depicted in figure
\ref{concave_example}. Let $f:\mathbb{R}^{2}\rightarrow\mathbb{R}$ be a function that
admits an antiderivative $F$.

\begin{figure}
\includegraphics[scale=0.9]{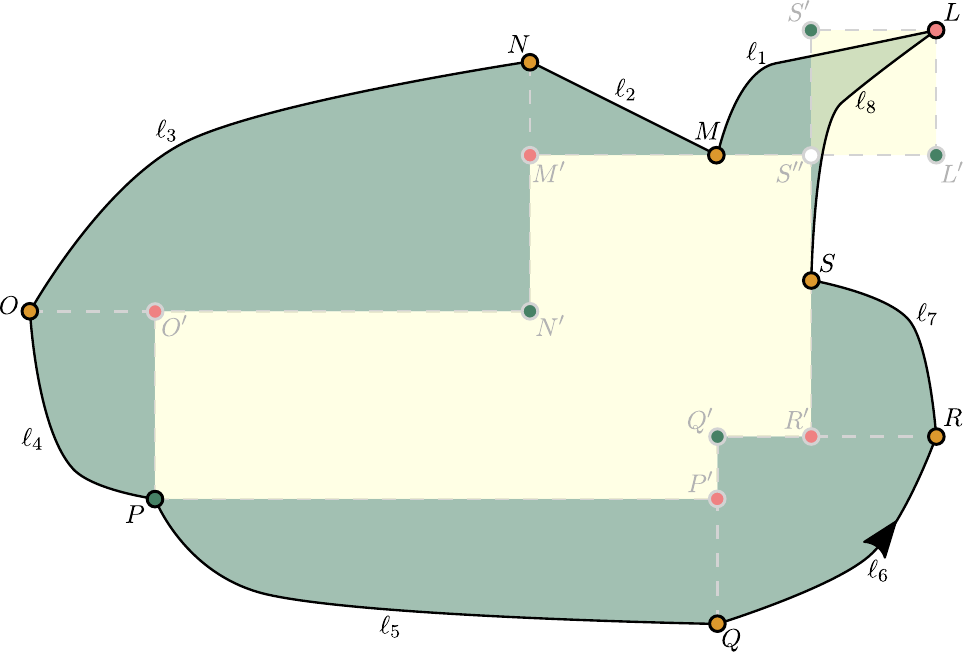}
\caption{An illustration of theorem \ref{ftc_theorem} with a domain that is partly convex and partly convex, where the domain’s edge is $\bigcup_{i=1}^{8}\ell_{i}$.}
\label{concave_example}
\end{figure}

It holds that:
\begin{align}
\begin{aligned}
\underset{\gamma_{1}}{\sqint}F & =\underset{D\left(\gamma_{1}\right)}{\iint}f\boldsymbol{dx}-\frac{1}{2}F\left(L\right)+F\left(L'\right),\\
\underset{\gamma_{2}}{\sqint}F & =\underset{D\left(\gamma_{2}\right)}{\iint}f\boldsymbol{dx}-F\left(M'\right),\\
\underset{\gamma_{3}}{\sqint}F & =\underset{D\left(\gamma_{3}\right)}{\iint}f\boldsymbol{dx}+F\left(N'\right),\\
\underset{\gamma_{4}}{\sqint}F & =\underset{D\left(\gamma_{4}\right)}{\iint}f\boldsymbol{dx}+\frac{1}{2}F\left(P\right)-F\left(O'\right),\\
\underset{\gamma_{5}}{\sqint}F & =\underset{D\left(\gamma_{5}\right)}{\iint}f\boldsymbol{dx}+\frac{1}{2}F\left(P\right)-F\left(P'\right),\\
\underset{\gamma_{6}}{\sqint}F & =\underset{D\left(\gamma_{6}\right)}{\iint}f\boldsymbol{dx}+F\left(Q'\right),\\
\underset{\gamma_{7}}{\sqint}F & =\underset{D\left(\gamma_{7}\right)}{\iint}f\boldsymbol{dx}-F\left(R'\right),\\
\underset{\gamma_{8}}{\sqint}F & =\underset{D\left(\gamma_{8}\right)}{\iint}f\boldsymbol{dx}-\frac{1}{2}F\left(L\right)+F\left(S'\right).
\end{aligned}
\end{align}
Adding up these equations, while considering the equality
\[
\underset{L'S'S''L}{\iint}f\boldsymbol{dx}=F\left(S\right)-F\left(S'\right)+F\left(S''\right)-F\left(L'\right)
\]
and applying theorem \ref{wang}, results with $\underset{\gamma}{\sqint}F=\underset{D}{\iint}f\boldsymbol{dx}$
as theorem \ref{ftc_theorem} states. Note that in the concave portion of the domain, the theorem subtracts the function's integral over the rectangular domain $LS'S''L'$. This is the reason that the detachments of the curve at the point $L$ and along the curves $\ell_1,\ell_8$ are flipped by construction.
\end{exm}

\begin{figure}[htp]

\subfloat{
  \includegraphics[clip,width=0.75\columnwidth]{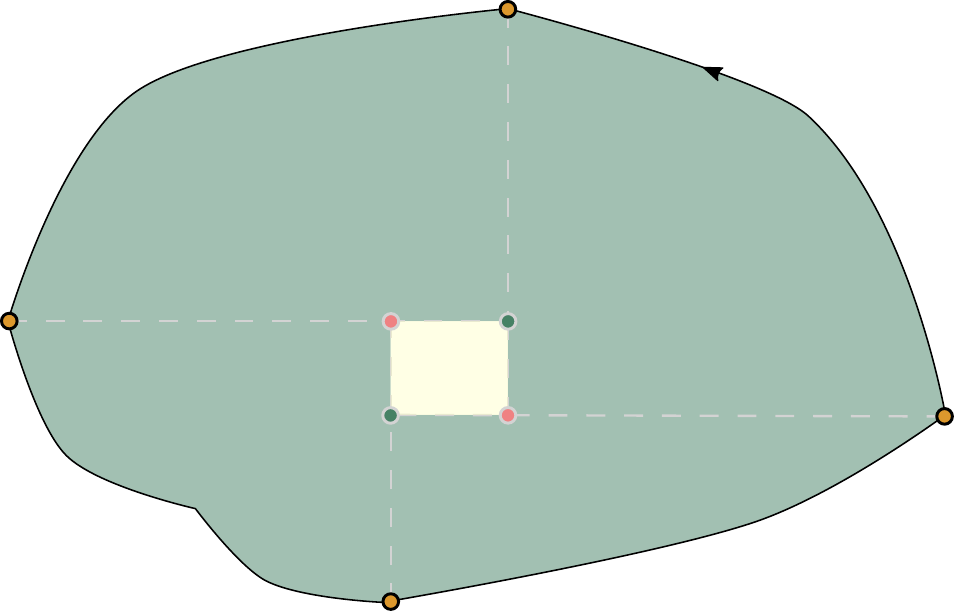}
}

\subfloat{%
  \includegraphics[clip,width=0.75\columnwidth]{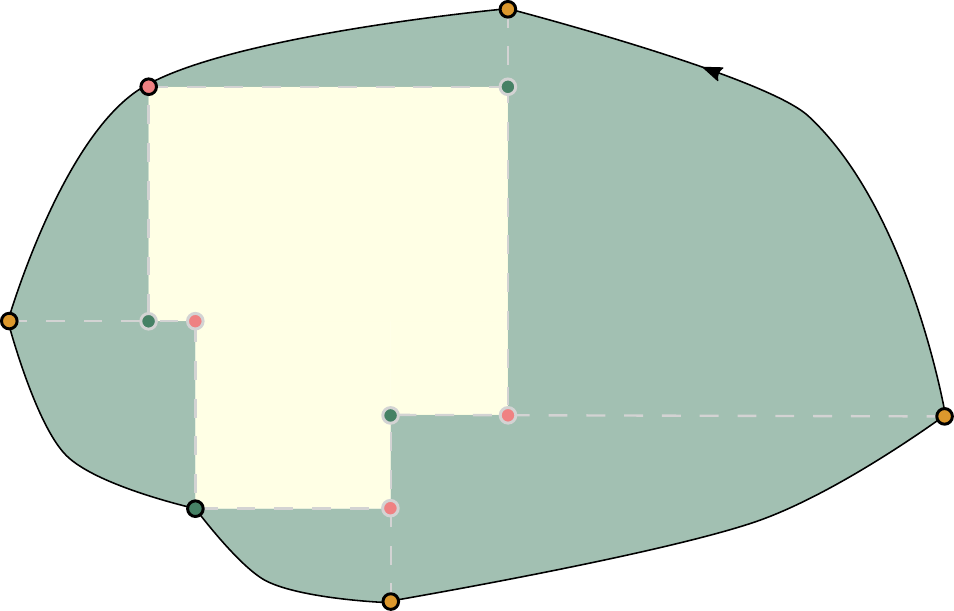}%
}

\subfloat{%
  \includegraphics[clip,width=0.75\columnwidth]{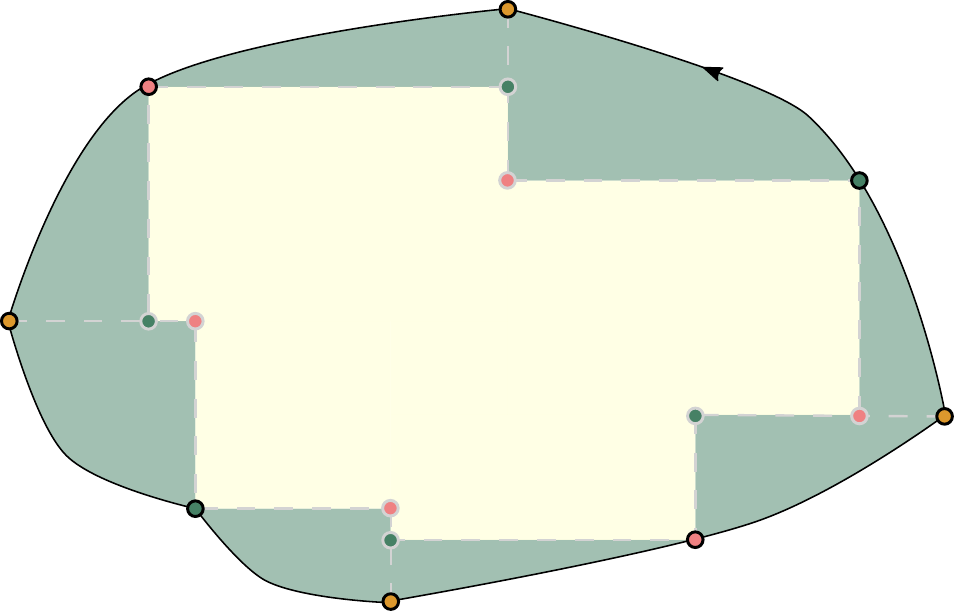}%
}

\caption{An illustration of different monotonic divisions of the same closed curve (the upper is the minimal such division) when applied in theorem \ref{ftc_theorem}.  The subtler the division is, the larger the internal rectangular domain becomes - over whom the integral is calculated by aggregating the antiderivative’s values at the endpoints of each monotonic curve’s (positive) local domain.}

\end{figure}

\chapterimage{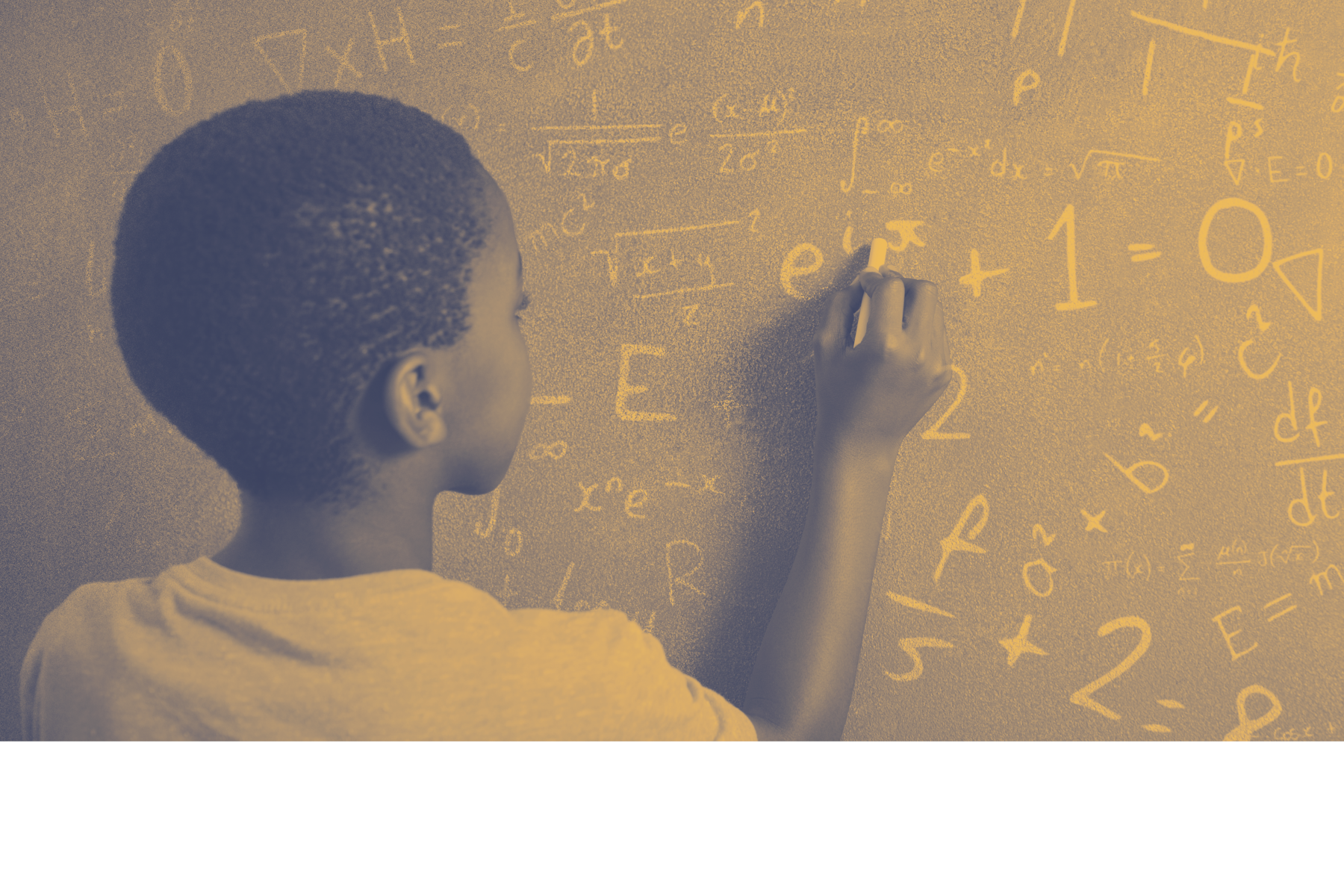}
\chapter{Semi-discrete Complex Analysis}

Can we apply the detachment operator in the Complex plane? As in Real Analysis, a theory based on the derivative sign would go through rates calculations and be trivial. But what if we can introduce a theory about the properties of “complex trends” while calculating them directly without going through the derivative? Would this theory apply to non-differentiable and even discontinuous functions, similar to the case in Real Analysis? And would it capture trends in the function's image more reliably and coherently than the derivative sign?

Complex Analysis is abundant with concepts that generalize their Real Analysis counterparts, such as continuity and the derivative. Inspired by its AI, natural sciences, and Real Analysis applications, let us consider a complex version of a function's detachment and analyze its basic mathematical properties.

\index{Detachment}
\section{Defining the Complex Detachment}
\begin{figure}
\includegraphics[scale=0.8]{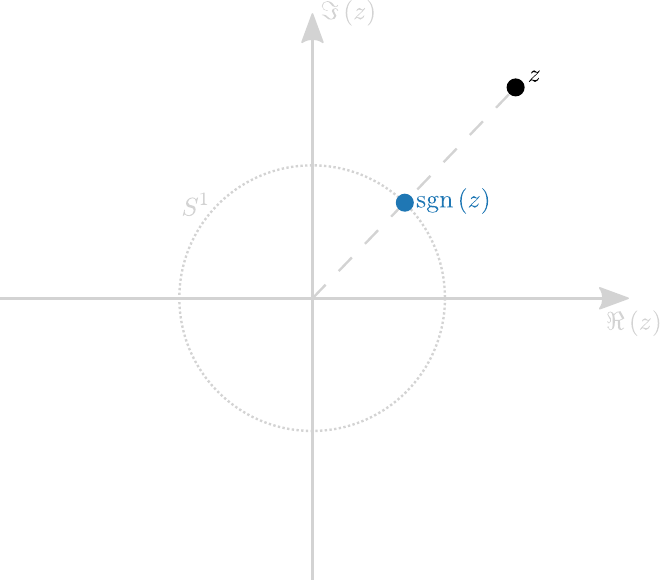}
\caption{An illustration of the complex signum function}
\label{complex_sign_fig}
\end{figure}

Recall that the signum of a complex number $z$ is defined as:

$$\text{sgn}\left(z\right)\equiv\frac{z}{\left|z\right|}=e^{iarg\left(z\right)},$$

where $arg\left(\cdot\right)$ is the complex argument function. Geometrically, this means that the sign function maps any point in the complex plane to its closest counterpart on the unit circle, as illustrated in Figure \ref{complex_sign_fig}.

Note that in Real Analysis, we suggested softening the definition of the detachment with respect to the derivative by considering merely one-sided detachments, defined by one-sided limits, and did not require them to agree. We naturally extend the notion of one-sided derivatives in higher dimensions with directional derivatives. The Complex Analysis literature conventionally focuses on differentiable and, more specifically, analytic functions. However, complex partial derivatives, also known as Wirtinger derivatives, are prevalent. They particularly serve to define the complex directional derivatives, which are useful as well, depending on the context. For example, they are applied in advanced research (see \cite{dructu2018geometric}), and also serve as a pedagogical and auxiliary thought tool (see \cite{shabat2003introduction}). Let us suggest an analogous natural way to define the directional detachment.

\begin{definition}[Detachment of a complex function]\label{complex_detachment_definition} Let us define the $\varphi$-detachment of a function $f:\mathbb{C}\longrightarrow\mathbb{C}$ at a point $z_{0}\in\mathbb{C}$ as follows:

$$\begin{array}{ccc}
& f_{\varphi}^{;}:\mathbb{C}\longrightarrow S^{1}\bigcup\left\{ 0\right\} \\
& f_{\varphi}^{;}\left(z_{0}\right)\equiv e^{-i\varphi}\underset{\begin{array}{c}
\left|z\right|\to\left|z_{0}\right|\\
\arg\left(\Delta z\right)=\varphi
\end{array}}{\lim}\text{sgn}\left[f\left(z\right)-f\left(z_{0}\right)\right],
\end{array}$$

where $S^{1}$ is the unit circle in the complex plane. We will say that $f$ is $\varphi$-detachable at $z_{0}$ if the limit exists.
\end{definition}

We may think of the $e^{-i\varphi}$ coefficient as a generalization of the $\pm 1$ coefficient from the one-sided detachment in Real Analysis. Both coefficients render the detachment more consistent with the derivative sign, as shown in the following statement.

\begin{figure}[H]
\includegraphics[scale=1.3 ]{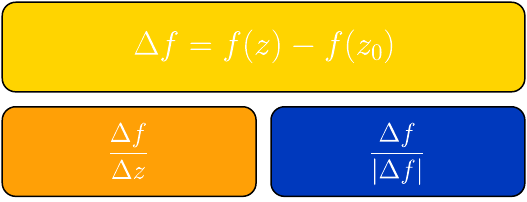}
\caption{In the complex detachment we divide $\Delta f$ by its norm (as in the definition of the signum function), rather than by $\Delta z$ as in the complex derivative.}
\end{figure}

The following claim shows that the detachment adds a calculation value with respect to the derivative only in cases where the derivative vanishes.

\begin{clm}If $f:\mathbb{C}\longrightarrow\mathbb{C}$ is both differentiable and $\varphi$-detachable at the point $z_{0},$ and $f'\left(z_{0}\right)\neq0,$ then:

$$f_{\varphi}^{;}\left(z_{0}\right)=\text{sgn}\left[\frac{\partial f}{\partial\varphi}\left(z_{0}\right)\right].$$
\end{clm}
\begin{proof}According to the definition:

\begin{align}
&\begin{aligned}
f_{\varphi}^{;}\left(z_{0}\right) &\equiv e^{-i\varphi}\underset{\begin{array}{c}
\left|z\right|\to\left|z_{0}\right|\\
\arg\left(\Delta z\right)=\varphi
\end{array}}{\lim}\text{sgn}\left[f\left(z\right)-f\left(z_{0}\right)\right]=\underset{\begin{array}{c}
\left|z\right|\to\left|z_{0}\right|\\
\arg\left(\Delta z\right)=\varphi
\end{array}}{\lim}\frac{\text{sgn}\left[f\left(z\right)-f\left(z_{0}\right)\right]}{e^{i\varphi}}\\
&=\underset{\begin{array}{c}
\left|z\right|\to\left|z_{0}\right|\\
\arg\left(\Delta z\right)=\varphi
\end{array}}{\lim}\frac{\text{sgn}\left[f\left(z\right)-f\left(z_{0}\right)\right]}{\text{sgn}\left(z-z_{0}\right)}
=\underset{\begin{array}{c}
\left|z\right|\to\left|z_{0}\right|\\
\arg\left(\Delta z\right)=\varphi
\end{array}}{\lim}\text{sgn}\left[\frac{f\left(z\right)-f\left(z_{0}\right)}{z-z_{0}}\right] \\
&=\text{sgn}\underset{\begin{array}{c}
\left|z\right|\to\left|z_{0}\right|\\
\arg\left(\Delta z\right)=\varphi
\end{array}}{\lim}\left[\frac{f\left(z\right)-f\left(z_{0}\right)}{z-z_{0}}\right]=\text{sgn}\left[\frac{\partial f}{\partial\varphi}\left(z_{0}\right)\right],
\end{aligned}
\end{align}
where the fifth transition is because the sign function is continuous in $\mathbb{C}\backslash\left\{ 0\right\}$.
\end{proof}

For brevity, we will denote the $\varphi$-detachment at the point $z_{0}$ by $f^{;},$ rather than $f_{\varphi}^{;}\left(z_{0}\right)$.

\section{Geometric Intuition}
Suppose that a function is $\varphi$-detachable at $z_{0}.$ Then, by the limit definition, given $\epsilon>0,$ for any sufficiently small environment of $z_{0},$ the function's values at a point $z$, satisfy that:

$$\left|\text{sgn}\left[f\left(z\right)-f\left(z_{0}\right)\right]-f^{;}\right|<\epsilon.$$

This means that $f\left(z\right)$ is bounded between two lines intersecting at $z_{0}.$ The angle between the spanning vector of each line and $f^{;},$ is $\epsilon.$

Recall that, in the realm of the real functions, the detachment definition is equivalent to the following statement:

$$\exists\dot{B}_{\pm}\left(x_{0}\right):\forall x\in\dot{B}_{\pm}\left(x_{0}\right):\,\,\,\,\,\text{sgn}\left[f\left(x\right)-f\left(x_{0}\right)\right]=f_{\pm}^{;}\left(x_{0}\right),$$

due to the discontinuity of the sign function at zero.

However, an analogous statement in the complex domain does not necessarily hold. Such a statement is more strict than the existence of the limit stated in equation \ref{complex_detachment_definition}. Geometrically, the former is defined merely for functions whose image near a point is on a straight line. The reason lies in the different geometric interpretations of the discontinuity of the sign function in one vs. two dimensional functions.

We know that the limit $\underset{z\to0}{\lim}\text{sgn}\left(z\right)$ does not exists because it depends on the different paths by which we let $z$ approach zero. Can this intuition help us establish a geometric interpretation for the detachment?

The detachability of a real valued function at a point is a synonymous with the function having a “local trend”. The detachability of a complex function points out the existence of a “local trend” in a broader sense.

Intuitively, a $\theta$-detachable function maps points along a line in the environment of the point, to points “approaching” a single line.

The differentiability of a function at a point allows us to calculate its values with the linear approximation formula;  it states that the function's value is linear in the difference between the points (up to the function $\rho$):
$$f\left(z\right)=f\left(z_{0}\right)+f'\left(z_{0}\right)\left(z-z_{0}\right)+\rho\left(z\right)\left(z-z_{0}\right),$$
where $\underset{z\to z_{0}}{\lim}\rho\left(z\right)=0$ and $z\neq z_{0}.$

In contrast, the detachment suggests a different kind of approximated linear relation of the function's image, where the formula itself does not involve information from the domain. Suppose $f$ is $\varphi$-detachable, and let $\epsilon>0.$ According to the limit definition, we have that for $z$ in a sufficiently small environment of $z_{0}$ such that $\arg\left(z-z_{0}\right)=\varphi$:

$$\left|\text{sgn}\left[f\left(z\right)-f\left(z_{0}\right)\right]-f^{;} e^{i\varphi} \right|<\epsilon.$$

Since $f^{;}$ is the limit of points on the (closed) unit circle, it has to be on the circle as well. The size of the chord between $\text{sgn}\left[f\left(z\right)-f\left(z_{0}\right)\right]$ and $f^{;}e^{i\varphi}$ is at most $\epsilon.$ Let us assume, without loss of generality, that $0<\arg\left(f^{;}e^{i\varphi}\right)<\frac{\pi}{2}.$

Let $\theta_{f}$ be the angle between $\text{sgn}\left[f\left(z\right)-f\left(z_{0}\right)\right]$ and $f^{;}e^{i\varphi}.$ The formula for the size of a chord given the angle $\theta_{f}$ and radius $1$ yields:

$$\left|\text{sgn}\left[f\left(z\right)-f\left(z_{0}\right)\right]-f^{;}e^{i\varphi}\right|=2\cdot1\cdot\sin\left(\frac{\theta_{f}}{2}\right),$$
hence:
$$\theta_{f}=2\arcsin\left(\frac{\left|\text{sgn}\left[f\left(z\right)-f\left(z_{0}\right)\right]-f^{;}e^{i\varphi}\right|}{2}\right)<2\arcsin\left(\frac{\epsilon}{2}\right).$$
Therefore, we show that the argument of the difference between the function's image at points in the environment of $z_{0}$ and its image at $z_{0}$ satisfies: $$arg\left(f\left(z\right)-f\left(z_{0}\right)\right)\in\left(\arg\left(f^{;}e^{i\varphi}\right)-2\arcsin\left(\frac{\epsilon}{2}\right),\arg\left(f^{;}e^{i\varphi}\right)+2\arcsin\left(\frac{\epsilon}{2}\right)\right).\label{arg_delta_f_bounds}$$

We will refer to the above analysis, which associates the bound and the chord length with these of the angle, as a \textbf{“chord-angle”} analysis.

\section{Computability and Consistency}
Similar to the case in Real Analysis, detachability at a point does not necessarily implies continuity or differentiability - and vice versa. Let us illustrate some basic examples to gain a better understanding of why this is the case.

\begin{exm}Let us consider the following function:

$$f\left(z\right)=\begin{cases}
\frac{1}{z}, & z\neq0,\\
0, & z=0.
\end{cases}$$

Let us calculate the $\varphi$-detachments by the definition:

$$f_{\varphi}^{;}\left(0\right)\equiv e^{-i\varphi}\underset{\begin{array}{c}
\left|z\right|\to\left|z_{0}\right|\\
\arg\left(\Delta z\right)=\varphi
\end{array}}{\lim}\text{sgn}\left[f\left(z\right)-f\left(z_{0}\right)\right]=e^{-i\varphi}\cdot\underset{\begin{array}{c}
\left|z\right|\to\left|z_{0}\right|\\
\arg\left(\Delta z\right)=\varphi
\end{array}}{\lim}\left[\frac{1}{\text{sgn}\left(z\right)}\right]=e^{-2i\varphi}.$$

This is an example of a non-differentiable and discontinuous function, for which the detachment is nevertheless able to provide monotony information.
\end{exm}

\begin{exm}Let $f\left(z\right)=z^{2}.$ The derivative is $f'\left(z_{0}\right)=2z_{0}.$ Particularly, at $z_{0}=0$ the derivative vanishes and we ought to calculate the detachment based on its definition:

$$f_{\varphi}^{;}\left(0\right)\equiv e^{-i\varphi}\underset{\begin{array}{c}
\left|z\right|\to\left|z_{0}\right|\\
\arg\left(\Delta z\right)=\varphi
\end{array}}{\lim}\text{sgn}\left[f\left(z\right)-f\left(z_{0}\right)\right]=e^{-i\varphi}\underset{\begin{array}{c}
\left|z\right|\to\left|z_{0}\right|\\
\arg\left(\Delta z\right)=\varphi
\end{array}}{\lim}\text{sgn}\left(z^{2}\right)=e^{-i\varphi}\cdot e^{2i\varphi}=e^{i\varphi}.$$

Thus, similarly to the case in Real Analysis, while the directional derivatives do not provide monotony information at the stationary point, the directional detachment does.
\end{exm}

\begin{exm}Let $f\left(z\right)=\sqrt{z}.$ The derivative is $f'\left(z_{0}\right)=\frac{1}{2\sqrt{z_{0}}}.$ At $z_{0}=0$, the derivative does not exist and we ought to calculate the detachment based on its definition:

$$f_{\varphi}^{;}\left(0\right)\equiv\underset{\begin{array}{c}
\left|z\right|\to\left|z_{0}\right|\\
\arg\left(\Delta z\right)=\varphi_{2}
\end{array}}{\lim}\text{sgn}\left[f\left(z\right)-f\left(z_{0}\right)\right]=\underset{\begin{array}{c}
\left|z\right|\to\left|z_{0}\right|\\
\arg\left(z\right)=\varphi_{2}
\end{array}}{\lim}\text{sgn}\left(\sqrt{z}\right)=e^{\frac{i\varphi}{2}}.$$

As such, similarly to the case in Real Analysis, the detachment may also provide monotony information at the derivative's singularities for continuous functions.
\end{exm}

\begin{exm}Let $$f\left(z\right)=\begin{cases}
z\sin\left(z\right), & z\neq0,\\
0, & z=0.
\end{cases}$$

Then $f$ is differentiable at $z=0,$ but the definition of its directional detachment yields:

\begin{align}
&\begin{aligned}
f_{\varphi}^{;}\left(0\right)&\equiv\underset{\begin{array}{c}
\left|z\right|\to\left|z_{0}\right|\\
\arg\left(\Delta z\right)=\varphi_{2}
\end{array}}{\lim}\text{sgn}\left[f\left(z\right)-f\left(z_{0}\right)\right]=\underset{\begin{array}{c}
\left|z\right|\to\left|z_{0}\right|\\
\arg\left(z\right)=\varphi_{2}
\end{array}}{\lim}\text{sgn}\left(z\sin\left(z\right)\right) \\
&=e^{i\varphi}\underset{\begin{array}{c}
\left|z\right|\to\left|z_{0}\right|\\
\arg\left(\Delta z\right)=\varphi_{2}
\end{array}}{\lim}\text{sgn}\left[\sin\left(z\right)\right],
\end{aligned}
\end{align}

which is undefined. Thus, similarly to the case in Real Analysis, the complex detachment may not provide monotony information at points of infinite oscillations.
\end{exm}

\begin{exm}Let $f\left(z\right)=\Re\left(z\right),$ whose image is the real line. The sign of its directional derivative satisfies:

$$\text{sgn}\left[\frac{\partial f}{\partial\varphi}\left(z_{0}\right)\right]\equiv \text{sgn}\left\{ \underset{\begin{array}{c}
\left|z\right|\to\left|z_{0}\right|\\
\arg\left(\Delta z\right)=\varphi
\end{array}}{\lim}\left[\frac{\Re\left(\Delta z\right)}{\Delta z}\right]\right\} =\text{sgn}\left[\cos\left(\varphi\right)e^{-i\varphi}\right]=\begin{cases}
e^{-i\varphi}, & \left|\varphi\right|<\frac{\pi}{2},\\
0, & \left|\varphi\right|=\frac{\pi}{2},\\
-e^{-i\varphi}, & \text{else},
\end{cases}.$$

Since the derivative may vanish, let us calculate the $\varphi$-detachment directly according to the definition:

\begin{align}
&\begin{aligned}
f_{\varphi}^{;}\left(z_{0}\right) &\equiv e^{-i\varphi}\underset{\begin{array}{c}
\left|z\right|\to\left|z_{0}\right|\\
\arg\left(\Delta z\right)=\varphi
\end{array}}{\lim}\text{sgn}\left[f\left(z\right)-f\left(z_{0}\right)\right] \\
&=e^{-i\varphi}\underset{\begin{array}{c}
\left|z\right|\to\left|z_{0}\right|\\
\arg\left(\Delta z\right)=\varphi
\end{array}}{\lim}\text{sgn}\left[\Re\left(\Delta z\right)\right]=\begin{cases}
e^{-i\varphi}, & \left|\varphi\right|<\frac{\pi}{2},\\
0, & \left|\varphi\right|=\frac{\pi}{2},\\
-e^{-i\varphi}, & \text{else}.
\end{cases}
\end{aligned}
\end{align}

This is an example where the detachment agrees with the derivative sign even in cases where it vanishes ($\left|\varphi\right|=\frac{\pi}{2}$ in this example).
\end{exm}

\section{Auxiliary Terminology}

Assumptions and abbreviations. We will apply the following notations below:

We focus on directional detachments in a given angle $\varphi.$ Therefore, by a $\delta$-“environment” of the point $z_{0}$, we refer to the $\delta$-lengthed line segment starting at $z_{0}$ and tilted by $\varphi$ with respect to $z_{0}.$
We assume, without loss of generality, that all the complex numbers' arguments are between $0$ and $\pi.$
For the two variables $z_{0}$ and $z$ in the definition domain of a function $f,$ we apply the abbreviation $\Delta f=f\left(z\right)-f\left(z_{0}\right).$
The value of an operator applied to a function is calculated at $z_{0}$, unless we specifically mention otherwise; for example, $\text{sgn}\left(f\right)=\text{sgn}\left(f\left(z_{0}\right)\right).$

We assume detachability, but not differentiability nor continuity, of the functions involved in the results below. As a tradeoff, we will apply the following intuitive and useful concepts.

\begin{definition}[Complex sign-continuity]
Let $f:\mathbb{C}\longrightarrow\mathbb{C}$ be a function, and let $z_{0}\in\mathbb{C}.$ Then, $f$ is sign-continuous at $z_{0}$ if $\text{sgn}\left(f\right)$ is continuous.
\end{definition}

\index{Bimodularity}
\begin{definition}[Bimodularity]
Let $f,g:\mathbb{C}\longrightarrow\mathbb{C}$ be a pair of functions, and let $z_{0}\in\mathbb{C}.$ Then $f,g$ are:

\begin{itemize}
\item Bimodular at $z_{0}$ if for each $z$ in its neighborhood: $\left|\Delta f\right|=\left|\Delta g\right|.$
\item Locally bimodular at $z_{0}$ if for each $z$ in its neighborhood, $\left|f\Delta g\right|=\left|g\Delta f\right|.$
\item Spatially bimodular at $z_{0} $ if for each $z$ in its neighborhood, $\left|f\left(z\right)\Delta g\right|=\left|g\left(z\right)\Delta f\right|.$
\end{itemize}
\end{definition}

\begin{exm}Any complex function is sign-continuous whenever the function's angle of change is continuous. This may happen at the point of non-differentiability, such as cusps. It may also happen at discontinuities, for example if the function's graph approaches a line in a discontinuous fashion.
\end{exm}

\begin{exm}Any function $g:\mathbb{C}\longrightarrow\mathbb{C}$ is everywhere bimodular to all its reflections with respect to the axes: $$g_{1}\left(z\right)=-f\left(z\right), g_{2}\left(z\right)=\overline{f\left(z\right)}, g_{3}\left(z\right)=-\overline{f\left(z\right)},$$ and to all its rotations by a unit vector, $g_{4}\left(z\right)=f\left(z\right)e^{i\xi}.$
\end{exm}

\begin{exm}Let $f,g:\mathbb{C}\longrightarrow\mathbb{C},$ and let $z_{0}\in\mathbb{C}.$ If there is an environment where $g$ is a constant multiplication of $f,$ then they are locally and spatially bimodular. More precisely, if $g\neq0$ and $\frac{f}{g}$ is continuous at $z_{0}$ and constant in its neighborhood, then $f,g$ are both locally and spatially bimodular at $z_{0}.$ Let us show local bimodularity. Given $z$ in an environment of $z_{0}, $ where $\frac{f}{g}$ is held constant:

$$\left|f\left(z_{0}\right)\Delta g\right| =\left|\frac{f\left(z_{0}\right)}{g\left(z_{0}\right)}g\left(z_{0}\right)g\left(z\right)-f\left(z_{0}\right)g\left(z_{0}\right)\right|=\left|\frac{f\left(z\right)}{g\left(z\right)}g\left(z_{0}\right)g\left(z\right)-f\left(z_{0}\right)g\left(z_{0}\right)\right|=\left|g\left(z_{0}\right)\Delta f\right|,$$

where the fourth equality is due to the continuity and constancy of $\frac{f}{g}.$
A similar argument holds for their spatial bimodularity. Note that this condition is sufficient but not necessary for bimodularity.
\end{exm}

\begin{remark}Let $f:\mathbb{C}\longrightarrow\mathbb{C}$ be a detachable function and $z_{0}\in C.$ Let $\epsilon>0.$ We previously showed with a chord-angle analysis, that the detachability of a function $f$ implies that there is an environment of $z_{0}$ wherein:

$$\arg\left(\Delta f\right)\in\left(\arg\left(f^{;}e^{i\varphi}\right)-2\arcsin\left(\frac{\epsilon}{2}\right),\arg\left(f^{;}e^{i\varphi}\right)+2\arcsin\left(\frac{\epsilon}{2}\right)\right).$$

Let us use the following notation for brevity:
$$\theta_{f}\equiv\arg\left(\frac{\Delta f}{f^{;}e^{i\varphi}}\right)<2\arcsin\left(\frac{\epsilon}{2}\right)\label{theta_f_bound},$$ where the equality can be rewritten as follows:

$$\text{sgn}\left(\Delta f\right)=f^{;}e^{i\left(\varphi+\theta_{f}\right)}\label{sgn_df_formula}.$$
\end{remark}

\begin{remark}Similarly, if $f$ is sign-continuous, then given $\epsilon>0,$ the limit definition ensures the existence of an environment such that for each $z$, $\left|\text{sgn}\left(f\left(z\right)\right)-\text{sgn}\left(f\right)\right|<\frac{\epsilon}{2},$ therefore $\text{sgn}\left(f\left(z\right)\right)\in\left(\text{sgn}\left(f\right)-\epsilon,\text{sgn}\left(f\right)+\epsilon\right).$ Let us use the following notation:

$$\theta\equiv\arg\left(f\left(z\right)\right)-\arg\left(f\left(z_{0}\right)\right),$$

which can be rewritten as $$\text{sgn}\left(f\left(z\right)\right)=\text{sgn}\left(f\left(z_{0}\right)\right)e^{i\theta}.\label{sgn_f_formula}$$

Let us cite the following known result without proof.
\end{remark}

\begin{lem}\label{argz1_z2}Let $z_{1},z_{2}\in\mathbb{C}.$ If $\left|z_{1}\right|=\left|z_{2}\right|$ such that $0<\arg\left(z_{1}\right),\arg\left(z_{2}\right)<\pi$ then: $$\arg\left(z_{1}+z_{2}\right)=\frac{\arg\left(z_{1}\right)+\arg\left(z_{2}\right)}{2}.\label{arg_sum}$$
\end{lem}

\begin{lem}\label{theta_fg_upper_bound}Let $f,g:\mathbb{C}\longrightarrow\mathbb{C}$ be detachable functions at $z_{0}\in\mathbb{C}$ and let $\epsilon>0.$ Then there is an environment of $z_{0}$ such that for each $z$ there:

$$\left|e^{\frac{i}{2}\left(\theta_{f}+\theta_{g}\right)}-1\right|<\epsilon.$$
\end{lem}
\begin{proof}By the following transitions:

$$\left|e^{\frac{i}{2}\left(\theta_{f}+\theta_{g}\right)}-1\right|=2\sin\left(\frac{\theta_{f}+\theta_{g}}{4}\right)<2\sin\left(\frac{2\arcsin\left(\frac{\epsilon}{2}\right)+2\arcsin\left(\frac{\epsilon}{2}\right)}{4}\right)=\epsilon,$$

where the first equality is due to a chord-angle analysis of the angle $\frac{\theta_{f}+\theta_{g}}{2}$, the inequality is due to the upper bound provided in formula ($\ref{theta_f_bound}$),
applied to both $\theta_{f},\theta_{g}$, combined with the monotony of the $\sin$ function near zero.
\end{proof}

\begin{lem}\label{theta_theta_fg_upper_bound}Let $f,g:\mathbb{C}\longrightarrow\mathbb{C}$ be detachable functions at $z_{0}\in\mathbb{C},$ where $f$ is sign-continuous. Let $\epsilon>0.$ Then there is an environment of $z_{0}$ such that for each $z$:

$$\left|e^{\frac{i}{2}\left(\theta+\theta_{f}+\theta_{g}\right)}-1\right|<\epsilon.$$
\end{lem}
\begin{proof}From the sign continuity of $f$ we know that for each $z$ close enough to $z_{0}$:

$$\left|\text{sgn}\left(f\left(z\right)\right)-\text{sgn}\left(f\right)\right|<\frac{\epsilon}{2},$$

therefore $\text{sgn}\left(f\left(z\right)\right)\in\left(\text{sgn}\left(f\right)-\epsilon,\text{sgn}\left(f\right)+\epsilon\right),$ and a chord-angle analysis yields that we can bound the angle $\theta=\arg\left(f\left(z\right)\right)-\arg\left(f\left(z_{0}\right)\right),$ from above, by $2\arcsin\left(\frac{\epsilon}{4}\right).$

Thus, it holds that:

\begin{align}
&\begin{aligned}
\left|e^{\frac{i}{2}\left(\theta+\theta_{f}+\theta_{g}\right)}-1\right| &=2\sin\left(\frac{\theta+\theta_{f}+\theta_{g}}{4}\right)<2\sin\left(\frac{2\arcsin\left(\frac{\epsilon}{4}\right)+2\arcsin\left(\frac{\epsilon}{4}\right)+2\arcsin\left(\frac{\epsilon}{4}\right)}{4}\right) \\
&=2\sin\left(\frac{3}{2}\arcsin\left(\frac{\epsilon}{4}\right)\right)
<2\sin\left(2\arcsin\left(\frac{\epsilon}{4}\right)\right) \\
&=4\sin\left(\arcsin\left(\frac{\epsilon}{4}\right)\right)\cos\left(\arcsin\left(\frac{\epsilon}{4}\right)\right)\leq4\sin\left(\arcsin\left(\frac{\epsilon}{4}\right)\right)=\epsilon
\end{aligned}
\end{align}
\end{proof}

\section{Algebraic Rules}

\begin{clm}\textbf{Constant multiple rule.} Let $f:\mathbb{C}\longrightarrow\mathbb{C}$ be detachable at the point $z_{0}\in\mathbb{C}$, and let $c\in\mathbb{C}$. Then $cf$ is also detachable and:

$$\left(cf\right)^{;}=\text{sgn}\left(cf^{;}\right).$$
\label{complex_constant_multiple_rule}
\end{clm}
\begin{proof}Directly from the definition of the detachment:

\begin{align}
\begin{aligned}
\left(cf\right)^{;}\left(z_{0}\right) \equiv\underset{z\to z_{0}}{\lim}\text{sgn}\left[\left(cf\right)\left(z\right)-\left(cf\right)\left(z_{0}\right)\right]
&=\underset{z\to z_{0}}{\lim}\text{sgn}\left[c\Delta f\right] \\
&=\underset{z\to z_{0}}{\lim}\text{sgn}\left(c\right)\text{sgn}\left(\Delta f\right) \\
&=\text{sgn}\left(c\right)\underset{z\to z_{0}}{\lim}\text{sgn}\left(\Delta f\right) \\
&=\text{sgn}\left(c\right)f^{;}=\text{sgn}\left(cf^{;}\right)
\end{aligned}
\end{align}
\end{proof}

\index{Sum Rule}
\begin{theorem}[Sum rule]Let $f,g:\mathbb{C}\longrightarrow\mathbb{C}$ be detachable at $z_{0}\in\mathbb{C}$ and $f,g$ be bimodular at $z_{0}.$

Then $f\pm g$ is also detachable and:

$$\left(f\pm g\right)^{;}=\text{sgn}\left(f^{;}\pm g^{;}\right).$$
\label{complex_sum_rule}
\end{theorem}
\begin{proof}Let $\epsilon>0.$ Without loss of generality, assume that $\arg\left(f^{;}\right)\leq\arg\left(g^{;}\right).$ Let us prove the formula for the sum, and leave the one for the difference as an exercise. There are $\delta$-environments of $z_{0}$, such that for each $z$ , the following conditions hold:

\begin{table}[H]
\centering
        \begin{tabular}{ccc}
        \toprule
        \textbf{Environment} & \textbf{Condition} & \textbf{Reason} \\
        \midrule
        $\delta_1$ & $\left|e^{\frac{i}{2}\left(\theta_{f}+\theta_{g}\right)}-1\right|<\epsilon$ & Lemma \ref{theta_fg_upper_bound}, since $f$ and $g$ are detachable \\
        $\delta_2$ & $\left|\Delta f\right|=\left|\Delta g\right|$ & $f,g$ are bimodular \\
        \bottomrule
        \end{tabular}
\end{table}

Then, for each $z$ in the $\min\left\{ \delta_{1},\delta_{2}\right\}$ -environment, it holds that:

\begin{align}
&\begin{aligned}
&\left|\text{sgn}\left(\left(f+g\right)\left(z\right)-\left(f+g\right)\left(z_{0}\right)\right)-\text{sgn}\left(f^{;}+g^{;}\right)e^{i\varphi}\right| \\
&=\left|\text{sgn}\left(\Delta f+\Delta g\right)-\text{sgn}\left(f^{;}+g^{;}\right)e^{i\varphi}\right| \\
&=\left|\text{sgn}\left(\left|\Delta f\right|f^{;}e^{i\theta_{f}}e^{i\varphi}+\left|\Delta g\right|g^{;}e^{i\theta_{g}}e^{i\varphi}\right)-\text{sgn}\left(f^{;}+g^{;}\right)e^{i\varphi}\right| \\
&=\left|\text{sgn}\left(\left|\Delta f\right|f^{;}e^{i\theta_{f}}+\left|\Delta g\right|g^{;}e^{i\theta_{g}}\right)-\text{sgn}\left(f^{;}+g^{;}\right)\right| \\
&=\left|e^{i\arg\left[\left|\Delta f\right|f^{;}e^{i\theta_{f}}+\left|\Delta g\right|g^{;}e^{i\theta_{g}}\right]}-e^{\frac{i}{2}\left[\arg\left(f^{;}\right)+\arg\left(g^{;}\right)\right]}\right| \\
&=\left|e^{\frac{i}{2}\left[\arg\left(f^{;}e^{i\theta_{f}}\right)+\arg\left(g^{;}e^{i\theta_{g}}\right)\right]}-e^{\frac{i}{2}\left[\arg\left(f^{;}\right)+\arg\left(g^{;}\right)\right]}\right| \\
&=\left|e^{\frac{i}{2}\left[\arg\left(f^{;}\right)+\theta_{f}+\arg\left(g^{;}\right)+\theta_{g}\right]}-e^{\frac{i}{2}\left[\arg\left(f^{;}\right)+\arg\left(g^{;}\right)\right]}\right| \\
&=\left|e^{\frac{i}{2}\left[\arg\left(f^{;}\right)+\arg\left(g^{;}\right)\right]}\left\{ e^{\frac{i}{2}\left(\theta_{f}+\theta_{g}\right)}-1\right\} \right| \\
&=\left|e^{\frac{i}{2}\left(\theta_{f}+\theta_{g}\right)}-1\right|<\epsilon,
\end{aligned}
\end{align}

where the second transition is due to the notation in formula \ref{sgn_df_formula}, the fifth transition is due to the third condition above and formula ($\ref{arg_sum}$). The inequality is due to the first condition.
\end{proof}

\begin{theorem}[Product rule]Let $f,g:\mathbb{C}\longrightarrow\mathbb{C}$ be spatially bimodular and detachable at the point $z_{0}\in\mathbb{C}.$ Assume that $f$ is sign-continuous, and $g,g(z_0)$ are bimodular, or vice versa, at $z_{0}.$ Then, $fg$ is also detachable at $z_{0},$ and:
$$\left(fg\right)^{;}=\text{sgn}\left[f^{;}\text{sgn}\left(g\right)+g^{;}\text{sgn}\left(f\right)\right].$$
\label{complex_product_rule}
\end{theorem}
\begin{proof}Let $\epsilon>0,$ and, without loss of generality, assume that $f$ is sign continuous and $g,g\left(z_{0}\right)$ are bimodular at $z_{0}.$ There are $\delta$-environments, such that for each $z$, the following conditions hold:

\begin{table}[H]
\centering
        \begin{tabular}{ccc}
        \toprule
        \textbf{Environment} & \textbf{Condition} & \textbf{Reason} \\
        \midrule
        $\delta_1$ & $\left|e^{\frac{i}{2}\left(\theta+\theta_{f}+\theta_{g}\right)}-1\right|<\epsilon$ & Lemma \ref{theta_theta_fg_upper_bound}, since $f$ and $g$ are detachable \\
        $\delta_2$ & $\left|g\left(z\right)\right|=\left|g\right|$ & $g,g\left(z_0\right)$ are bimodular \\
        $\delta_3$ & $\left|f(z)\Delta g\right|=\left|g(z)\Delta f\right|$ & $f,g$ are spatially bimodular \\
        \bottomrule
        \end{tabular}
\end{table}

Let $z$ in the $\min\left\{ \delta_{1},\delta_{2},\delta_{3}\right\}$-environment. Then:

\begin{align}
&\begin{aligned}
&\left|\text{sgn}\left[\left(fg\right)\left(z\right)-\left(fg\right)\left(z_{0}\right)\right]-\text{sgn}\left[f^{;}\text{sgn}\left(g\right)+g^{;}\text{sgn}\left(f\right)\right]e^{i\varphi}\right|\\
&=\left|\text{sgn}\left[f\left(z\right)g\left(z\right)-f\left(z\right)g\left(z_{0}\right)+f\left(z\right)g\left(z_{0}\right)-f\left(z\right)g\left(z_{0}\right)\right]-\text{sgn}\left[f^{;}\text{sgn}\left(g\right)+g^{;}\text{sgn}\left(f\right)\right]e^{i\varphi}\right|\\
&=\left|\text{sgn}\left[f\left(z\right)\Delta g+g\left(z_{0}\right)\Delta f\right]-\text{sgn}\left[f^{;}\text{sgn}\left(g\right)+g^{;}\text{sgn}\left(f\right)\right]e^{i\varphi}\right|\\
&=\left|\text{sgn}\left[\left|f\left(z\right)\right|\text{sgn}\left(f\left(z\right)\right)\left|\Delta g\right|\text{sgn}\left(\Delta g\right)+\left|g\left(z_{0}\right)\right|\text{sgn}\left(g\right)\left|\Delta f\right|\text{sgn}\left(\Delta f\right)\right]-\text{sgn}\left[f^{;}\text{sgn}\left(g\right)+g^{;}\text{sgn}\left(f\right)\right]e^{i\varphi}\right|\\
&=\left|\text{sgn}\left[\left|f\left(z\right)\right|\text{sgn}\left(f\right)e^{i\theta}\left|\Delta g\right|g^{;}e^{i\theta_{g}}e^{i\varphi}+\left|g\left(z\right)\right|\text{sgn}\left(g\right)\left|\Delta f\right|f^{;}e^{i\theta_{f}}e^{i\varphi}\right]-\text{sgn}\left[f^{;}\text{sgn}\left(g\right)+g^{;}\text{sgn}\left(f\right)\right]e^{i\varphi}\right|\\
&=\left|\text{sgn}\left[\text{sgn}\left(f\right)e^{i\theta}g^{;}e^{i\theta_{g}}+\text{sgn}\left(g\right)f^{;}e^{i\theta_{f}}\right]-\text{sgn}\left[f^{;}\text{sgn}\left(g\right)+g^{;}\text{sgn}\left(f\right)\right]\right|\\
&=\left|e^{i\left[\text{sgn}\left(f\right)e^{i\theta}g^{;}e^{i\theta_{g}}+\text{sgn}\left(g\right)f^{;}e^{i\theta_{f}}\right]}-e^{i\left[f^{;}\text{sgn}\left(g\right)+g^{;}\text{sgn}\left(f\right)\right]}\right|\\
&=\left|e^{\frac{i}{2}\left[\arg\left(f\right)+\theta+\arg\left(g^{;}\right)+\theta_{g}+\arg\left(g\right)+\arg\left(f^{;}\right)+\theta_{f}\right]}-e^{\frac{i}{2}\left[\arg\left(f\right)+\arg\left(g^{;}\right)+\arg\left(g\right)+\arg\left(f^{;}\right)\right]}\right|\\
&=\left|e^{\frac{i}{2}\left[\arg\left(f\right)+\arg\left(g^{;}\right)+\arg\left(g\right)+\arg\left(f^{;}\right)\right]}\left[e^{\frac{i}{2}\left(\theta+\theta_{f}+\theta_{g}\right)}-1\right]\right|=\left|e^{\frac{i}{2}\left(\theta+\theta_{f}+\theta_{g}\right)}-1\right|<\epsilon,
\end{aligned}
\end{align}
where the fourth transition is due to the notation in formula \ref{sgn_df_formula} and due to the second and third conditions above, the seventh transition is due to lemma \ref{argz1_z2}, and the inequality is due to lemma \ref{theta_theta_fg_upper_bound}.
\end{proof}

\begin{theorem}\textbf{Quotient rule.} Let $f,g\neq0:\mathbb{C}\longrightarrow\mathbb{C}$ be locally bimodular and detachable at the point $z_{0}\in\mathbb{C},$ where $g$ is sign-continuous at $z_{0}.$ Then $\frac{f}{g}$ is also detachable, and:

$$\left(\frac{f}{g}\right)^{;}=\text{sgn}\left[\frac{\text{sgn}\left(g\right)f^{;}-\text{sgn}\left(f\right)g^{;}}{g^{2}}\right].$$
\label{complex_quotient_rule}
\end{theorem}

\begin{proof}First, let us develop the expression by the definition of the limit:

\begin{align}
&\begin{aligned}
\left(\frac{f}{g}\right)^{;}
&\equiv e^{-i\varphi}\underset{z\to z_{0}}{\lim}\text{sgn}\left[\left(\frac{f}{g}\right)\left(z\right)-\left(\frac{f}{g}\right)\left(z_{0}\right)\right] \\
&=\underset{z\to z_{0}}{\lim}\text{sgn}\left[\frac{f\left(z\right)g\left(z_{0}\right)e^{-i\varphi}-f\left(z_{0}\right)g\left(z\right)e^{-i\varphi}}{g\left(z\right)g\left(z_{0}\right)}\right] \\
&=\underset{z\to z_{0}}{\lim}\frac{\text{sgn}\left[f\left(z\right)g\left(z_{0}\right)e^{-i\varphi}-f\left(z_{0}\right)g\left(z\right)e^{-i\varphi}\right]}{\text{sgn}\left(g\left(z\right)\right)\text{sgn}\left(g\right)} \\
&=\frac{1}{\text{sgn}\left(g\right)^{2}}\underset{z\to z_{0}}{\lim}\text{sgn}\left[g\left(z_{0}\right)\Delta fe^{-i\varphi}-f\left(z_{0}\right)\Delta ge^{-i\varphi}\right] \\
&=\frac{1}{\text{sgn}\left(g\right)^{2}}\underset{z\to z_{0}}{\lim}\text{sgn}\left[g\left(z_{0}\right)\left|\Delta f\right|f^{;}e^{i\theta_{f}}-f\left(z_{0}\right)\left|\Delta g\right|g^{;}e^{i\theta_{g}}\right] \\
&=\frac{1}{\text{sgn}\left(g\right)^{2}}\underset{z\to z_{0}}{\lim}\text{sgn}\left[\text{sgn}\left(g\right)\left|g\left(z_{0}\right)\right|\left|\Delta f\right|f^{;}e^{i\theta_{f}}-\text{sgn}\left(f\right)\left|f\left(z_{0}\right)\right|\left|\Delta g\right|g^{;}e^{i\theta_{g}}\right],
\end{aligned}
\end{align}

where the fourth transition is because $g$ is sign-continuous at $z_{0},$ and the fifth is due to the notation in formula \ref{sgn_df_formula}.

Next, let us show that:
$$\underset{z\to z_{0}}{\lim}\text{sgn}\left[\text{sgn}\left(g\right)\left|g\left(z_{0}\right)\right|\left|\Delta f\right|f^{;}e^{i\theta_{f}}-\text{sgn}\left(f\right)\left|f\left(z_{0}\right)\right|\left|\Delta g\right|g^{;}e^{i\theta_{g}}\right]=\text{sgn}\left[\text{sgn}\left(g\right)f^{;}-\text{sgn}\left(f\right)g^{;}\right],$$
with an $\epsilon-\delta$ analysis. Let $\epsilon>0.$ There are $\delta$-environments of $z_{0}$ such that for each $z$ there, the following conditions hold:

\begin{table}[H]
\centering
        \begin{tabular}{ccc}
        \toprule
        \textbf{Environment} & \textbf{Condition} & \textbf{Reason} \\
        \midrule
        $\delta_1$ & $\left|e^{\frac{i}{2}\left(\theta_{f}+\theta_{g}\right)}-1\right|<\epsilon$ & Lemma \ref{theta_fg_upper_bound}, since $f$ and $g$ are detachable \\
        $\delta_2$ & $\left|f\left(z\right)\Delta g\right|=\left|g\left(z\right)\Delta f\right|$ & $f,g$ are locally bimodular \\
        \bottomrule
        \end{tabular}
\end{table}

Then, for each $z$ in the $\min\left\{ \delta_{1},\delta_{2}\right\}$-environment, it holds that:

\begin{align}
&\begin{aligned}
&\left|\text{sgn}\left[\text{sgn}\left(g\right)\left|g\left(z_{0}\right)\right|\left|\Delta f\right|f^{;}e^{i\theta_{f}}-\text{sgn}\left(f\right)\left|f\left(z_{0}\right)\right|\left|\Delta g\right|g^{;}e^{i\theta_{g}}\right]-\text{sgn}\left[\text{sgn}\left(g\right)f^{;}-\text{sgn}\left(f\right)g^{;}\right]\right| \\
&=\left|\text{sgn}\left[\text{sgn}\left(g\right)f^{;}e^{i\theta_{f}}-\text{sgn}\left(f\right)g^{;}e^{i\theta_{g}}\right]-\text{sgn}\left[\text{sgn}\left(g\right)f^{;}-\text{sgn}\left(f\right)g^{;}\right]\right| \\
&=\left|e^{i\arg\left[\text{sgn}\left(g\right)f^{;}e^{i\theta_{f}}-\text{sgn}\left(f\right)g^{;}e^{i\theta_{g}}\right]}-e^{i\arg\left[\text{sgn}\left(g\right)f^{;}-\text{sgn}\left(f\right)g^{;}\right]}\right| \\
&=\left|e^{i\frac{\arg\left(gf^{;}e^{i\theta_{f}}\right)+\arg\left(-fg^{;}e^{i\theta_{g}}\right)}{2}}-e^{i\frac{\arg\left(gf^{;}\right)+\arg\left(-fg^{;}\right)}{2}}\right| \\
&=\left|e^{i\frac{\arg\left(g\right)+\arg\left(f^{;}\right)+\arg\left(e^{i\theta_{f}}\right)+\arg\left(f\right)+\arg\left(g^{;}\right)+\arg\left(e^{i\theta_{g}}\right)-\pi}{2}}-e^{i\frac{\arg\left(g\right)+\arg\left(f^{;}\right)+\arg\left(f\right)+\arg\left(g^{;}\right)-\pi}{2}}\right| \\
&=\left|e^{i\frac{\arg\left(g\right)+\arg\left(f^{;}\right)+\arg\left(f\right)+\arg\left(g^{;}\right)-\pi}{2}}\left(e^{i\frac{\left(\theta_{f}+\theta_{g}\right)}{2}}-1\right)\right|=\left|e^{i\frac{\left(\theta_{f}+\theta_{g}\right)}{2}}-1\right|<\epsilon,
\end{aligned}
\end{align}

where the first transition is due to the second condition above, and the inequality is due to lemma 3, which can be applied because of the first and second conditions.
\end{proof}

\part{Epilogue}
\chapterimage{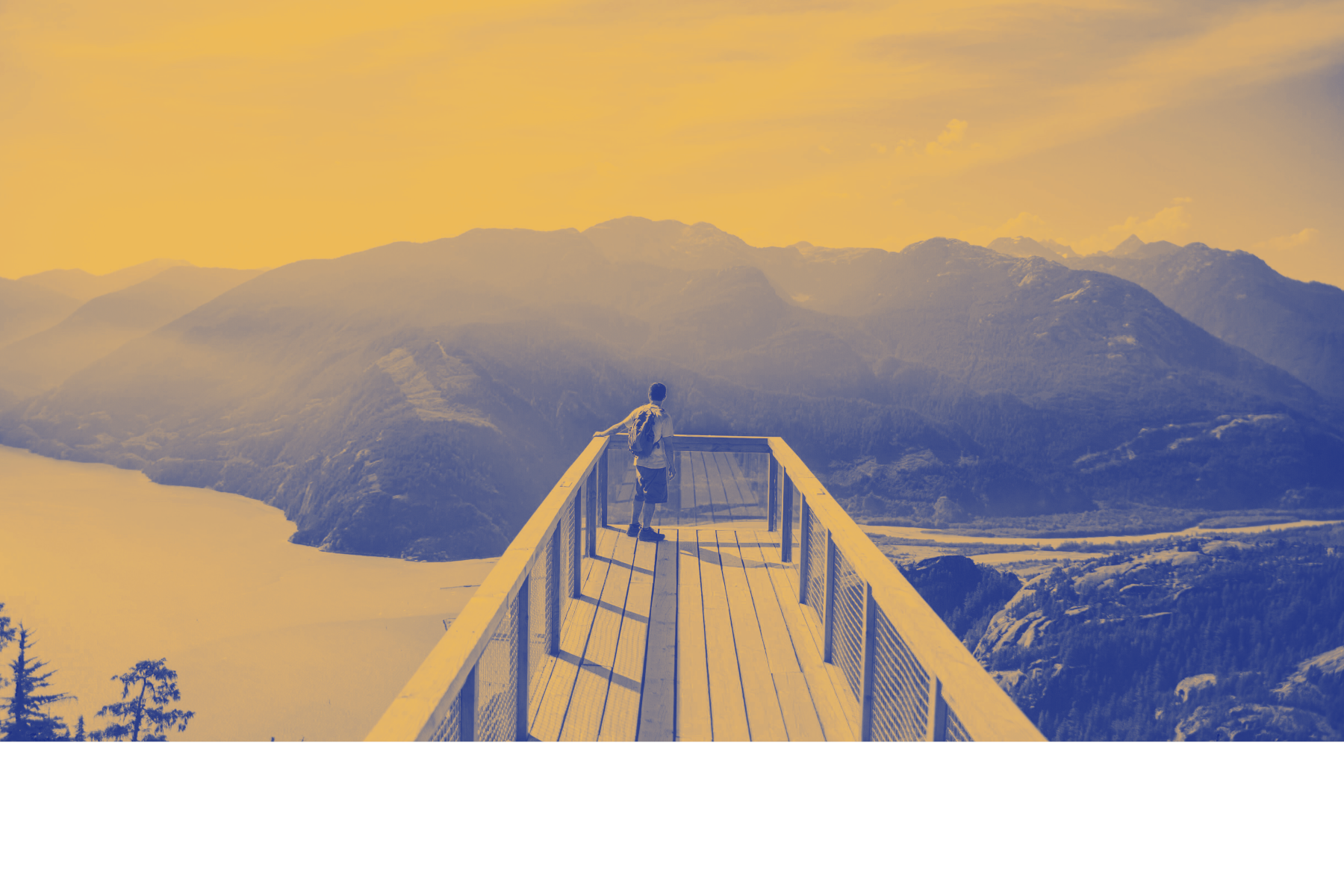}
\chapter{Summary}
Throughout this book, we explored examples from Trendland, an emerging set of applications across the scientific literature that leverage local trends. Let us cherry-pick some prominent examples.

Artificial Intelligence researchers find it lucrative to apply "sign" methods for efficient backpropagation, depending on the geometric setting. They enhance the "signed gradient descent" algorithm (RProp), thus forming another branch of optimization techniques on top of the rate-based methods built on gradient descent. In Image Processing and Computer Vision, various applications such as edge detection and deblurring apply images' derivatives signs.

Exploiting the trend is also prevalent in other branches of engineering. An example is Electrical Engineering, where Fault Analysis often applies the direction of the signal (where its accurate rate is redundant). Another example is Systems Engineering, where novel methods for Maximum power point tracking (MPPT) capture the derivative sign of the voltage. Finally, in Mechanical Engineering, Compensation formulas often incorporate friction information and specifically consider the direction of movement.

The trend is used in many other fields. Natural scientists, for example, often apply the derivative sign in qualitative analyses of natural phenomena and classify scenarios based on functions' trends. Biologists learn about the interactions between species with the sign of their "Community Matrix." Chemical Engineers apply the emerging field of Qualitative Trend Analysis to classify process trends according to their derivatives' signs across an interval. Physicists use the Banerjee criterion, based solely on the Arott plot's derivative sign, to find the order of the phase transition. It arguably comes as no surprise that mathematicians investigate functions' trends extensively; for example, in the theory of Locally Monotone operators. Statisticians apply the regression coefficient's slope sign to deduce the direction of the relationship between the variables and use the Mann-Kendall trend test to assess processes' trends. Social scientists use functions' partial derivative signs for comparative static analyses. 

Given the importance that exploiting the trend has in several fields, I find it relevant to investigate how functions' trends are taught. Several Science Education researchers suggest that students struggle with functions' trends when introduced as the derivative sign. It appears that the derivative sign is a confusing, non-intuitive notion and, sometimes, a conceptual challenge due to the use of two confusing concepts - velocity and speed. It may be helpful to introduce a dedicated tool that captures the trend independently of the rate.

In addition to a literature review of trends applications, we also surveyed other trends calculations approaches (on top of the derivative sign). These are helpful in scenarios where the rate information that the derivative captures is superfluous. They can also capture trends if the derivative is undefined or zeroed (at extrema points). While advanced mathematical tools such as the Dini derivative may address some of these scenarios, researchers often prefer other ad-hoc methods.

Surprisingly, it turns out that the "Detachment" operator, defined in Semi-discrete Calculus, concisely models the numerical tricks scientists have already been using, part of which we reviewed in this book; it further outperforms the derivative sign in modeling trends. Due to skipping the division operator, the Detachment is more numerically stable (less susceptible to overflow and gradient explosion) and efficient (up to 20\% faster in its discrete form). It is also computationally robust and consistent in continuous domains. Additionally, it meets the didactic requirement for a tool that separates the rate and the trend.

Since trend applications are a sign of the times, and the detachment operator serves to conduct them better in several senses, we explored a mathematical theory that relies on this point-wise operator, named "Semi-discrete Calculus." Results in this theory (see a summary in Table \ref{results_table}) often introduce a tradeoff between the information level provided by the theorem and the set of functions to which it is applicable, along with computational gains.

The Infinitesimal Calculus explores mainly two measurements: 1) Differential Calculus studies the instantaneous rates of change and the slopes of curves; 2) Integral Calculus studies the accumulation of quantities and areas under or between curves. This work shows that scientists, engineers, mathematicians, and teachers are increasingly using another change measurements tool: functions' local trends. While it seems to be a special case of the rate (via the derivative sign), this work proposes a separate and favorable mathematical framework for the trend, called Semi-discrete Calculus.

\begin{table}[h!]
\centering
        \begin{tabular}{ccccc}
        \toprule
        \textbf{Result} & \color[HTML]{FFA006} \textbf{$f'$} & \color[HTML]{7800cf} \textbf{$\int f$} & \color[HTML]{0039BD} \textbf{$f^;$} & \textbf{References} \\
        \midrule
        Constant multiple rule & \color[HTML]{FFA006} V & \color[HTML]{7800cf} V & \color[HTML]{0039BD} V & Claim \ref{constant_multiple_rule} \\
        Sum and difference rule & \color[HTML]{FFA006} V & \color[HTML]{7800cf} V & \color[HTML]{0039BD} V & Claim \ref{sum_rule} \\
        Product rule & \color[HTML]{FFA006} V & \color[HTML]{7800cf} V & \color[HTML]{0039BD} V & Theorem \ref{product_rule} \\
        Quotient rule & \color[HTML]{FFA006} V & \color[HTML]{7800cf} V & \color[HTML]{0039BD} V & Theorem  \ref{quotient_rule} \\
        Fermat's theorem & \color[HTML]{FFA006} V &  & \color[HTML]{0039BD} V & Theorems  \ref{fermat_for_derivative}, \ref{fermat_semi_discrete} \\
        Rolle's theorem & \color[HTML]{FFA006} V &  & \color[HTML]{0039BD} V & Theorems \ref{rolle_theorem}, \ref{rolle_theorem_detachment} \\
        Lagrange's theorem & \color[HTML]{FFA006} V & \color[HTML]{7800cf} V & \color[HTML]{0039BD} V & Theorems \ref{mvt_derivative_thm}, \ref{mvt_integral_thm}, \ref{mvt_detachment_thm} \\
        Fundamental theorem & \color[HTML]{FFA006} V & \color[HTML]{7800cf} V & \color[HTML]{0039BD} V & Theorems \ref{ftc_1d}, \ref{semi_discrete_fundamental} \\
        Chain rule & \color[HTML]{FFA006} V & \color[HTML]{7800cf} V & \color[HTML]{0039BD} V & Claims \ref{chain_rule_derivative}, \ref{chain_rule_detachment} \\
        Inverse function rule & \color[HTML]{FFA006} V & \color[HTML]{7800cf} V & \color[HTML]{0039BD} V & Claims \ref{inverse_function_rule_derivative}, \ref{inverse_function_rule_detachment} \\
        Functional power rule & \color[HTML]{FFA006} V & \color[HTML]{7800cf} V & \color[HTML]{0039BD} V & Claims \ref{functional_power_rule_derivative}, \ref{functional_power_rule_detachment} \\
        Taylor Series & \color[HTML]{FFA006} V & & \color[HTML]{0039BD} V & Theorem \ref{taylor_series}, Corollary \ref{semi_discrete_taylor} \\
        L'Hôpital rule & \color[HTML]{FFA006} V & & \color[HTML]{0039BD} V & Theorems \ref{lhopital_derivative}, \ref{lhopital} \\
        Extrema tests & \color[HTML]{FFA006} V & & \color[HTML]{0039BD} V & Claims \ref{first_derivative_test}, \ref{first_detachment_test}, \ref{second_derivative_test}, \ref{second_detachment_test} \\
        Green's theorem & \color[HTML]{FFA006} V & \color[HTML]{7800cf} V & \color[HTML]{0039BD} V & Theorems \ref{green_theorem}, \ref{ftc_theorem} \\
        Complex constant multiple rule & \color[HTML]{FFA006} V & \color[HTML]{7800cf} V & \color[HTML]{0039BD} V & Claim \ref{complex_constant_multiple_rule} \\
        Complex sum rule & \color[HTML]{FFA006} V & \color[HTML]{7800cf} V & \color[HTML]{0039BD} V & Theorem \ref{complex_sum_rule} \\
        Complex product rule & \color[HTML]{FFA006} V & \color[HTML]{7800cf} V & \color[HTML]{0039BD} V & Theorem \ref{complex_product_rule} \\
        Complex quotient rule & \color[HTML]{FFA006} V & \color[HTML]{7800cf} V & \color[HTML]{0039BD} V & Theorem \ref{complex_quotient_rule} \\
        \bottomrule
        \end{tabular}
\color{black}
\caption{A collection of Calculus results, some of them illustrated in this work, in which case the exact reference is provided. The second, third, and fourth columns indicate whether versions of the result or a rearrangement thereof that applies the derivative, the integral and the detachment are currently known, respectively.}
\label{results_table}
\end{table}

\section{The Author}
The author graduated with degrees in Mathematics and Computer Science from the Israel Institute of Technology (Technion), where he won the excellent teaching assistant award for teaching Advanced Calculus. Ever since, he has been leading AI research teams and invented several patents in Data Science, Statistics, and AI. One of his patents incorporates Semi-discrete Calculus (\cite{shacharpatent}).

\section{Acknowledgements}
The scientific editor Fabian Goguta proofread the book, and the Physical Chemist Dr. Erez Zemel contributed to the discussion held in subsection \ref{nowhere_differentiability}.

%----------------------------------------------------------------------------------------
%	BIBLIOGRAPHY
%----------------------------------------------------------------------------------------

\chapterimage{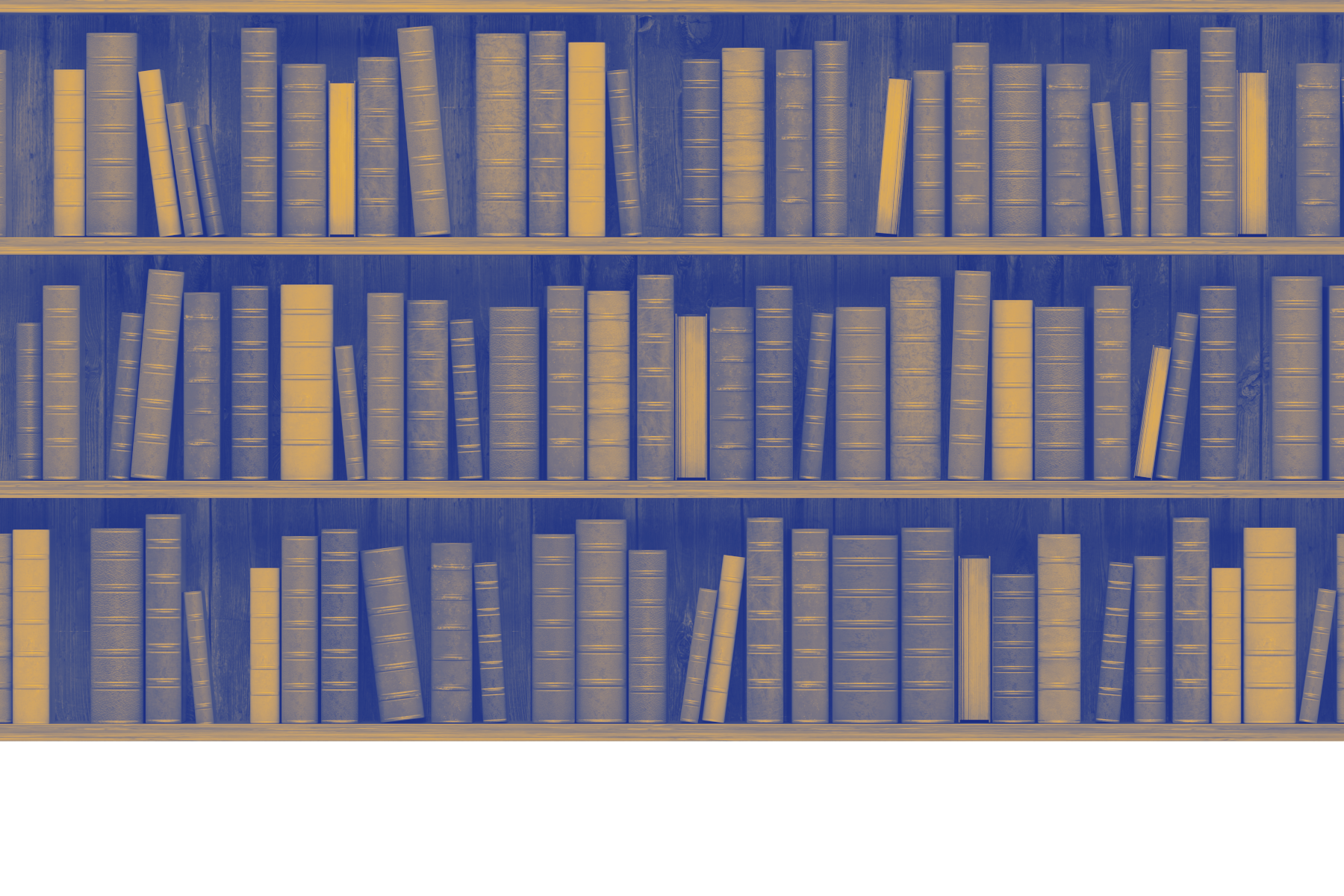}
\chapter{Bibliography}
\addcontentsline{toc}{chapter}{\textcolor{ocre}{Bibliography}} % Add a Bibliography heading to the table of contents

%------------------------------------------------

\section{Papers}
\addcontentsline{toc}{section}{Papers}
\defbibfilter{papers}{
  type=article or
  type=inproceedings or
  type=techreport or
  type=misc
}
\nocite{*}
\printbibliography[heading=bibempty,filter=papers]

%------------------------------------------------

\section{Theses}
\addcontentsline{toc}{section}{Theses}
\nocite{*}
\printbibliography[heading=bibempty,type=thesis]

%------------------------------------------------

\section{Books}
\addcontentsline{toc}{section}{Books}
\nocite{*}
\printbibliography[heading=bibempty,type=book]

%----------------------------------------------------------------------------------------
%	INDEX
%----------------------------------------------------------------------------------------
\chapterimage{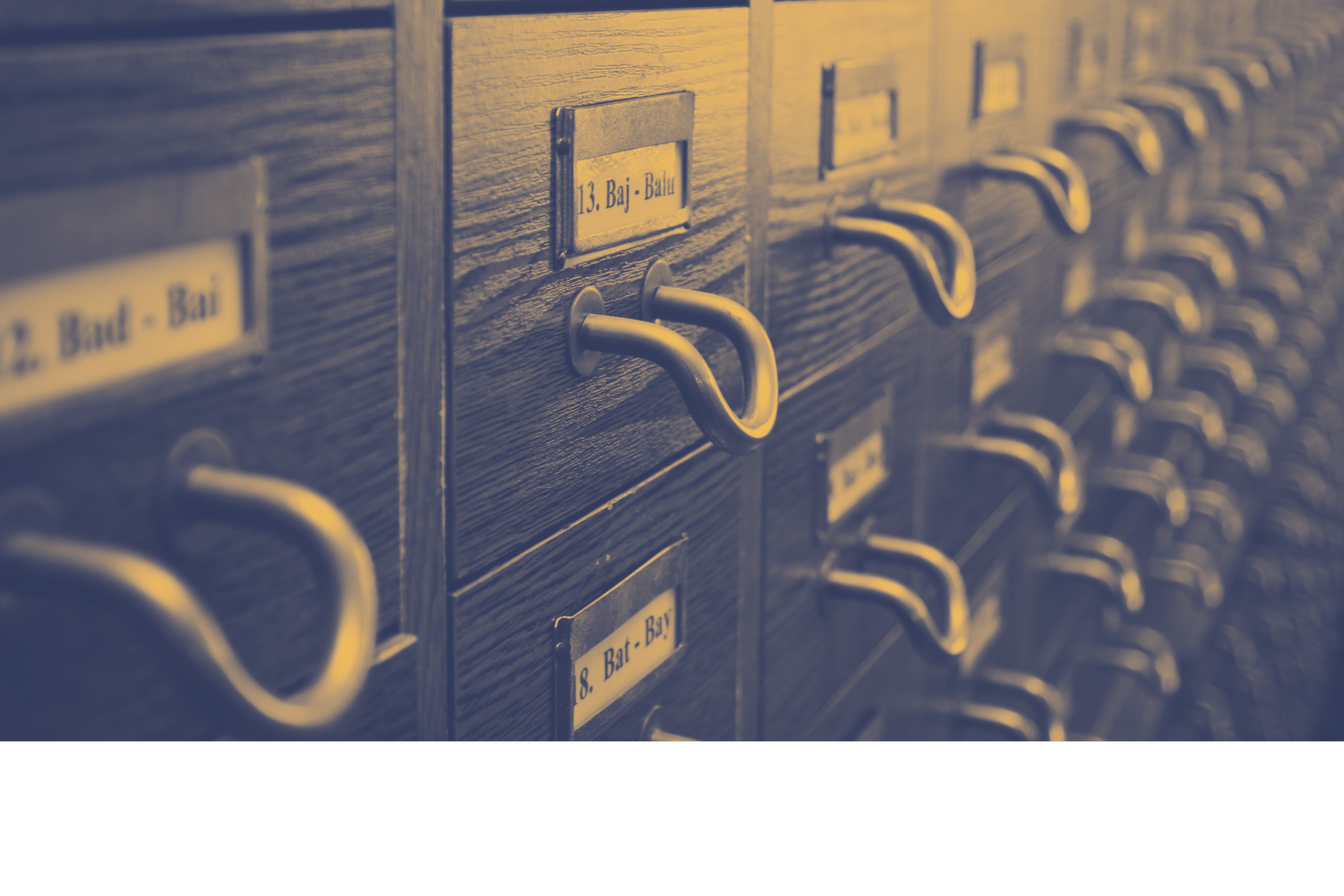} \cleardoublepage\phantomsection \setlength{\columnsep}{0.75cm}
% Space between the 2 columns of the index
\addcontentsline{toc}{chapter}{\textcolor{ocre}{Index}} % Add an Index heading to the table of contents
\printindex
\end{document}